\documentclass[12pt,a4paper]{report}
\usepackage[utf8]{inputenc}
\usepackage[francais,english]{babel}
\usepackage{CJKutf8}
\usepackage[pdftex]{graphicx} 
\usepackage[export]{adjustbox}
\usepackage[usenames,dvipsnames,svgnames,table]{xcolor}
\usepackage{amsmath}
\usepackage{mathtools}
\usepackage{amsfonts}
\usepackage{stackrel}

\usepackage{amssymb}
\usepackage{cases}
\usepackage{verbatim}
\usepackage{bm}%
\usepackage[absolute]{textpos} 
\usepackage{doi}
\usepackage{multicol} 
\usepackage{calc} 
\RequirePackage{geometry}
\usepackage{tikz} 
\usepackage{bibentry}
\usepackage{geometry}
\usepackage{hyperref}
\usepackage[acronyms]{glossaries}
\usepackage{color}
\usepackage{enumitem}
\nobibliography*
\definecolor{darkblue}{rgb}{0.0,0.0,0.27}
\hypersetup{colorlinks,breaklinks,
            linkcolor=darkblue,urlcolor=darkblue,
            anchorcolor=darkblue,citecolor=darkblue}

    \usetikzlibrary{calc}
       \usetikzlibrary{decorations.pathreplacing}
          \usetikzlibrary{decorations}
             \usetikzlibrary{arrows,decorations.markings} 
 \usepackage{subfiles}
\newcommand{\defeq}{\stackrel{\text{def.}}{=}}
\newcommand{\dif}{\mathrm{d}}
\newcommand{\R}{\mathbb{R}}
\newcommand{\C}{\mathbb{C}}
\newcommand{\N}{\mathbb{N}}
\newcommand{\Z}{\mathbb{Z}}

 \newcommand{\im}{\mathbf{i}}
  
 \newcommand{\abs}[1]{\left\vert#1\right\vert}
 \newcommand{\set}[1]{\left\{#1\right\}}
 
 \newcommand{\tBarnes}{\tilde{\mathbf{G}}}
 \newcommand{\Barnes}{\mathbf{G}}
\newcommand{\overlap}{\mathsf{q}}
\newcommand{\Overlap}{\mathsf{Q}}
 \newcommand{\vertex}{\mathcal{V}}
  \newcommand{\zmode}{\varphi_0}

	  \newcommand{\red}[1]{{\color{red} #1}}
	
	  \definecolor{blue}{rgb}{0,0,1}
	  \definecolor{green}{rgb}{0,.6,0}
	  \definecolor{red}{rgb}{0,0,0}
	  \definecolor{vio}{rgb}{1,0,1}
	  \definecolor{uv}{rgb}{0.5,0,0.5}
	  \definecolor{ama}{rgb}{0.3,0.3,0.3}

\graphicspath{{./images/}{./}} 

\newcommand{\PhDsumEN}{This thesis presents original results in two domains of disordered statistical physics: logarithmic correlated Random Energy Models (logREMs), and localization transitions in long-range random matrices.

In the first part devoted to logREMs, we show how to characterise their common properties and model--specific data. Then we develop their replica symmetry breaking treatment, which leads to the freezing scenario of their free energy distribution and the general description of their minima process, in terms of decorated Poisson point process. We also report a series of new applications of the Jack polynomials in the exact predictions of some observables in the circular model and its variants. Finally, we present the recent progress on the exact connection between logREMs and the Liouville conformal field theory.

The goal of the second part is to introduce and study a new class of banded random matrices, the broadly distributed class, which is characterid an effective sparseness. We will first study a specific model of the class, the Beta Banded random matrices, inspired by an exact mapping to a recently studied statistical model of long--range first--passage percolation/epidemics dynamics. Using analytical arguments based on the mapping and numerics, we show the existence of localization transitions with mobility edges in the ``stretch--exponential'' parameter--regime of the statistical models. Then, using a block--diagonalization renormalization approach, we argue that such localization transitions occur generically in the broadly distributed class.  } 

\makeglossaries
\newacronym{logrem}{logREM}{logarithmically correlated Random Energy Model}
\newacronym{rem}{REM}{Random Energy Model}
\newacronym{lft}{LFT}{Liouville field theory}
\newacronym{ope}{OPE}{Operator Product Expansion}
\newacronym{dpct}{DPCT}{directed polymers on the Cayley tree}
\newacronym{kpp}{KPP}{Kolmogorov–Petrovsky–Piscounov}
\newacronym{kpz}{KPZ}{Kardar--Parisi--Zhang}
\newacronym{rsb}{RSB}{replica symmetry breaking}
\newacronym[sort={RSB1}]{1rsb}{1RSB}{one--step replica symmetry breaking}
\newacronym{iid}{i.i.d.}{independent and identically distributed}
\newacronym{pdf}{pdf}{probability density function}
\newacronym{ir}{IR}{infra--red}
\newacronym{uv}{UV}{ultra--violet}
\newacronym{fpp}{FPP}{first--passage percolation}
\newacronym{bbm}{BBM}{Branching Brwonian motion}
\newacronym{gff}{GFF}{Gaussian Free Field}
\newacronym{ipr}{IPR}{inverse participation ratio}
\newacronym{ppp}{PPP}{Poisson point process}
\newacronym{sdppp}{SDPPP}{randomly shifted, decorated Poisson point process}
\newacronym{dozz}{DOZZ}{Dorn-Otto and Zamolodchikov-Zamolodchikov}
\newacronym{dos}{DoS}{density of states}
\newacronym[longplural={Power-law Banded Random Matrices}]{pbrm}{PBRM}{Power-law Banded Random Matrix}
\newacronym[longplural={conformal field theories}]{cft}{CFT}{conformal field theory}

\newacronym[longplural={Beta Banded Random Matrices}]{bbrm}{BBRM}{Beta Banded Random Matrix}

\begin{document}


\newgeometry{margin=2.cm}
\pagenumbering{arabic}

\begin{center}
{\Large \textbf{Disordered statistical physics in low dimensions \\ Extremes, glass transition, and localization}} \\
\vspace{0.5cm}
Xiangyu Cao \\
Ph.D. Thesis \\
\textit{Advisors}: Alberto Rosso, Raoul Santachiara \\
\textit{LPTMS, Univ. Paris-Sud, Université Paris--Saclay} \\
\end{center}
\vspace{1cm}

\textit{Abstract.} \PhDsumEN
\tableofcontents

\chapter*{\textit{Remerciements}}\label{ch:thank}

I am grateful to Jonathan Keating and Christopher Mudry who kindly accepted to be the Referees of the thesis manuscript. 

It is a pleasure to acknowledge Yan Fyodorov as a senior collaborator. His pioneering scientific works have been inspiring the original research reported in both parts of this thesis. Although we have only met once physically, I have learned a lot from him through our collaboration. Besides playing key rôles in our joint works (published and ongoing), he has been a careful reader of our other works, and shared his insights and suggestions.\vspace{.3cm}

Je remercie Bertrand Georgeot et Didina Serban d'avoir accepté d'être Membre du Jury de ma thèse.

Je suis reconnaissant envers M. Gilles Montambaux, qui a suivi mon doctorat en tant que responsable de l'\'Ecole Doctorale de Physique en Île--de--France. Son attention particulière et ses soutiens constants sont essentiels pour les résultats de mon parcours. J'ai apprécié particulièrement son écoute lors de nos entretiens annuels. 

Les travaux qui consituent cette thèse sont les fruits de mes séjours de 2015 à 2017 au sein du Laboratoire de Physique Théorique et Modèles Statistiques (LPTMS), qui m'a soutenu à tous les moments décisifs, et qui m'a accueilli dans son ambience vivante et travailleuse, indispensable pour mon emancipation scientifique. Je remercie tout particulièrement son directeur Emmanuel Trizac pour sa confiance et ses encouragements, sans lesquels j'aurais abandonné la recherche depuis longtemps. Mes gratitudes les plus spéciales vont à Claudine Le Vaou, la gestionnaire du laboratoire. C'est grâce à ses compétences et créativités exceptionnelles, et à son attention et bienveillence sans faille, que nous avons résoulu les nombreux problèmes administratifs hautement non--triviaux, qui sont les conditions \textit{sine qua non} du bon déroulement de ma thèse. Je remercie également Barbara Morléo, collègue de Claudine au CNRS, pour ses efforts aussi importants sur mon dossier. Je remercie les informaticiens du laboratoire, Vincent Degat et Qin Zhiqiang: tous les deux ont sauvé ma vie en sauvant mes données. D'ailleurs, le puissant \textit{cluster} du LPTMS est derrière toutes les études numériques de cette thèse. Enfin, un merci spécial va à Marie--Thérèse Commault pour sa gentillesse. 

Au LPTMS, j'ai eu l'occastion de discuter régulièrement avec plusieurs membre permanents. M'excusant d'être non--exlusif, je remercie particulièrement: Eugène Bogomolny pour ses nombreux conseils pointus, Silvio Franz pour m'avoir appris la théorie des verres de spin, et Christophe Texier, pour ses explications lucides sur des sujets difficiles comme la localisation d'Anderson; outre la recherche, les suggestions de Martin Lenz et Satya Majumdar ont été extrêmement utiles pour ma recherche de post--doc. Je suis heureux d'avoir connu mes camarades du bureau, et surtout mes co--disciples, Clélia de Mulatier, Inès Rodriguez Arias, et (de façon non-officielle) Timothée Thiery; je remercie tous les trois pour leur encouragement et sympathie. J'ai apprécié les discussions avec le brilliant et aimable post--doc Andrea De Luca. 

Mes travaux ont également été soutenus par \textit{Capital Fund Management}. Je remercie la générosité du groupe, et tout particulièrement son président Jean--Philippe Bouchaud, sans qui cela aurait été impossible. Je voudrais remercier une autre fois Jean--Philippe, en tant que physicien, parce que j'ai eu la chance de participer à la collaboration scientifique qu'il avait initiée, et qui est à l'origine des résultats du Chapitre \ref{ch:anderson} de cette thèse. Ses idées et ses grandes qualités intellectuelles m'ont toujours inspiré. Mon seul regret, auquel je compte remédier dans un futur proche, est de ne pas avoir suffisamment travaillé sur notre projet commun pendant la thèse. 

Je remercie Pierre Le Doussal d'avoir été un mentor et collaborateur important, bien qu'il ne soit pas mon directeur de thèse officiel. Il a porté son attention encourageante à mon travail avant le début de nos collaborations. En particulier, grâce à son soutien, j'ai fait mon premier voyage de recherche, à Santa--Barbara, pendant trois semaines riches et fructueuses. \`A risque de reprendre les mots de ses anciens élèves, les entretiens réguliers avec Pierre sont des moments précieux, qui font disparaitre toutes les angoisses du quotidien, grâce à l'enthousiasme scientifique infini de Pierre. De cela s'ensuivent son attention acharnée aux détails et son jusqu'au--boutism (dans le sens strictement positif du terme!) qui ne cesseront d'\^etre mes références.

Avant d'arriver au LPTMS, j'ai été étudiant à l'Institut de Physique Théorique (IPhT) à Saclay, sous la direction de Kirone Mallick, que je tiens à remercier. C'est Kirone, et Bernard Derrida, notre professeur de Master 2, qui m'avaient initié à la recherche en physique statistique. Au--delà du savoir--faire scientifique, Kirone m'a offert les leçons les plus précieuses que puisse recevoir un chercheur débutant: l'humilité, la préserverence, et l'ouverture d'esprit. Il a toujours été encourageant et de bons conseils, même quand j'avais l'intention de changer de sujet et de laboratoire. Cette transition n'aurait pas été possible sans sa bienveillence altruiste. Je remercie également Stéphane Nonnenmacher et Michel Bauer, directeurs de l'IPhT, d'avoir facilité ce transfert. 
Parmi les membres de l'IPhT, je voudrais remercier particulièrement: Jérémie Bouttier, pour les discussions scientifiques, et surtout pour m'avoir fait venir à Cargèse; Sylvain Ribault, pour les discussions, et pour ses codes \textit{open--source} et ses notes de cours en domaine public sans lesquels une partie importante de cette thèse serait absente.

Enfin, je ne remercierai jamais assez Alberto et Raoul, mes directeurs de thèse au LPTMS (d'ailleurs, Raoul a dirigé mon stage M2). La longueur de ce texte divergerait si je me mettais à énumerer leurs générosités, sans lesquelles la longueur du texte qui suit serait proche de zéro. Je leur dédie donc ce manuscrit comme je le fais à ma famille.


\vspace{.3cm}
\begin{CJK*}{UTF8}{gbsn}

致父母与爱妻佳骊。

\end{CJK*}
\vspace{.3cm}

The bibliography style of this thesis is provided by \href{https://scipost.org/}{SciPost}, an open--access online journal of Physics.
\pagebreak

\chapter*{Foreword}
With the invention of the microscope, a new world was discovered. It was considered so radically different from the macroscopic world that, according to Michel Foucault's account~\footnote{\textit{{Debate on Human nature}}, M. Foucault and N. Chomsky, in \cite{davidson1997foucault}.}, it led to the modern concept of \textit{life}. It was in terms of the latter that the irregular motion of pollens observed by Brown was understood, until Einstein (theory, 1905) and Perrin (experiment, 1908)  put on a firm footing the idea that all natural phenomena, beyond the realm of life, and as simple as the expansion of heated gas, have a microscopic origin. Unfortunately, that was too late to change the tragic life of the idea's father, Boltzmann, on whose tombstone was graved the following equation
\begin{equation*}
S = k \log W 
\end{equation*}
relating the macroscopic quantity, entropy ($S$) to a microscopic one, the number of microscopic configurations ($W$). The Boltzmann constant $k$ is as fundamental as the Planck's constant $\hbar$: its numerical value is only a consequence of choice of units. In this thesis, we use units so that $k = \hbar = 1$.

With this equation was born the subject of \textit{statistical} physics, \textit{i.e.}, the physics of counting large numbers ($W$ in the above equation). With quantum mechanics, it is one of the pillars of modern condensed matter theory. The standard paradigm of the latter was summarized by the Anderson's beautiful formula \textit{more is different}~\cite{anderson1972more} in 1972. When a large number of microscopic constituents organize themselves, some symmetry of the constituent law can be spontaneously broken, leading to phase transitions. About the same time, it was realized that spontaneous symmetry breaking and phase transitions can be described theoretically by quantum/statistical field theory and the renormalization group. The latter led to, in principle, a classification of the states of matter in terms of universality classes. Each of the latter is characterised by a small set of ``critical exponents'', which are independent of the microscopic nature of the constituents. The classification was carried out most successfully in two (or $1+1$) dimensions, thanks to the powers of \acrfull{cft}.

However, the success is largely limited to systems that reach rapidly enough their thermal equilibrium. Extending the paradigm to phenomena far from equilibrium (such as turbulence) is a largely open challenge. In this respect, a crucial intermediate is the \textit{disordered} systems: glasses, electronic systems with impurities, \textit{etc}. They are not strongly driven (like turbulent fluids) or active (like many biological systems), so the equilibrium formalism is still applicable. However, the existence of  a large hierarchy of time--scales brings an essential complication: the \textit{quenched randomness} in the microscopic Hamiltonian. As a consequence, the equilibration is only partial, and the system may visit a fraction of the complex \textit{energy landscape}. This led to new notions of phase transition and symmetry breaking, such as \textit{localization transition} and \textit{\gls{rsb}}. Their field theory description is not always known (\textit{e.g.}, the plateau transition in the integer quantum Hall effect), and the known ones have limited range of applications, and involve often technicalities such as super--symmetry (localization transitions) and/or functional renormalization group (manifolds in random media). 

In this respect, the statistical physics theory of disordered systems is largely driven by specific models, \textit{i.e.}, their analytic solutions and relations/mappings between them. The goal is to identify new universality classes, clarify the universal and \textit{non}--universal properties of the particular models, and seek their field theory description. This is the method of the present thesis. It has two subjects: \textit{\glspl{logrem}} (Chapter \ref{ch:logrem}) and \textit{localization transition with long--range hopping} (Chapter \ref{ch:anderson}). Although they are largely independent, their study shares a few common notions and methods, such as extreme value statistics and theoretical models known as polymers in random media. The purpose of Chapter \ref{ch:intro} is to introduce these basic notions, whereas Chapter \ref{ch:end} summarizes the thesis and discusses the perspective from a global point of view.

\section*{\Glspl{logrem}  \\ Overview of Chapter \ref{ch:logrem}}
\Glspl{logrem} are arguably the simplest, yet non--trivial, class of disordered statistical models. They can be defined as the problem of a thermal particle in a random potential. Since Sinai's study of the case where the random potential is a 1d Brownian motion, such problems have become prototypes of statistical physics with disorder. The \glspl{logrem} correspond to cases where the potential is Gaussian and \textit{logarithmically correlated}. From a theoretical point of view (reviewed in section \ref{sec:introPEL}), this is arguably the most interesting case, because as the result of competition between deep valleys of the potential and the entropic spreading of the particle, there is a \textit{freezing transition} at some finite temperature. The low--temperature, ``frozen'', phase is extremely glassy, in the sense that the thermal particle is caged in \textit{a few} deepest valleys of the potential. Remarkably, \glspl{logrem} are among the few cases where the method of \gls{rsb} is applicable in finite dimensional problems. The essential reason is that, the class of  \glspl{logrem} contains not only the problem of a thermal particle in log--correlated potentials, but also that of \gls{dpct}, and of \gls{bbm}. These are all mean--field statistical models defined on hierarchical (tree--like) lattices, where the \gls{rsb} is known to apply. 

This fundamental link between finite--dimensional and hierarchical models has an involved history going back to the \gls{rem} of spin glass, of which \glspl{logrem} are close cousins. The main driving force behind the discovery of the link are the study of \textit{2D} disordered systems: Dirac fermions in random magnetic field, and random--gauge XY model. The randomness behind these models reduces all to the 2D \gls{gff}, which is one of the most natural ways to construct log--correlated potentials. They lead either to 2D \glspl{logrem}, or to 1D \glspl{logrem}, by restricting the thermal particle to a 1D geometry in the plane where the 2D \gls{gff} is defined. To be fair, it must be mentioned that another way of generating 1D log--correlated potential comes from random matrices (through the logarithm of their characteristic polynomial) and number theory (Riemann $\zeta$ function on the critical line): this unexpected link to pure mathematics has stimulated many recent developments. 

Therefore, the domain of \glspl{logrem} has become a inter--disciplinary area attracting the attention of theoretical physicists and mathematicians alike. In this respect, the point of view and the original contributions of this thesis are focused on the 2D \gls{gff}--related aspects. Indeed, many questions that we will investigate can be asked on a single \gls{logrem}, the \textit{circular model}, and its variants. Put simply, it describes a thermal particle confined onto the unit circle, on which a random potential is defined by restricting a 2D \gls{gff} to the circle (from the random matrix point of view, one would define as the log of the characteristic polynomial of a random unitary matrix). Now, we may ask:
\begin{itemize}
\item Why, historically speaking, do theorists come to study this model? What is already known about it (section \ref{sec:logremintro} except \ref{sec:multifracintro})?  
\item What is the relation between the above continuum definition and its discrete definition\textit{s} (section \ref{sec:IRUV})?
\item What is the thermodynamics of the model, why does it have a freezing transition? Why and how can we study it using \gls{rsb}? What are the general consequences of this approach to \glspl{logrem} (sections \ref{sec:logREMdef} and \ref{sec:RSB})?
\item What is full distribution of the free energy (this is known and reviewed in section \ref{sec:treetoplane})? How does it depend on the specific definition of the model? What if we consider a \textit{realistic} 2D \gls{gff} on a finite domain, such as a disk with Dirichlet boundary condition (section \ref{sec:Dirichlet})? 
\item What other observables can we calculate for the circular model? Are there some analytical tools that can be systematically applied (a candidate is the Jack polynomials, see section \ref{sec:Jack})?
\item In the low temperature phase, the thermal particle is caged in a few deepest minima of the potential. How do they behave, \textit{i.e.}, what is the extreme order statistics? How can we study them using the \gls{rsb} approach? For instance, how can we predict the distribution of gap between the deepest and second minimal values (section \ref{sec:orderstat})?
\end{itemize}
As we will see, the sections \ref{sec:logremintro}, \ref{sec:logREMwhat}, \ref{sec:RSB} and \ref{sec:orderstat} are based on results which apply to general \glspl{logrem}, but we discuss them using the same example of circular model for the sake of concreteness. On the other hand, most of the results of section \ref{sec:Jack} are specific to the circular model and its variants. 

Therefore, using the circular model, we have provided an overview of Chapter \ref{ch:logrem}, excluding section \ref{sec:liouville} on the relation between \gls{lft} and \glspl{logrem}, and the sections \ref{sec:multifracintro}, \ref{sec:Gaussian} and \ref{sec:bindingetc}, whose essential purpose is to provide conceptional preparations for the relation. Relating precisely \gls{lft} and \glspl{logrem} is one of the main contributions of this manuscript. Although it does rely on some of the general results obtained in the previous sections (in particular, those in section \ref{sec:orderstat} concerning the full minima process), section \ref{sec:liouville} is quite detached from the rest of the Chapter: for example, we will start from a genuinely \textit{2D} \gls{logrem} (while the circular model is defined on a 1D geometry); also, we will be focused on the \textit{Gibbs measure} (the \textit{position} of thermal particles) instead of the free energy and minimal values. Therefore, we refer to section \ref{sec:lftoverview} for a more specific overview. The original results of section \ref{sec:liouville} are recent; they are impacting considerably how we view \glspl{logrem} in general, and raised many new questions that are under active investigation. These questions, as well as those raised by the previous sections,  are discussed informally in section \ref{sec:finallogrem}.

\section*{Localization transition with long--range hopping \\ Overview of Chapter \ref{ch:anderson}}
Localization transitions are arguably the most prototypical and most fascinating phenomena usually associated to disordered systems. With regard to the previous Chapter on \glspl{logrem}, freezing transition could also be viewed as a localization transition, in which the \textit{classical, thermal} particle is caged in few sites. However, properly speaking, the term ``localization transitions'' is reserved to the \textit{quantum mechanics} of particles in random potentials. At a formal level, this implies that the fundamental random object is not the partition function or the Gibbs measure, but the Hamiltonian, \textit{i.e.}, a \textit{random matrix} and its \textit{eigenvectors} (eigenstates). The latter can be localized or extended (with respect to the basis usually corresponding to the positions of the quantum particle). The sharp changes between the two possibilities (\textit{i.e.}, localization transitions), and the existence of other intermediate ones, are non--trivial questions with important physical applications \footnote{In fact, cast in this general mathematical form, localisation transitions' applications go beyond the realm of quantum mechanics in disordered medium; see section \ref{sec:loc-intro} for a brief and incomplete overview.}. 

Given the importance of localization transitions, there are many theoretical techniques to study them. The one mainly employed by Chapter \ref{ch:anderson} is classic and based on the following well--known lesson of Feynman: one can turn a quantum mechanics problem of a particle into a \textit{classical} statistical mechanics problem of a 1D extended object. In this thesis, this 1D object will be called a \textit{polymer}. Indeed, it is known that random matrices and their eigenvector localization can be studied by mapping to the problem of polymers in random media. In turn, the latter models found themselves in a well--known web of mappings, which relate them to models of \gls{fpp} and out--of--equilibrium growth. We will review these relations in section \ref{sec:introkpz}. The starting point of the original material in Chapter \ref{ch:anderson} is to extend this web of mappings to the context involving long--range hopping.  

The out--come of the new mappings is that, starting from the recently studied long--range \gls{fpp}/growth models (reviewed in section \ref{sec:fisher}), we defined a new ensemble of random matrices, the \glspl{bbrm}. They turn out to be \textit{superficially} comparable to the \glspl{pbrm}. As we will review in \ref{sec:loc-intro}, the \glspl{pbrm} were studied as 1D, long--range proxies of the standard Anderson model, which has no localization transition in 1D and 2D; however, the localization transition of the \glspl{pbrm} is pathologically simple and has in particular no \textit{mobility edges} (\textit{i.e.},  separation of the spectrum of one matrix into localized and extended eigenstates). Remarkably, as we will explore in section \ref{sec:bbrm}, the new \glspl{bbrm} turn out to have a different and richer phase diagram from \glspl{pbrm}, and have in particular localizations transitions with mobility edges. To show this, we will use the mapping that motivated the model (section \ref{sec:mappingAL}), along with other arguments and numerical evidences, in section \ref{sec:LT}.

Nonetheless, we would like to emphasize that the main interest of Chapter \ref{ch:anderson} is \textit{not} the results on the specific \gls{bbrm} model, but their generalization to a larger class of banded random matrices with \textit{broadly distributed} elements, in section \ref{sec:bbrmgen}. Roughly speaking, such matrices are \textit{sparse}: the moments of the matrix elements are much larger than their typical values, so most of the elements are small, but there are a few ``black swans''. This turns out to be the most important feature of the \gls{bbrm} model, and the one that is behind many of its localization properties. The essential point of our demonstration is that, although the exact mapping to the long--range \gls{fpp}/growth models is limited to the \glspl{bbrm}, the \textit{methods} used to study the statistical models can be adapted to the ``quantum'' (random matrix) context, with much looser constraints on the matrices' properties. Therefore, we shall conclude \ref{ch:anderson} by predicting the existence of localization transitions for a large class of \textit{new} banded random matrices. This raises many open questions, which we will discuss in section \ref{sec:finalanderson}.

\chapter*{Publication List}
The original research results on which this thesis is based are also available in the following peer-reviewed publications or preprints under review:
\begin{itemize}
\item[] \cite{cao15gff} \bibentry{cao15gff}. \\
\textbf{Reported in:} sections \ref{sec:jackintro} and \ref{sec:Dirichlet}.
\item[] \cite{cao16maxmin} \bibentry{cao16maxmin}. \\
\textbf{Reported in:} sections \ref{sec:minmax} and \ref{sec:EA}.
\item[]\cite{cao16loc} \bibentry{cao16loc}. \\
\textbf{Reported in:} sections \ref{sec:bbrm} and \ref{sec:bbrmgen}.
\item[] \cite{cao16order} \bibentry{cao16order}. \\
\textbf{Reported in:} sections \ref{sec:logREMdef}, \ref{sec:IRUV}, \ref{sec:EA}, \ref{sec:RSB}, and \ref{sec:orderstat}.
\item[]\cite{cao16liouville} \bibentry{cao16liouville}. \\ 
\textbf{Reported in:} section \ref{sec:liouville}. 
\end{itemize}

\printglossary[type=\acronymtype,style=index]

\chapter{Motivation and overview} \label{ch:intro}
\glsresetall
\section{Extreme value statistics}\label{sec:evs}
The statistical properties of extreme values among a large number of random variables are relevant in a wide range of disciplines, \textit{e.g.}, statistics, physics, meteorology, and finance. For instance, the knowledge of minimal and maximal stock prices over a period is obviously valuable in financial markets; estimating the most catastrophic flooding that would occur in 100 years is a crucial issue for big cities near water. The distribution of extreme eigenvalues is an important topic in random matrix theory, to which we will come back in Section \ref{sec:introkpz}. 

The basic question of extreme value statistics can be stated roughly as follows: let $V_1, \dots, V_M \in \mathbb{R} $ be $M$ random real variables, what is the distribution of 
\begin{equation} V_{\min} = \min_{i=1}^M V_i \,, \end{equation}
in the limit when $M$ is very large? More precisely speaking, as $M \to \infty$, one seeks to know, in increasing precision: how does the typical value $V_{\min}$ behave, as a function of $M$ ? how much does $V_{\min}$ fluctuate around its typical value? Finally, what is the probability distribution of the fluctuation in the $M\to \infty$ limit? This series of questions leads to the following Ansatz
\begin{equation}
V_{\min} \longrightarrow a_M + b_M y \,,\, M\to\infty \,. \label{eq:evsansatz}
\end{equation}
Here the convergence is in distribution.  $a_M$ the typical value, and $b_M$ the amplitude of the fluctuation, are both deterministic numbers that depend on $M$, whereas $y$ is a random variable whose distribution becomes $M$-independent in the $M\to \infty$ limit. The random variable $y = (V_{\min} - a_M) / b_M$ is also called the \textit{rescaled} minimum. The Ansatz \eqref{eq:evsansatz} is natural, widely applicable, and will apply to all the situations considered in this work. Therefore, in this framework, the problem of extreme value statistics reduces to determining $a_M$, $b_M$ and the probability distribution of $y$. 

\begin{figure}
\center
\includegraphics[scale=.5]{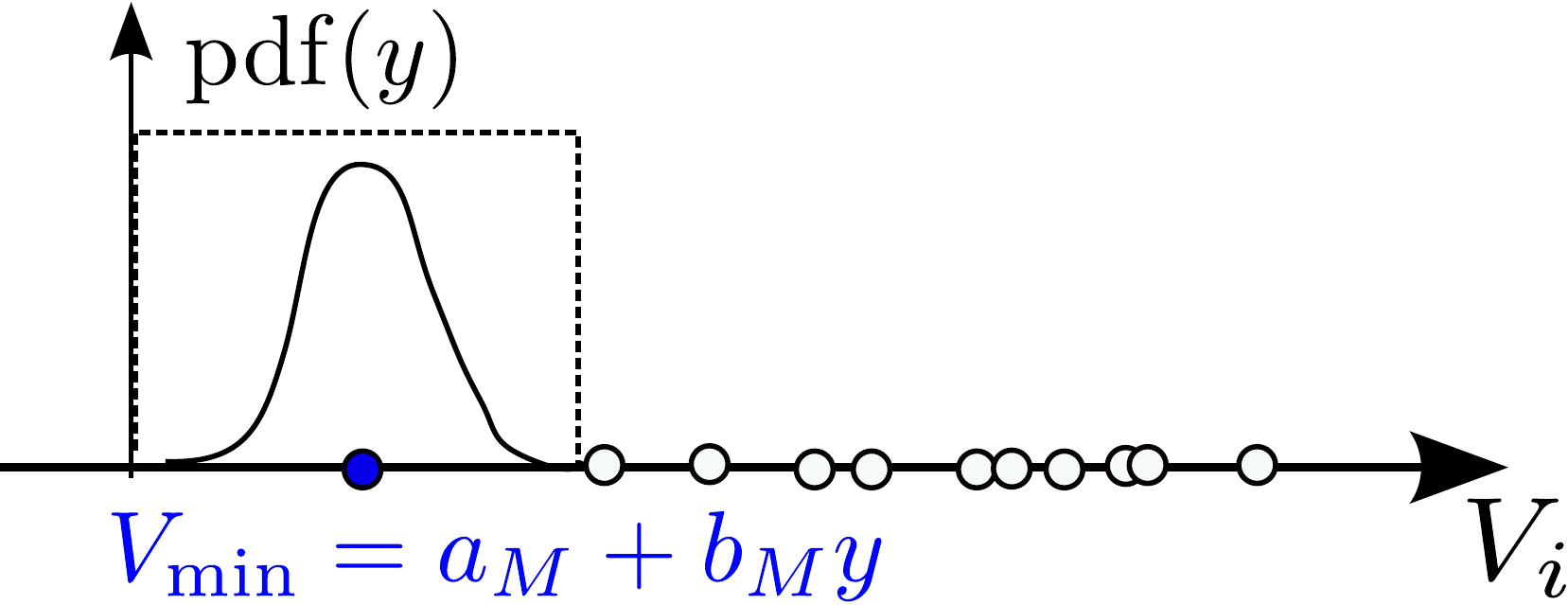}
\caption{An illustration of the extreme value statistics problem, and the Ansatz eq. \eqref{eq:evsansatz}.} \label{fig:evsAnsatz}
\end{figure}

Clearly, the answer depends on the statistical properties of the variables $V_1, \dots, V_M$ themselves, in particular, whether and how they are correlated. A whole chapter of this thesis (Chapter \ref{ch:logrem}) will be devoted to the case where $V_i$'s are \textit{logarithmically} correlated. 

Here, let us consider the simplest case where $V_i$ are independent and identically distributed (i.i.d.). According to the classical Fisher--Tippett--Gnedenko theorem, the distribution of $y$ belongs to either the Fréchet, Weibull, or the Gumbel family, depending on the asymptotic behaviour of $V_i$'s distribution at the $V_i \to -\infty$ limit (left tail):
\begin{enumerate}
\item When $V_i$ has an algebraic left tail, $y$ belongs to the Fréchet family;
\item When $V_i$ is bounded from below, $y$ belongs to the Weibull family;
\item Otherwise, $y$ has the Gumbel distribution.
\end{enumerate}
Let us further specialize to an (important) example of the Gumbel case, in which $V_i$'s are \gls{iid}. Gaussian variables defined by the following mean and covariance
\begin{equation}
\overline{V_i} = 0 \,,\, \overline{V_i V_j} = 2 \ln M \delta_{ij} \,,\, i,j = 1, \dots, M\,. \label{eq:REMdef}
\end{equation}
Then, $V_{\min}$ satisfies the Ansatz eq. \eqref{eq:evsansatz} with the following parameters:
\begin{subequations}\label{eq:REMzeroT}
\begin{align}
&a_M  = -2 \ln M + \frac12 \ln \ln M + \frac12 \ln(4\pi) \,,\, b_M = 1 \,,\, \\
&\mathrm{pdf}(y) = \exp\left(y - e^{y} \right) \stackrel{y\to-\infty}{\sim} e^{y}\,,
\end{align}
\end{subequations}
where $\mathrm{pdf}(y)$ denotes the \gls{pdf}.
A self-contained demonstration of this result will be given in section \ref{sec:REM}. The strategy will be that of disordered statistical physics: 
\begin{enumerate}
\item Regard the extreme value $V_{\min}$ as the \textit{ground state} of a statistical physics model whose energies are $V_1, \dots, V_M$. This means introducing a temperature $T = 1/\beta$, and writing down the canonical partition function and the free energy: 
\begin{equation}
\mathcal{Z} = \sum_{i=1}^M \exp(-\beta V_i) \,,\, \mathcal{F} = -\beta^{-1} \ln \mathcal{Z} \,. \label{eq:defZandF}
\end{equation}
\item Study the disordered statistical physics model at finite temperature. In particular, determine the limit distribution of the free energy $\mathcal{F}$, in the sense of the scaling Ansatz \ref{eq:evsansatz}. In the statistical physics context, the $M\to \infty$ is usually called the \textit{thermodynamic limit}. This is the hard core of the approach, since the model defined by eq. \eqref{eq:defZandF} is \textit{disordered}, and disordered statistical physics is in general difficult, both analytically and numerically. In the case of the example \eqref{eq:REMdef}, the resulting statistical model is the Derrida's famous \gls{rem}  \cite{derrida1980random}. It is one of the foundational toy model in disordered statistical physics; section \ref{sec:REM} will be devoted to it. Its importance can be illustrated by the remarkable fact that such a simple model has a phase transition, at $\beta = \beta_c = 1$ using the above definition. 
\item Take the zero temperature limit, in which $\mathcal{F}$ will tend to $V_{\min}$: 
\begin{equation}
V_{\min} = \lim_{\beta\to +\infty} \mathcal{F} \,.
\end{equation}
There is two technical assumptions behind the above statement. First, the \textit{absence} of zero-temperature phase transition, which assures that the $M \to \infty$ and $\beta\to \infty$ limits commute. Second, the ground should not be degenerated; in particular, the entropy should vanish as $T \to 0$. In the problems treated in this thesis, both assumptions turn out to be fulfilled. 
\end{enumerate}

The above relation between extreme value statistics and disordered statistical physics is of central importance and has far more applications. We mention two notable ones. The first is the study of disordered systems (\textit{e.g.}, spin glasses) in low temperatures. Extending the above reasoning, it is natural to expect that this problem is related to the statistical properties of the ground state, and the lowest excited states, which are the \textit{higher extreme order statistics} of the energies $V_i$, 
\begin{equation}
V_{\min} =V_{\min,0} \leq V_{\min,1} \leq V_{\min,2}  \leq \dots \,,
\end{equation}
noting that $V_{\min,k}$ is the $(k+1)$-th ordered minimal value. Even when $V_i$'s are independent, $V_{\min,k}$'s will be non-trivially correlated, and it is important to characterize the joint distribution of the \textit{extrema process}. For the Random Energy Model, defined by eq. \eqref{eq:REMdef}, in the $M\to\infty$ limit, the extrema process is known to be the \textit{Gumbel Poisson point process}. We will discuss this in section \ref{sec:REMorder}.

The second is the statistical physics of \textit{optimization problems} \cite{mezard87beyond,dotsenko1995introduction}. Formally, an optimization problem amounts to finding the minimum \textit{value} and \textit{position} of some \textit{cost function} of a (usually large) number of variables, $V(x_1, \dots, x_n)$. It is very fruitful to think of the latter as a $n$-dimensional \textit{potential energy landscape} in a $(n+1)$-dimension space, and the searched minimum is the deepest valley. Many optimization algorithms can be seen as the simulation of a thermal particle in that potential. To understand the behaviour and efficiency of these algorithms, it is the statistical physics of a thermal particle in such a potential. 

The notion of potential energy landscape is central in disordered statistical physics and its wide applications. The landscapes involved are often complex, and involve a large of number of parameters unknown \textit{a priori}. Therefore, their theoretical model is often \textit{random} potential energy landscapes. 

\section{A thermal particle in a random potential}\label{sec:introPEL}
Random potential energy landscapes studied in spin glass and optimization problems are often in high dimensions. In contrast, the problem of thermal particles in a low-dimensional random potential is easier to state, has its own applications, and remains highly non-trivial. 

As an instructive illustration, let us consider Gaussian potentials on a one-dimensional lattice, labelled by $j=0, \dots, M-1$, with periodic boundary condition. Assuming translation invariance, the potential can be generated by Fourier transform 
\begin{equation}
V_j = \Re \sum_{k=-M/2}^{M/2-1} \sqrt{\mu_{k}} \exp\left(2 \pi \im \frac{jk}{M}\right) \mathsf{N}_k \,.
\label{eq:FFTgen}	
\end{equation}
Here, $(\mathsf{N}_k)$ is a sequence of \gls{iid} standard complex Gaussian variables, and $\mu_k \geq 0$ control all the statistical properties of the potential, \textit{e.g.}, the variance and the correlations: 
\begin{align}
\overline{V_j^2} &=  \sum_{k=-M/2}^{M/2-1} \mu_k  \,,\,
\overline{V_0 V_j} = \sum_{k=-M/2}^{M/2-1} \mu_k \cos\left(\pi \frac{jk}{M}\right) \,,\, \nonumber  \\
 \overline{(V_0 - V_j)^2} &=  \sum_{k=-M/2}^{M/2-1} 4 \mu_{k} \sin^2\left(\pi \frac{jk}{M}\right)  \,, \label{eq:covariance-gen}
\end{align}
The last equation measures the \textit{roughness} of the potential, and is not affected by the \textit{zero-mode} $\mu_{k=0}$. Its effect is a trivial global shift to the potential, and will be set to $0$. 

\begin{figure} 
\center \includegraphics[scale=.5]{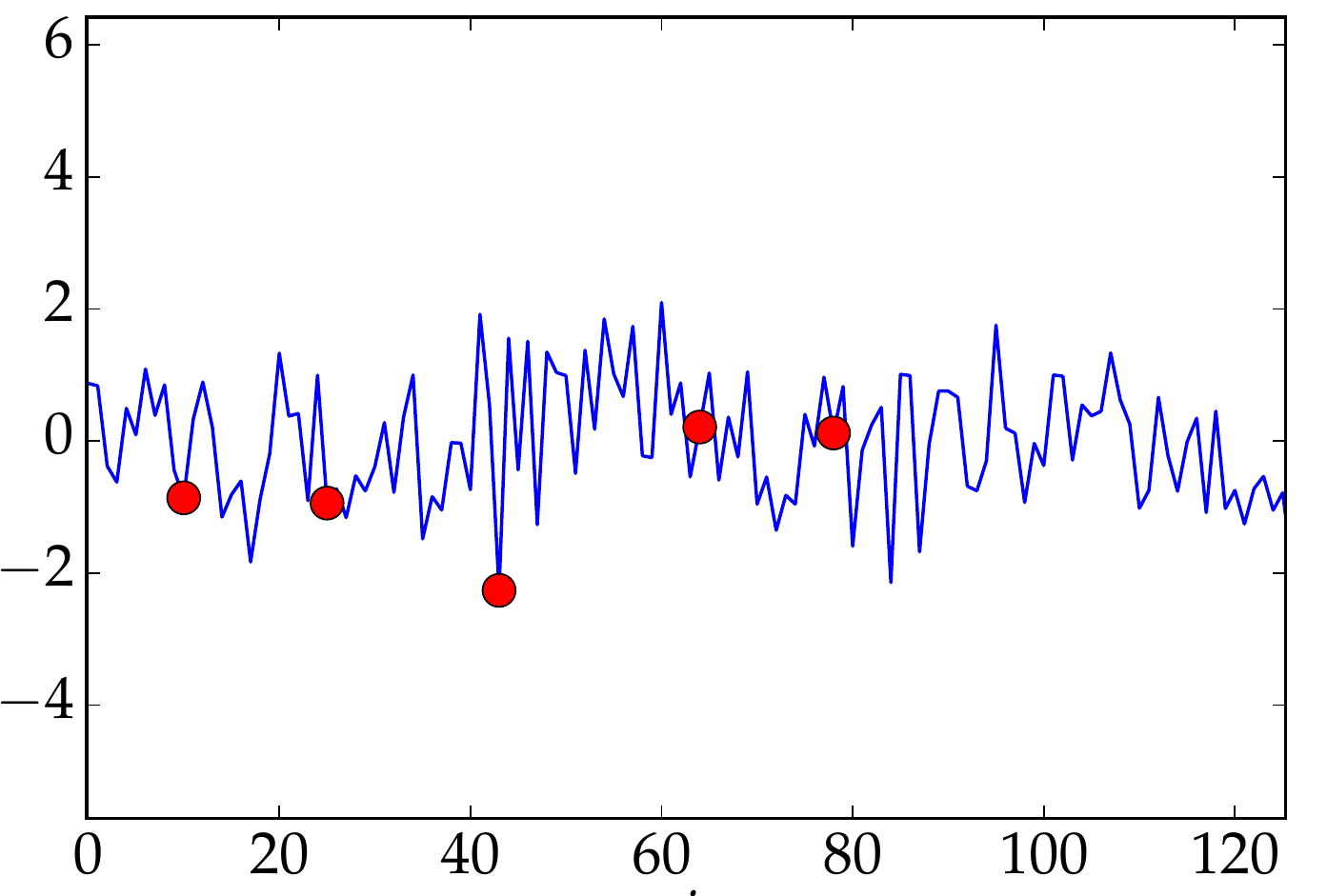} \includegraphics[scale=.5]{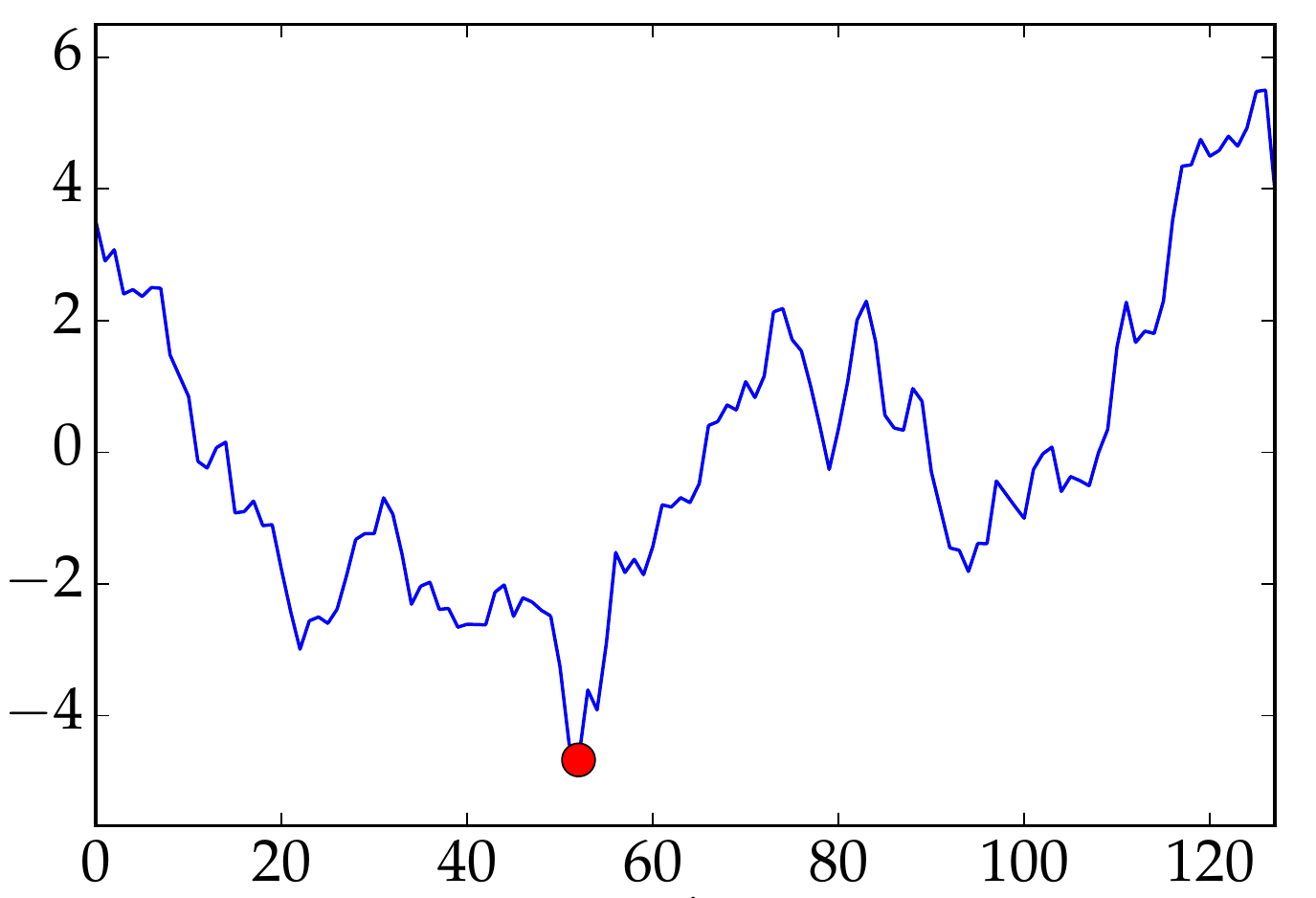}
\caption{Samples of random potentials defined by eq. \eqref{eq:FFTgen} and \eqref{eq:muk-gen}, with $\alpha = 0$ (white noise, left panel) and $\alpha = 2$ (Brownian motion, right panel). In the former case, a thermal particle visits freely the whole system, while in the latter, it is confined in the deepest minimum.} \label{fig:WNBM}
\end{figure}
An important class of potentials is given by amplitudes $\mu_k$ that depend algebraically on $\abs{k}$; more precisely,
\begin{equation}
\mu_0 = 0 \,,\, \mu_{k\neq 0} = \pi M^{-1} \sin^{-\alpha}\left( \frac{\pi \abs{k}}{M}\right) \stackrel{\abs{k} \ll M}\sim k^{-\alpha} M^{\alpha-1} \,. \label{eq:muk-gen}
\end{equation}
where $\alpha$ is a real parameter. Let us recall the morphology of the potential as function of $\alpha$:
\begin{itemize}[label=$\square$]
\item When $\alpha = 0$, $V_j$ are \gls{iid} standard Gaussian, with variance $\overline{V_j^2} = 1$ (notice the difference with eq. \eqref{eq:REMdef}). So the potential is a white noise. 
\item When $\alpha \in (0,1)$, the variance tends to finite constant in the $M\to \infty$ limit,  $\overline{V_j^2} \to \frac{\Gamma \left(\frac{1}{2}-\frac{\alpha }{2}\right)}{\sqrt{\pi } \Gamma \left(1-\frac{\alpha }{2}\right)}$. Off-diagonal correlations decay algebraically: when $1 \ll j \ll M$, by eq. \eqref{eq:covariance-gen} and \eqref{eq:muk-gen}, 
\begin{equation}
\overline{V_0 V_j} \sim \sum_k k^{-\alpha} M^{\alpha-1} \cos(\pi jk/M) \sim  \abs{j}^{\alpha-1} \,,
\end{equation} 
therefore, as $j, M \to \infty$, $\overline{(V_0 - V_j)^2}$ remains bounded, \textit{i.e.}, the potential is not rough. 
\item When $\alpha > 1$, the potential is rough:
\begin{equation}
\overline{V_j^2} \sim M^{2H}  \,,\, \overline{(V_0 - V_j)^2} \sim \abs{j}^{2 H} \,,\, 1 \ll \abs{j} \ll M \,,\, H = \frac{\alpha-1}{2} \,, 
\end{equation}  
where $H$ is the \textit{Hurst exponent} of the roughness. In particular, when $\alpha=2$, the potential can be compared locally to a Brownian motion in a suitable continuum limit, while general $\alpha \in (0,1)$ corresponds to a \textit{fractional} Brownian motion.
\end{itemize}

Now, what is the behaviour of a thermal particle of temperature $T = 1/ \beta$ in one of the above potentials? Note that the statistical model would be formally identical to eq. \eqref{eq:defZandF}, yet with $V_j$ non-trivially correlated.  A standard strategy of qualitatively understanding any such model is to compare the minimum energy, $V_{\min} = \min_{i=0}^{M-1}(V_i) $, with the entropy of visiting every one of the $M$ sites $S = \ln M$. This determines roughly whether the system is in an energy-dominating, or entropy-dominating, phase. Such an analysis was carried out in detail in \cite{carpentier2001glass}, section II, whose main message is the following:
\begin{itemize}[label=$\square$]
\item When $\alpha < 1$, $\abs{V_{\min}} \ll \ln M$. The model has only a high-$T$ phase;
\item When $\alpha > 1$, $\abs{V_{\min}} \sim M^{\alpha - 1}  \gg \ln M$. The model has only a low-$T$ phase. Nevertheless, we note the problem of a thermal particle in a random potential generated by a (fractional) Brownian motion is classic (dating back to Sinai \cite{sinai1983limiting}) and well--studied problem with wide applications, see \cite{schehr2014exact} for a review.
\end{itemize}

\begin{figure}
\center
\includegraphics[scale=.5]{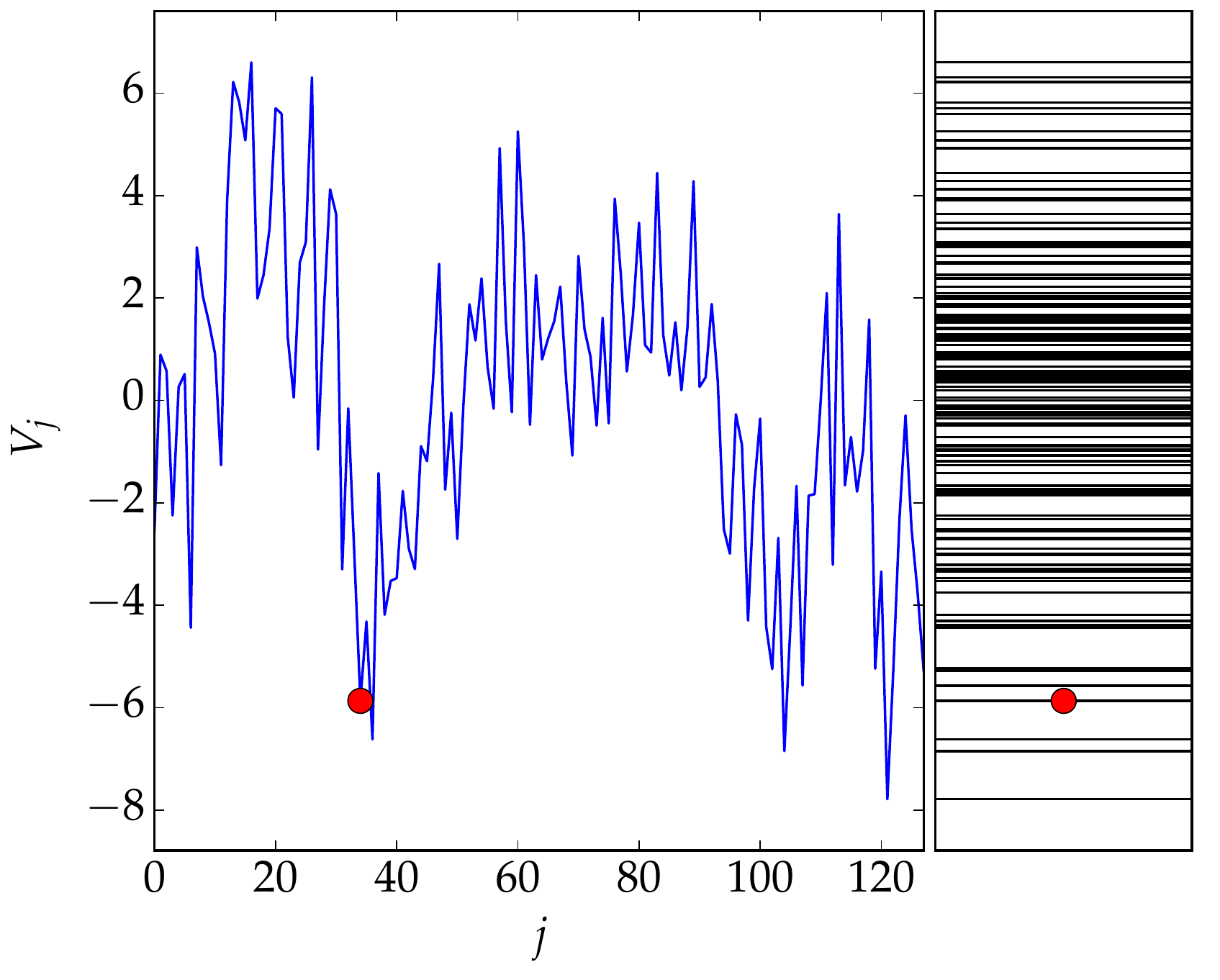}
\caption{A sample of the log-correlated random potential, generated by eq. \eqref{eq:FFTgen} and \eqref{eq:muk-gen}, with $\alpha = 1$. The potential is plotted in the left panel; all its energy values are plotted at the right panel. A thermal particle at finite temperature occupying a position with low--lying energy (but not the minimum energy, which is attained at $j = 120$) is also depicted.}  \label{fig:logREM1}
\end{figure}
These results point to the case $\alpha=1$, which was deliberately left out, and is in fact the most interesting from the thermodynamic point of view: there is a phase transition at some finite $\beta = \beta_c$ separating a high-$T$ phase from a low-$T$ phase. Indeed, when $\alpha=1$, equations \eqref{eq:muk-gen} and \eqref{eq:covariance-gen} imply that the variance $\overline{V_j^2} = 2 \ln M + O(1)$ , \textit{i.e.}, it is proportional to the (infinite-$T$) entropy $S = \ln M$, comparable to the \gls{rem} (eq. \eqref{eq:REMdef}). Therefore, the out-come of energy--entropy competition depends non--trivially on the temperature and can induce a phase transition. Yet, unlike the \gls{rem}, correlations do exist and are logarithmic: 
\begin{equation} \overline{V_j V_0} \sim 2 \ln (M / \abs{j}) \,,\, 1 \ll j \ll M \,. \end{equation} 
This is an example of \glspl{logrem}. Remarkably, the critical temperature of this \gls{logrem} $\beta_c = 1$, identical to the \gls{rem}. The \gls{rem}, the \glspl{logrem} and their phase transition (called \textit{freezing}) will be the subject of Chapter \ref{ch:logrem}. 

Let us anticipate that minimum behaviour of the \gls{logrem} defined above (eq. \eqref{eq:FFTgen}, \eqref{eq:muk-gen} with $\alpha=1$):
\begin{subequations}\label{eq:FBzeroT}
\begin{align}
&a_M = -2 \ln M + \frac32 \ln \ln M + O(1)  \,,\, b_M = 1 \,,  \label{eq:FBzeroTlead} \\
&\mathrm{pdf}(y) = 2 e^y K_0\left(2 e^{y/2}\right) \stackrel{y\to -\infty}{\sim} \abs{y}e^y \,, \label{eq:FBzeroTpdf}
\end{align}
\end{subequations}
where $K_n(x)$ is the Bessel $K$-function. One should compare \eqref{eq:FBzeroT} to the analogous result for the uncorrelated \gls{rem}, eq. \eqref{eq:REMzeroT}: we shall see that, the $\frac32 \ln \ln M$ correction in $a_M$ and the $\abs{y} e^y$ left-tail (in contrast to $e^y$ in the \gls{rem} case) are \textit{universal} features of \glspl{logrem}. On the other hand, the precise full distribution of $y$ depends on the specific  \gls{logrem} studied here, \textit{i.e.}, on the fact that it is a 1D potential, with periodic boundary condition, \textit{etc}. In fact, we will see in section \ref{sec:treetoplane} (around eq. \eqref{eq:FBgfreeze}) that this \gls{logrem} is the famous \textit{circular model of $\mathrm{1/f}$-noise} \cite{fyodorov2008statistical}. 

\section{Polymers in random media}\label{sec:introkpz}
 Let us come to another classic problem which illustrates the ideas above, and will play a rôle in both Chapter \ref{ch:logrem} and \ref{ch:anderson}. Although called polymers in random media, it is related to a range of seemingly different problems. Here, let us motivate it with the Eden model \cite{eden1961two,richardson73growth} of non-equilibrium growth, which was a simplified description of the expansion of microcosm colonies. 
 
\begin{figure}
  \center 
\includegraphics[scale=1]{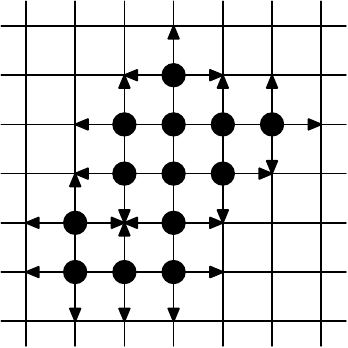}
\includegraphics[scale=1]{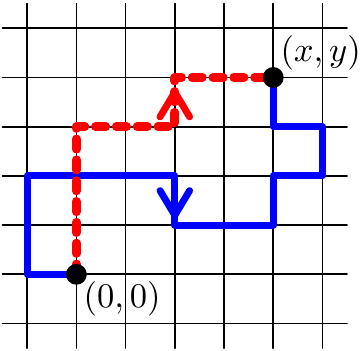}
  \caption{\textit{Left}: An illustration of the Eden model. The dots represent occupied sites, and the arrows represent the possible ways of new occupation. Each happens with a rate $\dif t$. \textit{Right}: Two polymers from $(0,0)$ to $(x,y) = (4,4)$. The dashed red one is a \textit{directed polymer}, while the blue one is undirected.}\label{fig:eden}
  \end{figure}

 The Eden model is a continuous-time Markov stochastic lattice model. It can be defined on any lattice; here, let us take the two-dimensional square lattice as example (see Figure \ref{fig:eden} for an illustration). Each lattice site $(x,y)\in\Z^2$ can be either empty or occupied; occupied sites remain occupied forever; during any infinitesimal time interval $\dif t$, any empty site $(x,y) \in \Z^2$ has probability $\dif P = n \dif t$ to become occupied, where $n$ is the number of occupied sites among its neighbours, $(x\pm1, y), (x, y \pm 1)$. In other words, every occupied site occupies each of its neighbours at rate $1$, and all the attempts occur independently. Initially ($t = 0$), only the origin $(0,0)$ is occupied. Following the dynamics for a long time ($t \gg 1$), the colony of occupied sites will acquire a macroscopic shape (which is unknown analytically!), whose linear size $L$ grows linearly in time $L \propto t$. Observed more closely, the surface of the colony is rough, and it is widely believed that the local dynamics at the intermediate scale $1 \ll \ell \ll L$ is believed to be described by the famous \gls{kpz} equation \cite{kardar1986dynamic,Halpin-Healy2015}
\begin{equation}\label{eq:kpz}
\partial_t h(u,t) = \nu \partial_u^2 h(u,t) + \frac{\lambda}{2} (\partial_u h(u,t))^2 + \eta(u,t) \,,\, 
\end{equation}
where $\eta(u,t)$ is a space--time white noise. A subtlety involved in this statement is that the \gls{kpz} equation describes the irreversible stochastic growth of \textit{simple} interfaces, \textit{i.e.}, those can be described as the graph of some height function $h(u)$ (with some choice of the coordinates $h$ and $u$), while the Eden--interface is not in general simple at the discrete level (\textit{e.g.}, it has ``holes'', see Figure \ref{fig:eden}, left panel). 
 
KPZ equation is known to describe a range of phenomenon, including the \textit{directed} polymer in random media model. Usually, seeing this involves a Cole-Hopf transform manipulation on the \gls{kpz} equation (see for example \cite{calabrese2010kpz,dotsenko10kpz,dotsenko10kpz1}). Here,  one may relate the Eden model to an \textit{undirected} polymer in random media model (the difference will be discussed below). For this, we consider the \textit{first passage time},
 \begin{equation}
 T(x,y) = \text{time at which } (x,y) \text{ becomes occupied.}
 \end{equation}   
 In particular, $T(0,0) = 0$; since every site changes its status only once, the above quantity is well-defined.

 To derive the formula for $T(x,y)$ (see eq. \eqref{eq:TxyZgen} below), which establishes the claimed relation between the Eden model and the polymer model,  consider a lattice edge $\mathsf{e}: (x,y)-(x',y')$ connecting the two sites, and assume $(x',y')$ is occupied before $(x,y)$. According to the dynamics rule, $(x,y)$ will be occupied \textit{from} $(x',y')$ at $T(x',y') + \tau(\mathsf{e})$, where $\tau(\mathsf{e})$ is the \textit{waiting time} assigned to $\mathsf{e}$. The waiting time for all edges is \gls{iid}. random variables, with standard exponential distribution $\mathrm{pdf}(\tau(\mathsf{e})) = e^{-\tau(\mathsf{e})}$ ($\tau(\mathsf{e}) \geq 0$). Now, Remembering that $T(x,y)$ can be occupied from any of its neighbours, we have
 $$  T(x,y) = \min_{\mathsf{e}:(x',y)-(x,y)} \left[ T(x',y') + \tau(\mathsf{e}) \right] \,.  $$
 Iterating this formula (\textit{i.e.}, writing $T(x',y')$ at the right hand side as a minimum over its neighbours, and so on), and combing with the initial condition, one concludes that $T(x,y)$ is a minimum over all the lattice paths $\mathfrak{p}: (0,0) - (x,y)$ made of a sequence of edges $e_1, \dots, e_s$
 \begin{equation}\label{eq:Txypolymer}
 T(x,y) = \min_{\substack{\mathfrak{p}: (0,0)\to(x,y)}}
  E[\mathfrak{p}] \,,\, E[\mathfrak{p} = (\mathsf{e}_1, \dots, \mathsf{e}_{s})]= \left[ \sum_{j=1}^s \tau(\mathsf{e}_j)  \right] \,,
 \end{equation} 
 where the length of the path  $s = s(\mathfrak{p})$ is \textit{not} fixed. What we just obtained is a \gls{fpp} model \cite{hammersley1966first,auffinger201550} corresponding to the Eden growth model. In section \ref{sec:fpp} we will see the another application of such a mapping.
 
 Equation \eqref{eq:Txypolymer} expresses $T(x,y)$ as a minimum. So, we can define the finite--temperature version of it.  For this, we regard $T(x,y)$ as the minimum energy of the ensemble of undirected polymers, modelled as \textit{any} lattice paths connecting $(0,0)$ and $(x,y)$ in a random medium, described by the random energies $\{\tau(\mathsf{e})\}$; see Figure \ref{fig:eden} (right panel) for illustration. The energy of a polymer configuration is the sum of its edge-energies. Such a model can be defined at finite temperature $T = 1/\beta$, in the canonical grand-canonical ensemble, by the partition function
 \begin{equation}\label{eq:polymerZgen}
 \mathcal{Z} = \sum_{\substack{\mathfrak{p}: (0,0)\to(x,y)}}
 \exp\left[ - \beta \sum_{j=1}^s (\tau(\mathsf{e}_j)  - \mu)\right] \,,
 \end{equation}
 where $\mu$ is the chemical potential coupled of a monomer. Then, it is not hard to see that
 \begin{equation} \label{eq:TxyZgen}
  T(x,y) = \left[ - \beta^{-1} \ln \mathcal{Z}\right]_{\mu = 0, \beta \to +\infty}  
  \end{equation}
 is retrieved as the zero temperature, zero chemical potential limit. This is the advocated relation between the Eden growth model and polymers in random media: the latter is the \textit{finite--temperature} version of the \gls{fpp} problem corresponding to the Eden model. Note that if the chemical potential $\mu\neq 0$, the zero--temperature limit gives a \gls{fpp} model with 
 \begin{equation} E[\mathfrak{p} = (\mathsf{e}_1, \dots, \mathsf{e}_{s})] =  \sum_{j=1}^s \left(\tau(\mathsf{e}_j) -\mu \right)  \label{eq:Ewithchemical}\end{equation} 
instead of eq. \eqref{eq:Txypolymer}. It is interesting to consider $\mu < 0$ and $\mu > 0$ separately:
\begin{itemize}
\item When $\mu > 0$, there can be some edge such that $\tau(\mathsf{e}) < \mu$. This is problematic because the polymer can visit this edge back and forth to lower its energy to $-\infty$. In statistical physics, this is well--known to be a signature of \textit{condensation}, \textit{e.g.} Bose--Einstein condensate. Although $\mu > 0$ is forbidden in statistical physics, it does have an application when we relate the polymer model to quantum mechanics, as we shall see in sections \ref{sec:mappingAL} and \ref{sec:LT}. 
\item When $\mu < 0$, this energy function punishes longer paths. In particular, when $\mu \to -\infty$, one obtains the \textit{directed} polymers in random media model, whose partition function sums over only lattice paths of minimal length $s = s_{\min} = \abs{x} + \abs{y}$ (see the dashed line in Figure \ref{fig:eden}, right panel for an example):  
   \begin{equation}\label{eq:ZDP}
   \mathcal{Z}_{\text{DP}} = \sum_{\substack{\mathfrak{p}: \mathsf{e}_1, \dots, \mathsf{e}_{s_{\min}} \\ (0,0)\to(x,y)  }}
    \exp\left[ - \beta \sum_{j=1}^{s_{\min}} \tau(\mathsf{e}_j)\right] \,.
   \end{equation} 
 Moreover, it is also well-known that (see \textit{e.g.}, \cite{calabrese2010kpz}), with the time coordinate $t = x+y$ and space coordinate $u = x-y$, the free energy satisfies the \gls{kpz} equation, in an appropriate scaling regime. Now, we can now argue the undirected polymer model defined by eq. \eqref{eq:polymerZgen} is also described by \gls{kpz} in the continuum limit, since the dominant contribution to $\mathcal{Z}$ comes from polymers that are directed in the large scale. 
\end{itemize}

 The major concrete consequence of ``belonging to the \gls{kpz} class'' is the following prediction. For $(x,y) = (r\cos\theta, r\sin\theta)$ far from $(0,0)$ ($r \to \infty$),  $T(x,y)$ satisfies the following version of Ansatz eq. \eqref{eq:evsansatz}
 \begin{equation}\label{eq:TW}
 T(r\cos\theta, r\sin\theta) \stackrel{r\to \infty}{\longrightarrow} r a(\theta) +  r^{\frac{1}{3}} b(\theta) \chi \,,
 \end{equation}
 where $\chi$ obeys a Tracy--Widom distribution \cite{Tracy1994} \footnote{More precisely, the GUE one, since the Eden model is defined with droplet initial condition.}. Another universal signature is the scaling $r^{\frac{1}{3}}$. On the other hand, the pre--factors $a(\theta)$ and $b(\theta)$ are not universal and depend on the microscopic details of the model. In particular, $a(\theta)$ is directly related to the limit shape of the Eden model, so is unknown. The finite--temperature  energy  $-\beta^{-1} \mathcal{Z}$ (for $\mu \geq 0$) is also expected to have the scaling form eq. \eqref{eq:TW} with different $a(\theta)$ and $b(\theta)$ but the same exponent $r^{1/3}$ and distribution $\chi$. 
 
 Remarkably, eq. \eqref{eq:TW} describes another problem of extreme value statistics in \textit{random matrix theory} \cite{tracy2002distribution}: for a large class of random matrices of si $N \times N$, as $N \to \infty$, the largest eigenvalue $\lambda_{\min}$  limiting distribution  $\lambda_{\min} = a \sqrt{N} + b N^{1/6} \chi$; it can be matched with eq. \eqref{eq:TW} by setting $N = r^2$. This is not a coincidence, but there is a deep connection. In fact, using ingenious combinatorial methods, it has been shown \cite{Johansson2000kpz} that, $\lambda_{\min}$ could be closely related to a cousin to the first--passage percolation problem, the \textit{last}--passage percolation, whose finite--temperature version is in turn the directed polymer in random media model (eq. \eqref{eq:ZDP})!

The fact that the same universality governs seemingly unrelated phenomena makes \gls{kpz} a fascinating subject, and the various related models are extensively studied. Although this thesis contains no direct contribution to \gls{kpz} in finite dimensions, let us mention the following:
\begin{itemize}
\item The directed polymer model can be defined in any dimension, and also on a Cayley tree. That case turns out to be an important case , for several reasons: it can be exactly analysed \cite{derrida1988polymers}; it corresponds to \gls{kpz} in $d = \infty$ dimension; moreover, it is also a \gls{logrem} (see section \ref{sec:DPCT})! In fact it was the first \gls{logrem} ever studied, before those defined by log-correlated random potentials in Euclidean spaces (\textit{e.g.}, the one in section \ref{sec:introPEL}). As we shall see in Chapter \ref{ch:logrem}, the remarkable close relation between a $d = \infty$ model and finite dimensional ones is the essential reason behind the applicability of \gls{rsb}, a disordered--systems method usually limited to mean-field models, to finite-dimension \glspl{logrem}. 
\item 
 \gls{kpz} universality also appears in \textit{Anderson localization}, which we study in Chapter \ref{ch:anderson}. It is a well--known quantum mechanical phenomenon of a particle in a random potential. Its relation to the \textit{classical} statistical physics of the directed polymer in random media problem was realized during the early days of \gls{kpz} \cite{medina89anderson,medina92anderson}, and turned out quite fruitful up to now. We shall revisit this in section \ref{sec:LT}.
\end{itemize}

\glsresetall
\chapter{Log-correlated Random Energy Models}\label{ch:logrem}
\section{History and background}\label{sec:logremintro}
This section reviews the theoretical backgrounds that lead to the field of \glspl{logrem}. Section \ref{sec:REM} is a quite detailed introduction to the \gls{rem}, covering its freezing transition and free energy distribution. Section \ref{sec:DPCT} introduces the \gls{dpct} model, which is the first \gls{logrem}. Section \ref{sec:treetoplane} reviews the key developments of \glspl{logrem} on Euclidean spaces, including works of Mudry \textit{et. al.} and of Carpentier--Le Doussal, the circular model of Fyodorov--Bouchaud, and the freezing duality conjecture. Section \ref{sec:multifracintro} introduces the basic multi--fractal properties of \glspl{logrem}, and the transitions associated. 

\subsection{Random Energy Model (REM)} \label{sec:REM}
The \acrlong{rem} has been introduced in Chapter \ref{ch:intro} as the finite temperature version of a simple extreme value statistics problem. The genuine motivation of Derrida \cite{derrida1980random,derrida1981random} was to design a simple toy model of spin glass. Since the classic work of Edwards and Anderson \cite{edwards1975theory}, spin glass is associated with the Ising model with random interaction, whose partition function (at zero magnetic field) is
\begin{equation}
\mathcal{Z} = \sum_{\sigma_1 = \pm 1} \dots  
\sum_{\sigma_N = \pm 1} \exp(-\beta H[\sigma_1, \dots, \sigma_N]) \,,\, 
H[\sigma_1, \dots, \sigma_N] = \sum_{<ij>} J_{ij} \sigma_i \sigma_j \,,\,   \label{eq:RBIsing}
\end{equation}
where $\beta$ is the interaction, $\sum_{<ij>}$ sums over neighbouring pairs in a lattice, and $J_{ij}$'s are \textit{random} couplings: usually, they are taken as \gls{iid} Gaussian with variance $1 / N$, so that 
\begin{equation}
\overline{H[\sigma_1, \dots, \sigma_N]^2} \propto N\,. \label{eq:Hvariance}
\end{equation}
 The model defined by eq. \eqref{eq:RBIsing} turned out to be significantly more difficult than its non--disordered cousin. It is still poorly understood in three dimensions. The solution of the mean field version, defined on a fully connected network, was a heroic effort, and involved the invention of new theoretical insights and methods: the \textit{replica trick} initiated by Sherrington and Kirkpatrick \cite{sherrington75spinglass}, the de Almeida--Thouless stability \cite{dealmeida78stability}, which led to the discovery of \textit{\acrlong{rsb}}, accumulating in the Parisi's \textit{tour--de--force} exact solution \cite{parisi79spinglass,parisi80sk}. The complete story, as well as the foundational papers, can be found in Part one of classic book \cite{mezard87beyond} (other textbooks include \cite{dotsenko1995introduction}; mathematicians may prefer \cite{Talagrandbook}). 

A striking feature of the spin--glass theory is its non-rigorousness: both the replica trick and the \acrfull{rsb} involve manipulations which would shock any mathematician. The \gls{rem} was the first step in the mathematical development of spin--glass theory. For this, Derrida tremendously simplified the model eq. \eqref{eq:RBIsing}. It retains two features of eq. \eqref{eq:RBIsing} and \eqref{eq:Hvariance}: the partition function sums over an exponential number $M = \exp(cN)$ of configurations, and the energy of each configuration has variance $\propto N \propto \ln M$. However, it neglects the correlations between energy levels. After some rescaling, one obtains  the \acrfull{rem} partition function that we have seen in section \ref{sec:evs}:
\begin{equation}
\mathcal{Z} = \sum_{j=1}^M \exp(-\beta V_j) \,,\, (V_j) \text{ Gaussian, }  \overline{V_j} = 0\,,\, \overline{V_i V_j} =  2 \ln M \,  \delta_{ij}\,. \label{eq:REMdef1}
\end{equation}
By the above reasoning, we expect that the thermodynamic quantities of the \gls{rem} are proportional to $\ln M$, which is the system volume of the spin glass model being mimicked.

\subsubsection*{Freezing transition}
The most important fact about the \gls{rem} is that it has a phase transition at $\beta = \beta_c$, where $\beta_c = 1$ with the normalization of eq. \eqref{eq:REMdef1}. The simplest way to see it is to work in the micro--canonical ensemble (as in \cite{derrida1980random}). For this, let us fix an energy $E$. Since each energy level is a Gaussian with \gls{pdf} $P(V_j) = (4\pi\ln M)^{-\frac12} \exp(-V_j^2 / 4 \ln M)$, $j = 1, \dots, M$, the (mean) number of configuration with energy $E$ is $\overline{\mathcal{N}(E)} = P(E) M$. As a consequence, $\overline{\mathcal{N}(E)} \ll 1$ when $\abs{E/\ln M}>2$. When $E / \ln M \in  (-2 , 2 )$, $ \overline{\mathcal{N}(E)} \sim M^{1-\frac{E^2}{4\ln M}} (4\pi\ln M)^{-\frac12} \gg 1$. Since the energies are independent, $\mathcal{N}(E)$ has a binomial distribution, which has no algebraic (fat) tail. For such distributions, the mean value is representative of the typical value; therefore, the entropy can be estimated by
\begin{equation}
\frac{S(E)}{\ln M} \to \frac{\ln \overline{\mathcal{N}(E)}}{\ln M}  = 1 - \frac{1}{4} \left[\frac{E}{\ln M}\right]^2  \,,\, \frac{E}{\ln M} \in (-2, 2) \,,\, \label{eq:REMentropy}
\end{equation}
Using standard thermodynamics formulas, we calculate the inverse temperature $\beta = \frac{\partial S}{\partial E} = - E / (2\ln M) \in (-1, 1)$, and the free energy 
\begin{equation} \mathcal{F} / \ln M = (E - \beta^{-1} S) / \ln M \to -\left(\beta + \beta^{-1} \right) \,,\, \abs{\beta} < 1\,. \label{eq:FhighT} \end{equation} 
Remark that as $\beta \to 1$, $E / \ln M \to -2$, $S / \ln M \to 0$: the entropy becomes sub-extensive at a non--zero temperature. This is a key signature of disordered systems: it is called the \textit{entropy crisis}, and is responsible for their glassy behaviour and slow dynamics. What happens at lower temperatures? Since the entropy cannot further decrease and cannot increase either, the only possibility is that $S = 0$, and the free energy becomes temperature independent:
\begin{equation}
\mathcal{F} / \ln M \to - 2 \,,\,  \beta  \geq 1 \,. \label{eq:Ffreeze}
\end{equation}
From now on we restrict to $\beta \geq 0$ by default. Equations \eqref{eq:FhighT} and \eqref{eq:Ffreeze} imply a second--order transition at $\beta = \beta_c = 1$, called the \textit{freezing} transition. In particular, as $\beta \to \infty$, we recovers the leading term $V_{\min} = -2\ln M + \dots$ of the extreme value statistics, eq. \eqref{eq:REMzeroT}. To recover the correction terms and the Gumbel law of the fluctuation, one may use the \acrlong{1rsb} method, as we will discuss in section \ref{sec:REMRSB} below. Here, we give an elegant replica-free method (provided in \cite{cao16order}, last appendix) to calculate the distribution of $\mathcal{F}$ at any temperature. This allows also introducing some standard formalism that will be used recurrently in this chapter.

\subsubsection*{Free energy distribution}
A central quantity in disordered statistical physics is the (exponential) generating function of partition function
\begin{equation}
G_\beta(x) = \overline{\exp\left(- e^{\beta x} \mathcal{Z} \right)} \,. \label{eq:defGgen}
\end{equation}
It can be interpreted in several ways: 
\begin{enumerate}
\item As the exponential generating series
\begin{equation}
G_\beta(x) = \sum_{n=0}^{\infty}  \frac{(-\mu )^n}{n!} \overline{\mathcal{Z}^n}  \,,\, \mu = e^{\beta x}  \label{eq:Gasseries}
\end{equation}
of the \textit{replicated partition sums} $\overline{\mathcal{Z}^n}$, $n=0, 1, 2, \dots$: they are the object that is calculated in the replica approach. However, the series, as a function of $\mu$, has usually zero convergence radius around $\mu = 0$, so eq. \eqref{eq:Gasseries} is only useful for formal manipulations, for example, the differentiation
\begin{equation}
\partial_x G_\beta(x) =  \sum_{n=0}^{\infty}  \frac{(-\mu )^n}{n!} n \beta  \overline{\mathcal{Z}^n}  \,.
\end{equation}
\item As a cumulative distribution function. This is the most important interpretation, and is in fact the proper definition of eq. \eqref{eq:defGgen}. To explain it, let us introduce $g$, a random variable independent of $\mathcal{Z}$ which has the standard Gumbel distribution. That is, its cumulative distribution function and Laplace transform are respectively \footnote{In this thesis, if $P(x)$ is the \gls{pdf} of a random variable, both $\int_x^{\infty} P(x') \dif x'$ and $\int_{-\infty}^x P(x') \dif x'$ can be called the cumulative distribution function (the former occurs more often). When it matters, we will always be precise about which one is considered.}
\begin{equation}\label{eq:Gumbeldef}
\overline{\theta(x - g)} = \exp(-e^{-x}) \Leftrightarrow \overline{e^{t g}} = \Gamma(1 - t) \,,\, \Re(t) < 	1 \,.
\end{equation}
where $\theta$ is the Heaviside step function and $\Gamma$ is Euler's Gamma function. Now, eq. \eqref{eq:defGgen} implies 
\begin{equation}
G_\beta(x) = \overline{\exp\left(- e^{\beta (x - \mathcal{F} )}\right)} 
= \overline{\theta(- \beta (x - \mathcal{F}) - g )} = \overline{\theta(\left[\mathcal{F} - g/\beta\right]- y)} \,, \label{eq:Gasconvolution}
\end{equation}
\textit{i.e.}, the probability that the sum of $\mathcal{F}$ and an independent rescaled Gumbel $-g/\beta$ is larger than $y$. Therefore $-\partial_x G_\beta(x)$ is the \gls{pdf} of the convolution $(\mathcal{F} - g/\beta)$:
\begin{equation}
\overline{\delta(\mathcal{F} - g / \beta - x)} = -\partial_x G_\beta(x) \,. \label{eq:partialGipdf}
\end{equation}
 As a consequence, we have the Fourier--Laplace transform relations 
\begin{subequations}
\begin{align}
& \int_\R  (-\partial_x G_\beta(x)) e^{tx}  \dif x = \Gamma(1 + t / \beta) \, \overline{\exp(t\mathcal{F})} \,, \label{eq:Lap}\\
&\int_{r + \im \R} \frac{\dif t}{2 \pi \im } e^{- t x} \, \beta^{-1} \Gamma(t / \beta) \, \overline{\exp( t \mathcal{F})} = G_{\beta}(x) \,, \label{eq:inverseLap}
\end{align}
\end{subequations}
where $r$ can be chosen such that the vertical contour $r + \im \R$ is at the right of all the poles of the integrand. Equations \eqref{eq:Gasconvolution} through eq. \eqref{eq:inverseLap} are fundamental to follow and understand various technical aspects of this Chapter. For the engaged Reader, the best way to get familiar with them is to follow the \gls{rem} case, especially, working through the details from eq. \eqref{eq:GbetaREMhighT} to eq. \eqref{eq:REMFhighT} and from eq. \eqref{eq:GbetaREMfreeze} to eq. \eqref{eq:etFREMlowT}. 

Remark that, applying the residue theorem to eq. \eqref{eq:inverseLap}, and assuming that $\overline{\exp( t \mathcal{F})}$ has no poles, we would obtain a formal series coming from the poles of $\Gamma(t/\beta)$ at $t/\beta = -n, n = 0, 1, 2, \dots$:
$$ G_{\beta}(x) = \sum_{n=0}^{\infty} \frac{(-1)^n}{n!} e^{\beta n y}  \overline{\mathcal{Z}^n} + \dots \,,$$
comparable to eq. \eqref{eq:Gasseries}. Therefore, the integral eq. \eqref{eq:inverseLap} can be used as a non--rigorous re--summation of the series eq. \eqref{eq:Gasseries}, \textit{if} we can continue $\overline{\mathcal{Z}^n}$ to \textit{complex} $n$. Such manipulations are behind the \textit{replica approach}, as we will see in section \ref{sec:RSB}.
\item Finally, it is important to note that
\begin{align}
&G_\beta(y) = \overline{\prod_{j=1}^M \theta_\beta(V_j - x)} \,,\, \text{where} \label{eq:Gasproduct}\\
&\theta_\beta(x) = \exp\left(-e^{-\beta x}\right) \stackrel{\beta\to \infty}{\longrightarrow} \theta(x) \,,
\end{align}
\textit{i.e.}, $\theta_\beta$ is a finite temperature smearing of the Heaviside step function $\theta$. As a consequence:
\begin{subequations}\label{eq:Ginfty}
\begin{equation}
G_\beta(x)  \stackrel{\beta\to \infty}{\longrightarrow} \overline{\prod_{j=1}^M \theta(V_j - x)} = \overline{\theta(V_{\min} - x)} \,, 
\end{equation}
\textit{i.e.}, $G_{\infty}(x)$ is one minus the cumulative distribution function of $V_{\min}$. The \gls{pdf} of $V_{\min}$ is then obtained by derivation:
\begin{equation}
-\partial_y G_\infty (x) = \overline{\delta(V_{\min} - x)} \,.
\end{equation}
\end{subequations}
\end{enumerate}
The above discussion is completely general, and applies to any disordered statistical physics model.

For the REM (see \cite{cao16order}, last appendix), for which $V_j$'s are independent, eq. \eqref{eq:Gasproduct} simplifies to 
\begin{align}
G_\beta(x) &= \left[ \gamma_\beta(x) \right]^{M} \stackrel{M\to\infty}{\longrightarrow}
e^{\hat\gamma_\beta(y)} \,,\, \hat{\gamma}_\beta = \lim_{M\to \infty} \left(M (\gamma_\beta -1)\right)\,, \label{eq:GbetaREM} \\ 
\gamma_\beta(x) &= \overline{\theta_\beta(V_j - x)} = 
\int_\R \frac{\dif v e^{-\frac{v^2}{4\ln M}}}{\sqrt{4\pi \ln M}} \exp(-e^{\beta(x-v)}) \,. \label{eq:gammaREM}
\end{align}
The last quantity can be calculated by a Hubbard-Stratonovich transform: \textit{i.e.}, we insert into the above equation the identity 
\begin{equation}
\frac{e^{-\frac{v^2}{4\ln M}}}{\sqrt{4\pi \ln M}} = \int_{\R-\im \epsilon} \frac{\dif p}{2\pi} e^{- p v  + p^2 \ln M}\,, \,  \epsilon  >  0 \,, \label{eq:HStransform}
\end{equation}
and then integrate over $v$ in terms of the Gamma function, 
\begin{align}
\gamma_\beta(x) &=  \int_{\epsilon + \im \R} \frac{\dif p}{2\pi \im } 
e^{p^2 \ln M - x p} \int_\R \dif v e^{- p (v - x)} \exp\left(- e^{\beta(x-v)} \right) \label{eq:REMGammaint}
 \\& =
  \int_{\epsilon + \im \R} \frac{\dif p}{2\pi \im } e^{p^2 \ln M - x p} \beta^{-1} \Gamma(p / \beta) 
 \\ &= 1 +  \int_{-\epsilon + \im \R} \frac{\dif p}{2\pi \im } e^{p^2 \ln M - x p} \beta^{-1} \Gamma(p / \beta) \label{eq:REMgamma1}
\end{align}
In eq. \eqref{eq:HStransform}, the contour is at the right of the imaginary axis so that the $v$-integral in eq. \eqref{eq:REMGammaint} converges at $v \to + \infty$. In eq. \eqref{eq:REMgamma1} we move the contour across the imaginary axis, picking up the residue of the Gamma pole at $p = 0$, which gives $1$ by the Cauchy's formula. 

The remaining integral shall be analysed by the saddle point/steepest descent method, the saddle point of \eqref{eq:REMgamma1} being at $p_* =  x / (2 \ln M)$. For this, \eqref{eq:GbetaREM} implies that one should consider the regime of $y$ where $\hat\gamma_\beta \sim O(1)$, or $\gamma_\beta \sim 1/M$. Expectedly, such regimes are in agreement with eq. \eqref{eq:FhighT} and \eqref{eq:Ffreeze}. So the analysis is different in the two phases (see Figure \ref{fig:REM} for an illustration): 

\begin{figure}
\center
\includegraphics[scale=1]{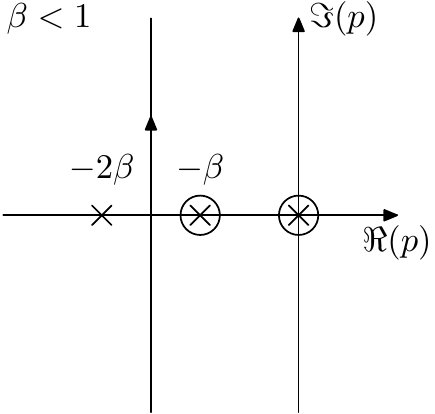}
\includegraphics[scale=1]{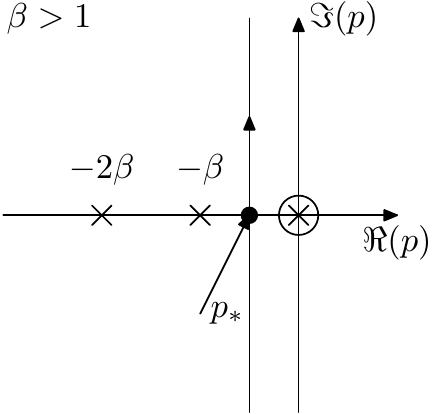}
\caption{Illustrations of the deformed contour integral. The Gamma poles at indicated with a cross. The residue of the pole at $0$ gives the term $1$ in eq. \eqref{eq:REMgamma1}. \textit{Left}: In the $\beta < 1$ phase, the dominant contribution comes from the pole at $-\beta$.  \textit{Right}: In the $\beta > 1$ phase, the dominant contribution comes from the saddle point at $p_* > -\beta$.}\label{fig:REM}
\end{figure}
\begin{enumerate}
\item $\beta < 1$. Eq. \eqref{eq:FhighT} implies the relevant regime should be $x = -(\beta + \beta^{-1}) \ln M + y$, $y \sim o(\ln M)$. But then $p_* / \beta < - 1$, so to deform the contour to cross the saddle point, one has to pick up residues of some Gamma poles. It is not hard to check that the pole at $p = - \beta$ has dominant contribution: 
\begin{align}
& \hat\gamma_\beta (x) = - e^{\beta y} \stackrel{\eqref{eq:GbetaREM}}\Rightarrow G_\beta(y) \stackrel{M\to \infty}{=}  \exp\left(- e^{\beta y}\right) \,,\, 
x = y - (\beta + \beta^{-1}) \ln M \,. \label{eq:GbetaREMhighT}
\end{align}
By  eq. \eqref{eq:partialGipdf}, this means that 
$$ \overline{\theta(\mathcal{F} - g / \beta - x)} = \exp(-\exp(\beta (x + (\beta + \beta^{-1})\ln M))) \,,$$
\textit{i.e.}, the \gls{pdf} of $\mathcal{F} - g / \beta$ is equal to that of $-g/\beta - (\beta + \beta^{-1})\ln M$ (where we recall that $g$ is a standard Gumbel random variable independent of $\mathcal{F}$). As a consequence, the free energy becomes deterministic in the thermodynamic limit:
\begin{equation} \mathcal{F} \equiv -(\beta + \beta^{-1}) \ln M  \,.
\label{eq:REMFhighT}\end{equation} 
\item $\beta > 1$. Eq. \eqref{eq:Ffreeze} points to the regime $x = - 2 \ln M + o(\ln M)$, so $ p_* / \beta > -1$, thus no pole-crossing is needed to deform the contour to the saddle point. Yet, $x = -2 \ln M$ would give $\gamma_\beta = 1 + M^{-1} ( 4 \pi \ln M )^{-\frac12}$, with an extra correction from the Jacobian; so the correct regime of $x$ should be $\sim -2 \ln M + \frac12 \ln\ln M$. More precisely, one can check the following:
\begin{align}
G_\beta(x) \stackrel{M\to \infty}{\longrightarrow}  \exp\left(- e^{y}\right) \,,\, 
x =  y -  2 \ln M + \frac12 \ln \ln M  +  c_\beta \,, \label{eq:GbetaREMfreeze}
\end{align}
where $c_\beta = \frac12 \ln (4\pi) - \ln \Gamma (1-\beta^{-1})$. It diverges at $\beta\to 1_+$, ensuring the matching with the $\beta < 1$ phase. At the zero--temperature $\beta \to \infty$ limit, we retrieve eq. \eqref{eq:REMzeroT} in Chapter \ref{ch:intro}, by recalling eq. \eqref{eq:Ginfty}. For the sake of later comparison, we note that eq. \eqref{eq:GbetaREMfreeze} has an exponential tail at $y\to-\infty$: 
\begin{equation}
1 - G_\beta(x) \sim  e^{y} + O(e^{2y}) \,,\, y \to -\infty \,. \label{eq:REMtail}
\end{equation}

In terms of eq. \eqref{eq:partialGipdf}, the equation \eqref{eq:GbetaREMfreeze} means that the \gls{pdf} of the convolution $(\mathcal{F} - g/\beta)$ is \textit{temperature independent} in the $\beta > 1$ phase, up to a translation. This is a remarkable fact that further justifies calling the $\beta=1$ transition \textit{freezing}: not only the extensive free energy freezes, but also its fluctuation, \textit{after convolution with $g/\beta$}. It should be stressed that the free energy distribution does not freeze, because it is obtained from that of $\mathcal{F} - g/\beta$ by undoing the convolution with $-g/\beta$; in terms of moment generating function, by applying eq. \eqref{eq:Lap}, it is not hard to see
\begin{equation} 
\overline{e^{t\mathcal{F}}} \stackrel{M\to \infty}{\longrightarrow} \frac{\Gamma(1 + t)}{\Gamma(1 + t/\beta)} 
M^{-2t} \left(\ln M\right)^{\frac{t}2} e^{tc_\beta} \,. \label{eq:etFREMlowT}
\end{equation} 
\end{enumerate}
A curious mathematical message of the above analysis is the following: in the $\beta < 1$ phase, $\hat{\gamma}_\beta$ is dominated by a residue (\textit{discrete term}), while in the $\beta > 1$ phase, it is dominated by a saddle point integral (\textit{continuous term}), which induces the $\ln \ln M$ correction. We shall observe strikingly similar patterns in the \acrlong{lft}, see section \ref{sec:liouville}. 

\subsubsection*{Large deviation and near-critical scaling}
The remainder of this section, rather technical and detached from the rest of the manuscript, is devoted to a closer look at the \gls{rem}'s \textit{near-critical regime} and \textit{large deviation} associated. We use the temporary notations 
\begin{equation} \beta = 1 - \tau \,,\, \abs{\tau} \ll 1 \,,\, x = -2 N + \delta \,, N = \ln M \,. \end{equation}
We will allow $\delta$ to be potentially large (so it is to be differentiated with $y$), so as to explore the effect of atypical free energy fluctuations. As indicated above, we should focus on the competition between the saddle point at 
\begin{equation} p_* = \frac{\delta}{2N} - 1  \end{equation} 
and the pole at $p = - \beta$ (we shall assume $0 > p_* > - 2 \beta$ to avoid the effect of other $\Gamma$-- poles). We will also concentrate on $\hat\gamma_\beta = \ln G_\beta$, by eq. \eqref{eq:GbetaREM}. There are three cases:
\begin{enumerate}
\item $2 N \tau  > \delta$, $p_* < -\beta$, and we have the contribution of both the saddle point and the pole (the latter is sub-dominant, but will be kept here)
\begin{align}
& \hat{\gamma}_\beta(x) = - e^{\beta \delta + N \tau^2} + 
e^{\delta - \frac{\delta^2}{4N}}\int_{s\in \R} \frac{\dif  s}{2\pi} \beta^{-1} \Gamma(p_* / \beta +  \im s ) e^{-s^2 \beta^2 N}  \\
 & = - e^{\beta \delta + N \tau^2} - e^{\delta - \frac{\delta^2}{4N}} \left( \frac{1}{\sqrt{4 \pi N}} \frac{\Gamma(2 + p_*/\beta)}{(p_* + \beta)} +  O(N^{-\frac32}) \right)  \,. \label{eq:highTREMlargedev}
\end{align}
This means that conditioning a left large deviation of free energy drives the system into a discrete term dominating phase, even if $\beta > 1$. More generally, as $\delta \to -\infty$, one can drive the system into phases where there are more and more Gamma poles. 
\item  $2N\tau < \delta$,  $p_* > -\beta$ so the pole does not contribute at all:
\begin{align}
\hat{\gamma}_\beta(x) &= e^{\delta - \frac{\delta^2}{4 N}}\int_{s\in \R} \frac{\dif  s}{2\pi} \beta^{-1}  \Gamma(p_* / \beta +  \im s ) e^{-s^2 \beta^2 N}   \\
&=   e^{\delta - \frac{\delta^2}{4 N}} \left(\frac{1}{\sqrt{4 \pi N \beta^2}} p_*^{-1} \Gamma(1 + p_*/\beta) + O(N^{-\frac32}) \right) \,. \label{eq:lowTREMlargedev}
\end{align}
This means that right large deviation $\delta > 2 N \tau$ drives the system into the phase where continuous term dominates, even if $\tau > 0$, that is, the typical system is in the high-$T$ phase.
\item   $2 N \tau = \delta$, the saddle point is exactly at the $\Gamma$-pole, giving a novel type of contribution
\begin{align}
\hat{\gamma}_\beta(x) &= e^{\delta - \frac{\delta^2}{4N}}\int_{s\in \R - \im \epsilon} \frac{\dif  s}{2\pi} \Gamma(-1 +  \im s ) e^{-s^2 \beta^2 N}  \label{eq:REMcritical0} \\
& = e^{\delta - \frac{\delta^2}{4N}}\int_{s\in \R - \im \epsilon} \frac{\dif  s}{2\pi} \left[ \frac{\im}{s} + (\gamma_E - 1) + O(s)\right] e^{-s^2 \beta^2 N}  \\
& =  e^{\delta - \frac{\delta^2}{4N}} \left[-\frac{1}{2} + \frac{(\gamma_E - 1)}{\sqrt{4\pi N \beta^2}}  + O(N^{-\frac32})\right] \label{eq:critical} \,.
\end{align}
\end{enumerate}
Observe that equations \eqref{eq:highTREMlargedev}, \eqref{eq:lowTREMlargedev} and \eqref{eq:critical}, written as such, cannot be matched together: this is because they concern the \textit{large deviations} of the \gls{rem} free energy. This is reflected by the fact that $p_* =  \frac{\delta}{2N} - 1$ is kept constant in the above cited equations. The results in the different phases diverge as $p_* \to 1$, and cannot be used to describe the critical regime. Denoting
\begin{align} 
\alpha = 1  + p_* / \beta =  \frac{\delta - 2 N \tau}{2\beta N} =  \frac{\delta}{2N} - \tau + O(\delta\tau N^{-1} , \tau^2, \delta^2 N^{-2}) \,, \end{align}
the critical regime corresponds to $\abs{\alpha} \ll 1$. In this regime, $\hat\gamma_\beta$ is given by the following integral (compare to \eqref{eq:REMcritical0})
\begin{align}
\hat{\gamma}_\beta(x)e^{-\delta + \frac{\delta^2}{4N}} = &\int_{s\in \R - \im \epsilon} \frac{\dif  s}{2\pi}  \Gamma(-1 + \alpha +  \im s ) e^{-s^2 \beta^2 N}  \\
=&  \int_{s\in \R - \im \epsilon} \frac{\dif  s}{2\pi} \left[- \frac{1}{\alpha + \im s} + (\gamma_E - 1) + O(\alpha + \im s) \right]  e^{-s^2 \beta^2 N} \\
 = &
 -f\left(\alpha  \beta  \sqrt{N}\right)  + \frac{\gamma_E-1}{\sqrt{4\pi N \beta^2}} + O(N^{-3/2}) \\
f(y)& = \frac{1}{2}e^{y^2}(\text{sgn}(y) -\text{erf}(y)) = 
\begin{cases}
\frac{1}{2} + O(y^2) & y \to 0_+ \\
-\frac{1}{2} + O(y^2) & y \to 0_- \\
\pm \frac{1}{2\sqrt{\pi}y} + O(y^{-3})  & y \to \pm \infty
\end{cases} 
\end{align} 

Note that the jump at $y = 0$ compensates exactly the appearance/disappearance of the discrete term in \eqref{eq:highTREMlargedev}/\eqref{eq:lowTREMlargedev}. Indeed, near $p_* = \beta$, all the three formulas \eqref{eq:highTREMlargedev}, \eqref{eq:lowTREMlargedev} and \eqref{eq:critical} can be matched in the following scaling description of the \gls{rem}'s near critical free energy distribution:
\begin{align}
&\hat{\gamma}_\beta(x) = - e^{\delta - \frac{\delta^2}{2N}}\left[\tilde{f}(\alpha \beta \sqrt{N})  + \frac{1-\gamma_E}{\sqrt{4\pi N\beta^2}} + O(N^{-3/2}) \right] \,,\, \tilde{f}(y) = \frac{1}{2}e^{y^2}(1 - \text{erf}(y)) \,. \label{eq:scaling}
\end{align}
The divergence of $\tilde{f}(y)$ at $y\to -\infty$ matches exactly the appearance of the discrete term in \eqref{eq:highTREMlargedev}, since $y = \alpha \beta \sqrt{N} \Rightarrow y^2 = \beta \delta + N \tau^2 - (\delta - \delta^2/(4N)) $ is exactly the difference of the exponents. Equation \eqref{eq:scaling} reveals also a scale $\sqrt{t} \alpha = 1 \Leftrightarrow \abs{\delta - 2 N\tau} \sim \sqrt{N}$, so $\delta = 0$ (centre of the distribution) is in the critical regime if $N < \tau^{-2}$, or $M  < \exp(\tau^{-2})$: the finite size effect of the freezing transition is very strong.  

\subsection{Directed polymer on the Cayley tree}\label{sec:DPCT}
The \gls{rem} introduced in the previous section is characterized by the total absence of correlations. They were subsequently incorporated in the \textit{generalized} \glspl{rem} \cite{derrida1985generalization,derrida1986solution}. The generalized \glspl{rem} have played decisive rôles in the mathematical development of spin glass theory; in particular, it led to the Ruelle cascade \cite{ruelle1987mathematical}. The latter describes the structure of low energy states of the mean field spin glass model, and is a pillar of its modern rigorous treatment (see \cite{Talagrandbook}).   

\begin{figure}[h]
\center \includegraphics[scale=.6]{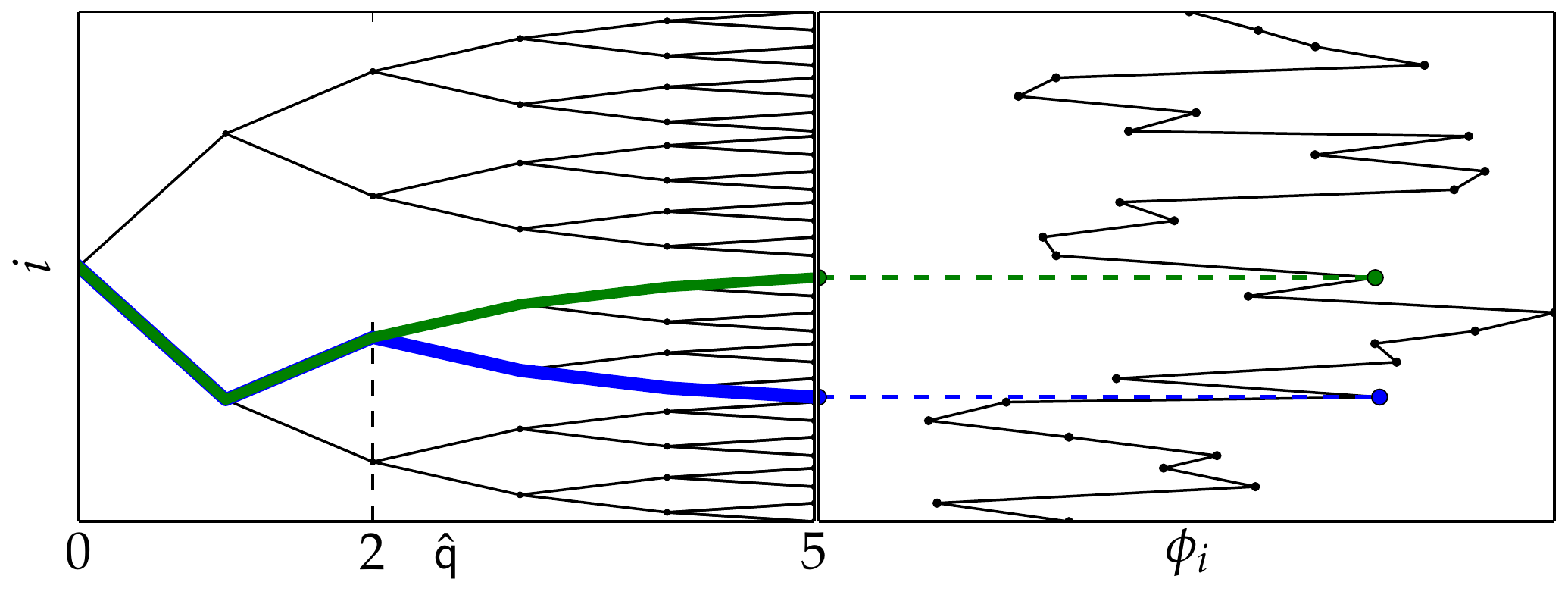}
\caption{A Cayley tree of branching number $\kappa = 2$ and $n = 5$. Two directed polymers are drawn in bold and different colours. They have common length $\hat\overlap = 2$, and overlap $\overlap = .4$. On the right panel, a sample of the energies of the directed polymers is plotted.} \label{fig:tree}
\end{figure}
Another important correlated variant of \gls{rem} is the \acrfull{dpct} model, mentioned in section \ref{sec:introkpz}. It is defined on a Cayley tree (see Figure \ref{fig:tree} for an illustration), described by its branching number $\kappa$ and the number of generations $n \in \N$, so that the total number of leaves is $M = \kappa^n$. To each edge, we associate an independent energy, which is a centred Gaussian variable of variance $2 \ln \kappa$.  Directed polymer (DP) are by definition the simple path from the tree root to one of the leaves, and its energy $V_i, i = 1,\dots,M$ is the sum of those of its edges. Therefore, the energy of each individual DP is a centred Gaussian with variance $n \times 2\ln \kappa = 2 \ln M$, identical to the REM, eq. \eqref{eq:REMdef1}; moreover, the correlation of any two DP's is 
\begin{equation}
\overline{V_i V_j} = 2  {\hat\overlap}_{ij}  \ln \kappa \,, \label{eq:covDPCT}
\end{equation}
where $\hat q_{ij}$ is the common length of the DP's $i$ and $j$. For example, the matrix for $\kappa = 2$ and $n = 2$ is 
\begin{equation}
\left(\hat{\overlap}_{ij}\right) = \begin{pmatrix}
2 & 1 & 0 & 0 \\
1 & 2 & 0 & 0 \\
0 & 0 & 2 & 1 \\
0 & 0 & 1 & 2 \\
\end{pmatrix} \,.
\end{equation}
A closely related quantity is the \textit{overlap}, defined by a simple rescaling:
\begin{equation} 
\overlap_{ij} \defeq \frac{\overline{V_i V_j}}{2 \ln M} = \frac{ \hat \overlap_{ij} }{n} \in [0,1] \,. \label{eq:defoverlap}
\end{equation}
The overlap is an important notion of the spin glass theory: its definition depends on the model, so as to measure the ``similarity'' of two configurations (in the Ising model eq. \eqref{eq:RBIsing}, it is defined as $q_{\sigma, \sigma'} = N^{-1} \sum_{j} \sigma_j \sigma'_j \in [-1,1]$). For \glspl{logrem}, the first equality of eq. \eqref{eq:defoverlap} will be used  in general, while the second equality is specific to the \gls{dpct} model. 

\subsubsection{\gls{bbm} and \gls{kpp} equation}
\begin{figure}
\center
\includegraphics[scale=.5]{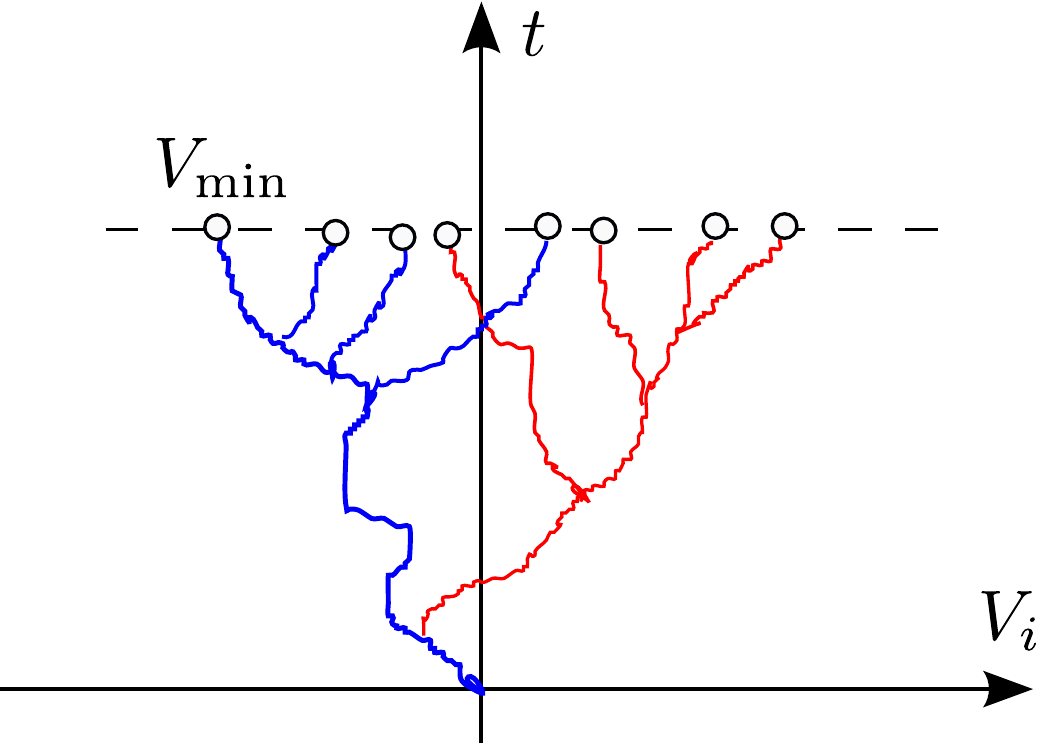}\quad
\includegraphics[scale=.5]{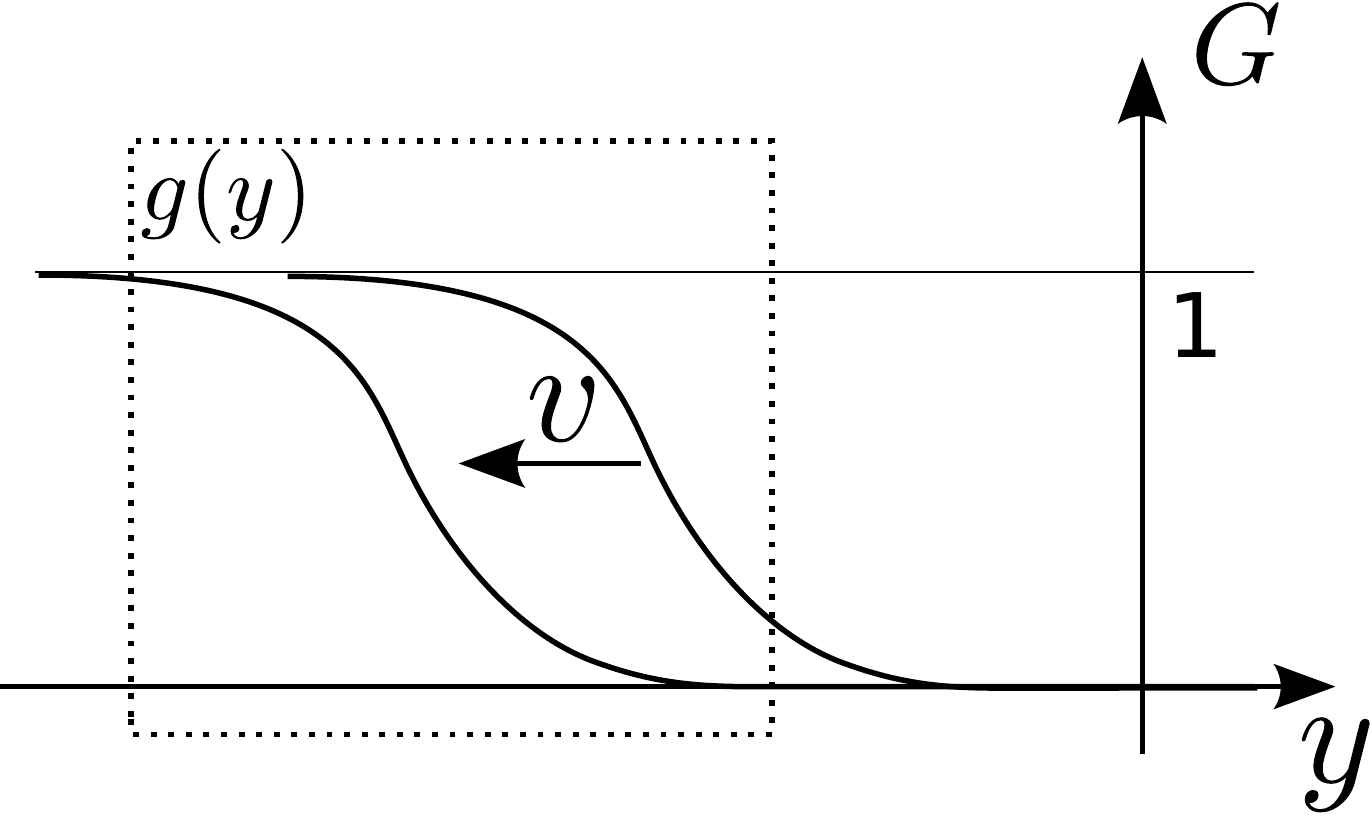}
\caption{\textit{Left}: An illustration of the \gls{bbm} model. The energy levels $V_i$ split at rate $1$ and do independent Brownian motions such that $(\dif V_i)^2 = 2 \dif t$. We indicate by different colours the two families of energy levels separated at the first split. \textit{Right}: Illustration of the \gls{kpp} equation. The solution is a travelling wave of velocity $v<0$ and of a profile $g(y)$ given by eq. \eqref{eq:ODEwaveKPP}. }\label{fig:BBMKPP}
\end{figure}
The \gls{dpct} is considerably more involved to study analytically than the \gls{rem}. A natural idea  \cite{derrida1988polymers}, which is particularly efficient for problems defined on trees, is to consider the evolution of the directed polymer's energy levels $V_j$ as one adds a generation $n \to n + 1$. In one ``time--step'', every level $V_j$ splits into $\kappa$ degenerate ones, and then each new level is incremented by an independent centred Gaussian of variance $2 \ln \kappa$. It is not hard to see that the resulting energies are a sample of the \gls{dpct} with $(n+1)$ generations. This description makes clear the affinity of the \gls{dpct} to the \gls{bbm} model, which we defined now. In this continuous space-time model, each energy level does an independent 1d Brownian motion with $\dif V_i \dif V_j = 2 \dif t \delta_{ij}$, and branches into two identical copies with rate $\dif t$. At $t = 0$, there is only one energy level at $0$; therefore, at time $t$, there are in average 
\begin{equation} M = \overline{M(t)} = e^t  \label{eq:MofBBM}
\end{equation}
energy levels. So for \gls{bbm}, the thermodynamics quantities should be proportional to $\ln M = t$. At time $t$, each energy level is a centred Gaussian of variance 
\begin{equation} \overline{V_j^2} =  2 t \sim 2 \ln M \,, \label{eq:varianceofBBM}
 \end{equation}
and $\overline{V_{i} V_j}$ is $2$ times the branching moment of their common ancestor.  Therefore, defining $\mathcal{Z} = \sum_{j=1}^{M(t)} e^{-\beta V_j}$ for the \gls{bbm}, it is reasonable to expect that its statistical physics is close to that of \gls{dpct}, defined by $\mathcal{Z} = \sum_{j=1}^{M = \kappa^n} e^{-\beta V_j}$.

The key finding of \cite{derrida1988polymers} is that for \gls{bbm}, the function $G(x,t) = G_\beta(x) = \overline{\prod_{i=1}^{M(t)} \theta_\beta(V_j - x)}$ (eq. \eqref{eq:Gasproduct}, depending now on $t$) satisfies the Fisher-\gls{kpp} equation \cite{kolmo91kpp} equation:  
\begin{subequations}\label{eq:KPPs}
\begin{eqnarray}
&\partial_t G = \partial_{x}^2 G + G^2 - G \,,\,  \label{eq:KPP}\\
& G(x,0) = \theta_\beta(-x) = \exp(-e^{\beta x}) \,. \label{eq:ICKPP}
\end{eqnarray}
\end{subequations}
We outline the reasoning leading to this. To calculate $\dif G = G(x, t + \dif t) - G$, we enumerate what can happen during $(0, \dif t)$ to the only energy level:
\begin{itemize}
\item[(BM)] it moves to $\pm \sqrt{\dif t}$ with probability $\dif t$ (for each sign), in which case $\dif G = \mp \sqrt{\dif t} \partial_x G + \frac{1}{2} \dif t \partial_{xx} G$ (by It\^o's calculus);
\item[(B)] it splits into two copies with probability $\dif t$, in which case, by independence of the two sub--trees, $\dif G({t + \dif t},y) = (G^2 - G)\dif t$.
\end{itemize} 
Summing all the contributions gives eq. \eqref{eq:KPP}. In fact, the above reasoning shows the well known fact that for \textit{any} function $\phi(y)$, the observable of the \gls{bbm}
\begin{align}
G_\phi(y) = \overline{\prod_{j=1}^{M(t)} \phi(V_{j}(t) - y)}  \label{eq:gphi}
\end{align}
satisfies eq. \eqref{eq:KPP}. Only the initial condition depends on the particular form of $\phi = \theta_\beta$, which is the only place where $\beta$ enters in into the \gls{kpp} equation. Finally, remark that a similar recursion reasoning can be applied to \gls{dpct}. The result would be a discrete version of eq. \eqref{eq:KPPs}, to which the discussion below still applies \textit{mutatis mutandis}, yet much less elegantly.

\subsubsection{\gls{rem} \textit{vs.} \gls{bbm}}
The Fisher--\gls{kpp} equation is one of the best understood non-linear partial differential equations. It is one of the first examples of a \textit{reaction--diffusion} model. More concretely, we will interpret \gls{kpp} as a population dynamics model, in which $\rho = 1 - G$ is the local population density. The dynamics is a combination of \texttt{(i)} a uniform spatial diffusion of individuals and \texttt{(ii)}  a local Logistic--equation dynamics $\dot{\rho} = \rho(1 - \rho)$ describing an exponential reproduction which saturates at the stable fixed point $\rho = 1$ (the other fixed point $\rho = 0$ is unstable). In the initial condition eq. \eqref{eq:ICKPP}, the right half-line is populated. It is known that the colony will invade the left-half in a \textit{travelling wave} fashion, \textit{i.e.}, the solution in the long--time limit is of the form
\begin{equation}
G_\beta (x) = G(x,t) = g(x - a(t)) \,,\, a(t) = vt + o(t) \,, \label{eq:travallingwave}
\end{equation}
where $a(t)$ gives the leading, $t$-depending, behaviour of the free energy distribution, $v < 0$ is the velocity of the travelling wave, and the wave profile $g(x)$ satisfies the ordinary differential equation, which is obtained by plugging the above eq. \eqref{eq:travallingwave} into eq. \eqref{eq:KPP}
\begin{equation}
v g' + g'' + g^2 - g = 0 \,, g(-\infty) = 1 \,,\, g(+\infty) = 0 \,. \label{eq:ODEwaveKPP}
\end{equation} 
It is not hard to show that the above equation with the limit conditions determines $g$ uniquely up to a global translation (a way to proceed is to interpret eq. \eqref{eq:ODEwaveKPP} as Newtonian dynamics). A crucial and non--trivial part of the \gls{kpp} theory is the \textit{velocity selection}, \textit{i.e.}, determining $v$ by the initial condition. In the \gls{bbm} context, by eq. \eqref{eq:ICKPP} and eq. \eqref{eq:Gasconvolution}, this amount to determining the free energy density $v = \lim_{t \to\infty}\mathcal{F} / t = \lim_{M\to\infty} \mathcal{F} / \ln M$ as a function of temperature $\beta$. The result is however surprisingly simple: it is identical to the \gls{rem}, eq. \eqref{eq:FhighT} and \eqref{eq:Ffreeze}:
\begin{equation}
 \frac{\mathcal{F}}{\ln M}  \to v =\begin{cases}
-(\beta + \beta^{-1}) \,,\,&  \beta < 1 \,,\\ -2 \,,\, & \beta \geq 1 \,.
\end{cases} \label{eq:FleadingBBM}
\end{equation}
From a \gls{kpp}--travelling wave point of view, behind the two cases above are two distinct velocity selection mechanisms. In fact, $v$ is selected by the left tail of the initial condition alone. Now, recall from eq. \eqref{eq:ICKPP} that, $1 - G(x, t=0) = \rho(x,t = 0) = 1 - \exp(-e^{\beta x})=e^{\beta x} + O(e^{2\beta x})$ as $x  \to -\infty$. The dynamics in the regime where $\rho(x,t) \ll 1$ can be approximated by the linearization of \gls{kpp} eq. \eqref{eq:KPP}: $\rho_t = \rho_{xx} + \rho + O(\rho^2)$, which can be solved by Fourier transform (just as the text-book solution of the diffusion equation). The result is
\begin{align}
1 - G(x,t) = \rho(x,t) \approx& - e^{t}\int_{ \im \R-\epsilon } \frac{\dif p}{2 \pi \im} e^{- p x + p^2 t} \beta^{-1} \Gamma(p/\beta) + O(\rho^2)  \,, \label{eq:rhoxtpertube}
\end{align}
which is strikingly similar to the \gls{rem} analysis, \textit{e.g.}, eq. \eqref{eq:REMgamma1}. Therefore, eq. \eqref{eq:FleadingBBM} can be understood by repeating the discrete (residue) \textit{vs.} continuous (saddle point) argument used for the \gls{rem}. This is not only a coincidence: a \gls{kpp} treatment of the \gls{rem} has been done in \cite{carpentier2001glass}, section III.D.2, and the resulting equation is \textit{exactly} eq. \eqref{eq:rhoxtpertube}, with no further non--linearity.  

Going back to statistical physics, eq. \eqref{eq:FleadingBBM} implies that the \gls{bbm} and the \gls{rem} share the same thermodynamics (leading behaviour of free energy); in particular, both have a freezing transition at $\beta = 1$. Moreover, eq. \eqref{eq:FleadingBBM}, \eqref{eq:ODEwaveKPP} and \eqref{eq:travallingwave} implies that the $t \to \infty$ limit profile of $G_\beta(x)$ freezes (is $\beta$-independent) in the whole $\beta > 1$ phase: recall that \gls{rem} has also this feature, see eq. \eqref{eq:GbetaREMfreeze}. 

Then, what are the differences between \gls{bbm} and \gls{rem}? Concerning the free energy distribution, there are two essential ones: the \textit{left tail of the limit shape} and the \textit{log--correction} to the extensive behaviour. As we shall see in section \ref{sec:treetoplane}, they are important \textit{universal} signatures of the \gls{logrem} class. So a more detailed review is in order.

\subsubsection{$\abs{y} e^y$ left tail of the limit shape}
For the \gls{bbm}, the limit profile of $G_\beta(x)$, up to a translation, is given by \eqref{eq:ODEwaveKPP} (with $v$ given by \eqref{eq:FleadingBBM}), and thus different from \gls{rem}'s (minus) Gumbel profile, given by eq. \eqref{eq:GbetaREMhighT} and \eqref{eq:GbetaREMfreeze}. In particular, when $\beta < 1$, the \gls{bbm}'s free energy has a $O(1)$ non--trivial fluctuation, in contrast to \gls{rem}, eq. \eqref{eq:GbetaREMhighT}. When $\beta > 1$, the left tail of $G_\beta$ is 
\begin{equation}
1 - G_\beta(x) = 1 -  g(y) \sim A \abs{y} e^{y} + \dots \,,\, x = y + a(t) \,,\, \beta > 1 \,. \label{eq:BBMtail}
\end{equation}
which is different from the \gls{rem} analogue $e^{y}$, see eq. \eqref{eq:REMtail}. To understand this, we can look at the ODE \eqref{eq:ODEwaveKPP} around $g(y)\sim 1, y \to -\infty$. The linearised equation is $v (1-g)' + (1-g)'' + (1-g) = 0$, whose general solution is $A y e^{y} + B e^y$ when $v = -2$ (which is the case for all $\beta \geq 1$). Since the differential equation eq. \eqref{eq:ODEwaveKPP} has a non--linear part, it is reasonable that the solution is always perturbed out of the 1d manifold $\set{A = 0}$ and must have $A \neq 0$. This gives us eq. \eqref{eq:BBMtail}.

\subsubsection{Log--correction}
The \textit{sub--leading correction} to eq. \eqref{eq:travallingwave} is also different from the \gls{rem}. Indeed, a classic result in \gls{kpp} theory is:
\begin{equation}
G_\beta(y + a(t)) \stackrel{t\to\infty}\longrightarrow g(y) \,,\,  a(t)  = \begin{cases} (-\beta - \beta^{-1})t + c_\beta \,,\, & \beta < 1 \,, \\  -2 t + \frac12 \ln t + c_\beta  \,,\,& \beta = 1 \,,\, \\ -2t + \frac32 \ln t +  c_\beta \,,\, & \beta > 1 \,, \end{cases} \label{eq:GbetaBBM}
\end{equation}
where $c_\beta$ denotes some $t$--independent function of $\beta$, which is different from the \gls{rem} one. Eq. \eqref{eq:GbetaBBM} holds not only for \gls{bbm}, but also for \gls{dpct}, upon replacing $t = \ln M$ (the $O(1)$ correction $c_\beta$ varies also and depends on the Cayley tree's branching number). 
The reason behind the $\frac32$-correction in \gls{bbm}/\gls{dpct} is non--trivial. Indeed, it is the subject of a very recent study \cite{schmidt2015}, which constructed a class of models that interpolates the \gls{rem} ($\frac12$ correction) and the \gls{bbm}/\gls{dpct} ($\frac32$ correction, respectively). While the original and classic treatment is Bramson's memoirs \cite{Bramson1983memoirs}, there is a short and elegant explanation found recently by \cite{hamel2013short}. The key point is that the solution to the linearised \gls{kpp}, eq. \eqref{eq:rhoxtpertube}, is qualitatively wrong when for $x > a(t)$. For the true solution, $\rho = 1 - G$ becomes a constant in that regime (see Figure \ref{fig:BBMKPP}). To reproduce this qualitatively while keeping the approximating equation linear, one considers the same diffusion equation, but with \textit{Dirichlet boundary condition} along the line $\rho(-2t + C,t) = 0$ (with $C > 0$ large but fixed). Such an equation can be solved by combining a shift of the reference frame and a mirror image trick; one can check that the following is a general solution:
\begin{align}
&\rho(-2t + C + y)  = \int_{\im \R - 1} \frac{\dif p}{2 \pi \im} e^{(p+1)^2 t - y p} f(p) \,,\, f(-1+p) = -f(-1 - p) \nonumber  \\
 \approx& \int_{\im \R - 1} \frac{\dif p}{2 \pi \im} e^{(p+1)^2 t - y p} a_1 \times(p+1)  = 
\frac{a_1 y e^{y-\frac{y^2}{4t}}}{4 \sqrt{\pi} t^{\frac{3}{2}}} \to  
c \abs{y} e^{ y - \frac32 \ln t  } \,, \label{eq:threehalf}
\end{align}
as $t \to \infty$. In the first equation of the second line, we expanded $f(p)$ around $p = -1$, $f(p) = a_1 (p+1) + O((p+1)^3)$, to get the leading term of the saddle point approximation. By doing this we get both the $\frac32 \ln t$ shift in the $\beta > 1$ phase and the $\abs{y}e^{y}$ tail. At the critical temperature $\beta= 1$, the initial condition has an exponential tail $e^{y}$ corresponding to a pole $f(p\to-1) \sim (1 + p)^{-1}$: this pole cancels the $(p+1)$ factor in eq. \eqref{eq:threehalf}, and we get the $\frac12 \ln t$ correction.

As a recapitulation of the above discussions, we deduce that at the zero temperature $\beta \to \infty$ limit, the extreme value statistics problem associated to \gls{bbm}/\gls{dpct} fits into the Ansatz eq. \eqref{eq:evsansatz} as follows:
\begin{equation}
V_{\min} = -2\ln M + \frac32 \ln \ln M + c_\beta + y \,,\, P(y) \sim A \abs{y} e^{y} \,,\, y \to -\infty \,. \label{eq:evsBBM}
\end{equation}
The two distinguishing features discussed above are both manifest.

To conclude, we note the arguments leading to eq. \eqref{eq:BBMtail} and \eqref{eq:GbetaBBM} apply for a large class of \gls{kpp}-type equations; for example, $G_t = G_{xx} + G^k - G$ ($k \geq 2$), which models the a \gls{bbm}'s variant in which energy levels split into $k$ identical copies.  Therefore, these equations are \textit{universal} features of all the imaginable models similar to the \gls{bbm}/\gls{dpct}.

\subsection{From Cayley tree to 2D Gaussian Free Field}\label{sec:treetoplane}
\red{The idea of relating \gls{bbm}/\gls{dpct} to a thermal particle in a log--correlated potential appeared in the study of certain disordered systems in 2D, both classical and quantum. An important classical case is the disordered XY model \cite{rubinstein83xy,carpenter98XY,mudry99XY}, while a representative in quantum mechanics is the 2D Dirac fermions in random magnetic field \cite{chamon1996localization,castillo97dirac,kogan96prelocalised,horovitz2002freezing}. In both cases, the log--correlated potential in question is the massless 2D \gls{gff}, which encodes essentially the quenched disorder. In turn, the freezing transition of the \gls{logrem} defined by the 2D \gls{gff}, upon further ramification, has important consequences in these models. We will briefly explain this for the XY model at the end of this section, around eq. \eqref{eq:XY}. The quantum applications are beyond the scope of this thesis, yet they are important subjects of future study, since freezing phenomena are present in a few non--conventional symmetry classes of 2D localisation transitions \cite{gade1993anderson,ludwig94qhe,mudry96cft,motrunich02particle,mudry03DoS}, see \cite{evers2008anderson} for a review.}

Let us come to define the 2D \gls{gff}, the 2D plane will be most often identified to the complex plane, with coordinates $z = x + \im y, z^* = x - \im y,  x, y\in \R$. In statistical field theory, the 2D \gls{gff} $\phi(z) = \phi(x,y)$ defined by the massless quadratic action
\begin{equation}
\mathrm{P}[\phi(z)]  \propto \exp\left(- \int_\C \frac{(\nabla \phi)^2}{4\pi \sigma^2}  \dif^2 z  \right) \,,\, \label{eq:GFFaction}
\end{equation} 
where $\dif^2 z = \dif x \wedge \dif y$ is the area element, and $(\nabla \phi)^2 = (\partial_x \phi)^2 + (\partial_y \phi)^2 = \partial_z \partial_{z^*} \phi$ is the kinetic term of the massless free field action. The resulting covariance (Green function in field theory language) is logarithmic in real space
\begin{equation}
\overline{\phi(z) \phi(w)} = - \sigma^2 \ln \abs{z - w} \,, \label{eq:GFF2p}
\end{equation}
where $\sigma^2$ is the coupling constant ($\sigma^{-2}$ is also known as the stiffness) that will be fixed later. 

Some digression is helpful here to put the 2D \gls{gff} in larger physical contexts. Massless Gaussian Free Fields in general dimensions are the foundation of perturbative quantum/statistical field theory and a trivial fixed point of the renormalization group. Its Green function obeys power-law $\propto r^{2-d}$ except at $d=2$, where it is log--correlated. On the other hand, log--correlated Gaussian potentials can be defined on any spatial dimension, since they occur most naturally in $d=2$ as 2D \gls{gff}, or in $d = 1$, either as the restriction of 2D \gls{gff} to some curve (a circle or an interval), or as the $1/f$-noise. 

The 2D \gls{gff} is therefore at the intersection of log--correlated potentials and statistical field theory, and plays a central rôle in the critical phenomena in 2D statistical physics, the history of which is too long to review fairly here. With hindsight, it is clear that the log--correlation of the 2D \gls{gff} underlies the long--range order in the Berezinsky-Kosterlitz-Thouless transition \cite{berezinskii1971destruction,berezinskii1972destruction,kosterlitz73kt}, which concerns the super--fluidity in 2D. The Coulomb--gas picture that emerged have found wide applications in 2D critical models \cite{nienhuis822D}, and their continuum description: 2D conformal field theories \cite{belavin19843cft,dotsenko1984conformal,di1987relations}. When mathematicians began to turn these physical predictions into theorems, the 2D \gls{gff} became their favourite tool \cite{Sheffield07}: the notable cases are the Schramm (stochastic)--Loewner evolution \cite{cardy2005sle,duplantier06dual} and the geometry of random surfaces (2D quantum gravity) \cite{duplantier09kpz}. 

Beyond the \textit{static} statistical models, 2D \gls{gff} is also the stationary state of the $(2+1)$-d random interface growth described by the Ewdards-Wilkinson equation \cite{edwards82wilkinson} and the anisotropic \gls{kpz} equation \cite{wolf91anikpz,prahofer1997kpz,borodin2014anikpz,borodin09anikpz}, which found recent application in out--of--equilibrium super--fluidity \cite{altman15ani}.
 
\begin{figure}[h]
\center
\includegraphics[scale=.8]{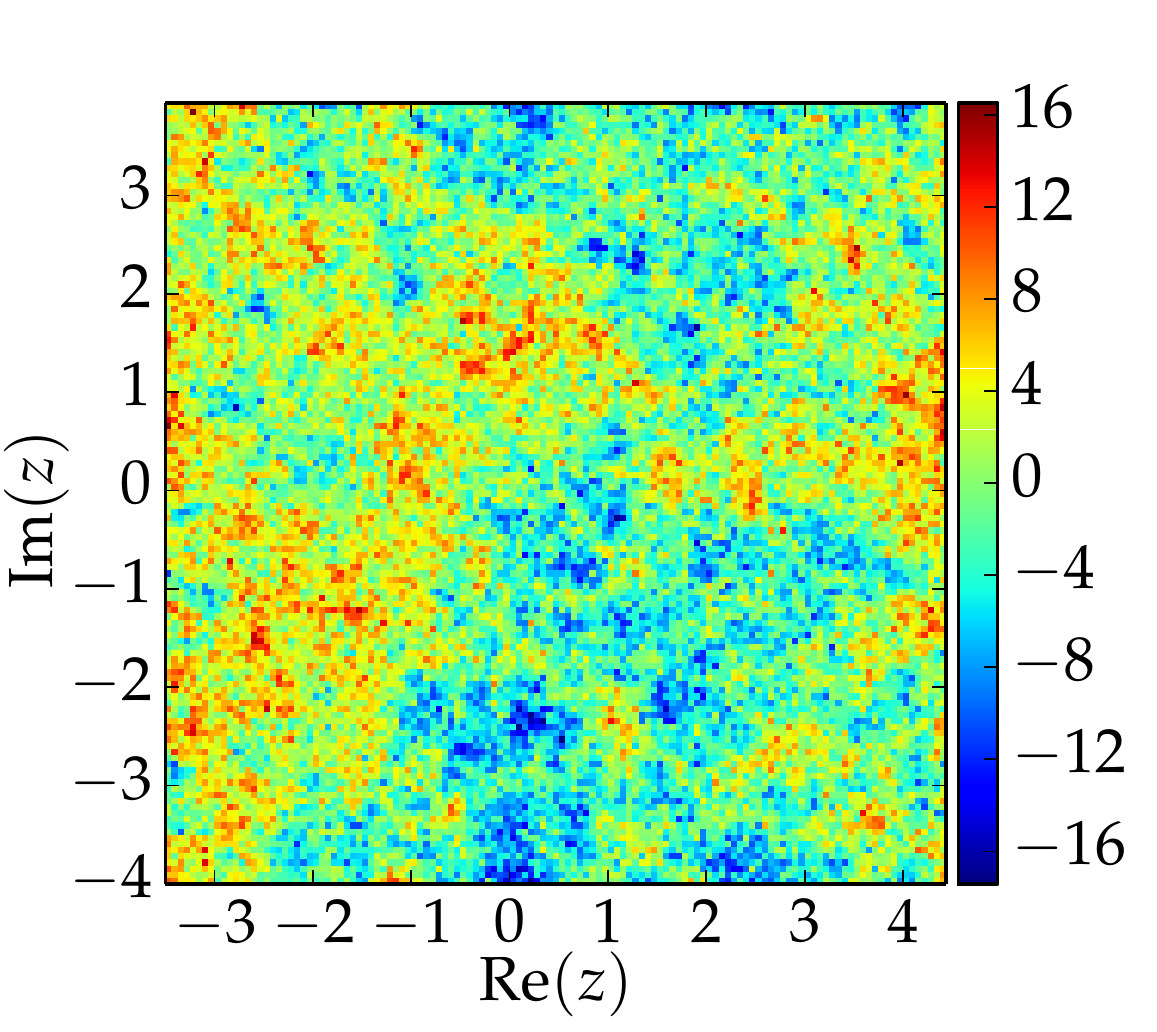}
\caption{A sample of the 2D \gls{gff}, regularized on a lattice on a square torus (periodic boundary condition), with $M = 2^{16}$; the image is compressed so there are less visible pixels.}
\label{fig:2Dgff}
\end{figure}

\subsubsection{2D \glspl{logrem}}
Let us now come back to the problem of a thermal particle in a 2D \gls{gff} potential. One can write down the partition function
\begin{equation}
Z = \int_\C \dif^2 z e^{-\beta \phi(z)} \,,\,
\end{equation}
but it is only formal, mainly because it is well-known that the 2D \gls{gff}, eq. \eqref{eq:GFF2p}, has divergences at both short distances and long distance. Both should be regularized to make the model properly defined. The simplest way to do this is to discretize the 2D \gls{gff} on a toroidal square lattice of size $R\times R$, and of lattice spacing $\epsilon$. Then the field can be generated by a 2D analogue of eq. \eqref{eq:FFTgen}
\begin{align}
& \phi(x + \im y) = \Re \sum_{q=-L/2}^{L/2-1}\sum_{p=-L/2}^{L/2-1} \exp\left[2\pi \mathbf{i} \left( \frac{q x}{R}+\frac{p y}{R}\right)\right] \mathsf{N}_{p,q} \sqrt{\mu_{p,q}} \,,\,  \label{eq:FFT2D}  \\  
& \mu_{p,q} =  \frac{2}{\pi(p^2 + q^2)} \text{ if } (p^2 + q^2 > 0) \,,\,  \mu_{0,0} = 0  \,.
\end{align}
where $\mathsf{N}_{p,q}$ are \gls{iid} independent complex standard normal variables, and $x, y = 0,\epsilon, 2\epsilon, \dots, R - \epsilon$ are the coordinates of the lattice points, and there are $M = (R/\epsilon)^2$ of them. The formula eq. \eqref{eq:FFT2D} is used to simulate numerically 2D \gls{gff} in this thesis. A sample is plotted in Figure \ref{fig:2Dgff}. By eq. \eqref{eq:FFT2D}, the covariance of such a field is
\begin{equation}\label{eq:2DGFFdis}
\overline{\phi(z) \phi(w)} = 4 \ln \abs{\frac{R}{z-w}} \,, \epsilon \ll \abs{z-w} \ll R \,,
\end{equation}
while  $\overline{\phi(z)^2} = 2 \ln M + O(1)$ (a heuristic way to remember this is noting that $\overline{\phi(z)^2} = \overline{\phi(z) \phi(z + \epsilon)} + O(1)$). So it differs from eq. \eqref{eq:GFF2p} (with $\sigma^2 = 4$) by a \textit{diverging} constant $4\ln R$. This is the \gls{ir} divergence of 2D \gls{gff}; its rôle in \glspl{logrem} will be discussed in section \ref{sec:Gaussian}. Now that we have generated a discrete potential, the discrete partition function $\mathcal{Z}$ eq. \eqref{eq:defZandF} can be defined; the thermodynamic limit is achieved by taking $M\to\infty$, \textit{i.e.}, $\epsilon\to0$ \textit{and/or} $R\to \infty$: since there are no other scales in the problem, the only dimensionless parameter is $R/\epsilon$.

It turns out that \textit{the resulting thermodynamics is identical to that of \gls{dpct}/\gls{bbm}}. In particular, the particle goes through also a freezing transition. This correspondence between \gls{dpct}/\gls{bbm} and 2D \gls{gff} was first put forward (partially) in \cite{chamon1996localization} by Chamon, Mudry and Wen; In fact, the \gls{dpct}--2D \gls{gff} relation extends to finer details free energy fluctuations, and can be summarized as follows:  \vspace{.3cm} 

\fbox{\begin{minipage}{.96\textwidth}
Upon fixing the normalization $\sigma^2 = 4$, the free energy $\mathcal{F}$ of a thermal particle in a 2D \gls{gff} has the same leading and sub--leading behaviour as \gls{bbm} and \gls{dpct}: 
\begin{subequations}\label{eq:GbetaGFF}
\begin{equation}
G_\beta(y + F_M) \stackrel{M\to\infty}{\longrightarrow} g(y) \,,\, F_M = \begin{dcases} -\left( \beta + \beta^{-1} \right) \ln M  \,,\, & \beta < 1 \,, \\  -2 \ln M + \frac12 \ln \ln M   \,,\,& \beta = 1 \,,\, \\ -2  \ln M + \frac32 \ln \ln M \,,\, & \beta > 1\,,\end{dcases} \label{eq:GbetaGFFleading}
\end{equation}
where $g(y)$ is some $M$-independent limit shape, which depends on the regularization procedure of the model (so in general different from the \gls{bbm} one), as well as $\beta$. Moreover, identically to \gls{bbm}, in the $\beta > 1$ phase, $g(y)$ becomes $\beta$-independent and maintains its $\beta = 1$ value (freezes):
\begin{equation}
g(y)\vert_{\beta > 1} = g(y)\vert_{\beta = 1} \,. \label{eq:GbetaGFFshape}
\end{equation}
In particular, we have the asymptotic left tail
\begin{equation}
g(y)\vert_{\beta \geq 1} \sim A \abs{y} e^{y} \,,\, y \to -\infty \,. \label{eq:GbetaGFFtail}
\end{equation} 
\end{subequations}
 \end{minipage}}
 \vspace{.3cm}

The above statement and a physical demonstration thereof was the content of Carpentier and Le Doussal's work \cite{carpentier2001glass}. This paper applied a real-space, Kosterlitz--Thouless type renormalization group (RG) analysis to the 2D \gls{gff} problem. The similar RG was designed for the random gauge XY model in \cite{carpenter98XY,carpentier00XYlong} (see below for more history). The outcome is that $G_\beta(y)$ satisfies approximatively the a \gls{kpp} equation, to which the general result \eqref{eq:GbetaBBM} applies. This explains why eq. \eqref{eq:GbetaBBM} is valid almost \textit{verbatim} in the 2D \gls{gff} problem. As a consequence, the extreme value statistics of 2D \gls{gff} is also governed by eq. \eqref{eq:evsBBM}. \red{We mention also that \gls{kpp}--type renormalization group equations were later obtained in related quantum mechanics problems \cite{mudry03DoS}.}

Another important point revealed by the analysis \cite{carpentier2001glass} is that \textit{the dimension is irrelevant} for the problem of log--correlated potentials: eq. \eqref{eq:GbetaGFF} and associated properties (freezing of the profile) hold for a log--correlated potential in any dimension, provided the potential is normalized such that the freezing temperature $\beta_c = 1$ (see section \ref{sec:logREMdef}, eq. \eqref{eq:logdecay}). The question of dimension dependence of \glspl{logrem} is also discussed in \cite{fyosom2007}.

The affinities between log--correlated potentials and \gls{bbm}/\gls{dpct} go beyond free energy distribution, and suggest strongly that all these models belong to a same class, which we call \glspl{logrem}, and which is characterized by universal properties such as eq. \eqref{eq:GbetaGFF}. The latter, in the general context of \glspl{logrem}, is known as the \textit{freezing scenario}. 

\subsubsection{Circular model}
The above results lead to an outstanding question: can we calculate the limit distribution $g(y)$ in \eqref{eq:GbetaGFF}? For the \gls{bbm}, this problem is exactly (if not explicitly enough) solved by the ODE \eqref{eq:ODEwaveKPP}. For the 2D \gls{gff}, regularized on a 2D domain, no exact answer is known!

Nonetheless, since we know that the freezing scenario  (eq. \eqref{eq:GbetaGFF}) holds also for log--correlated dimensions in other dimensions than $2$, we can hope to make progress in 1D. 
As mentioned before, a common way to construct them is to restrict the 2D  \gls{gff} to some 1D curve, such as the unit circle or the interval $[0,1]$. It turns out that, in these two geometries, the limit distribution $g(y)$ can be exactly calculated, as was shown by Fyodorov--Bouchaud \cite{fyodorov2008statistical} and Fyodorov--Le Doussal--Rosso \cite{fyodorov2009statistical}. We now review the former case, called the \textit{circular model}, at a formal level, with the goal of exposing the main idea of the Fyodorov--Bouchaud's solution. 

For this, let us write down the (continuous, formal) partition function of the circular model:
\begin{equation}
Z = \int_0^{2\pi} \frac{\dif \theta}{2 \pi} \exp\left({- \beta \phi(e^{\im \theta})}\right) \,, 
\overline{\phi(z)\phi(w)} = -2 \abs{z-w} \,,\, \overline{\phi(z)} = 0 \,. \label{eq:ZofFB}
\end{equation}
Notice that, since the dimension changes, the normalization of eq. \eqref{eq:GFF2p} is fixed to $\sigma^2 = 2$, so that $\beta_c = 1$. Then, we proceed by the \textit{replica trick}, which starts by calculating integer moments of $Z$:
\begin{align}\label{eq:FBreplica}
\overline{Z^n} &= \int_0^{2\pi} \prod_{a=1}^n \frac{\dif \theta_a}{2 \pi}  
\overline{\exp\left({-\beta \sum_{a=1}^n \phi(z_a)}\right)} \,,\, z_a = e^{\im \theta_a}   \,.
\end{align}
To compute this, we use the Wick theorem, which holds any Gaussian variable $V$ 
\begin{equation}
\overline{\exp(V)} = \exp\left(\overline{V} + \frac12\overline{V^2}^c\right) \,,\, \overline{V^2}^c = \overline{V^2} - \left(\overline{V}\right)^2 \label{eq:WickThm}
\end{equation}
applied to $V = -\beta \sum_{a=1}^n  \phi(z_a)$:
\begin{align*}
\overline{\exp(V)}& = \exp\left( \frac12 \sum_{a=1}^n \beta^2 \overline{\phi(z_a)^2} + 
\beta^2 \sum_{a<a'} \overline{\beta \phi(z_a) \beta \phi(z_{a'})} \right)  \\
& = \prod_{a=1}^n 
 e^{\frac12 \beta^2 \overline{\phi(z_a)^2} } \prod_{1\leq a<b \leq n} \abs{z_a - z_b}^{-2\beta^2} \,.
\end{align*}
In the second equation we applied \eqref{eq:ZofFB}. Plugging into eq. \eqref{eq:FBreplica}, we have
\begin{equation}
 \overline{Z^n}= \int_0^{2\pi} \prod_{a=1}^n \frac{\dif \theta_a}{2 \pi} \prod_{a=1}^n 
 e^{\frac12 \beta^2 \overline{\phi(z_a)^2} } \prod_{1\leq a<b \leq n} \abs{z_a - z_b}^{-2\beta^2} \,,\,  \label{eq:ZnCoulombgas}
 \end{equation}
Eq. \eqref{eq:ZnCoulombgas} is a Coulomb gas integral: the $n$ replicas interact by a power law \textit{attractive} force coming from exponentiating the log correlation. Now, the ``self--interaction'' $\overline{\phi(z)^2}$ is formally infinity by eq. \eqref{eq:ZofFB}: this is a \gls{uv} divergence, which should be regularized. Nevertheless, we do not need to discuss explicitly this issue here, since in \eqref{eq:ZnCoulombgas}, so long as the regularized value of $\overline{\phi(z)^2}$ is independent of $z$, the term involving it in \eqref{eq:ZnCoulombgas} can be absorbed into a re--normalization of $Z$, which is equivalent to a shift in the free energy, immaterial to the calculation of its limit distribution. So, we shall set $\overline{\phi(z)^2} = 0$, \textit{i.e.}, $\phi$ is ``normal ordered'' in field theory term. 

Now, eq. \eqref{eq:ZnCoulombgas}, with $\overline{\phi^2(z)} = 0$, is known as the Dyson integral \cite{dyson1962statistical1} (see \cite{forrester2010log} and \cite{forrester2008importance} for a review on more general exactly solvable Coulomb gas integrals), and its value is exactly known: 
\begin{equation} \label{eq:Dyson}
\overline{Z^n} = \int_0^{2\pi} \prod_{a=1}^n \frac{\dif \theta_a}{2 \pi} \prod_{1\leq a<b \leq n} \abs{z_a - z_b}^{-2\beta^2} = \frac{\Gamma(1-n\beta^2)}{\Gamma(1-\beta^2)^n} \,,\, 
\end{equation}
whenever the integral converges, \textit{i.e.}, when $n \beta^2 < 1$. Therefore, the moments $\overline{Z^n}$ of the \textit{continuous} partition function only exist for $n < \beta^{-1/2}$: for any $\beta < 1$, only a finite number of moments exist; for $\beta > 1$, none of them exists. This leaves too little information to determine the distribution of $Z$. The key \textit{non-rigorous} step of the replica trick is to analytically continue eq. \eqref{eq:Dyson} to generic value of $n \in \C$. By doing this, we can obtain (guess) the moment generating function of the (continuous) free energy $F = -\beta^{-1} \ln Z$ (which differs from $\mathcal{F}$ by a constant shift):
\begin{subequations} \label{eq:FBhighT}
\begin{align} 
&\overline{\exp(tF)} = \overline{Z^{-t / \beta}} = \Gamma(1 + t\beta) \Gamma(1-\beta^2)^{t/\beta} \,. \label{eq:etfFB} \\
\Rightarrow\, & \overline{\exp(t(F - g/\beta))} =  \Gamma(1 + t\beta) \Gamma(1+t/\beta)  \Gamma(1-\beta^2)^{t/\beta} \,. \label{eq:etyFB}
\end{align}
The last equation is useful because by the analogue of eq. \eqref{eq:inverseLap}, $\overline{\exp(t(F - g/\beta))}$ the Laplace transform of the limit distribution $-g'(y)$; 
\begin{equation}
g(y) = \int_{\im \R + \epsilon} \frac{\dif t}{2\pi \im t} \, e^{-ty} \, \overline{\exp(t(F - g/\beta))} \,. \label{eq:FBghighT}
\end{equation}
\end{subequations}

Now, the key point is that equations \eqref{eq:FBhighT} hold only in the phase $\beta < 1$ (and for $\Re(t)>-1/\beta$). The basic reason behind this is that the naïve continuum formalism presented just now is invalid in the $\beta > 1$ phase; we shall understand this better using the \acrlong{1rsb} approach in section \ref{sec:RSB}. Fortunately, for our purpose here, we \textit{do not need} to compute $\overline{\exp(t(F - g/\beta))}$ for $\beta > 1$, because the answer will be given by the \textit{freezing scenario}, eq. \eqref{eq:GbetaGFFshape}, once we know $g(y)$ (or $\overline{\exp(t(F - g/\beta))}$) at $\beta = 1$. However, The point $\beta = 1$ is tricky, as the factor $\Gamma(1-\beta^2)^{t/\beta}$ in eq. \eqref{eq:etfFB} and \eqref{eq:etyFB} diverges. Yet, this factor is of form $e^{c_\beta t}$, so corresponds to a first moment shift of the distribution of $F$ (and $F - g/\beta$), and can be discarded if our only goal is to determine $g(y)$ up to a translation. Doing this, and applying \eqref{eq:GbetaGFFshape} and then inverse Laplace--Fourier transform, we have
\begin{equation}
g(y)\vert_{\beta > 1} =  g(y)\vert_{\beta = 1} = \int_{\im \R + \epsilon} \frac{\dif t}{2\pi \im t}
 \Gamma(1 + t)^2 e^{-ty} = 2 e^{y/2} K_1(2e^{y/2}) \,, \label{eq:FBgfreeze}
\end{equation}
$K_n$ denoting the Bessel $K$-function. In particular, since the pole at $t = -1$ is a double one, we have the left tail $g(y) \sim A \abs{y}e^y, y \to -\infty$, verifying the freezing scenario prediction eq. \eqref{eq:GbetaGFFtail}. Taking derivative of eq. \eqref{eq:FBgfreeze} we have $-g'(y) = 2 e^y K_0\left(2 e^{y/2}\right)$, just as eq. \eqref{eq:FBzeroTpdf} in section \ref{sec:introPEL}. This is not a coincidence, since the \gls{logrem} considered in section \ref{sec:introPEL} (more precisely, eq. \eqref{eq:muk-gen}  with $\alpha = 1$) is one way of defining properly the circular model at the discrete level. The calculation above, combined with eq. \eqref{eq:GbetaGFFleading}, essentially demonstrates eq. \eqref{eq:FBzeroT}, leaving however numerous subtle points, which will be treated in a more systematic manner in this chapter. 

By observing the above derivation, we can already be convinced that the particular limit distribution, eq. \eqref{eq:FBgfreeze}, is specific to the circular model, as it comes from the Dyson integral eq. \eqref{eq:Dyson}. The analogous Coulomb gas integral for the interval model would be the Selberg integral 
$$  \int_0^1 \prod_{a = 1}^n \dif x_a  \prod_{a < b }\abs{x_a - x_b}^{-2\beta^2} \,,$$
whose solution and analytical continuation are different \cite{fyodorov2009statistical}, leading to a distinct limit distribution. While we will not study it in this thesis, we will study how the limit distribution of the circular model is modified if the 2D \gls{gff} is put on a finite disk, in section \ref{sec:Dirichlet}. 

\subsubsection{Freezing--duality conjecture}
The authors of \cite{fyodorov2009statistical} observed that, eq. \eqref{eq:etyFB}, with the factor $\Gamma(1-\beta^2)^{t/\beta} = e^{c_\beta t}$ discarded, is \textit{dual}. It means that $\Gamma(1 + t\beta)\Gamma(1 + t/\beta)$ would be invariant under the formal change of variable $\beta \to  1/\beta$; equivalently, it is a function of the {variable} $\beta + \beta^{-1}$. At the same time, the same quantity freezes in the $\beta > 1$ phase, by Laplace transform of eq. \eqref{eq:GbetaGFFshape}. Furthermore,  \cite{fyodorov2009statistical} showed that the same co-existence of freezing and duality  prevails when the 2D \gls{gff} is restricted on the interval $[0,1]$ instead of a circle. The analysis of the \textit{interval model} requires analytically continuing the Selberg integrals \cite{forrester2008importance}, which are more complex cousins of the Dyson integral eq. \eqref{eq:Dyson}, and we will not do it here. 

Nevertheless, we have already encountered such co-existences. The first, quite trivial, example is the free energy density of the \gls{rem} and the \glspl{logrem}, 
\begin{equation} -  \lim_{M\to\infty}\mathcal{F} / \ln M  \to \begin{cases}
\beta + \beta^{-1} \,,\,& \beta < 1 \,,\,  \\
2 \,,\, & \beta > 1 \,.
\end{cases} = \includegraphics[scale=.2,valign=c]{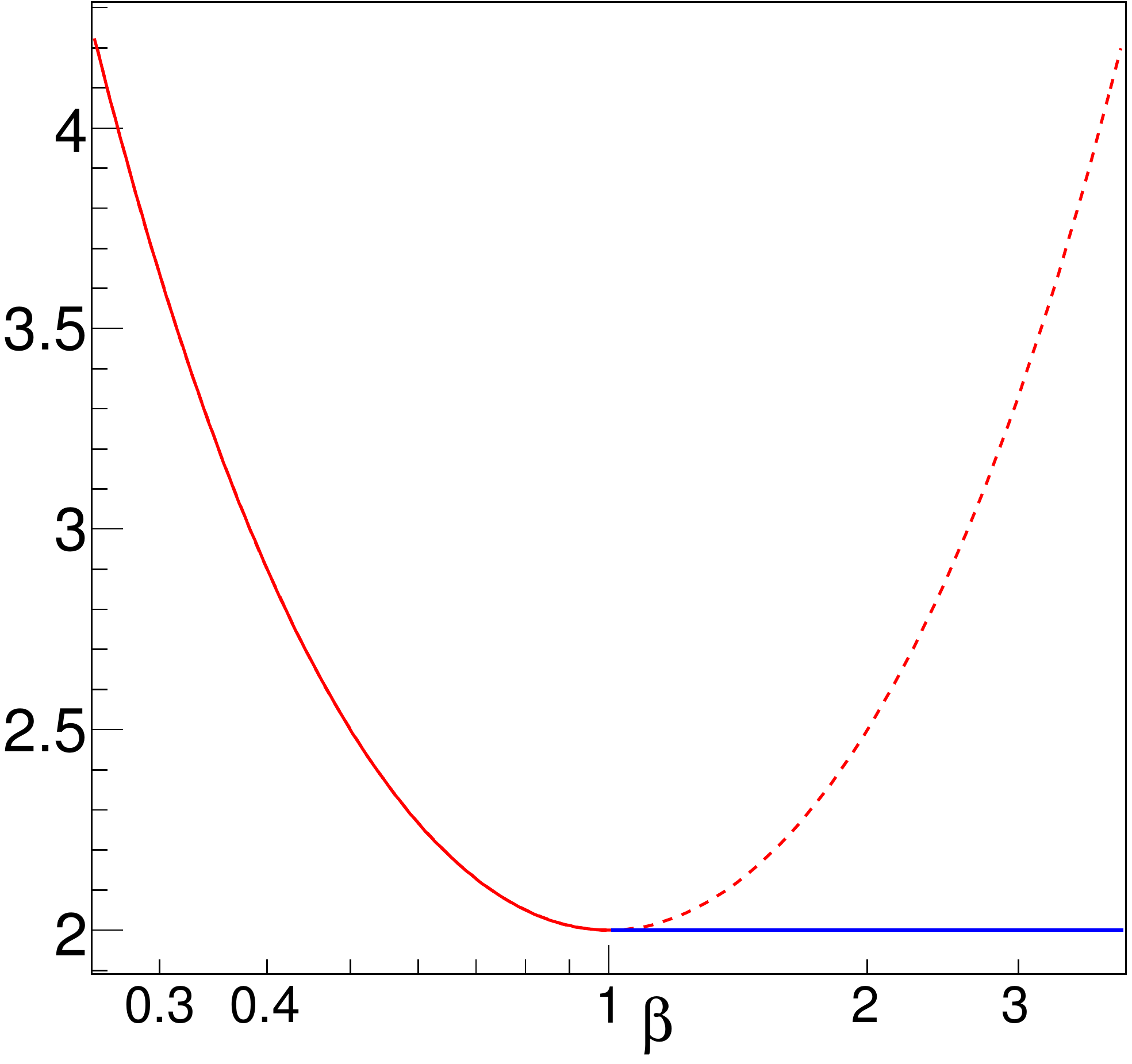} \,, \label{eq:freezedual}
 \end{equation}
It is dual in the $\beta < 1$ phase and freezes in the $\beta > 1$ phase. Moreover, the same thing can be said for the limit shape $g(y)$ in \gls{bbm}, which is determined by the equation \eqref{eq:ODEwaveKPP}: $g'' + v g' + g^2 - g = 0, g(-\infty) = 1, g(+\infty) = 0$. This equation depends only on the selected travelling wave velocity $v = \lim_{M\to\infty}\mathcal{F} / \ln M$. So the duality and freezing of $g(y)$ follow directly from that of $v$, eq. \eqref{eq:freezedual}. On the other hand, notice that the same \textit{cannot} be said about the uncorrelated \gls{rem}: indeed, eq. \eqref{eq:GbetaREMhighT} implies that $g(y) = \exp\left( -e^{\beta y} \right)$, which is not dual. 

Based on the above evidences, it was put forward in \cite{fyodorov2009statistical} the \textit{freezing-duality conjecture} (this name came from \cite{fyodorov2015moments}), which we can succinctly phrase as:  \vspace{.2cm}

\fbox{
\begin{minipage}{.95\textwidth}
\center \textit{In \glspl{logrem}, dual quantities in the $\beta  < 1$ phase freeze in the $\beta > 1$ phase.}
\end{minipage}
} \vspace{.2cm}

To this day, this conjecture is verified in all exact solvable \glspl{logrem}, but the reason behind has not been understood.  We shall provide some rationale for it in section \ref{sec:freezingRSB}, around eq. \eqref{eq:Zbcopy} and eq. \eqref{eq:FDC}, and further special cases where it is checked in section \ref{sec:Jack}. Finally, we note that the duality observed in \glspl{logrem} echoes those in $\beta$--random matrix theory \cite{dumitriua2002matrix} (which has been applied to study \glspl{logrem}, see \cite{fyodorov2010freezing,fyodorov2015moments}) and in 2D conformal field theory \cite{dotsenko1984conformal,duplaniter09dual}. The latter is not merely a superficial reminiscence: as we shall see in section \ref{sec:liouville}, \glspl{logrem} are closely related to the \acrfull{lft}.

\subsubsection{\red{Application: disordered XY model}}
With the hindsight of the freezing transition in 2D \gls{gff}, let us briefly review its application to the disordered XY model. This model was studied by Rubinstein \textit{et. al.} \cite{rubinstein83xy} long ago, who obtained only a partially correct phase diagram. An extensive literature was devoted to its correction \cite{korshunov93xy,cha95crystal2d,tang96xy,korshunov96xy,scheidl97vortex,carpenter98XY,mudry99XY,carpentier00XYlong}, in which the freezing transition in 2D \gls{gff} played a central role. Let us explain this point in the simplest possible approximation (adapted from \cite{carpentier2001glass}, appendix D), and refer to the above original works for full treatments.

\begin{figure}
\center 
\includegraphics[scale=1.]{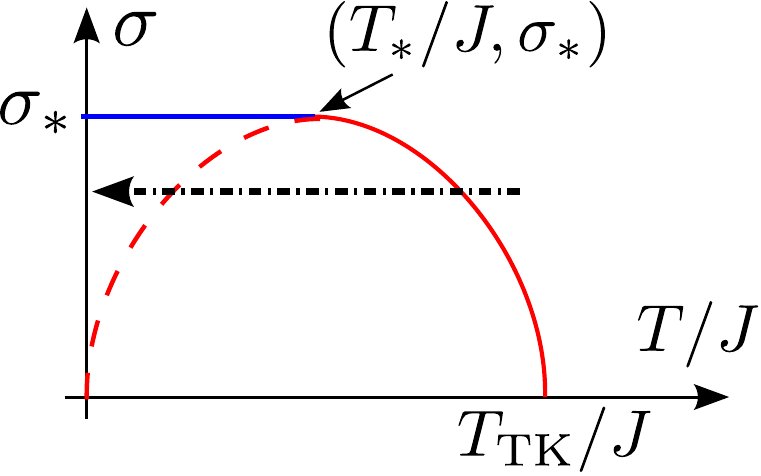}
\caption{A sketch of the phase diagram of the disordered XY model. The solid curves (red and blue), corresponding to $\eta = 0$ in eq. \eqref{eq:XYphase}, separates the low temperature phase (the bounded region below it) and the high temperature phase. The red dashed curve is the wrong prediction that misses the freezing transition, and continues analytically the red solid curve beyond its validity. In particular, it would imply a re--entrance transition indicated by the dashed arrow. The correct phase separation for $T < T_*$ is instead given by the blue straight line, making the re--entrance disappear.}\label{fig:XYphase}
\end{figure}
As a warm--up, we recall that the XY model \textit{without} disorder can be defined by the Hamiltonian
\begin{equation}  H[(\theta_j)] = \sum_{<jk>} J \, V(\theta_j - \theta_k) \,,\, V(\theta \to 0) \sim \theta^2 \,,  \label{eq:XY}\end{equation}
where $(\theta_j)$ is a (discrete) field of $O(2)$ spins on a 2D lattice, and $V$ is a $2\pi$--periodic function that describes the interaction between neighbouring spins (denoted by $\left<ij\right>$). This model has a Kosterlitz--Thouless transition at temperature $T = T_{\text{KT}}$: when $T > T_{\text{KT}}$, there is a  proliferation of vortices (topological defects of the XY field); when $T < T_{\text{KT}}$, the vortices form bound pairs and annihilate each other, and there is a line of fixed points enjoying an algebraic quasi--long range order. A qualitative estimate of $T_{\text{KT}}$ can be obtained by comparing the energy cost of one vortex $E \sim 2 J \ln (R/\epsilon)$ ($J$ is the coupling constant of XY model) and the entropy gained, $S = \ln (R^2/\epsilon^2)$ (contributed by the position of the vortex). Equating $E / T_{\text{KT}} \approx S$ gives $T_{\text{KT}} \approx J$. 

Now we consider the disordered XY model defined as follows,
\begin{equation}  H[(\theta_j), (A_{jk})] = \sum_{<jk>} J \, V(\theta_j - \theta_k - \sigma A_{jk}) \,,\,   \label{eq:XY1} \end{equation}
where $\sigma$ is the disorder strength, and $A_{jk}$ is a quenched random gauge field which is uncorrelated, and whose corresponding magnetic field is $B(z)$. As a consequence, if $\Delta \phi(z) = \sigma B(z)$ ($\Delta$ is the discrete Laplacian), $\phi(z)$ will be a 2D \gls{gff} with coupling constant $\sigma$, see eq. \eqref{eq:GFF2p}. Moreover, it is known that the presence of gauge field modifies the vortex energy as follows:
\begin{equation*}
E \leadsto E' = E \pm J \phi(z) = 2 J \ln (R/\epsilon) \pm J \phi(z) \,,
\end{equation*}
where the sign depends on the orientation of the vortex. Therefore, the partition function of configurations with one (say, positive) vortex is very close to that of a 2D \gls{logrem}:
$$ \mathcal{Z} =  e^{-E / T} \sum_{j=1}^M \exp (- J \phi(z_j) / T) \,. $$
Its leading behaviour can be easily obtained by the freezing scenario eq. \eqref{eq:GbetaGFF}, upon proper rescaling: 
\begin{equation} \mathcal{Z} = \left(\frac{R}{\epsilon}\right)^\eta \times O(1) \,,\, \eta = \begin{cases}
-2J / T + \frac12 \sigma^2 J^2 / T^2 + 2 & T > J\sigma /2 \,,\\
- 2J / T + 2 \sigma J / T  &  T < J \sigma / 2\,,
\end{cases}  \label{eq:XYphase}\end{equation}
where the second case corresponds to the frozen phase. 
The disordered XY model is in the high temperature phase if and only if vortices are favourable, \textit{i.e.}, $\eta > 0$. The line of phase transition is thus given by $\eta = 0$, and sketched in Figure \ref{fig:XYphase}. It is non--analytic at the point $\sigma =  \sigma_* = 1, T = T_* = J/2 < T_{\text{KT}}$. Below $T_*$, the critical line becomes flat and given by $\sigma = \sigma_*$, independently of $T_*$: this is the manifestation of freezing transition in the XY context (a similar phase diagram can be found in the binding transition of \glspl{logrem}, see section \ref{sec:bindingetc} and in particular, Figure \ref{fig:binding}). If the freezing transition had been overlooked, one would have continued the $T > J\sigma / 2$ expression in eq. \eqref{eq:XYphase} beyond its scope of validity, and predict \textit{wrongly} a re--entrance transition for $\sigma < \sigma_*$, as shown in Figure \ref{fig:XYphase}. Therefore, identifying the free transition is essential to the correct \textit{qualitative} understanding of the disordered XY model's phase diagram.

We end this discussion by warning the Reader that the above explanation does \textit{not} represent accurately the rich physical questions involved the disordered and non--disordered XY models. A faithful treatment of either would require the formalism of Kosterlitz--Thouless renormalization group \cite{kosterlitz73kt}, and its disordered extension \cite{carpenter98XY,carpentier00XYlong}. From a \gls{logrem} point of view, the disordered XY model can be seen as that of \textit{interacting} thermal particle\textit{s} in a log--correlated potential, and is thus considerably more involved than \glspl{logrem}!

\subsection{Multi-fractality of logREMs}\label{sec:multifracintro}
Another universal feature believed to be shared by \glspl{logrem} (and also by the \gls{rem}, but to a limited extent), is their multi--fractal properties. In fact, these are the main focus of the pioneering work \cite{chamon1996localization,castillo97dirac}. Multi--fractality appears in different contexts of physics, \textit{e.g.}, turbulence, random geometry, and critical wave-functions in Anderson transitions \cite{evers2008anderson}. For the \glspl{logrem}, multi--fractality will refer to that of the normalized \textit{Gibbs measure}:
\begin{equation}
p_{\beta,j} = \frac{1}{\mathcal{Z}} e^{-\beta V_j}  \,,\,  j=1, \dots, M \,,\, \mathcal{Z} = \sum_{j=1}^M e^{-\beta V_j}  \,. \label{eq:defGibbs}
\end{equation}

\begin{figure}[h]
\center 
\includegraphics[scale=.4]{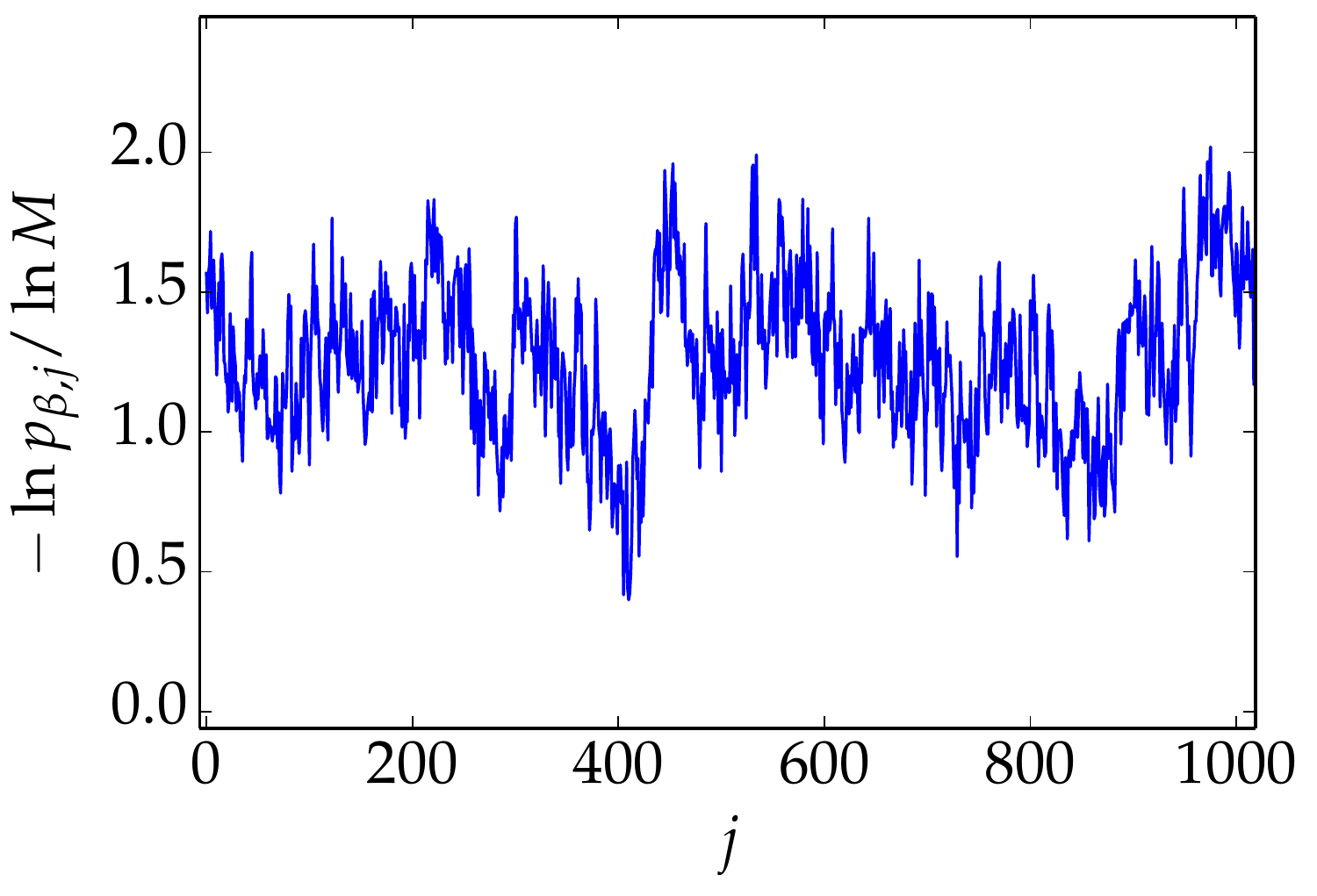}
\includegraphics[scale=.4]{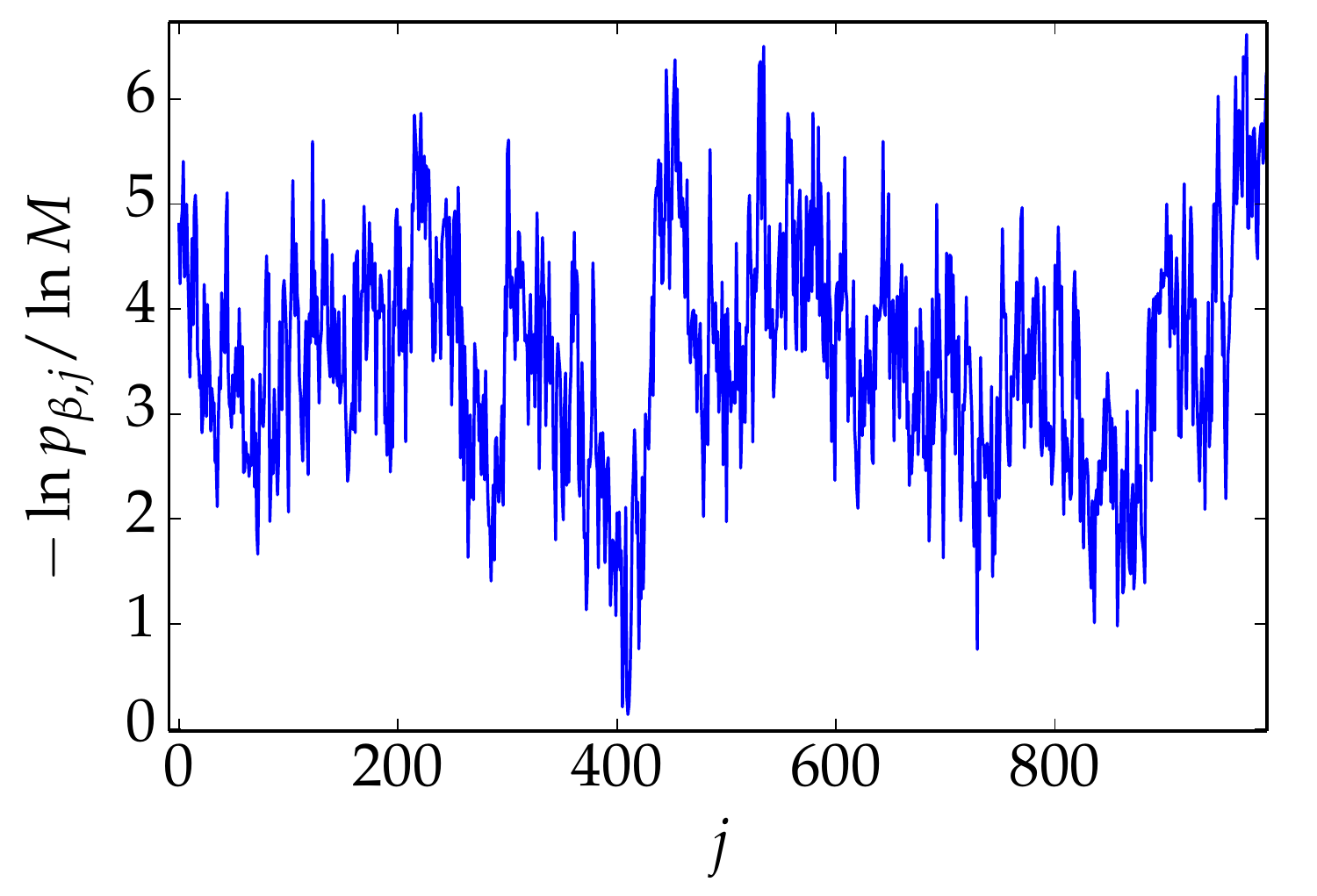}
\includegraphics[scale=.6]{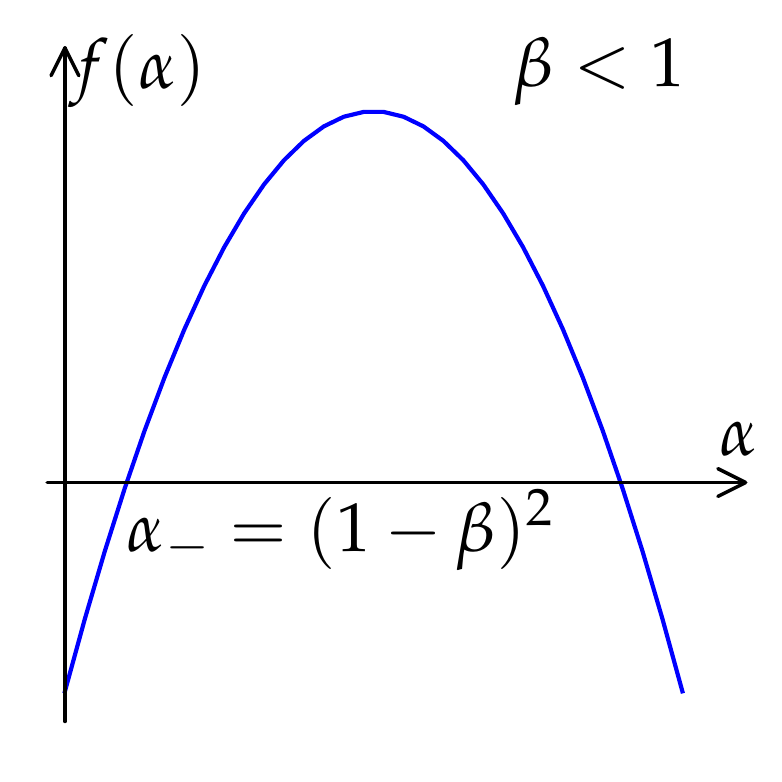} 
\includegraphics[scale=.6]{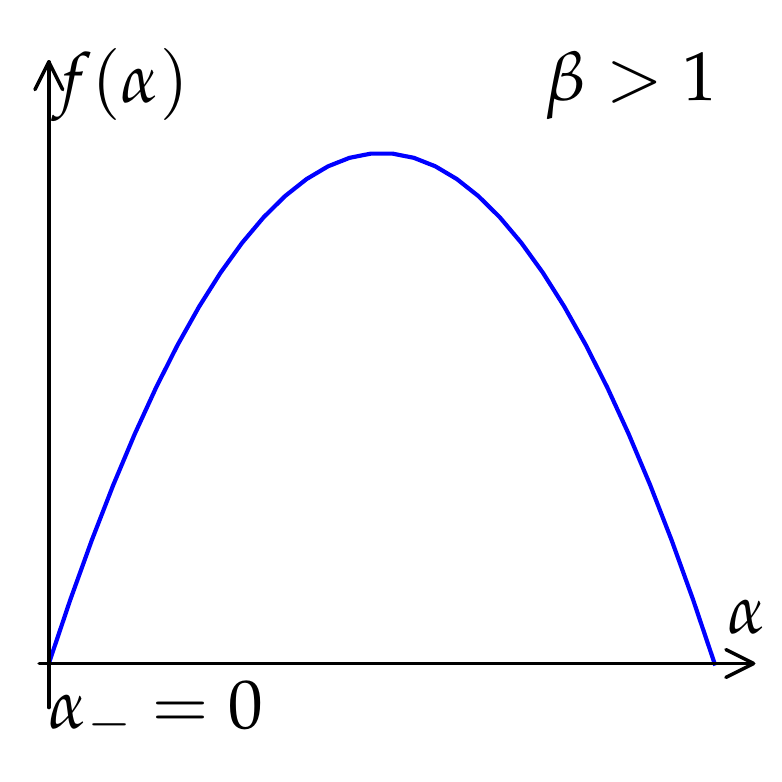} 
\caption{Left: The log-scale colour plot of the Gibbs measure of a \gls{logrem} of 2D\gls{gff} on a torus.  } \label{fig:multifracdef}
\end{figure}
As we can see in figure \ref{fig:multifracdef}, $p_{\beta,j}$'s magnitude spans a large range. The \textit{multi--fractal spectrum} of $p_{\beta,j}$ is defined by making a histogram of $-\frac{\ln p_{\beta,j}}{\ln M}$, and then ``plot'' it the log scale:
\begin{subequations} \label{eq:defmultispec}
\begin{align}
&f(\alpha) =  \frac{1}{\ln M \dif \alpha} \ln \abs{\set{j : -\frac{\ln p_{\beta,j}}{\ln M} \in [\alpha, \alpha + \dif \alpha]} }  \,, \alpha \geq 0 \,  \label{eq:defmultispeca} \\
\Leftrightarrow &\abs{\set{j: p_{\beta,j} \sim M^{-\alpha}}}  = M^{f(\alpha)} \times \text{corrections,}  \label{eq:defmultispecb}
\end{align}
\end{subequations}
where $\abs{X}$ denotes the number of elements in the set $X$. In the above equations, $f(\alpha)$ is defined for any sample of disorder $(V_j)$. However, as $M \to \infty$, $f(\alpha)$ is expected to become deterministic, and is called the multi--fractal spectrum. Note that the definitions eq. \eqref{eq:defmultispec} applies to any long sequence of positive numbers $p_j, j = 1, \dots, M$. In quantum mechanics applications, they are usually the occupation probabilities of a wave function on a lattice, $p_j = \abs{\langle j \vert \psi \rangle}^2$;  in general, the multi--fractal spectrum is an important measure of \textit{critical} wave functions, \textit{i.e.}, those appearing at Anderson/quantum Hall transitions \cite{evers2008anderson}. 

For the \gls{rem} and \glspl{logrem},  $f(\alpha)$ is quadratic \cite{chamon1996localization}:
\begin{equation}\label{eq:falphaREM}
f(\alpha) = \frac{4(\alpha_+ - \alpha)(\alpha - \alpha_-)}{(\alpha_+ - \alpha_-)^2} \,, 
\alpha_{-} = \begin{cases}
(1-\beta)^2 & \beta < 1 \,, \\ 0 & \beta \geq 1 \,,
\end{cases} \; \alpha_+ = \alpha_- + 4\beta \,,
\end{equation}
see figure \ref{fig:multifracdef} for plots in the two phases. 
Note that the freezing transition manifests itself also in the non-analytic $\beta$-dependence of $f(\alpha)$ at $\beta = 1$. 

 A simple derivation of eq. \eqref{eq:falphaREM} goes as follows. By eq. \eqref{eq:defGibbs}, $\alpha_j = \beta (V_j - \ln \mathcal{F})/\ln M$. Now since $\mathcal{F} / \ln M$ is deterministic as given by eq. \eqref{eq:freezedual}, and $V_j$ is a centred Gaussian of variance $2 \ln M$. So when $M \to \infty$, $\alpha_j$ is a Gaussian of mean $\beta^2 + 1$ and variance $2 \beta^2 / \ln M$. Therefore, the normalized probability distribution of $\alpha$ is
\begin{equation} \ln \mathrm{P}(\alpha) = \begin{dcases} 
-\frac{(\alpha - \beta^2 - 1)^2}{4 \beta^2} + o(\ln M) \,,\, \\
-\frac{(\alpha - 2 \beta)^2}{4 \beta^2} + o(\ln M) \,.
\end{dcases}
\end{equation}
Plugging this into $f(\alpha) = \frac{1}{\ln M} \ln \left( M \mathrm{P}(\alpha) \right)$ gives eq. \eqref{eq:falphaREM} after some algebra. The tacit assumption behind the above reasoning is that the correlation between $(V_j)$ and $\mathcal{F}$ has no effect on $f(\alpha)$. Indeed, as we will see later (section \ref{sec:LFTmultifrac}), the effect of the correlations are limited to the sub--leading order. 

\subsubsection{\Acrfull{ipr}}
The (generalized) \glspl{ipr} are important observables closely related to the multi--fractal spectrum. They are widely used in quantum mechanics to detect localization transitions (see section \ref{sec:LT} below). For a statistical model, the \glspl{ipr} are defined in terms of its Gibbs measure $p_{\beta,j}, j = 1, \dots, M$, as follows:
\begin{equation}
P_q = \sum_{j=1}^M p_{\beta,j}^q  \,,\, q \geq 0 \,. \label{eq:defIPR}
\end{equation}
\Glspl{ipr} for $q < 0$ are seldom considered. This definition is as general as the multi--fractal spectrum: in particular, it can be calculated for each disorder realization. In the large system $M \to \infty$ limit,  we can compute its leading behaviour using the definition of the multi--fractal spectrum eq. \eqref{eq:defmultispec}, and the saddle point approximation
\begin{align}
P_q = \int_I \dif \alpha \,  M^{-q \alpha + f(\alpha)} &=  M^{-\tau_q} \times \text{corrections} \,,\, \label{eq:IPRtau} \\
\text{where } \tau_q &= \min_{\alpha \in I} \left[ q \alpha - f(\alpha) \right] \,. \label{eq:Legendre}
\end{align}
Here, $\tau_q$ is called the multi--fractal \textit{exponent}. As a function of $q$, it is calculated as the Legendre transform of the multi--fractal spectrum $f(\alpha)$, eq. \eqref{eq:Legendre}. In general, it is calculated by solving the equation $\partial_\alpha f(\alpha_q) = q$ for $\alpha_q$. The solution is unique if Since $f(\alpha)$ is convex. Thi is the case for \glspl{logrem}, see eq. \eqref{eq:falphaREM}, and we have 
\begin{equation}
a_q = \begin{cases}
-2 q \beta^2 + \beta^2 + 1 \,,\,& \beta \leq 1 \\ - 2q\beta^2 + 2 & \beta > 1 \,.
\end{cases} \label{eq:alphaqLegendre}
\end{equation}
In general, $\alpha_q$ has the following interpretation: the dominating contribution to $P_q$ comes from those points $p_{\beta,j} \sim M^{-\alpha_q}$. When $q$ increases, $\alpha_q$ decreases. In terms of $\alpha_q$, the multi--fractal exponent is given as $\tau_q = q \alpha_q - f(\alpha_q)$. After simple algebra, one finds for the \glspl{logrem}:
\begin{equation}
\tau_q = q \beta \left( Q - q \beta \right) - 1 \,,\, Q = \begin{cases}
\beta + \beta^{-1} & \beta \leq 1 \,,\\ 2 & \beta > 1 \,.
\end{cases} \label{eq:tauqlogrem}
\end{equation}
The above notation is suggested by the \acrlong{lft}, see section \ref{sec:liouville}.

The careful reader must have noticed that a crucial point has been left unexplained: what is the domain $I$ of the integral and the minimum? Indeed, if $\alpha_q \notin I$, eq. \eqref{eq:tauqlogrem} would be wrong! In fact, $I$ depends on whether $P_q$ is calculated in the \textit{typical} or \textit{annealed} ensemble (\cite{mudry96cft}, see \cite{evers2008anderson}, section II.C.7 for a review). Such difference is omnipresent in disordered systems, and arises for quantities like $P_q$, which can fluctuate strongly from one disorder realization to another, and has a disorder--induced \gls{pdf} $P(P_q)$. The latter has often fat (algebraic) tails, coming from \textit{rare} samples; as a result, its mean value may be different from its typical value (another example for which this is the case is the partition function $\mathcal{Z}$). Therefore, it is crucial to discuss which observable of the distribution of $P_q$ is being calculated, \textit{i.e.}, specifying the ensemble:  
\begin{figure}[h]
\center 
\includegraphics[scale=.5]{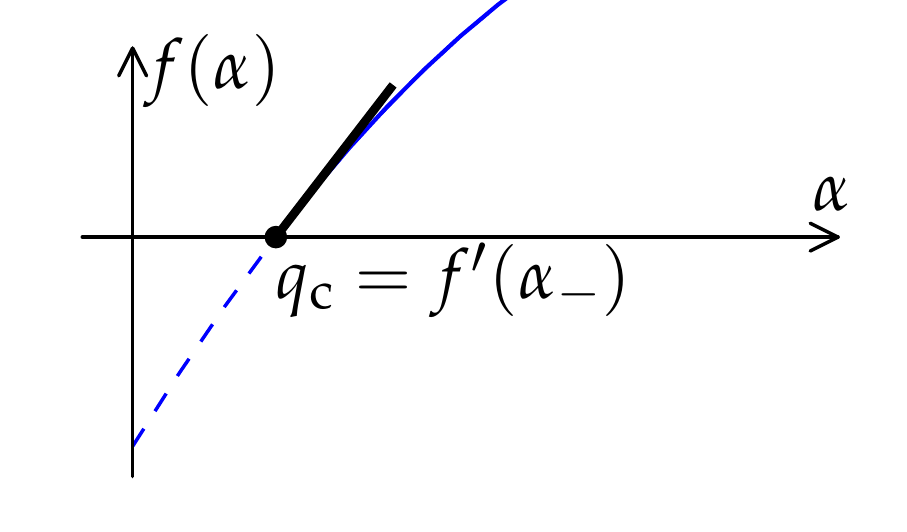}
\includegraphics[scale=.5]{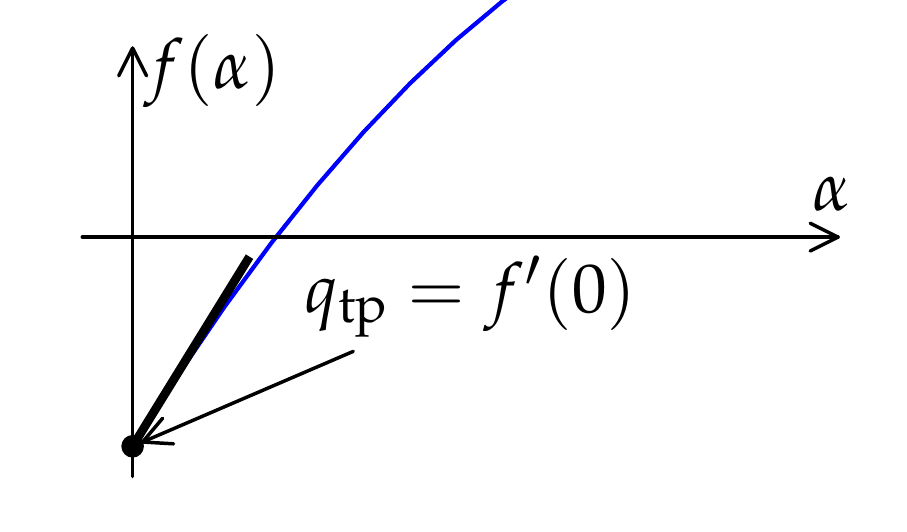}
\caption{\textit{Left}: illustration of the $\alpha_-$--dominance transition in the typical ensemble, see eq. \eqref{eq:typicalensemble}. The multi--fractal spectrum is not defined when $f(\alpha) < 0$. \textit{Right}: illustration of the termination point transition in the annealed ensemble, see eq. \eqref{eq:IPRann}. The multi--fractal spectrum is defined until $\alpha = 0$. } \label{fig:multifractals}
\end{figure}
\begin{itemize}[label=$\square$]
\item The \textit{typical} ensemble (Figure \ref{fig:multifractals}, left) concerns the typical value of $P_q$; in most cases, it is equivalent to the \textit{quenched} ensemble $\exp \overline{\ln P_q}$. But here it is more helpful to think of \textit{one} large typical sample of $(p_{\beta,j}), j = 1,\dots, M$; their logarithms are distributed as described by $f(\alpha)$. The definition eq. \eqref{eq:defmultispec} implies that  when $f(\alpha) < 0$, there are $M^{f(\alpha)} \ll 1$ points $j$ such that  $p_{\beta,j} \sim M^{-\alpha}$: such points are absent in one large typical sample; so $I =\{\alpha: f(\alpha) \geq 0 \} = [\alpha_-, \alpha_+]$, see eq. \eqref{eq:falphaREM}. Eq. \eqref{eq:tauqlogrem} holds in the typical ensemble only when $\alpha_q \geq \alpha_-$ in eq. \eqref{eq:alphaqLegendre}, \textit{i.e.}, $ q \leq q_c  = f'(\alpha_-) = \beta^{-1}$ (in the two phases). When $q > \beta^{-1}$, the  value $\alpha_q$ used in eq. \eqref{eq:Legendre} should not  replaced by $\alpha_q^{\text{typ}} =  \alpha_-$. In summary, the typical ensemble exponents are
\begin{equation}
\tau_q^{\text{typ.}} = \begin{dcases}
q \beta \left( Q - q \beta \right) - 1 \,,   &  q < \beta^{-1} \,, \\
q \alpha_- \,, & q \geq \beta^{-1} \,.
\end{dcases}\label{eq:typicalensemble}
\end{equation}
The non-analyticity at $\tau_q^{\text{typ.}}$ is referred to differently in the literature: ``freezing'' in \cite{evers2008anderson}, ``pre--freezing'' in \cite{carpentier2001glass}, ``$\alpha_-$--dominance'' in \cite{fyodorov2009pre}. We will use the last name since the other two will refer to other transitions in this work.
\item The \textit{annealed} ensemble (Figure \ref{fig:multifractals}, right) concerns the \textit{mean value} $\overline{P_q}$, averaged over \textit{all} disorder samples. So, even if $f(\alpha) < 0$, Gibbs measure values $p_{\beta,j} \sim M^{-\alpha}$ will be present in some rare samples, and the above transition is absent: eq. \eqref{eq:tauqlogrem} is true beyond $q = \beta^{-1}$. However, since $p_{\beta, j} \leq 1$ by normalization, so  $\alpha$ can never be negative. Therefore, when $q > q_{tp} = f'(0) = Q / (2\beta)$ (in the two phases, with $Q$ defined in eq. \eqref{eq:logremdef}), the value $\alpha_q$ given by eq. \eqref{eq:alphaqLegendre} is non-physical and should be replaced by $\alpha_q^{\text{ann.}} = 0$. So the annealed ensemble exponents are
\begin{equation}\label{eq:IPRann}
\tau_q^{\text{ann.}} = \begin{dcases}
q \beta \left( Q - q \beta \right) - 1 \,,   &  q < Q/(2\beta) \,, \\
Q^2 / 4 - 1 \,, & q \geq Q/(2\beta) \,.
\end{dcases}
\end{equation}
The non-analyticity of $\tau_q^{\text{ann.}}$ has also different names: ``termination point transition'' in \cite{evers2008anderson} and ``pre--freezing'' in \cite{fyodorov2009pre} (beware of the confusion!). We will use the term \textit{termination point transition} in this work. 
\end{itemize} 
The above discussion is limited to the leading behaviours (exponents). In section \ref{sec:LFTmultifrac} we will explore the sub--leading corrections: in particular, in the annealed ensemble, they will be predicted for the first time using \acrlong{lft}. 

\section{What are logREMs?}\label{sec:logREMwhat}
\subsection{General characterization} \label{sec:logREMdef}
In section \ref{sec:DPCT} and \ref{sec:treetoplane}, we have reviewed \glspl{logrem} defined on hierarchical lattices and on Euclidean spaces of different dimensions.  Despite apparent differences, they share many key properties, which invite the following questions: what is common in their definition? Can we give a unifying characterization? Clearly, all \glspl{logrem} are defined by Gaussian energy levels $V_j, j = 1,\dots, M$. Therefore, they are completely determined by the mean values $\overline{V_j}, j = 1, \dots, M$ and the covariance matrix 
\begin{equation} \overline{V_j V_k}^c \defeq \overline{V_j V_k} - \overline{V_j} \times \overline{V_k} \,,\, j, k = 1, \dots, M \,.\end{equation} 
This matrix cannot be arbitrary, and must reflect the ``logarithmic correlations''. How to translate this term into a quantitative language?  A simple proposal, given in \cite{cao16order}, section II, is to look at histograms of the covariances $\overline{V_j V_k}^c$ with $j$ fixed, similarly to the definition of multi--fractal spectrum, eq. \eqref{eq:defmultispec}. 

In this thesis, we normalize all the \glspl{logrem} so that the freezing temperature $\beta_c = 1$. Then, for both hierarchical \glspl{logrem} and Euclidean space ones, the following holds: 
\begin{align} 
\forall j\,,\, \abs{\set{k : \overline{V_j V_k}^{c} \geq  2 \overlap \ln M  }} \approx M^{1 - \overlap}  \,,\, \overlap \in (0,1) \,. \label{eq:logremdef}
\end{align}
Recall from eq. \eqref{eq:defoverlap} that $\frac{\overline{V_j V_k}^{c}}{2 \ln M }$ is the general definition of the \textit{overlap} in \glspl{logrem}; so the above equation says that for any fixed energy level $j$, there are $\sim M^{1-\overlap}$ levels that have overlap $\geq \overlap$ with it \footnote{Remark that he same statement holds had we considered $=\overlap$ in the place of $\geq \overlap$}. In particular, setting $\overlap = 1$, we require the variance $\overline{V_j^2}^c \approx 2 \ln M $. 

Let us check that eq. \eqref{eq:logremdef} is satisfied by the different \glspl{logrem}: 
\begin{itemize}[label=$\square$]
\item For the \gls{dpct} defined on a Cayley tree of branching number $\kappa$ and depth $n$ so that $M = \kappa^n$, by \eqref{eq:covDPCT}, $\overline{V_j V_k}^c \geq 2 \overlap \ln M$ if and only if the common length $\hat \overlap_{jk}$ is larger than $ \hat{\overlap} = \overlap n$, where $n$ is the depth of the tree. For fixed $j$, this is satisfied for all $k$ belonging to a sub--tree of size $\kappa^{n-\overlap n} = M^{1 - \overlap}$. The reasoning for the \gls{bbm} is similar.
\item For a log--correlated potential defined on a $d$-dimensional lattice made of points $\mathbf{x}_j \in \R^d, j = 1,\dots, M$, the correct normalization is
\begin{equation}
\overline{V_j V_k} \approx 2d  \ln \frac{R}{\abs{\mathbf{x}_j - \mathbf{x}_k}} \,,\, \epsilon \ll \abs{\mathbf{x}_j - \mathbf{x}_k}  \ll R \,. \label{eq:logdecay}
\end{equation}
where $\epsilon$ and $R$ are the \gls{uv} cut--off (lattice spacing) and \gls{ir} cut--off (linear size of the lattice), respectively, so that $(R/\epsilon)^d = M$. Then, for fixed $j$, $\overline{V_j V_k} \geq 2 \overlap \ln M$ for $k$ in a $d$-dimensional ball of radius $ r = R^{1-\overlap} \epsilon^\overlap$, containing $\sim(r/\epsilon)^d = M^{1-\overlap}$ points.
\end{itemize} 
On the other hand, the uncorrelated \gls{rem} \textit{does not} satisfy eq. \eqref{eq:logremdef}, because their the overlap $\overline{V_i V_j}^c / (2 \ln M)$ can only assume two values $0$ and $1$. The right hand side eq. \eqref{eq:logremdef} would be modified to a step function $\theta(-q)$ for \gls{rem}. 

In section \ref{sec:RSB}, we will see that eq. \eqref{eq:logremdef} is an essential input to the \acrlong{rsb} approach to \glspl{logrem}. Nevertheless,  we should be careful enough not to claim the unifying characterization a definition of \glspl{logrem} in a mathematical sense. The reason is that there is an important additional property of \glspl{logrem}, called \textit{ultra--metricity}. We will discuss this in section \ref{sec:RSB}.

\subsection{IR and UV data} \label{sec:IRUV}
The general characterization \eqref{eq:logremdef} specifies only the structure of covariance matrix in the regime $0 < \overlap < 1$, which is responsible for the universal properties of the \glspl{logrem}, such as the free scenario, eq. \eqref{eq:GbetaGFF}. However, many other observables, such as the full limit distribution of the free energy, are not universal but model dependent; for example, the minimum distribution for the circular model, eq. \eqref{eq:FBgfreeze} is different from that of \gls{bbm}, given by eq. \eqref{eq:ODEwaveKPP} (with $v = -2$). Moreover, in the sequel, we will be interested in other observables, \textit{e.g.}, second minimum and gap distribution (section \ref{sec:orderstat}), and the distribution of minima positions (section \ref{sec:liouville}). These observables will all be model dependent; yet, in the $M\to \infty$ limit, they should not depend on all the details of the discrete model definition, \textit{i.e.}, $\overline{V_j}$ and $\overline{V_j V_k}^c, j, k = 1, \dots, M$. So what is the information relevant for calculating the above observables in the thermodynamic limit?

This question was already considered in \cite{fyodorov2009statistical}. We gave a systematic treatment of the issue in \cite{cao16order}, concentrating on the \glspl{logrem} defined on Euclidean spaces. It turns out that the relevant information can be organized into two groups, called the \textit{\gls{ir} and \gls{uv} (limit) data}, which we describe in a non--technical manner below (the complete technical description can be found in \cite{cao16order}, section 2.1, and will be illustrated for the circular model below): 
\begin{itemize}[label=$\square$]
\item The \textit{\gls{ir} data} specify what is the geometric manifold $X$ (circle, interval, 2D torus, \dots) on which the \gls{logrem} potential $\phi(z), z\in X$ is defined, what is the mean value $\overline{ \phi(z)}$, and the covariance $\overline{\phi(z) \phi(w)}$ for $z,w \in X$ are two distinct points; note that in the unit of lattice spacing, the distance between $z$ and $w$ go to infinity in the thermodynamic limit. 

We will show in section \ref{sec:RSB} that the \gls{ir} data determine the continuum Coulomb gas integral in the replica approach of the \gls{logrem}, which determines in turn the limit distribution (modulo a translation) of the free energy. 
\item The \textit{\gls{uv} data} concerns the covariance $\overline{V_i V_j}$ between lattice points separated by a finite number of lattice spacing in the thermodynamic limit. 

We will show in section \ref{sec:orderstat} that the \gls{uv} data determine the distribution of the gaps (\textit{e.g.}, the difference between the second and first minima).
\end{itemize} 
The fact that only the two limiting scales are relevant is quite intuitive, since the scales in between are governed by the logarithmic decay eq. \eqref{eq:logdecay} and have no more freedom of model dependence. In section \ref{sec:RSB}, we shall give further support to this intuition, using the replica approach. 
 
\subsubsection{Example: circular models}
\begin{figure}[h]
\center
\includegraphics[scale=.5]{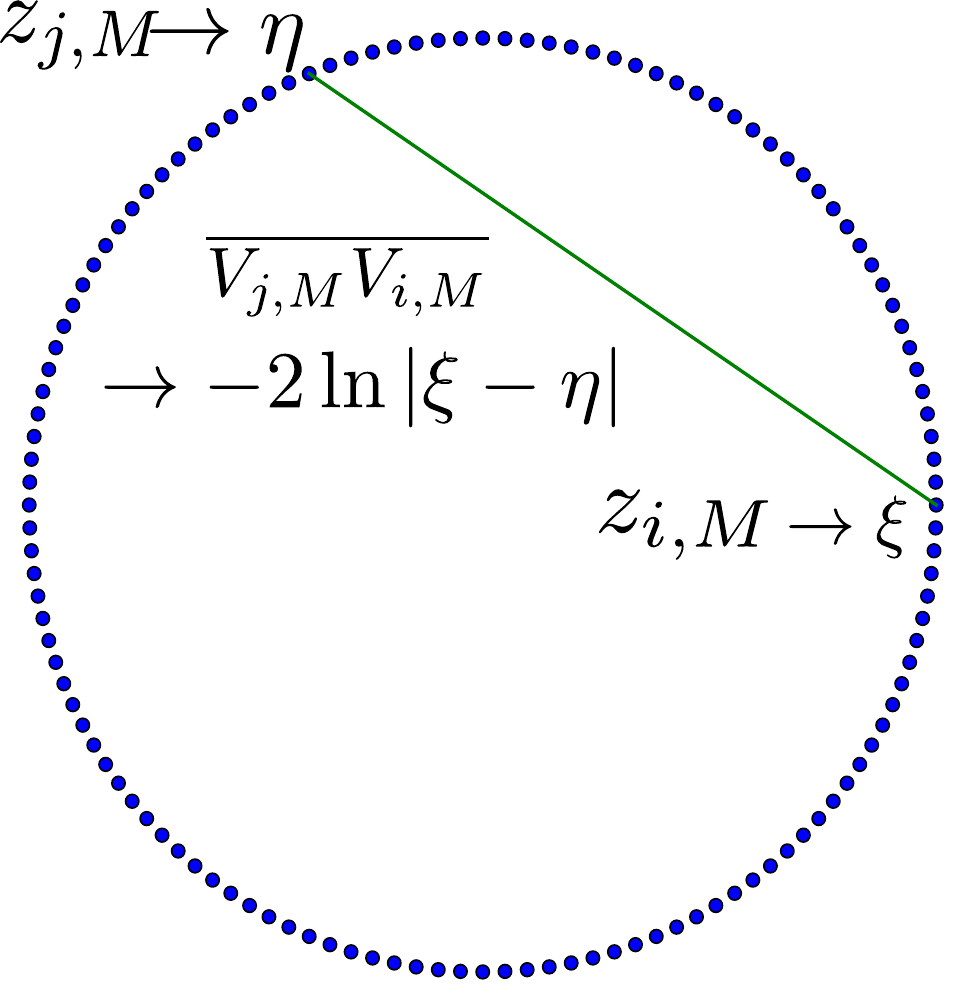}
\includegraphics[scale=.5]{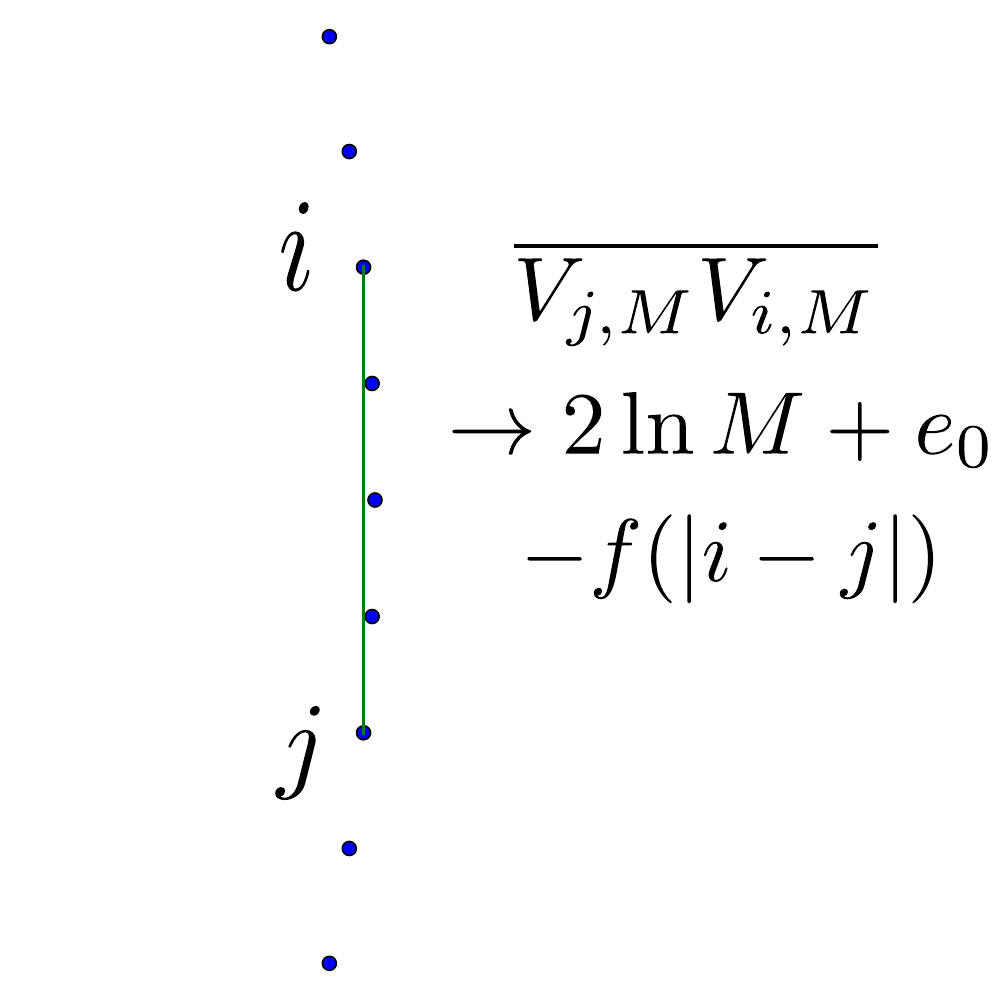}
	\caption{An illustration of the \gls{ir} (left) and \gls{uv} (right) covariances in the circle model. The formula in the \gls{ir} limit corresponds to the $1/f$--noise case in eq. \eqref{eq:circularIR}. } \label{fig:uvir}
\end{figure}
We now illustrate more technically the above ideas in the case of the circular model\textit{s}. We emphasized the plural form because several different definitions can be found in the literature \cite{fyodorov2008statistical,fyodorov2009statistical,rosso12counting,cao15gff,cao16maxmin}, and our goal is to organize them in terms of \gls{uv} and \gls{ir} data. For the clearness's sake, we shall denote $V_{j,M}, j = 1, \dots, M$ the potential values in a system of size $M$; any of the circular models can be defined by discrete Fourier transform (see also section \ref{sec:introPEL})
\begin{align}
&V_{j,M} = \Re \sum_{k=-M/2}^{M/2-1} \sqrt{\mu_{\abs{k},M}} \exp\left(2 \pi \im \frac{jk}{M}\right) \mathsf{N}_k \,,\, \mu_0 = 0 \,. \label{eq:FFT} \\
& \overline{V_{j,M}} = 0 \,,\, \overline{V_{i,M}V_{j,M}} = \sum_{k=-M/2}^{M/2-1} \mu_{\abs{k},M}
 \exp\left[2 \pi \im \frac{(i -j) k}{M}\right] \,.   \label{eq:circularVar}
\end{align}
As shown in section \ref{sec:introPEL}, for the model to be a \gls{logrem} (with the correct normalization for 1D), the Fourier modes $\mu_k$ should be $\sim 1 / k$ for $1 \ll k \ll M/2$. Again, some freedom is allowed at the two limiting scales. Here are some choices:
\begin{subequations}
\begin{align}
\text{$1/f$-noise model: }  & \mu_{k,M} = k^{-1}   \,,  \label{eq:muk1overf}\\
\text{Dirichlet model: } &  \mu_{k,M} = k^{-1}(1 - q^{k}) \,,\, \abs{q} < 1 \label{eq:mukDirichlet}\\
\text{Long range model: } & \mu_{k,M} = \pi M^{-1} \sin^{-1}\left( \frac{\pi k}{M}\right) \,. \label{eq:mukLR}
\end{align}
\end{subequations}
The first choice will be the default one that we refer to as the \textit{circle model} of $1/f$-noise. The second is related to the 2D \gls{gff} with Dirichlet boundary condition, see below and section \ref{sec:Dirichlet}; it reduces to the $1/f$-noise model at $q = 0$. The last one is also what is considered in section \ref{sec:introPEL} (the name comes from \cite{fyodorov2009statistical}, section 6). 

Let us consider the \textit{\gls{ir} data} of these models. For all of them, the underlying manifold is the unit circle, endowed with the uniform lattice $z_{j,M} = \exp\left(2 \pi \im j / M \right)$. The \gls{ir} limit of covariance is obtained by letting $i$ and $j$ depend on $M$, such that $z_{i,M}\to \xi$ and $z_{j,M} \to \eta$, $\xi \neq \eta, \abs{\xi} = \abs{\eta} = 1$, then eq. \eqref{eq:circularVar} implies 
\begin{align} \label{eq:circularIR}
\overline{V_{i,M}V_{j,M}} \stackrel{M\to\infty}{\longrightarrow} 
\sum_{k = -\infty}^{\infty} (\xi \eta^*)^k \lim_{M\to\infty}\mu_{\abs{k}, M} 
= \begin{dcases}
-2 \ln \abs{\xi - \eta} \,,\, & \text{$1/f$-noise},  \\
-2 \ln \abs{\frac{\xi - \eta}{1 - q \xi \eta^*}} \,,\, & \text{Dirichlet},  \\
-2 \ln \abs{\xi - \eta} \,,\, & \text{long range}. \\
\end{dcases} 
\end{align}
Therefore, the $1/f$-noise model and the long range model have the \textit{same} \gls{ir} data, corresponding to the restriction of the planar 2D \gls{gff}, eq. \eqref{eq:GFF2p} to the unit circle. As a consequence, the long range and $1/f$-noise have the same limit distribution of free energy (modulo a $O(1)$ translation). On the other hand, the Dirichlet model has different \gls{ir} data, and we shall discuss in section \ref{sec:Dirichlet} how that affects the free energy distribution. 

We now come the \textit{\gls{uv} data}. For this, we need to repeat the above calculation, but keeping $i-j$ constant as $M \to \infty$. We shall separate the case $i = j$ (variance) from the others: \begin{equation}\label{eq:variancelogrem} \overline{V_j^2} = \sum_{k=M/2}^{M/2-1} \mu_k \stackrel{M\to\infty}{\longrightarrow} 2 \ln M + e_0 \,, \end{equation}
where $e_0$ is a constant that is different for the three models; \textit{e.g.}, for the Dirichlet model, it is $e_0 =  \gamma_E + \ln 4 - 2 \ln (1 - q)$. We will see in section \ref{sec:RSB} that $e_0$ contributes only a shift to the free energy distribution. More interesting is the difference between the covariances and the variance, which will affect the gap between the minima (see section \ref{sec:orderstat}). An explicit calculation leads to:
\begin{align} \label{eq:UVlimit}
&\overline{V_j^2} - \overline{V_i V_j} \stackrel{M\to\infty}\longrightarrow f(\abs{i-j})  =
\int_{0}^{\frac12}  2(1 -   \cos(2 \pi x \abs{i-j} )  M \mu_{M x , M}  \dif x \nonumber \\ 
&f(n) =  \begin{dcases}
2 (-\text{Ci}( \pi n )+\log (n)+\gamma +\log \pi) \,,\, &  \text{$1/f$-noise and Dirichlet},\\
\int_0^{\frac\pi2} \frac{\sin^2 (n \theta) }{\sin(\theta)} \dif \theta  & \text{long range}
\end{dcases}
\end{align}
where $\text{Ci}(x) = - \int_{x}^{\infty} \cos t / t \dif t$ is the cosine integral. So the Dirichlet and the $1/f$-noise models have the same gap distributions, which are different from the long range model. 

\subsection{IR divergence in logREMs}\label{sec:Gaussian}
Log--correlated potentials, such as the 2D \gls{gff}, have both \gls{uv} and \gls{ir} divergences, which both need to be regularized to construct well-defined \glspl{logrem} (or to have any statistical physical application). In this thesis, the \gls{uv} divergence is always tamed by a discrete lattice, on which the variance (``self energy'') $\overline{V_i^2}^c = 2 \ln M + O(1)$, eq. \eqref{eq:variancelogrem}. Another way of \gls{uv}-regularization present in the mathematical literature is the Gaussian multiplicative chaos \cite{kahane1985chaos}, see \cite{rhodes2014gaussian} for recent review; roughly speaking, this approach stays in the continuum but suppresses the small wave--length fluctuations in order to make the regularized potential smooth. In the limit where the regularization is removed, one can retrieve most of the thermodynamic results of discrete \glspl{logrem}, \textit{e.g.}, freezing transition \cite{madaule2013glassy,duplantier2014critical}, and relation to \acrlong{lft} \cite{david2016liouville}, see also section \ref{sec:liouville} below. The only disadvantage is that it is not possible to define higher order statistics (\textit{e.g.}, second minimum) in this framework. In all cases, it is fair to say that the \gls{uv} regularization of \glspl{logrem} is well understood. 

The \gls{ir} regularization deserves more discussion. The circular models are defined by restricting the 2D \gls{gff} on the unit circle, which is a finite geometry, making the \gls{ir} divergence of the 2D \gls{gff} irrelevant. The same can be said for the interval model \cite{fyodorov2009statistical}. For both models, using the methods outlined in section \ref{sec:treetoplane}, one can make predictions about the free energy distribution and compare them with numerical simulations.

\subsubsection{Gaussian model}
Another way of \gls{ir} regularization, considered in \cite{fyodorov2009statistical,fyodorov2010freezing}, is to taking an infinite geometry, and confining the thermal particle with a deterministic potential in addition to the log potential. The example considered in \textit{op. cit.} is the ``Gaussian model''. Its continuum partition function is
\begin{equation}
Z = \int_\R \frac{\dif x}{\sqrt{2\pi}} e^{-\beta (\phi(x) + x^2/2)} \,,\,
\end{equation}
where $\phi(x)$ is the 2D \gls{gff} in eq. \eqref{eq:GFF2p} (with $\sigma^2 = 2$). Namely, the potential is the restriction of 2D \gls{gff} on the infinite line \textit{plus} a deterministic harmonic potential. One can proceed by mimicking the solution to the circular model in section \ref{sec:treetoplane}. The replicated partition function is the Mehta integral, also exactly solved:
\begin{equation}
\overline{Z^n} = \int_\R \prod_{a=1}^{n} \left[ \frac{\dif x_a}{\sqrt{2\pi}} e^{-\beta x_a^2 / 2} \right] \prod_{a<b} \abs{x_a - x_b}^{-2\beta^2} = \beta^{-n+\beta^2 n(n-1)} \prod_{j=1}^{n}\frac{ \Gamma(1 - j \beta^2)}{\Gamma(1-\beta^2)} \,, \label{eq:Mehta}
\end{equation}
The analytical continuation of the product into $n\in\C$ can be also done, using the (generalized) Barnes function $\tilde{G}_\beta(z)$ (see the appendix below, around eq. \eqref{eq:Barnes}, for its basic properties); the resulting Laplace transform of the free energy is 
\begin{align}
\overline{\exp(tF)}& = \exp(C_0 + C_1 t + C_2 t^2) / \tilde{G}_\beta(\beta^{-1} + t) \, \label{eq:etFGauss},
\end{align}
where $C_i$'s depend only on $\beta$ but not on $t$. However, as observed in \cite{fyodorov2009statistical}, this equation is not physical, Indeed, since the Barnes function has the asymptotic behaviour:
\begin{equation}
\ln \tilde{\Barnes}_\beta(x) \sim  x^2 \ln x + O(x^2)  \,,\,  x \to + \infty \,, \label{eq:Barnesasymptotics}
\end{equation} 
$\ln \overline{\exp(tF)}$ fails to be convex on the interval $t \in [0,\infty)$ for any choice of $C_i$ (it is analytically defined on that interval). So its inverse Laplace transform cannot be a positive probability distribution, and the prediction eq. \eqref{eq:etFGauss} is not physical!

This pathology is an illustration of the \gls{ir} divergence of \glspl{logrem}. It is not restricted to the Gaussian model, but appear in all exactly solved Coulomb gas integrals on infinite geometries, plaguing attempts of turning into predictions of the free energy distribution of some \gls{logrem}. In particular, this prevented such interpretations of the 2D Dotsenko--Fateev integrals \cite{dotsenko1984conformal} in terms of \gls{logrem} observables (in all fairness, there do exist proposals, \textit{e.g.} in \cite{david2016liouville}, which are unfortunately not physically significant enough from the \gls{logrem} viewpoint; see section \ref{sec:liouville} for further discussion). The origin of such pathologies is a rather intricate question from an analytical point of view  (\textit{i.e.}, properties of analytical continuations of Coulomb gas integrals). We will briefly comment on it below. 

Nevertheless, from a \gls{logrem} point of view, the pathology should be expected. To explain this, we recall that any computer simulation and physical realization of the Gaussian model would come with a \gls{ir} cut-off $L$ and \gls{uv} cut--off $\epsilon$. Concretely, we may take a discrete lattice with $M = L/\epsilon$ points, with positions 
$$ x_j = -L/2 + j \epsilon \,,\, j = 1,\dots, M \,. $$
The mean values and covariances of the \gls{logrem} potential are
$$ \overline{V_j} = \frac{x_j^2}{2} \,,\, \overline{V_j V_k}^c = \begin{dcases} 2 \ln \frac{L}{\abs{x_j - x_k}} = 2\ln \frac{M}{\abs{j-k}}\,, &  j \neq k \,, \\ 
2 \ln \frac{L}{\epsilon} = 2 \ln M  \,,& j = k \,. \end{dcases} $$
The numerator $L$ cannot be absent, otherwise the self-variance would not be $\sim 2 \ln M$, and the covariance matrix would not be positive-definite. Yet another way to see this necessity is to observe that, the Gaussian model with \gls{ir} cut--off $L$ must be defined by a 2D\gls{gff} with at least the same \gls{ir} cut-off, and such a 2D\gls{gff} has covariance $\propto \ln (L/\abs{x - y})$, see eq. \eqref{eq:2DGFFdis}. 

As $L \to \infty$ (with $\epsilon$ fixed), the thermal particle is caged by the harmonic potential in the region $\abs{x} \leq \ell_*$. Its size $\ell_*$ is determined by minimizing $\ell^2 / 2 + \mathcal{F}(\ell)$ with respect to $\ell$, where $\mathcal{F}(\ell) = -Q \ln (\ell/\epsilon)$ is the leading behaviour of the free energy of a \gls{logrem} of size $\ell / \epsilon$, see eq. \eqref{eq:freezedual}. The minimization gives $\ell_* =  \sqrt{Q} \sim O(1)$, independently of the cut--offs $L$ and $\epsilon$. Now the free energy fluctuation can be studied using the truncated replicated partition function
\begin{align}
\overline{Z^n} \approx &\int_{-\ell_*}^{\ell_*} \prod_{a=1}^n \left[ \frac{\dif x_a }{\sqrt{2 \pi}}  e^{-\beta x_a^2/2} \right] \prod_{a < b}  \frac{L^{2\beta^2}}{\abs{x_a - x_b}^{2\beta^2}} = L^{(n^2 -n)\beta^2} \, \overline{\widehat{Z}^n} \,,\,  \label{eq:ZnGaussian} \\
\widehat{Z} = & \int_{-\ell_*}^{\ell_*} \frac{\dif x }{\sqrt{2 \pi}}  e^{-\beta (x^2/2 + \phi(x)) } \,,\, 
\end{align}
In terms of free energy distribution, 
\begin{align}
\overline{\exp(tF)} = \exp\left(t^2 \ln L  + \beta t \ln L \right) \overline{\exp(t\widehat{F})} \,,\, \label{eq:etFGaussian2}
\end{align}
The Coulomb gas integral $\overline{\widehat{Z}^n}$ is defined on a finite interval, so we expect it can be continued analytically and be interpreted as the Laplace transform $\overline{\exp(t\widehat{F})}$, where $\widehat{F} = \beta^{-1}\ln \widetilde{Z}$ has a distribution independent of $L$. Now eq. \eqref{eq:etFGaussian2} implies that, modulo a first moment shift, the distribution of $F$  is obtained by a convolution of $\widehat{F}$ and a Gaussian distribution of variance $2 \ln L$: $F =  \mathsf{N} \times 2\ln L + \widehat{F}$, where $\mathsf{N}$ is a standard Gaussian independent of $\widehat{F}$. Therefore, in the Gaussian model with $L$ large but finite, the free energy fluctuation has a $L$--dependent variance $2 \ln L + O(1)$. The rescaled distribution $F / \sqrt{2\ln L} = \mathsf{N} + O(1/\sqrt{2\ln L})$ is a standard Gaussian in the $L\to \infty$ limit. 

We remark that the same result was obtained in a more accurate manner for a specific setting in \cite{fyodorov2009statistical}, section 5. The approach there is equivalent to  modifying the parabolic deterministic potential to  
$$\frac{x^2}{2} \leadsto - \frac{L^2}{8} \ln \left(1 - \frac{4 x^2}{L^2} \right) \,,$$
which tends to $\frac{x^2}{2}$ as $L\to \infty$ for each fixed point $x$ and diverges to $+\infty$ at the \gls{ir} cut--off $x = \pm L / 2$. This reduces the problem to a specific case of the interval model, solved in \cite{fyodorov2009statistical}. The main result is that, as $L\to \infty$, the variance of the free energy (denoted $F_L$) diverges, and all its higher cumulants have a finite limit:
\begin{align}
\overline{F_L^2}^c = 2 \ln L + O(1) \,,\, 
\overline{F_L^k}^c \stackrel{L\to\infty}{\longrightarrow}  \overline{F^k}^c \,,\, k = 3,4,5\dots \nonumber \\
\text{where } \overline{F^k}^c = \left[-\frac{\dif^k}{\dif t^k} \ln \tilde{\Barnes}_\beta(\beta^{-1} + t)\right]_{t=0} \,, \label{eq:FkGaussian}
\end{align}
\textit{i.e.}, $\overline{F^k}^c$ are the higher ``cumulants'' of the non--physical distribution predicted in eq. \eqref{eq:etFGauss}. Therefore, $F_L$ can be seen as a Gaussian with $\sim 2 \ln L$ variance, convoluted with an $O(1)$, yet non--positive correcting distribution. We believe that the pathology of the latter is closely related to the fact that the Mehta integral is defined on the infinite line; although we have no general demonstration of this relation, in section \ref{sec:Dirichlet}, we will comment on a case in which the correcting distribution is well--defined in probability terms. 

\subsubsection{\gls{ir} divergent \glspl{logrem}: free energy \textit{vs} Gibbs measure}
The above discussion shows that the Gaussian model is qualitatively different from \glspl{logrem} defined on finite domains, because it does \textit{not} satisfy the universal scaling behaviour of the free energy, eq. \eqref{eq:GbetaGFF}. In particular, in the zero temperature limit, the minimum distribution of the Gaussian model satisfies the Ansatz $V_{\min} \longrightarrow a_{L,\epsilon}  + b_{L, \epsilon} y$, where $b_{L,\epsilon} = \sqrt{2 \ln L}$ (as ${L\to\infty,\epsilon\to0}$), and $y$ has a standard Gaussian distribution. In contrast, for \glspl{logrem}, we would have $b_{L, \epsilon} = 1$ and $y$ non--Gaussian.

Moreover, we observe in retrospect that, our discussion of the Gaussian model does not depend on the precise shape of the confining potential $V(x) = \frac12 x^2$, as long as it diverges fast enough to $+\infty$ as $x \to \pm \infty$, in order to confine the thermal particle in an $O(1)$ region as the \gls{ir} cut-off is removed, $L \to \infty$. Let us call such models \textit{\gls{ir} divergent \glspl{logrem}}. Therefore, in all \gls{ir} divergent \glspl{logrem}, the free energy has a diverging variance, and its rescaled distribution tends to an uninteresting Gaussian.

This being said, \gls{ir} divergent \glspl{logrem} have still an interesting class of observables: the probability distribution of the \textit{position} of the thermal particle, or more generally, correlation functions of their normalized Gibbs measure (\textit{Gibbs measure statistics}). Unlike the free energy, the Gibbs measure has a non--trivial limit as $L \to \infty$, $\epsilon \to \infty$. Moreover, we can study it using the replica trick directly in that limit regardless of the ill--defined nature of the free energy distribution, because its diverging variance comes from long wave--length fluctuations ($\sim L$) of the log--correlated field, which do not affect the normalized Gibbs measure.

For the Gaussian model, such a study was carried out in \cite{fyodorov2010freezing}, which was motivated by the decaying Burgers turbulence with log--correlated initial flow potential, and which applied the method of $\beta$-random matrix theory. In section \ref{sec:liouville}, we will study an \gls{ir} divergent \gls{logrem} defined on the infinite plane, and relate its Gibbs measure statistics to the \acrlong{lft} on the infinite plane.

\subsubsection{Appendix: Barnes function}
We collect some basic facts of the generalized Barnes functions. They are entire functions of $z \in\C$, which depend on a parameter $\beta$. There are two normalizations, $\tBarnes_\beta(z)$ (\cite{fyodorov2015moments}, section 13.2) and $\Barnes_\beta(z)$ (\cite{fateev2000boundary}, eq. 3.17, used in \cite{fyodorov2009statistical,fyodorov2010freezing}). Both of them have simple zeros at $z = -n \beta - m\beta$, $n, m=0,1,2,\dots$, and are analytic everywhere. They differ by a trivial but cumbersome normalization factor 
\begin{equation}\label{eq:Barnes}
\tBarnes_\beta(z) = \Barnes_\beta(z) \beta^{\frac{z^2}2-\frac{z}2(\beta+ 1/\beta)} (2\pi)^{z(1/(2\beta)-1/2)}
\end{equation}
and are convenient for different situations. 

$\tBarnes$ is useful for analytically continuing products of Gamma's appearing in Selberg integrals because of the recursion relations:
\begin{equation}
\frac{\tBarnes(z + \beta)}{\tBarnes(z)}= \Gamma(z \beta) \,,\, \frac{\tBarnes(z + 1/\beta)}{\tBarnes(z)} = \Gamma(z/\beta) \,,\, \label{eq:Barnesrecursion}
\end{equation}
Iterating the first relation, we have:
\begin{equation}
\prod_{j=1}^n \Gamma(\beta z - j \beta^2) = \frac{\tBarnes_\beta(z)}{\tBarnes_\beta(z-n\beta)} \,.  \label{eq:barnesdef}
\end{equation}
which leads to eq. \eqref{eq:etFGauss}. Either of the relations \eqref{eq:Barnesrecursion}, together with the Stirling asymptotic formula $\ln \Gamma(x) \sim x \ln x + O(x) \,,\, x\to + \infty$, implies \eqref{eq:Barnesasymptotics}. $\Barnes$ is more closely related to \acrlong{lft}, see \ref{sec:liouville}, eq. \eqref{eq:UppBarne}.

We mention that a rich theory of Barnes functions applied to probability (closely related to 1D \glspl{logrem}) has been recently developed by Ostrovsky, see for example \cite{ostrovsky2009mellin,ostrovsky2013theory,ostrovsky2013selberg,ostrovsky2016barnes,ostrovsky2016gff,ostrovsky2016riemann}.

\section{Replica symmetry breaking (RSB)}\label{sec:RSB}
Most of the new results on \glspl{logrem} that will be discussed in this thesis are obtained using the \textit{replica approach}.  Section \ref{sec:treetoplane} has given a flavour of the method, by discussing the Fyodorov--Bouchaud's solution \cite{fyodorov2008statistical} of the circular model. We recall that the solution can be decoupled into two parts:
\begin{enumerate}
\item \textit{Algebraic/integrability}: the exact solution and analytic continuation of the Coulomb gas integral, such as eq. \eqref{eq:Dyson}; 
\item \textit{Physical/thermodynamics}: interpret that the previous part is correct only in the $\beta < 1$ phase, and use the freezing scenario for the $\beta > 1$ phase.
\end{enumerate}
In general, the existing literature and our contribution to Euclidean--space \glspl{logrem} can be classified into the above two parts. In this respect, sections \ref{sec:Jack} and \ref{sec:liouville} fall into the first part, while this section and section \ref{sec:orderstat} contribute to the second part. Its main object is to connect the discrete \gls{logrem} in the thermodynamic limit to the continuum results obtained in the first part. The discrete--continuum relation is complicated by the existence of phase transitions. In particular the freezing transition makes the relation not at all obvious in the $\beta > 1$ phase. Our goal in this section and section \ref{sec:orderstat} is to discuss that relation using the method of \acrfull{1rsb}. 

The basic object of any \gls{rsb} analysis is the replicated partition sum. For any random partition function $\mathcal{Z} = \sum_{j=1}^M e^{-\beta V_j}$, where $V_j$ are Gaussian, we have the following by applying Wick theorem eq. \ref{eq:WickThm}:
\begin{align}
\overline{\mathcal{Z}^n} = \sum_{j_0, \dots, j_n = 1}^M \prod_{a=1}^{n} \exp(-\beta \overline{V_{j_a}}) \prod_{a,b=1}^n \exp(\beta^2 \overline{V_{j_a} V_{j_b}}^c / 2) \,. \label{eq:Wick}
\end{align}
This equation holds thus for both the \gls{rem} and \glspl{logrem}. Throughout this section, we shall restrict to cases where $\overline{V_j} = 0$, with the exception in section \ref{sec:bindingetc}, where non--trivial mean values will be necessary to study the \textit{binding transition}.

\subsection{RSB for REM}\label{sec:REMRSB}
The  \gls{rsb} approach for the \gls{rem} and  \glspl{logrem} has been developed along side these models themselves, and is still being developed (see \textit{e.g.}, \cite{mottishaw15REM,derrida16kppfinitesize}). We have chosen to refrain from presenting this line of research in section \ref{sec:REM}, because the \gls{rem} was designed to be solvable \textit{with and without} replicas, and therefore a test ground for the replica theory. Nevertheless, as argued in \cite{mottishaw15REM} and as we will discuss below, the replica--free treatment of the \gls{rem} in section \ref{sec:REM} can be closely compared to its replica solution, which we review now. The goal is to introduce the basic ideas and twists of the replica approach, and try to provide some rationale for it.

For the \gls{rem}, in eq. \eqref{eq:Wick}, the mean values vanish $\overline{V_{j_a}} = 0$, and the covariances $\overline{V_{j_a} V_{j_b}} = 2 \ln M \delta_{j_a, j_b}$: they are non-zero only between replicas in the same \textit{group}, \textit{i.e.}, those occupying the same position. Suppose that the $n$ replicas form $k$ groups, of sis $m_1, \dots, m_k$, such that $m_1 + \dots + m_k  = n$.  The way of a partitioning $n$ replicas into these groups is given by the multinomial factor:
$$ \begin{pmatrix}
n \\ m_1; \dots; m_k
\end{pmatrix} = \frac{n!}{m_1!\dots m_k!} \,.$$
The corresponding  disorder--averaged Boltzmann factors is
$$ \prod_{a,b=1}^n \exp(\beta^2 \overline{V_{j_a} V_{j_b}}^c / 2)  =  \prod_{g=1}^k \exp\left( m_g^2 \beta^2 \ln M \right) \,. $$
Finally, we have to sum over the positions of the groups $1 \leq J_1< J_2 < \dots  < J_k \leq M$; for later applications, it is more convenient to regard the groups as distinguishable and divide by the symmetry factor: 
$$  \sum_{J_1 < \dots < J_k} [\dots] = \frac{1}{k!} \sum_{J_1\neq \dots \neq J_k} [\dots] \,, $$ 
Gathering the above considerations, we have the following 
\begin{align} \overline{\mathcal{Z}^n}  =&  \sum_{k=1}^{\infty}  \sum_{ m_1 + \dots+ m_k = n } \frac{n!}{m_1! \dots m_k!}\frac{1}{k!}  \sum_{J_1 \neq \dots \neq J_k = 1}^M  \prod_{g=1}^k \exp\left( m_g^2 \beta^2 \ln M \right) \nonumber \\
= & \sum_{k=1}^{\infty}  \sum_{ m_1, \dots, m_k } \frac{n!}{m_1! \dots m_k!}\frac{1}{k!} 
M(M-1) \dots (M-k+1) \prod_{g=1}^k M^{\beta^2 m_g^2}  \label{eq:REMZnreplica0} \\
\to &  \sum_{k=1}^{\infty}  \sum_{ m_1, \dots, m_k } \frac{n!}{m_1! \dots m_k!}\frac{1}{k!} 
\prod_{g=1}^k \exp\left( (m_g^2 \beta^2 + 1) \ln M \right) \label{eq:REMZnreplica} \,,
\end{align}
where in the second line we performed the (trivial) sum over $J_1, \dots, J_M$, and in the last line we took (naïvely) the limit $M \to \infty$ in which $M(M-1) \dots (M-k+1) \leadsto M^k$.

The crucial point of \gls{rsb} is to identify the dominant grouping configurations in eq. \eqref{eq:REMZnreplica} in the $M\to \infty$ limit, and \textit{in the $n \to 0$ limit}. Since eq. \eqref{eq:REMZnreplica} is not defined for non-integer $n$, this is the point from which the replica approach loses all its mathematical sense (eq. \eqref{eq:REMZnreplica0} is still rigorous) and relies on a set of (well-tested) conventions: \texttt{i.} $m_1 = m_2 = \dots = m_k = m = n / k$. So the sum over $k$ can be seen as one over $m$:
$$  \overline{\mathcal{Z}^n} = \sum_{m} \frac{\Gamma(1 + n)}{\Gamma(1 + m)^{n/k}}\frac{1}{\Gamma(1 + n/m)} 
\exp\left( \frac{n}{m} (m^2 \beta^2 + 1) \ln M \right) $$
where we used the Gamma functions to rewrite the factorials. The other conventions are: \texttt{ii.} $m$ is a parameter to be optimized in eq. \eqref{eq:REMZnreplica}, 
 \texttt{iii.} The optimization range is $m \in (n,1] \to (0, 1]$ (as $n \to 0$), and \texttt{iv.} the optimization is a \textit{minimization} of $\overline{Z^n}$ in the $M \rightarrow \infty$ limit:
\begin{align} \overline{\mathcal{Z}^n}& \approx \min_{m \in (0,1]}  \frac{\Gamma(1 + n)}{\Gamma(1 + m)^{n/k}}\frac{1}{\Gamma(1 + n/m)}
 \exp\left( \frac{n}{m} (m^2 \beta^2 + 1) \ln M \right) \,. \label{eq:ZnREM1RSB}
  \end{align}
where the sense of $\approx$ will be commented later. In the $M\to\infty$ limit, the minimization acts on the factor in front of $\ln M$, and gives 
\begin{align}
m = \begin{cases}
1 & \beta \leq 1 \,,\\
1/\beta & \beta >  1 \,.
\end{cases}  \label{eq:moptimumREM}
\end{align}
The non-analyticity at $\beta=1$ is the  \gls{rsb} signal of the freezing transition. Recalling that $m = m_1 = \dots = m_k$ is the number of replicas per group, let us discuss the above solution in the two phases:
\begin{enumerate}
\item In the $\beta < 1$ phase, $m = 1$: different replicas do not group together. For this reason the $\beta < 1$ phase is also called the ``replica symmetric'' phase. Eq. \eqref{eq:ZnREM1RSB} gives $\overline{\mathcal{Z}^n} \approx M^{(1 + \beta^2)n} \Rightarrow \overline{\exp(t\mathcal{F})} \approx M^{-t(\beta + \beta^{-1})}$, \textit{i.e.}, meaning that the free energy $\mathcal{F} = - ( \beta + \beta^{-1}) \ln M$ is deterministic. This result agrees \textit{exactly} in the $M\to \infty$ limit with eq. \eqref{eq:GbetaREMhighT}. 
\item In the $\beta > 1$ phase, $m = 1/\beta < 1$. The fact that the size of groups is non--integer is counter--intuitive, but should be seen as a consequence of the $n\to 0$ limit. Moreover, the fact it is between $n \to 0$ and $1$ is interpreted as non-trivial grouping of replicas (In fact, $m$ has a probability interpretation, as we shall see later in section \ref{sec:RSB1}). So the $\beta > 1$ is called the (one--step) replica symmetry breaking phase. Eq. \eqref{eq:ZnREM1RSB} gives 
\begin{align} 
 &\overline{\mathcal{Z}^n} \approx M^{2n\beta}  \frac{\Gamma(1+n)}{\Gamma(1+n\beta)} \Gamma(1 + 1/\beta)^{-n\beta} \,,
 \end{align}
 which should be compared to the exact ($M\to\infty$) result, eq. \eqref{eq:etFREMlowT} (see also texts after eq. \eqref{eq:GbetaREMfreeze}), which translates to 
 \begin{align}
 \overline{\mathcal{Z}^n} = M^{2n\beta} \left(4 \pi \ln M \right)^{-\frac{n\beta}2} 
  \frac{\Gamma(1-n\beta)}{\Gamma(1-n)} \Gamma(1 - 1/\beta)^{n\beta} \,. 
 \end{align}
 We see that there are two differences: the log correction is not present in the replica solution, and the Gamma functions would match exactly with the replacement 
 \begin{equation}
 \Gamma(1 +x) \leadsto \frac{1}{\Gamma(1 - x)} \,, \label{eq:Gammawierd}
 \end{equation}
 as if one applied the reflection formula but discarded the factor $x \pi / \sin(\pi x)$. Such phenomenon is well documented, \textit{e.g.}, in \cite{bouchaud1997universality}. 
\end{enumerate}
The above is the standard \gls{rsb} analysis of \gls{rem}. We see that it reproduces correctly the leading free energy. The sub--leading correction $\frac{1}{2} \ln (4 \pi \ln M)$ of free energy in the $\beta > 1$ phase should be added by hand; the limit distribution is also reproduced correctly provided the replacement eq. \eqref{eq:Gammawierd}.

At the level of \gls{rem}, it is possible to relate more closely the \gls{rsb} analysis with the exact treatment in section \ref{sec:REM}. For this, we shall take a different route from the exact eq. \eqref{eq:REMZnreplica0}. Rather than following the \gls{1rsb} recipe, we consider the moment generating function $G_\beta(x)$, eq. \eqref{eq:Gasseries}. This allows to lift the constraint $m_1 + \dots + m_k = n$:
\begin{align}
G_\beta(x) &= \sum_{n=1}^{\infty} \frac{(-e^{\beta x})^n}{n!} \sum_{k=0}^{\infty} \binom{M}{k} \sum_{m_1 + \dots  + m_k = n} n! \prod_{g=1}^k  \frac{M^{\beta^2 m_g^2}}{m_g!} \nonumber \\
           &= \left(1 + \sum_{m=1}^{\infty} \frac{(-e^{\beta x})^m }{m!} M^{\beta^2 m^2} \right)^M 
           \label{eq:gammaREMseries}
\end{align}
Let us compare it to the known exact term in the parenthesis, given in eq. \eqref{eq:gammaREM}:
$$ \int_{\im \R + \epsilon} \frac{\dif p}{2 \pi \im p} e^{-xp + p^2 \ln M} \Gamma(1 + p/\beta) \,.  $$
We see that eq. \eqref{eq:gammaREMseries} is obtained by summing the residues of the Gamma poles in the above equation, with the identification $p / \beta = -m$. The solution $m = 1/ \beta$ in the $\beta < 1$ phase corresponds to $p = - 1$, which is the saddle point that governs the frozen phase, see eq. \eqref{eq:GbetaREMfreeze}. The minimization--maximization inversion (convention \texttt{iv}) and the log-correction now become standard in the saddle point context. The optimization range limit $m \leq 1$ (convention \texttt{iii}) corresponds to the existence of Gamma pole at $p = - \beta$, whose residue will dominate in the $\beta < 1$ phase, see eq. \eqref{eq:GbetaREMhighT}. 
These observations are the starting points of a dictionary between the replica--free approach in section \ref{sec:REM} and the replica approach described above. In \cite{mottishaw15REM}, the authors went much further in this direction in order to calculate finite--size corrections in the \gls{rem}. Extending such a dictionary to \glspl{logrem} is an important open question. The difficulty lies in the fact that the \gls{rsb} approach of \glspl{logrem} is qualitatively distinct from that of the \gls{rem}, as we will see in the next section.
 
\subsection{RSB for logREMs and beyond} \label{sec:RSB1}
Now we come back to  \glspl{logrem}. An important scepticism that should be addressed before proceeding further is the \textit{applicability} of replica symmetry breaking to Euclidean  \glspl{logrem}. Indeed, in spin glass theory, \gls{rsb} has been only convincingly applied to mean field models, and its relevance in finite dimension is at least debatable. The same can be said for directed polymer in random media: no evidence of \gls{rsb} is found in the well--understood $d = 1$ \cite{calabrese2010kpz} case, while the mean field \gls{dpct} model is known since \cite{derrida1988polymers} to be amenable to an \gls{rsb} analysis. 

\subsubsection{Ultra--metricity of logREMs}
\begin{figure}[h]
\center
\includegraphics[scale=.5]{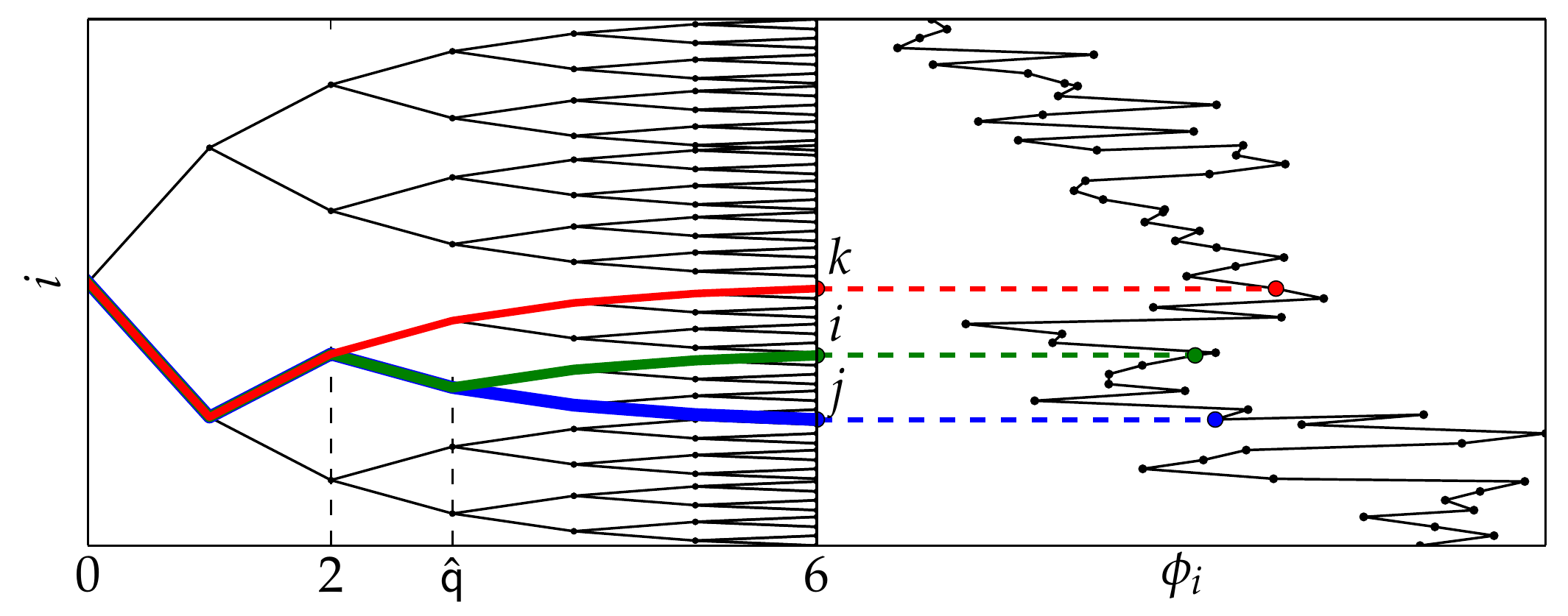}
\caption{Illustration of the ultra--metric triangle inequality eq. \eqref{eq:ultrametcity} for the \gls{dpct} model. Indeed, for the three directed polymers drawn in bold, named $i,j,k$ (as labelled in the figure), we have $ \overlap_{ik} = \overlap_{jk} < \overlap_{ij}$. In the energy landscape in the right panel, $\phi_i$ and $\phi_j$ are indeed more correlated than the other pairs.} \label{fig:ultrametric}
\end{figure}

It is generally accepted that \gls{rsb} is applicable if the Gibbs measure in the thermodynamic limit satisfies the \textit{ultra--metricity}. This notion is most easily explained in the \gls{dpct} model, where it simply refers to the following ``ultra--metric triangle inequality'':
\begin{align}\label{eq:ultrametcity}
\overlap_{ij} \geq  \min(\overlap_{jk}, \overlap_{ik}) \,,\,
\end{align}
for any triple of configurations $i,j$ and $k$. Recall from eq. \eqref{eq:defoverlap} that $\overlap_{ij}$ is proportional to their common length, so it is easy to verify eq. \eqref{eq:ultrametcity} on a Cayley tree (see Figure \ref{fig:ultrametric}). To explain the terminology here, we mention that the \textit{metric} in question is $D = 1 - \overlap$, so that it measures the distance (while $\overlap$ stands for the affinity) between two configurations. In terms of $D$, \eqref{eq:ultrametcity} is equivalent to $D_{ij} \leq \max(D_{jk}, D_{ik})$, which is stronger than the usual triangle inequality $D_{ij} \leq D_{jk} + D_{ik}$. 

Ultra--metric spaces are quite counter--intuitive. To describe how it looks like, let us fix any distance $D$, and consider the ``balls'' of radius $D$ centred at different points, $B(k, D) = \{j: D_{jk} \leq D \}$; using the ultra--metric triangle inequality, it is not hard to show that these balls are either disjoint or identical. In contrast, on the Euclidean space, it is very easy to imagine two intersecting but non-identical balls. Since this is true for any $D$, one concludes that a ball of radius $D$ is in turn made of disjoint balls of radius $D' < D$, and so on. This means that one should think of an ultra--metric space as hierarchical, \textit{i.e.}, its points are the leaves on a tree, with the overlap $\overlap_{ij}$ between $i$ and $j$ being the common length of the simple paths connecting them to the tree root. 

In mean field spin glass theory, the story is more involved than eq. \eqref{eq:ultrametcity}, because the latter is not satisfied for any triple of configurations; it is only true with $\to 1$ probability for any triple of configurations sampled from the Gibbs measure (at finite temperature), in the thermodynamic limit. The emergence of the hierarchical structure is a key discovery in the mean field spin glass theory (\cite{parisi79spinglass,parisi80sk,rammal1986ultrametricity,mezard87beyond}, for mathematical developments, see \textit{e.g.} \cite{panchenko2013parisi,bolthausen13spinglass}). 

For Euclidean  \glspl{logrem}, we claim that eq. \eqref{eq:ultrametcity} is satisfied in the $M \to \infty$ limit, without Gibbs measure sampling. To show this, recall that, the overlap in Euclidean  \glspl{logrem} has the same definition as hierarchical ones:
\begin{align} \overlap_{jk} = \overline{V_{j} V_{k}}^{c} / (2\ln M) = - \frac{d}{\ln M} \ln \abs{\frac{\mathbf{x}_j}{R} - \frac{\mathbf{x}_k}{R}} \,, \label{eq:overlapdis} \end{align}
where we used eq. \eqref{eq:logdecay}; $R$ and $\epsilon$ are  \gls{ir} and \gls{uv} cut-off's so that $(R/\epsilon)^d = M$. Then, for any triple $i,j,k$, we can use the usual triangle identity on the RHS of eq. \eqref{eq:overlapdis}:
\begin{align}
\overlap_{jk} &\geq - \frac{d}{\ln M} \ln \left(   \abs{\frac{\mathbf{x}_i}{R} - \frac{\mathbf{x}_j }{R}}+\abs{\frac{\mathbf{x}_i}{R} - \frac{\mathbf{x}_k}{R}} \right) = -\frac{d}{\ln M} 
\ln \left( M^{-\overlap_{ij} / d } +  M^{-\overlap_{ik}/ d} \right) \nonumber \\
& \stackrel{M\to\infty}{\longrightarrow} -\frac{d}{\ln M}   \ln M^{-\min(\overlap_{ij},\overlap_{ik})/d} = 
\min(\overlap_{ij},\overlap_{ik}) \,, \label{eq:ultrametric-proof}
\end{align}
which is our claim. Intuitively, the above calculation expresses simply the fact that any metric space can be seen as an ultra--metric space if we measure the distance in log--scale, with a diverging base ($R/\epsilon$ here). In the context of general Gaussian potential energy landscapes, the log--decay of the covariance is essential for the overlap defined by eq. \eqref{eq:overlapdis} to be ultra--metric: had we algebraically rough potentials (see section \ref{sec:introPEL}), we would have $\overlap_{ij} \sim 1 - \abs{(\mathbf{x}_i - \mathbf{x}_j)/R}^{\alpha}$, which does not satisfy \eqref{eq:ultrametcity} in any limit. 

Now we have gathered the two characteristics of  \glspl{logrem}: \textit{log--correlation}, quantified by the unifying counting criterion eq. \eqref{eq:logremdef}, and \textit{ultra--metricity}, eq. \eqref{eq:ultrametcity}. They are the theoretical input of the \gls{rsb} analysis of \glspl{logrem} (remark that for the \gls{rem}, ultra--metricity is trivially satisfied). 

\subsubsection{Multi-scale \glspl{logrem}}
We believe that the \gls{rsb} analysis for  \glspl{logrem} is better appreciated in a more general setting, that of \textit{multi--scale}  \glspl{logrem}. Their Euclidean realizations were considered in \cite{fyodorov2007explicit,fyodorov2008multiscale}, see also \cite{cao16order}, the first  appendix, which we follow here. 

To dilute the discussion, let us restrict to a one--dimensional lattice $j = 1, \dots, M$. The multi-scale \gls{logrem}'s covariance is written as follows:
\begin{equation}
\overline{V_{j}V_{k}}^c = - \int_0^1 \dif s \, 2 \rho(s) \ln \left(M^{-s} + \frac{\abs{j-k}}{M} \right) \,. \label{eq:covariancekstep}
\end{equation} 
Namely, it is a linear combination, weighted by a density $\rho(s) \geq 0$, of logarithmic correlations like eq. \eqref{eq:logdecay}, but which are regularized at a scale $\abs{i-j}/M \sim M^{-s}$. \Glspl{logrem} are retrieved with $\rho(s) = \delta(s-1).$  In the $M\to \infty$ limit (with $\abs{j-k} \sim M^{1-s}$, and $s$ fixed in the limit), eq. \eqref{eq:covariancekstep} implies 
\begin{align}
& \overline{V_{j}V_{k}}^c =  2 \overlap(s) \ln M \,,\, s = 1 - \frac{\ln (\abs{j-k}+1)}{\ln M} \label{eq:covmulti}  \\
& \overlap(s) = \int_0^1 \dif s' \, \min(s, s') \rho(s') = \int_0^s \rho(s')s'  \dif s' + s \int_s^1  \rho(s') \dif s' \,. \label{eq:Qofsmultiscale} 
\end{align}
The last equation implies the following general properties of $\overlap(s)$:
\begin{align}
&  \overlap(0) = 0 \,,\ \overlap'(s) \geq 0 \,,\, \overlap''(s) \leq 0 \,. \label{eq:Qofsproperties}
\end{align}
In addition, we fix the normalization as usual:
\begin{equation}
\overline{V_{j}^2} = 2 \ln M + o(\ln M)  \Leftrightarrow \int_0^1 \rho(s')s'  \dif s' = 1 \,. \label{eq:normmultiscale}
\end{equation}
Using the overlap definition eq. \eqref{eq:overlapdis}, and eq. \eqref{eq:covmulti}, we can show that multi--scale  \glspl{logrem} satisfy the following counting criterion generalizing eq. \eqref{eq:logremdef}:
\begin{align}
& \abs{\{ \, j \,: \overlap(j,k) > \overlap \}} \sim M^{1 - \Psi(\overlap)},\,\,  \overlap \in [0,1]\,,\label{eq:logremdeffull} \,,\, \Psi(\overlap(s))  =  s \,,\,
\end{align}
\textit{i.e.}, $\Psi(\overlap)$ is the inverse function of $\overlap(s)$ in eq. \eqref{eq:Qofsmultiscale}. Therefore, eq. \eqref{eq:Qofsproperties} and eq. \eqref{eq:normmultiscale} imply the following properties for  $\Psi(\overlap)$: 
\begin{align}
\Psi(0) = 0,\, \Psi(1) = 1, \quad \Psi'(\overlap) \geq 0 \,,\, \Psi''(\overlap)\geq0 \,. \label{eq:conditionPsiQ}
\end{align}
In particular,  \glspl{logrem} are retrieved by setting $\Psi(\overlap) = \overlap.$ 

The multi--scale \glspl{logrem} satisfy ultra--metricity, with the overlap definition $$\overlap_{jk} = \overline{V_{j} V_{k}}^{c} / (2\ln M) \,,$$ identical to eq. \eqref{eq:overlapdis}. The demonstration can be done along the same line as eq. \eqref{eq:ultrametric-proof}. Indeed, denoting $s_{ij} = 1 - \ln (\abs{i-j} + 1)/\ln M$, the triangle inequality implies $s_{jk} \geq 1 -  \ln (\abs{i-j} + \abs{j-k} + 1)/\ln M = 1 - \max(\ln (\abs{i-j} + 1)/\ln M, \ln (\abs{i-k} + 1)/\ln M) = \min (s_{ij}, s_{jk})$. Since $\overlap_{ij} = \overlap(s_{ij})$ is an increasing function of $s_{ij}$, we have the ultra--metricity inequality $\overlap_{jk} \geq \max(\overlap_{ij}, \overlap_{ik})$. Therefore, the \gls{rsb} analysis can be applied also to multi--scale  \glspl{logrem} defined on finite dimensional Euclidean spaces. This is quite remarkable with regard to the common belief that \gls{rsb} analysis would be only valid in the infinite-dimensional limit. In the context of generalized \glspl{logrem}, this restriction was suggested by the authors of \cite{fyodorov2008multiscale,fyodorov2007explicit,fyosom2007}. Here, in virtue of the arguments presented above, we propose the hypothesis that for (multi--scale) \glspl{logrem}, the \gls{rsb} analysis can be applied in \textit{any} dimension. 

\subsubsection{Full \gls{rsb} analysis}
The basic object of \gls{rsb} analysis in general is the replicated partition function $\overline{\mathcal{Z}^n}$, written as a sum over $n$ replica positions $j_a, a = 1, \dots, n$. It is common to organize the configurations by their \textit{overlap matrix}
\begin{equation}
\left[ \Overlap_{ab} \right]_{1 \leq a,b \leq n} \,,\, \Overlap_{ab} = \overlap_{j_a, j_b} = \frac{\overline{V_{j_a} V_{j_b}}}{2 \ln M} \,. \label{eq:overlapFull}
\end{equation}
Now we can write the replicated partition sum as an integral over the values of the overlap matrix:
\begin{align}
\overline{\mathcal{Z}^n} =\sum_{(j_a)}
 \overline{\exp\left( - \sum_{a} \beta V_{j_a} \right)} &=  
 \int  \prod_{a < b} \dif \Overlap_{ab} \exp(\ln M E[\Overlap] + \ln M S[\Overlap])   \nonumber  \,, \\
\text{ where } S[(\overlap_{ab})] &= \frac{1}{\ln M}  \ln \left[\sum_{(j_a)} \prod_{a<b} \delta\left( \overlap_{j_a, j_b} - \Overlap_{ab}  \right)\right] \,, \\
\text{ and } E[(\Overlap_{ab})] &=\left. \frac{1}{\ln M} \, \ln  \overline{\exp\left( - \sum_{a} \beta V_{j_a} \right)}\right\vert_{\overlap_{j_a, j_b} = \Overlap_{ab}}  \nonumber  \\
 & =  \frac{1}{\ln M} \, \ln  \exp\left( \beta^2 \ln M \sum_{a,b=1}^n  \Overlap_{ab} \right) \nonumber \\
 & =  \beta^2 \sum_{a,b=1}^n \Overlap_{ab} \,,\label{eq:EofQ0}
\end{align}
where we used eq. \eqref{eq:overlapFull} and the Wick theorem eq. \eqref{eq:Wick} (with $\overline{V_j} = 0$) in the calculation of $E[(\Overlap_{ab})]$. The notations $S$ and $E$ stand for ``entropy'' and ``energy'' respectively. By defining the ``Hamiltonian'' as the sum of the two contributions,
\begin{equation}
H[(\Overlap_{ab})] = -E[(\Overlap_{ab})] - S[(\Overlap_{ab})] \,, \label{eq:defHofES}
\end{equation}
we can write formally $\overline{\mathcal{Z}^n}$ as an integral over matrices, and apply the saddle point approximation to it:
\begin{align}
\overline{\mathcal{Z}^n} = \int \prod_{a < b} \dif \Overlap_{ab}  \exp(-\ln M H[(\Overlap_{ab})])
 \approx  \min_{\Overlap_{ab}, a<b} \exp(-\ln M H[(\Overlap_{ab})]) \,, \label{eq:ZnHogQmin}
  \end{align}
Note also that $H[(\Overlap_{ab})]$ describes the leading $M$-dependence of $\overline{\mathcal{Z}^n}$: only this part is necessary to determine the thermodynamics of the model. The integral in eq. \eqref{eq:ZnHogQmin} is over all symmetric matrices $\Overlap_{ab} = \Overlap_{ba}$. In the $M\to \infty$ limit, the normalization eq. \eqref{eq:normmultiscale} constraints the diagonal elements $\Overlap_{aa} = 1, a = 1, \dots, n$. To obtain the thermodynamics, it suffices to apply the saddle point approximation at the leading order and optimize $H[(\Overlap_{ab})]$ with respect to $\Overlap_{ab}$. As we have seen for the \gls{rem}, the optimization should be done in the formal $n \to 0$ limit and is in fact a minimization of $\mathcal{Z^n}$ (maximization of $H$). 

Important constraints on the overlap matrix are imposed by the ultra--metricity that holds in the $M\to \infty$ limit. In fact, in replica theory, ultra--metricity imposes that the overlap matrix assumes the Parisi Ansatz \cite{parisi79spinglass} of \textit{full} \gls{rsb}, which is a  generalization of the one-step \gls{rsb} applied to the \gls{rem}: the $n$ replicas form groups which are themselves divided into sub-groups, \textit{etc}. The group size depends only on the radius $D = 1 - \overlap$ of the ultra-metric ball that it occupies. As a consequence, up to a permutation of the indices $a = 1, \dots, n$, the overlap matrix is determined by a function
\begin{align}
 m(\overlap) = \abs{\{ b: \Overlap_{ab} \geq  \overlap \}} \,,\, \overlap \in [0,1] \,. \label{eq:mofqdef}
\end{align}
That is, $m(\overlap)$ is the size of groups that occupy (each) a ultra-metric ball of radius $D = 1 - q$. The above equation holds for any fixed replica $a$. In the $n \to 0$ limit, similarly to the ``order inversion'' ($n \leq m \leq 1$) that we have seen in the \gls{rem}, the group size \textit{increases} with $\overlap$, \textit{i.e.}, 
\begin{equation} 
1 \geq m(\overlap') \geq m(\overlap) \geq n \to 0 \,,\, 1 \geq \overlap' > \overlap \geq 0 \,.
\end{equation}
Since $\Overlap$ is determined by $m(\overlap)$ (up to permutation), the ``Hamiltonian'' in eq. \eqref{eq:ZnHogQmin} is also a functional of $m(\overlap): [0,1] \to [0,1]$. Let us  consider its explicit form. According to eq. \eqref{eq:defHofES}, the task can be divided into two parts, 
\begin{align}
H[m(\overlap)] &=  -E[m(\overlap)] - S[m(\overlap)]  \,. \label{eq:defHofES2}
\end{align} 
The energy part follows from eq. \eqref{eq:EofQ0}: 
\begin{align}
E[m(\overlap)] &= \beta^2 \sum_{a,b=1}^n \overlap_{a,b}  =  \beta^2 \int_0^1  \dif \overlap \sum_{a,b=1}^n \theta(\Overlap_{ab} - \overlap) =   \beta^2  n \int_0^1  m(\overlap)   \dif \overlap \,. \label{eq:EofQ2}
\end{align}
Note that it does not depend on $\Psi(\overlap)$, he characterizing function of the multi-scale \gls{logrem}, eq. \eqref{eq:logremdeffull}. On the other hand, $\Psi(\overlap)$ will affect The other, entropy, term $S[m(\overlap)]$, which comes from the sum over replica positions:
\begin{align}
 S[m(\overlap)] = -\frac{1}{\ln M} \ln \sum_{(j_a)} \prod_{a < b} \delta(\overlap(j_a, j_b) - \Overlap_{ab}) = - \int_0^1  \Psi'(\overlap) \frac{n}{m(\overlap)} \dif \overlap  \,, 
 \label{eq:SQ}
\end{align}
A brief explanation of the above identity is as follows (for more detailed explanation, see \cite{cao16order}, appendix A.3). For each scale $\overlap$, the number of groups corresponding to it is given by the total number of replicas over the groups size, $n / m(\overlap)$. For each such group, its position is determined up to a ball of radius $D = 1 - \overlap$, but is contained in another ball of radius $D = 1 - \overlap + \dif \overlap$; so, by eq. \eqref{eq:logremdeffull}, its entropy is $\ln M (\Psi(\overlap) - \Psi(\overlap - \dif \overlap)) = \ln M \Psi'(\overlap) \dif \overlap$. Summing over all $\overlap \in [0,1]$ gives eq. \eqref{eq:SQ}. 

Summarizing eq. \eqref{eq:EofQ2}, \eqref{eq:SQ} and \eqref{eq:defHofES2}, we have the following expression for $H[m(\overlap)]$:
\begin{align}
H[m(\overlap)] = - n \int_0^1 \left( \beta^2 m(\overlap) + \Psi'(\overlap) m(\overlap)^{-1} \right)  \dif \overlap \,. \label{eq:HQsimple}
\end{align}
Therefore, the maximization problem is quite trivial: one has only to remember $m(\overlap) \in [0,1]$, to get the following result:
\begin{equation}
m(\overlap) = \begin{dcases} \beta^{-1} \sqrt{\Psi(\overlap)} \,, &  \Psi'(\overlap) \leq \beta^2   \\
 1  \,,  & \Psi'(\overlap) > \beta^2 \,. \end{dcases}
\label{eq:moptimum1gen}
\end{equation}
Note the convexity of $\Psi(\overlap)$, eq. \eqref{eq:conditionPsiQ}, guarantees that the above solution is increasing. Plugging into eq. \eqref{eq:ZnHogQmin} (in the $n \to 0$ limit), we have the leading free energy
\begin{equation}
\frac{\overline{\mathcal{F}}}{\ln M} \longrightarrow  \lim_{n\to 0} \frac{1}{n \beta} H[m(\overlap)] = 
- \int_0^{1} \left(\beta m(\overlap)  + (\beta m(\overlap))^{-1} \Psi'(\overlap) \right)  \dif \overlap \,. \label{eq:FleadingMulti}
\end{equation}

A physical interpretation of $m(\overlap)$ is well known in terms of the \textit{overlap distribution} of two independent thermal particles in one random potential:
\begin{align}
& P(\overlap) \defeq \overline{\sum_{j,k} p_{\beta,j} p_{\beta, k} \delta(\overlap(j,k) - \overlap) } \,, \text{ where } p_{\beta,j} = \mathcal{Z}^{-1} e^{-\beta V_j} 
\end{align}
is the normalid Gibbs measure. $P(\overlap)$ turns out equal to the derivative of $m(\overlap)$:
\begin{equation}
P(\overlap) = m'(\overlap) \,,\, \overlap \in [0,1] \,. \label{eq:Pismprime}
\end{equation}
This result is quite standard in the replica theory \cite{mezard87beyond,dotsenko1995introduction}. Note that, to apply eq. \eqref{eq:Pismprime} correctly at $\overlap = 0$ and $1$, one has to assume that $m(q < 0) = 0$ and $m(q > 1) = 1$, so that $P(\overlap)$ is supported in the interval $[0,1]$ and is correctly normalized: $\int P(\overlap) \dif \overlap = 1.$ 

\begin{figure}
\center \includegraphics[scale=.7]{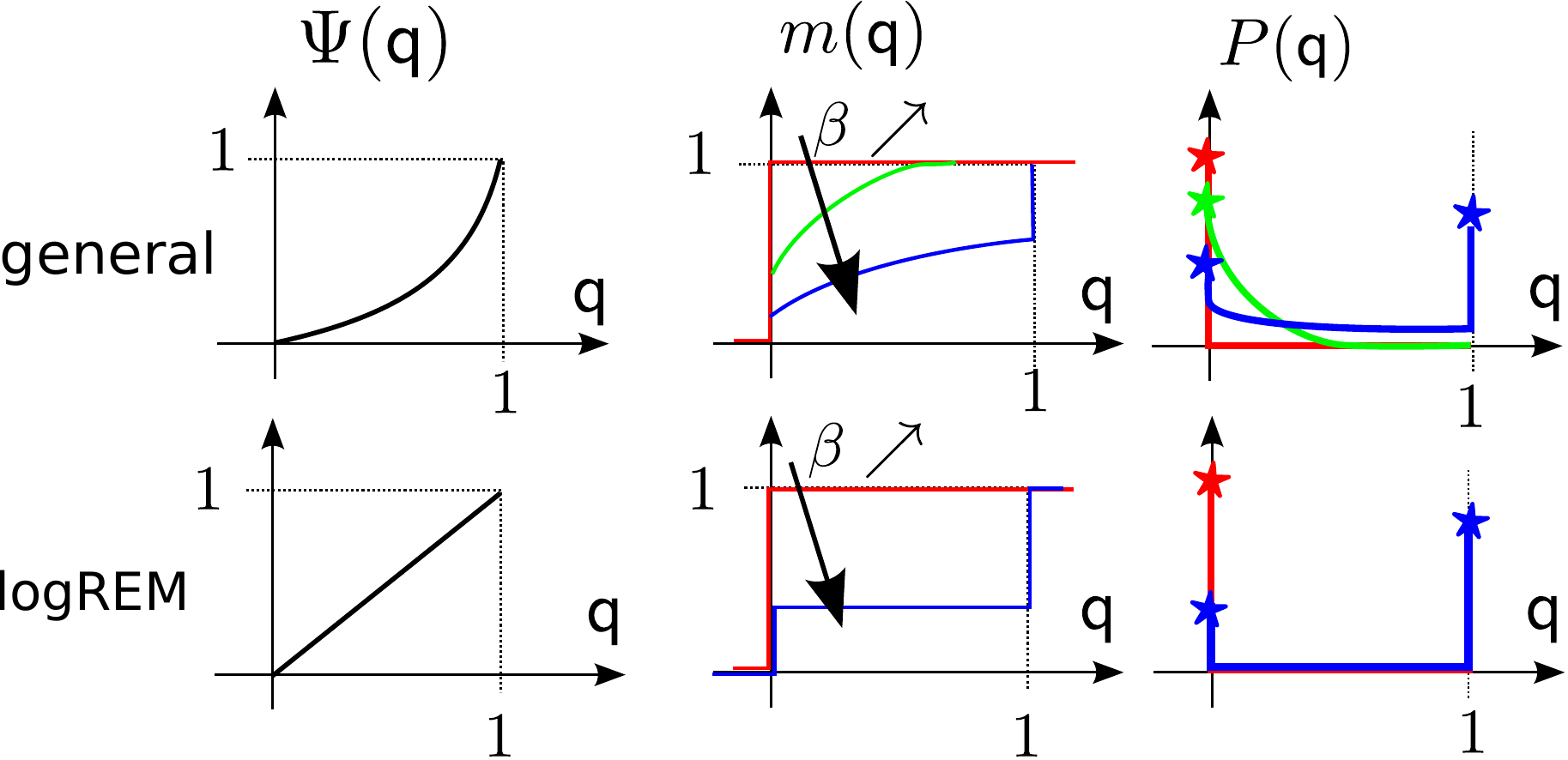}
\caption{A sketch of the functions $\Psi(\overlap)$ (left), $m(\overlap)$ (centre) and $P(\overlap)$ (right) of a generic multi--scale \gls{logrem} (top) and a \gls{logrem} (bottom). In the right panel, the stars represent delta peaks which generically occur at $\overlap = 0, 1$ For a generic multi--scale \gls{logrem}, $\Psi(\overlap)$ satisfies eq. \eqref{eq:conditionPsiQ}, and is a convex function. Then, by eq. \eqref{eq:moptimum1gen}, $m(\overlap) = 1$ for $\overlap \in (0,1)$  in the high temperature phase $\beta<\sqrt{\Psi'(0)}$, and $P(\overlap) = m'(\overlap) = \delta(\overlap)$ by eq. \eqref{eq:Pismprime} (red curves). After an interval of continuous phase transitions $\beta \in (\sqrt{\Psi'(0)},\sqrt{\Psi'(1)})$ (green curves), the system enters the frozen phase $\beta > \sqrt{\Psi'(1)}$ characterized by  a delta peak of $P(\overlap)$ at $\overlap = 1$. For the \gls{logrem}, $\Psi(\overlap) = \overlap$ is linear; as a consequence, the interval of continuous phase transitions shrinks to a point $\beta = \beta_c = 1$, and the overlap distribution  in the $\beta > 1$ is the sum of two deltas, $P(\overlap) = \beta^{-1} \delta(\overlap) + (1 - \beta^{-1}) \delta(\overlap - 1)$ (eq. \eqref{eq:PoverlaplogREM}).}\label{fig:RSB}
\end{figure} 

In light of this relation, let us look at the solution eq. \eqref{eq:moptimum1gen}; we refer to Figure \ref{fig:RSB} (upper row) for illustration. At high enough temperature $\beta < \sqrt{\Psi'(\overlap)}$, $m(\overlap) = 1$ for any $\overlap \in (0, 1)$. This is the replica symmetry phase: the $n$ replicas do not form non-trivial groups, and $P(\overlap) = \delta(\overlap)$, \textit{i.e.}, the overlap between two thermal particles is always ro (in the $M\to \infty$ limit), as if the random potential were non-existent. When the system is cooled down, the thermal particles become caged in smaller portions of the system, and two thermal particles can be in the same potential well, making the $P(\overlap)$ become non-ro for $\overlap > 0$. Generically, to each $\overlap$ corresponds a critical temperature $\beta_\overlap = \sqrt{\Psi(\overlap)}$ at which $m(\overlap)$ is non-analytical. The existence of a continuum of critical temperatures is a general feature of full \gls{rsb}. Finally, at low enough temperature, $\beta > \sqrt{\Psi'(1)}$ (whenever $\sqrt{\Psi'(1)} < \infty$), there is no more transitions, the thermal particles are caged into the deepest potential wells. This low temperature phase is a frozen phase because the free energy is also $\beta$--independent. Indeed eq. \eqref{eq:FleadingMulti} and \eqref{eq:moptimum1gen} (in which $m(\overlap)  =  \beta^{-1} \sqrt{\Psi(\overlap)}$ for all $\overlap$) imply that 
\begin{equation}
\frac{\overline{\mathcal{F}}}{\ln M} =2 \int_0^1  \sqrt{\Psi'(\overlap)}   \dif \overlap \,.
\end{equation}
This is the generalization (to multi--scale \glspl{logrem}) of the frozen ($\beta > 1$) phase of \glspl{logrem}, to which we come back below. 

\subsubsection{Back to \glspl{logrem}}
Now let us restrict to the \glspl{logrem} case, $\Psi(\overlap) = \overlap$, so $\Psi'(\overlap) = 1$ for $\overlap \in [0,1]$ (refer to Figure \ref{fig:RSB}, bottom row, for illustration).  Thus, eq. \eqref{eq:moptimum1gen} simplifies to 
\begin{equation}
\forall \overlap \in (0,1) \,,\, m(\overlap) = m =  \begin{dcases} 1 \,,\, &  \beta \leq 1 \,, \\
 \beta^{-1} \,,& \beta > 1 \,.\end{dcases} \label{eq:moptimum}
\end{equation}
In the context of replica trick, the \gls{logrem} is usually referred to as \acrfull{1rsb}, because if eq. \eqref{eq:moptimum} and eq. \eqref{eq:PoverlaplogREM}.  However, the \gls{1rsb} of \gls{logrem} should be distinguished from that of the \gls{rem}, as we will further comment below.  

By eq. \eqref{eq:Pismprime}, eq. \eqref{eq:moptimum} means that the overlap distribution obeys the ``zero--one'' law:
\begin{equation}
P(\overlap) = \begin{dcases}
\delta(\overlap)\,,\, &  \beta < 1 \,, \\
\beta^{-1} \delta (\overlap) + (1 - \beta^{-1}) \delta(1 - \overlap) \,, & \beta \geq 1 \,.
\end{dcases} \label{eq:PoverlaplogREM}
\end{equation}
Recalling the relation between the overlap and the (Euclidean) distance (eq. \eqref{eq:overlapdis}), this means that two thermal particles in a log--correlated potential are either of system size scale ($\overlap = 0$), or of lattice spacing scale ($\overlap = 1$); the $\beta > 1$ phase is characterized by the non-vanishing probability of the latter case.

\begin{figure}[h]
\center \includegraphics[scale=.5]{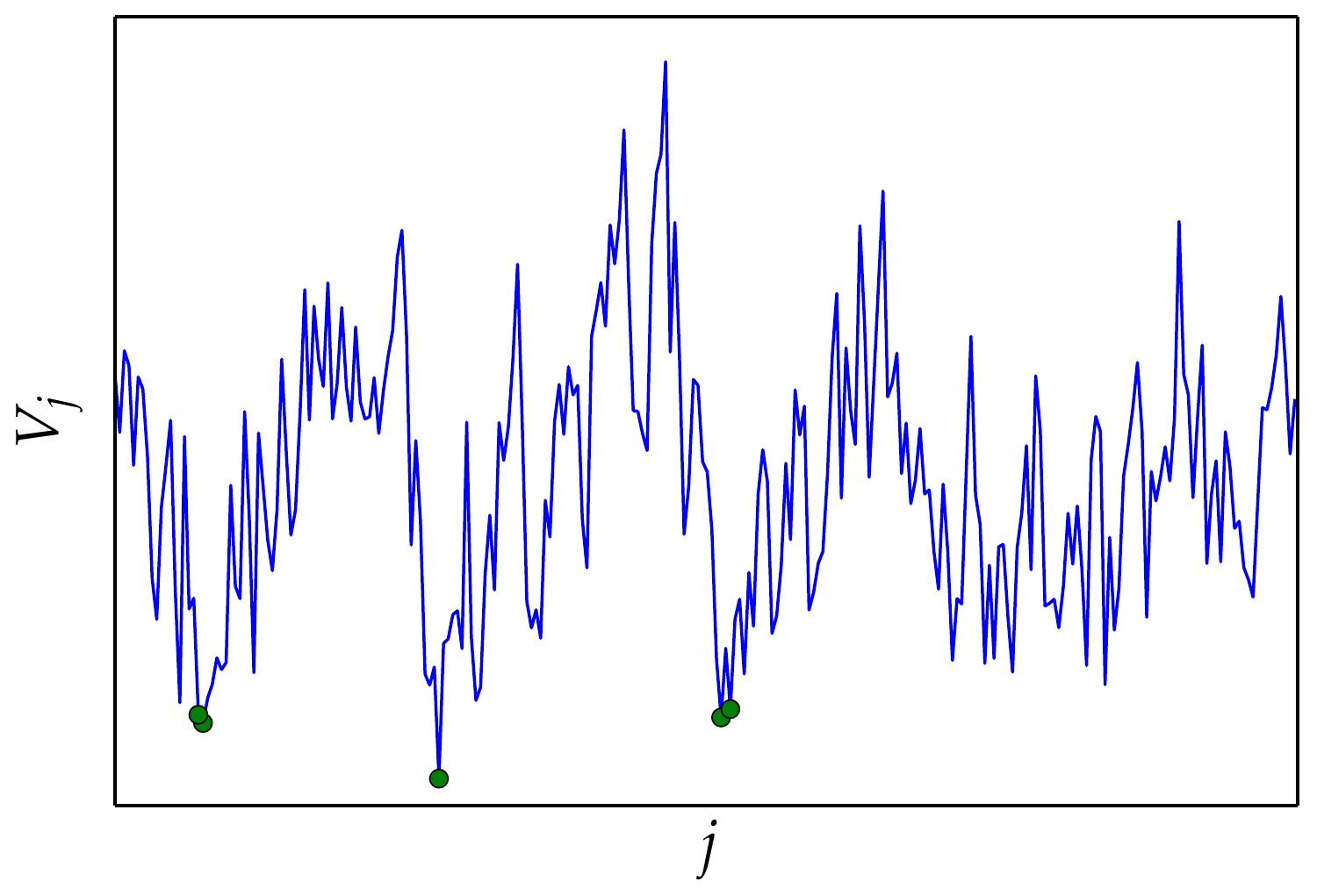}
\caption{The position (horizontal axis) and value (vertical axis) of the five deepest minima of a sample of the circular model. The clustering structure is apparent. } \label{fig:minimaposlogREM}
\end{figure}

A simple illustration of the above picture can be provided by the positions of deepest minima in a log-correlated potential, see Figure \ref{fig:minimaposlogREM} for a 1D example. We can see that they form clusters of lattice spacing size, while different clusters are separated by a system-scale distance. It is intuitive but not quantitatively clear how the overlap is related with minima positions; this will be treated in section \ref{sec:orderstat}.

Finally it is important to compare the \glspl{logrem} to the \gls{rem} in light of the analysis we just provided. Indeed, The formula of $m$ is identical to the \gls{rem} one, eq. \eqref{eq:moptimumREM}. The interpretation in terms of replica grouping is also similar: replicas do not form groups in the $\beta < 1$ (replica symmetry) phase, and form groups of size $m = 1/\beta$ (in the $n \to0$ limit) in the $\beta > 1$ phase. Each of such groups occupies a region of \gls{uv} cut--off ($\overlap = 1$) scale. The similarity of \glspl{logrem}'s and the \gls{rem}'s \gls{rsb} solution is not very surprising knowing the identity of their thermodynamics. Both of them are commonly referred to as \acrfull{1rsb} models. However, the \gls{1rsb} of \glspl{logrem} is more subtle than that of \gls{rem}. For the \gls{rem}, the overlap $\overlap_{ij} = \delta_{ij} \in \set{0,1}$ \textit{a priori}; while for the \gls{logrem}, $\overlap_{ij}$ can take any value $\in [0,1]$, and the zero-one law \eqref{eq:PoverlaplogREM} is achieved in the thermodynamics limit, and with respect to Gibbs measure sampling. Also, note that we retrieve the unique critical temperature $\beta_c = 1$ of \glspl{logrem} because for all $\overlap$, $\beta_{\overlap} = 1$, so the interval of critical temperatures shrinks to a single point. Therefore, the \gls{1rsb} of \glspl{logrem} is a \textit{degenerate} case of the \textit{full} \gls{rsb} solutions of multi-scale  \glspl{logrem}. This point has been also emphasized in another \gls{rsb} analysis of \glspl{logrem} in \cite{fyosom2007} (see also \cite{FyoLec2010}): these works revealed further the marginally stable modes generally associated to full \gls{rsb} solutions.  
 
\subsection{Freezing by 1RSB} \label{sec:freezingRSB}
The \gls{rsb} analysis of the previous section was performed at a thermodynamics level. The goal of this section is to go further and discuss how the full distribution of the free energy, and in particular the freezing scenario for \glspl{logrem}, can be understood using the replica approach.  We should use the \gls{1rsb} picture of \glspl{logrem} just obtained, and recast it in a form that allows us to take into account the \gls{ir} and \gls{uv} data (section \ref{sec:IRUV}) of  \glspl{logrem}. They were totally ignored in the previous \gls{rsb} analysis and now becomes crucial because the two extremal distances are the ones that the replicas have between one another. Such a \gls{1rsb} analysis of \glspl{logrem} looks quite different from the standard formalism. It first appeared in \cite{fyodorov2010freezing}, and was worked out in greater detail in \cite{cao16order}, section III.
\begin{figure}[h]
\center
\includegraphics[scale=.6]{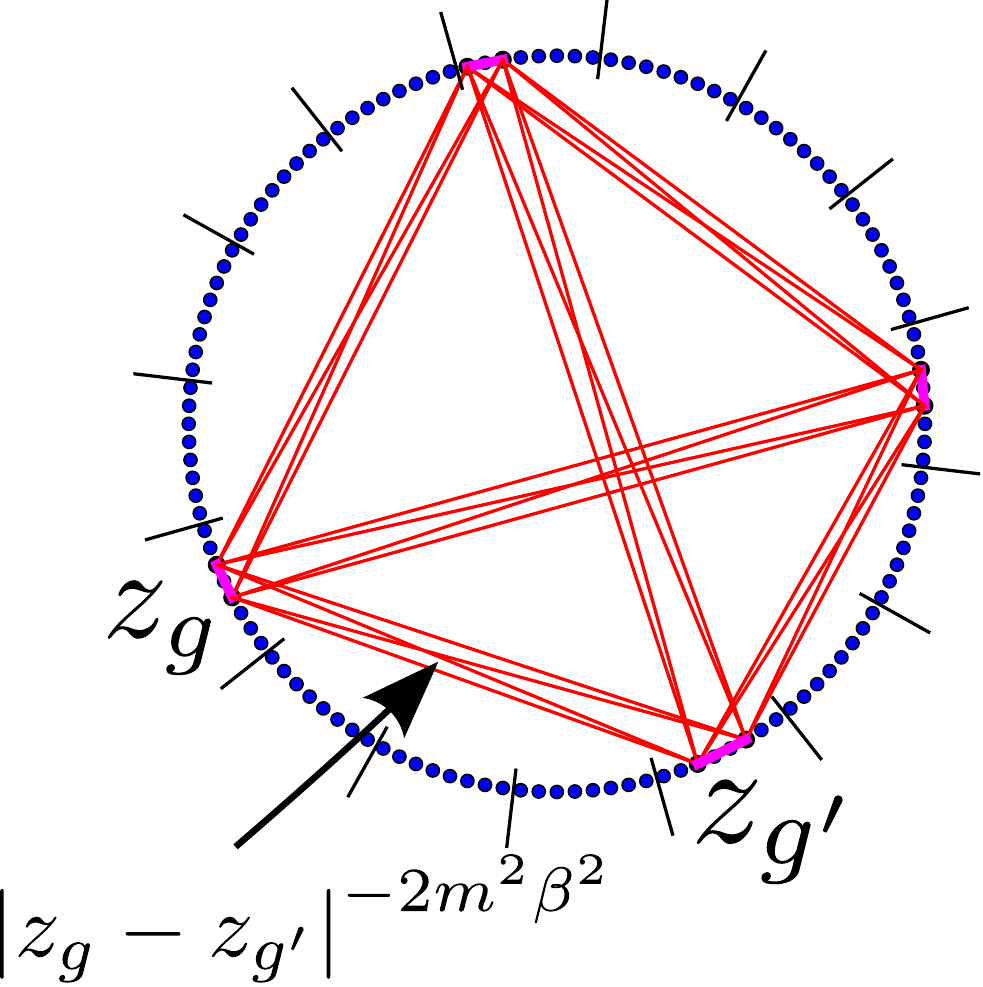}
\includegraphics[scale=.6]{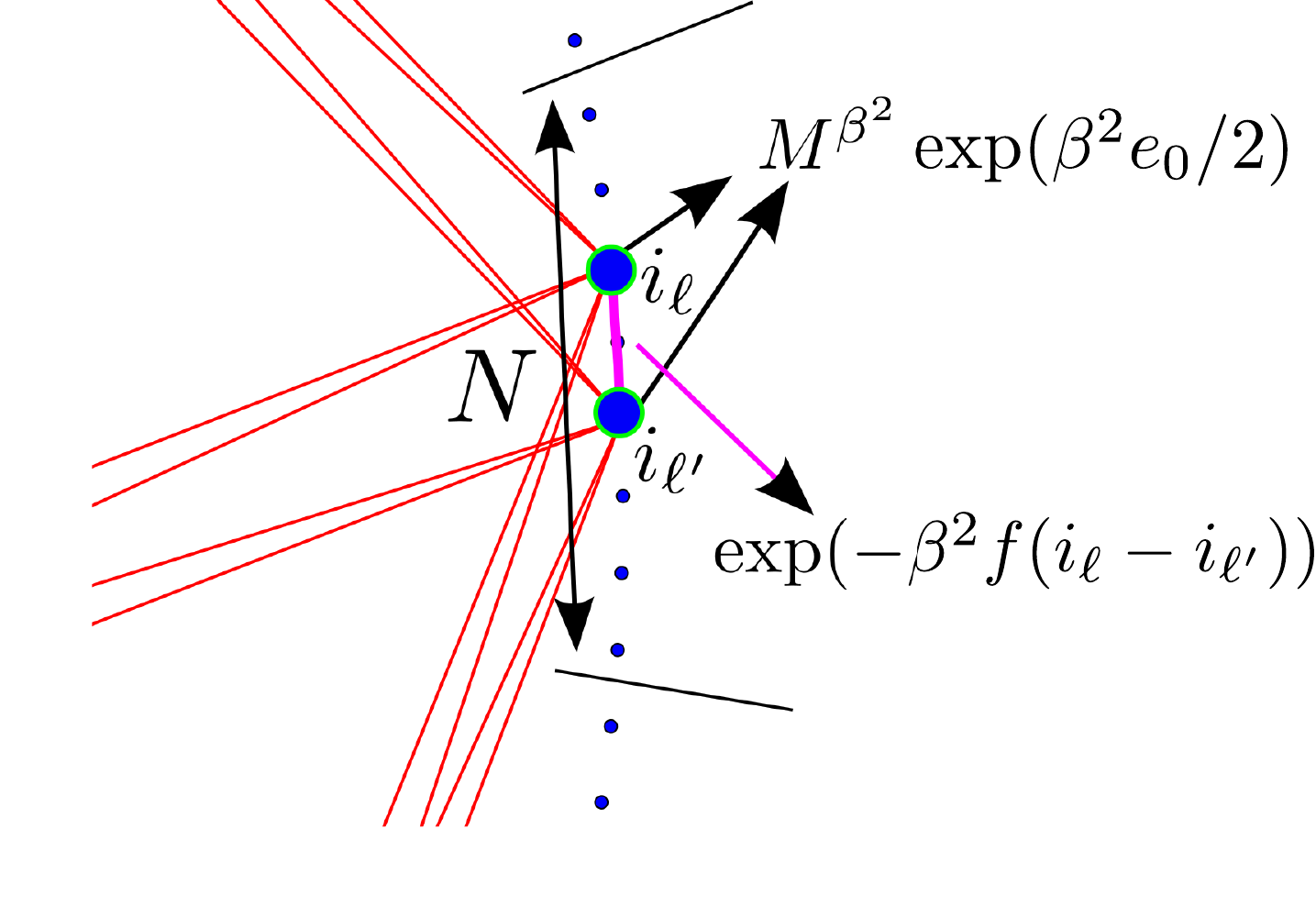}
\caption{An illustration of the \gls{1rsb} Ansatz. The circular model is divided into $M/N = 16$ blocks. The $n = 8$ replicas partitioned into groups of size $m = 2$. The right figure zooms into the block occupied by one group. The lines indicate the Wick contractions. The red (\gls{ir}) ones should be calculated using eq. \eqref{eq:IRdata1RSB} and the violet (\gls{uv}) ones should be calculated with eq. \eqref{eq:UVdata1RSB}. See also Figure \ref{fig:uvir}.} \label{fig:rsbcircle}
\end{figure}

In this section we explain that the above \gls{1rsb} Ansatz implies the freezing scenario in \gls{logrem}.
For the sake of concreteness, let us still take the circular model of $1/f$-noise as example, see section \ref{sec:IRUV} for definition. The basic idea of replica grouping is depicted in Figure \ref{fig:rsbcircle}. We divide the system into \textit{blocks} of $N$ sites, where $N$ is any intermediate scale such that both $N$ and $M / N$ go to $+\infty$ in the thermodynamic ($M\to \infty$) limit. Each block will be labelled by their position $\theta\in [0, 2\pi)$, or $z = e^{\im \theta}$, since all the lattice points in a same block will converge to the same continuum position as $M/N \to \infty$; positions inside a block will be labelled $1, \dots, N$; in other words, we have a one-to-one correspondence $j(\theta, i)$ between the ``global position'' $j$ and the hierarchical one $(\theta, i)$.

Now, in the replicated partition function
$$  \sum_{j_1, \dots, j_n} \overline{\exp\left( -\beta \sum_{a = 1}^n V_{j_a} \right)}  \,,$$
the $n$ replicas form $n/m$ groups of size $m$; when $n$ is continued to complex values we will have $m = \min(1, 1/\beta)$ by eq. \eqref{eq:moptimum}. Each group will occupy a block $\theta_1, \dots, \theta_{n/m}$, while distinct groups will be separated by system-size scale distances. This justifies that the sum over group positions can be replaced by continuum integral in the $M\to \infty$ limit. In addition to that, we need to sum over the ``micro-positions'' of the replicas their respective blocks. At last, one should not forget the combinatorial factor ($C_{n,m}$ below, see eq. \eqref{eq:REMZnreplica}) corresponding to assigning the $n$ replicas into $m$ (\textit{distinguishable}) groups. In summary, the sum over replica positions is rewritten as follows:
\begin{align}
&  \sum_{j_1, \dots, j_n} [\dots] \longrightarrow  C_{n,m}  \left( \frac{M}{N} \right)^{\frac{n}{m}} \int_{0}^{2\pi} \prod_{g = 1}^{n/m}\frac{\dif \theta_g}{2\pi} \sum_{i_1, \dots, i_n} [\dots] \,,\, \label{eq:sum2integral} \\
&  C_{n,m} = \frac{\Gamma(1 + n)}{\Gamma(1 + m)^{n/m} \Gamma(1 + n/m)} \,,\, \label{eq:Cnm}
\end{align}
where $[\dots]$ denotes the disorder--averaged Boltzmann weights, which, by Wick theorem, is a product of ``contractions'' $\exp\left(\beta^2 \overline{V_{j_a} V_{j_b}} \right), a,b=1, \dots, n$. We will also split them into two types, according to whether $a,b$ are in the same group or not. If they are not, we should use the  \gls{ir} limit of the $1/f$-noise circular model (see eq. \eqref{eq:circularIR}): 
 \begin{equation}   \overline{V_{j_a} V_{j_b}} \longrightarrow 2 \ln \abs{z_{g(a)} - z_{g(b)}} \,, \label{eq:IRdata1RSB} \end{equation}
where $g(a)$ is the group to which $a$ belongs. When $a,b$ are in same group, we use the \gls{uv} data (see eq. \eqref{eq:variancelogrem} and \eqref{eq:UVlimit}):
\begin{equation}
 \overline{V_{j_a} V_{j_b}} \longrightarrow 2 \ln M + e_0 - f(\abs{i_a - i_b}) \,.  \label{eq:UVdata1RSB}
\end{equation}
Collecting the equations eq. \eqref{eq:sum2integral} through \eqref{eq:UVdata1RSB}, we have the following: 
\begin{align}
&\overline{\mathcal{Z}^n} \, \rightarrow \,  M^{\frac{n}{m}(\beta^2 m^2 + 1)}  C_{n,m} \,  \overline{Z_{m\beta}^{n/m}} \, E_{m, \beta}^{\frac{n}{m}}  \label{eq:ZnRSB}  \\
&\overline{Z^{k}_b}  = \int_0^{2\pi} \prod_{g = 1}^{k} \left[ \frac{\dif \theta_g}{2\pi}  \right] \prod_{g < g'} \abs{e^{\im \theta_g} - e^{\im \theta_{g'}} }^{-b^2} = \frac{\Gamma(1 - k b^2)}{\Gamma(1 - b^2)^k} \label{eq:DysonRSB}\\
&E_{m,\beta} =  e^{\beta^2 m^2 e_0/2} \frac{1}{N} \sum_{i_1, \dots, i_m = 1}^N  \prod_{l,l' = 1}^m \exp(- \beta^2 f(i_l-i_{l'}) ) \,. \label{eq:Exi}
\end{align}
Here, $E_{m,\beta}$ gathers the sum over inter-block positions and the intra--group contractions (using eq. \eqref{eq:UVdata1RSB}); while $\overline{Z^k_b}$ is the Dyson integral (see eq. \eqref{eq:Dyson}) that results from the  \gls{ir} limit data eq. \eqref{eq:IRdata1RSB} and continuum integrals in eq. \eqref{eq:sum2integral}. It is continued to generic $k$ and $b$ Note that the Dyson integral appears in eq. \eqref{eq:ZnRSB} with ``renormalized'' temperature and number of charge numbers $b = \beta m, k = n / m$, as a result of grouping of replicas. Now let us discuss the result in the two phases. 

\subsubsection{High temperature phase}
In the $\beta < 1$ phase, by \eqref{eq:moptimum}, $m = 1$, so $b = \beta, k =  n / m = n$, and $C_{n,m} = 1$. $E_{m,\beta}$ becomes quite easy to evaluate: $E_{1,\beta}  =  e^{\beta^2 e_0 / 2}$. With all that, eq. \eqref{eq:ZnRSB} implies 
\begin{align}
\overline{\mathcal{Z}^n} =&  \overline{Z_{\beta}^{n}}  \left( M^{\beta^2 + 1} e^{\beta^2 e_0 / 2} \right)^n 
\Leftrightarrow \overline{\exp(t \mathcal{F})} = e^{t F_0} \Gamma(1 + t \beta)  \label{eq:etF1RSBhighT}
\end{align}
where the second equation is obtained with the change of variable $n = -t/\beta$ (since $\mathcal{F} = -\beta^{-1} \ln \mathcal{Z}$), and the first moment shift part of the free energy is:
\begin{align}
F_0 =& -\left(\beta + \beta^{-1}\right) \ln M - \beta e_0 / 2 + \beta^{-1} \ln \Gamma(1 - \beta^2) \,, \beta < 1 \,. \label{eq:FshifthighT}
\end{align}
The last factor comes from the denominator $\Gamma(1 - \beta^2)^n$ in the Dyson integral eq. \eqref{eq:DysonRSB}. Equation \eqref{eq:etF1RSBhighT} means the re-shifted free energy $f = \mathcal{F} - F$ satisfies $\overline{\exp(tf)} = \Gamma(1 + t\beta)$, \textit{i.e.}, it has the same distribution as $-\beta$ times a Gumbel variable. This determines completely the free energy distribution in the $M \to \infty$ limit. The result confirms what we have found in section \ref{sec:treetoplane}, eq. \eqref{eq:etfFB}. 

\subsubsection{Low temperature phase: freezing of distribution}
The $\beta > 1$ case is more interesting because it is directly related to the freezing scenario; it also involves a few subtleties that are not completely understood. 
 
Let us begin with the combinatorial factor $C_{n,m}$. By analogy to what we have seen in the \gls{rem}, when $n$ and $m$ become non-integer, they should be ``continued'' by applying the rule eq. \eqref{eq:Gammawierd}:
\begin{align}
C_{n,m} = \frac{\Gamma(1 + n)}{\Gamma(1 + m)^{\frac{n}{m}}\Gamma\left(1 - \frac{n}{m}\right)}  
\leadsto \frac{\Gamma(1 - m)^{\frac{n}{m}} \Gamma\left(1 + \frac{n}{m} \right)}{\Gamma(1 - n)} \,.
\end{align}
This manipulation has no satisfactory justification to the best of our knowledge. Carrying it out in eq. \eqref{eq:ZnRSB}, and recalling that in this phase, eq. \eqref{eq:moptimum} implies $m = 1/ \beta$, so $b = 1, n / m = n \beta$, we obtain:
\begin{align}
\overline{\mathcal{Z}^n} &= M^{2n} \frac{\Gamma(1 - m)^{\frac{n}{m}} \Gamma\left(1 + \frac{n}{m} \right)}{\Gamma(1 - n)} \left.\frac{\Gamma(1 - n\beta)}{\Gamma(1 - b^2)^n}\right\vert_{b \to 1}  E_{m,\beta}^{n\beta}  \\
\Rightarrow \overline{\exp(t \mathcal{F})} &= e^{t F_0} \frac{\Gamma(1 + t)^2}{\Gamma(1 + t / \beta)}   \label{eq:etF1RSBlowT}
\end{align} 
where the free energy shift is now \textit{formally}
\begin{equation}
F_0 = -2 \ln M - \beta e_0 / 2 + \beta^{-1} \Gamma(1-b^2) \vert_{b\to1} - \ln E_{1/\beta,\beta} \,.\, 
\beta > 1 \,,  \label{eq:Fshift1RSB}
\end{equation}
which is a quite problematic expression. Ignoring it for the moment, let us relate immediately eq. \eqref{eq:etF1RSBlowT} to the freezing scenario; indeed, that equation implies that, denoting $g$ a standard Gumbel random variable independent of $\mathcal{F}$, we have
\begin{equation}
\overline{\exp(t(\mathcal{F}-\beta^{-1} g))} =  \overline{\exp(t \mathcal{F})} \Gamma(1 + t / \beta) =  e^{tF_0} \Gamma(1+t)^2 \,, \label{eq:ety}
\end{equation}
\textit{i.e.}, up to a translation, the distribution of $\mathcal{F} - \beta^{-1}g$ maintains its $\beta=1$-form in the whole $\beta > 1$-phase. By inverse Laplace transform eq. \eqref{eq:Gasconvolution} (or eq. \eqref{eq:inverseLap}), this means $G_\beta(x) = \overline{\exp(-e^{\beta x} \mathcal{Z})}$ is also temperature-independent (up to a global shift) in the $\beta > 1$ phase. In summary, by the \gls{1rsb} analysis, we recovered the freezing of the limit shape (eq. \eqref{eq:GbetaGFFshape}) that we took for granted in section \ref{sec:treetoplane} (see discussion around eq. \eqref{eq:FBgfreeze}). Although we obtained this result by the \gls{rsb} approach, we note that recent mathematical developments in multiplicative chaos and in discrete 2D \gls{gff} have led to rigorous proofs of analogous results in respective settings \cite{madaule2013glassy,biskup2016extreme}.

At last, we shall comment on the term $E_{m,\beta}$. The alert Reader should have noticed that we have already analytically continued it to generic complex $m \in \C$, without indicating how this is possible  from the definition eq. \eqref{eq:Exi}. We shall postpone this discussion to section \ref{sec:orderstat}, around eq. \eqref{eq:Koffd}. 

\subsubsection{Log--correction and duality}
In the \gls{bbm} model, we explained the $\ln \ln M$ corrections by an argument based on the \gls{kpp} equation (eq. \eqref{eq:threehalf}); for \glspl{logrem} defined on the Euclidean plane, these properties are part of the general prediction (eq. \eqref{eq:GbetaGFF}) that was put forward in \cite{carpentier2001glass}, which showed that $G_\beta(x)$ still satisfy approximately \gls{kpp} equations (see discussion after eq. \eqref{eq:GbetaGFF}). To reconcile the current \gls{1rsb} formalism and the \gls{kpp} approach is the subject of ongoing joint investigation. Therefore, we cannot present a general and systematic derivation of these universal features in the \gls{rsb} framework. In fact, a major drawback of the \gls{rsb} methods in general is the difficulty to access sub--leading corrections: even for the \gls{rem} and \gls{dpct}/\gls{bbm}, such progress is made only recently \cite{mottishaw15REM,derrida16kppfinitesize} and relies on replica--free methods, as in section \ref{sec:REM}. 

However, as the list of exactly solved \glspl{logrem} extends, a pattern has emerged for the analytically continued integrals. Let us take again the example of the circular model; by Dyson integral eq. \eqref{eq:DysonRSB}, we have 
\begin{equation}\label{eq:Zbcopy}
\overline{Z_b^{-t/b}} \Gamma(1 + t /b) =  \Gamma(1 -b^2)^{\frac{t}{b}} \times \Gamma(1 + t b) \Gamma(1 + t / b) \,.
\end{equation}
Notice two features of the right hand side: \texttt{i} the first factor vanishes as $\propto (1 - b)^{t}$ as $b \to 1$. \texttt{ii} the remainder is invariant under the $b \leftrightarrow b^{-1}$ \textit{duality} transform. In \cite{rosso12counting,fyodorov2015high}, feature \texttt{i} was related to the $\frac32 \ln  \ln M$ in the $\beta > 1$ phase, especially in the ro--temperature limit. We shall not review the argument, which requires a formalism (\textit{counting statistics}) not introduced in this thesis. Nevertheless, we can already see that in eq. \eqref{eq:etF1RSBlowT} and \eqref{eq:Fshift1RSB}, feature \texttt{i} induces a diverging ($\to  + \infty$) shift of the free energy with respect to $-2 \ln M$. The non--trivial task (undertaken in \textit{op. cit.}) is to associate this divergence to si--dependent $\ln \ln M$ corrections.

The feature \texttt{ii}, known as the ``duality invariance'', has been associated to freezing by the \textit{freezing--duality conjecture}, see discussion around eq. \eqref{eq:freezedual}. What can be said about this conjecture in the  \gls{1rsb} framework? Although the latter does not demonstrate this conjecture, some instructive observation can be made. Indeed, recall that eq. \eqref{eq:Zbcopy} enters into the freezing distribution, that of $y := \exp(\mathcal{F}-g- F_0$, eq.  \eqref{eq:ety}, with $b = \beta  m = 1$, since $m = 1/ \beta$ in the $\beta > 1$ phase. Now, recall that in the \gls{rsb} formalism, $m$ is a parameter that was optimized, in the full \gls{rsb} analysis, see eq. \eqref{eq:HQsimple} and \eqref{eq:moptimum1gen}. There, only the leading behaviour of free energy was considered. In \cite{cao16order} we argued that the same $m$ is optimal (stationary) also for all the higher cumulants of the free energy. Indeed, if we regard $m$ as a free parameter in eq. \eqref{eq:etF1RSBlowT}, we would have $$\overline{\exp(t (\mathcal{F} - F_0)}  =  \Gamma(1 + t / \beta)^{-1} \Gamma(1 + t m \beta ) \Gamma(1 + t / (m \beta)) \,,\,$$ 
so by duality invariance, 
\begin{equation} \left[  \frac{\partial}{\partial m} \overline{\exp(t (\mathcal{F} - F_0))}  \right]_{m=1/\beta} = 0 \quad \Rightarrow  \quad  \left.\frac{\partial}{\partial m} \overline{\mathcal{F}^k}^c \right\vert_{m = 1/\beta} = 0 \,,\, k = 2, 3, \dots \,.  \label{eq:FDC}
\end{equation}
Heuristically, this means that, just as its extensive free energy, which corresponds to the selected velocity in \gls{kpp} equation, the whole limit distribution of the free energy is also ``selected''. Turning this intuition into precise predictions is a goal of ongoing research. 

To conclude this section, we emphasize that the quantitative understanding of the minimum distribution of the circular model (and other \glspl{logrem} on Euclidean spaces) is not complete: the first moment is still not determined to an $O(1)$ precision. 

\subsection{Pre--freezing and binding transition} \label{sec:bindingetc}  
The freezing transition is the only critical behaviour when we consider the extensive free energy in standard  \glspl{logrem} such as the circular model. Other transitions can occur when more involved observables are considered, and/or in enriched  \glspl{logrem}. Two examples have already appeared in the context of multi--fractality (section \ref{sec:multifracintro}). In this section, we will first review the pre--freezing transition, and then study the binding transition, which is necessary for constructing the relation to \acrlong{lft} in section \ref{sec:liouville}.

\subsubsection{Pre--freezing transition}
In this work, we use the term \textit{pre--freezing transition} to refer to the transition associated with the divergence of the continuum Coulomb gas integrals; so, the term has a \textit{different} meaning than that of the work \cite{fyodorov2009pre}, in which the same term refers to the \textit{termination point transition} in this thesis; see section \ref{sec:multifracintro} for a clarification. 

Recall that the Dyson integral eq. \eqref{eq:Dyson},  
$$ \overline{Z^n} = \int_0^{2\pi} \prod_{a=1}^n \frac{\dif \theta_a}{2 \pi} \prod_{1\leq a<b \leq n} \abs{z_a - z_b}^{-2\beta^2} \,. $$ 
is convergent if and only if
\begin{equation}\label{eq:prefreezing}
n \beta^2 < 1  \,.
\end{equation}
Indeed, this condition applies to general Coulomb gas integrals representing continuum replica averages of  \glspl{logrem}, because the divergence comes from configurations where the replicas $\theta_1, \dots, \theta_n$ are very close. The power--counting of the integrand (and measure) is $$ \dif (\theta_2-\theta_1) \dots \dif (\theta_n-\theta_1) \prod_{a<b}\abs{\theta_a - \theta_b}^{-2 \beta^2} \sim [\theta ]^{n-1 - \beta^2 n(n-1) } = [\theta]^{(n-1)(1 - n \beta^2)} \,, $$ 
where we note that due to translation invariance, the measure's contribution is $[\theta]^{n-1}$. The integral diverges if the exponent $(n-1)(1 - n \beta^2) > 0$, which is eq. \eqref{eq:prefreezing}. Note that when $\beta \geq 1$, eq. \eqref{eq:prefreezing} means $n < \beta^{-2} \leq 1$. So the freezing transition $\beta = 1$ happens when $\overline{Z^{n\to1}}$ ceases to exist. When $n > 1$, eq. \eqref{eq:prefreezing} gives a higher \textit{pre--freezing} critical temperature $\beta_{c,n} = n^{-1/2} < 1$: they can be heuristically considered as precursors of the freezing temperature $\beta_c = \beta_{c,1}$. However, what is the physical interpretation of pre--freezing? The answer, given in \cite{fyodorov2008statistical}, is that it corresponds to a transition in the \textit{negative large deviation} of the free energy $\mathcal{F}$, and we review their treatment here, restricting to the $\beta < 1$ phase. 

For this, we recall from eq. \eqref{eq:etF1RSBhighT} the Laplace transform of the shifted free energy
$$ \overline{\exp(t f)} = \Gamma(1 + t\beta) =  \frac{1}{1 + t\beta} + \dots \,,\, f = \mathcal{F} + (\beta + 1/\beta) \ln M + O(1) \,,  $$
where $\dots$ denotes poles  at $t = -2/\beta, -3/\beta$. By inverse Fourier transform, this corresponds to the exponential tail \begin{equation}  P(f) = e^{f/\beta} + O(e^{2f / \beta}) \,,\, f_* \ll f \ll 0 \,. \label{eq:ftailhighT} \end{equation} 
The crucial point is that eq. \eqref{eq:etF1RSBhighT} is not valid for $t < -1/\beta$, so eq. \eqref{eq:ftailhighT} will cross over to some other distribution when $f < f_*$ for some $f_*$. To determine that distribution and $f_*$, we need to compute $ \overline{\exp(t f)}$ for $t < -\beta^{-1}$, or $\overline{\mathcal{Z}^n}$ for $n = -t/\beta > \beta{-2}$: this is exactly the opposite condition of eq. \eqref{eq:prefreezing}. As argued in \cite{fyodorov2009statistical}, this is a new phase with another type of  \gls{rsb}: all the $n$ replicas are in the same block; that is, we plug $m = n$ into \eqref{eq:ZnRSB}, so the Coulomb gas integral eq. \eqref{eq:DysonRSB} becomes trivial, and we obtain:
\begin{align}
&\overline{\mathcal{Z}^n}  =  M^{\beta^2 n^2 + 1}  E_{n,\beta}  \,,\, n > 1 / \beta^2
\\ \Rightarrow\quad & \overline{\exp(t \mathcal{F})} = M^{t^2 + 1} E_{-t/\beta,\beta}  \,,\, t < -1/\beta \,.
\end{align}
By inverse Fourier transform this gives a Gaussian of variance $2\ln M$
\begin{equation}\label{eq:Pfprefreezing}
P(f) = \frac{M}{\sqrt{4 \pi \ln M}} \exp\left( -\frac{(f + F)^2}{4 \ln M}\right) \,,\, f < f_* \,.
\end{equation}
where $E_{-t/\beta,\beta}$ gives an $O(1)$ convolution which we have omitted. Note that $P(f)$ is not a normalized distribution on the whole real line: $\int_\R P(f) \dif f = M$, so it can be only valid for $f < f_*$. To determine $f_*$ (up to order $\ln M$), we match eq. \eqref{eq:Pfprefreezing} with eq. \eqref{eq:ftailhighT}:
\begin{equation} \label{eq:prefreezingf}  f_* / \beta \approx  \ln M - \frac{(f_* + F)^2}{ 4 \ln M } \Rightarrow f_* = -(\beta^{-1} - \beta) \ln M  + O(\ln \ln M) \,. \end{equation}

In summary, the exponential tail of eq. \eqref{eq:ftailhighT} crosses over to a Gaussian tail in the large deviation regime of atypically small free energy $\mathcal{F} = -(\beta + 1 / \beta) \ln M  + f_* =- 2 \beta^{-1} \ln M$. By eq. \eqref{eq:prefreezingf}, \eqref{eq:Pfprefreezing} and \eqref{eq:ftailhighT}, we compute the (negative) large deviation function of free energy: 
\begin{equation}
 \mathcal{L}(x) \defeq  \frac{-1}{\ln M} \ln \overline{\left(\frac{\mathcal{F}}{\ln M} - x \right)} 
 \stackrel{\beta<1}{=} \begin{dcases}
    - x/\beta -1 - 1/\beta^2 \,, &  -2\beta^{-1}  <  x  < -(\beta + \beta^{-1}) \,, \\
  x^2 / 4 - 1 \,, &   x < -2\beta^{-1} \,.
\end{dcases}
\end{equation}
The transition at $-2 \beta^{-1}$ is the physical interpretation of the pre--freezing transition. Note that although the \gls{1rsb} solution jumps from $m=1$ (high temperature phase) to $m=n$ (pre--freezing phase) at the transition, $ \mathcal{L}(x) $ has continuous derivative at the transition. Approaching the freezing temperature $\beta\to 1_-$, the pre--freezing point $-2 /\beta \to -2$ comes close to the typical free energy density $-(\beta + \beta^{-1})$. So the same analysis in the $\beta > 1$ phase requires carefully accounting for the $\ln \ln M$--order corrections.  

\subsubsection{Binding transition}
Let us consider a concrete model that displays the binding transition, the \textit{Morris circular model} (another example is the interval model, considered in \cite{fyodorov2009statistical}). Recall that, the potential of the circular model can be seen as restricting the 2D GFF $\phi$ (eq. \eqref{eq:GFF2p}) to the unit circle. In this respect, the random potential of the Morris circular model can be seen as $\phi(z) + 2 \alpha \ln \abs{z - 1}$. That is, we add a \textit{deterministic} part of the potential which has a \textit{logarithmic singularity}, here at $z = 1$. To define it properly on a lattice, let $C_{jk}, j,k = 1, \dots, M$ be the covariance matrix of the circular model (defined by eq. \eqref{eq:circularVar} and \eqref{eq:muk1overf}). Then, that of the Morris circular model is defined by:
\begin{equation} \label{eq:Morrisdef}
\overline{V_j} = -\alpha C_{j 1} \,,\, \overline{V_j V_k}^c = C_{jk} \,.
\end{equation}
Here we assume that the lattice points are given by $z_{j,M} = e^{2 \pi j / M}$. Then by eq. \eqref{eq:circularIR}, it is clear that $\overline{V_j} \to 2 \alpha \ln \abs{z-1}$ provided $z_{j,M} \to z \neq 1$ as $M\to \infty$: the mean values in eq. \eqref{eq:Morrisdef} reproduce the continuum potential $2 \alpha \ln\abs{z-1}$. On the other hand, its divergence at $z = 1$ is regularized by $\overline{V_1} = - 2 \alpha \ln M + O(1)$. 
 
\begin{figure}
\center
\includegraphics[scale=.5]{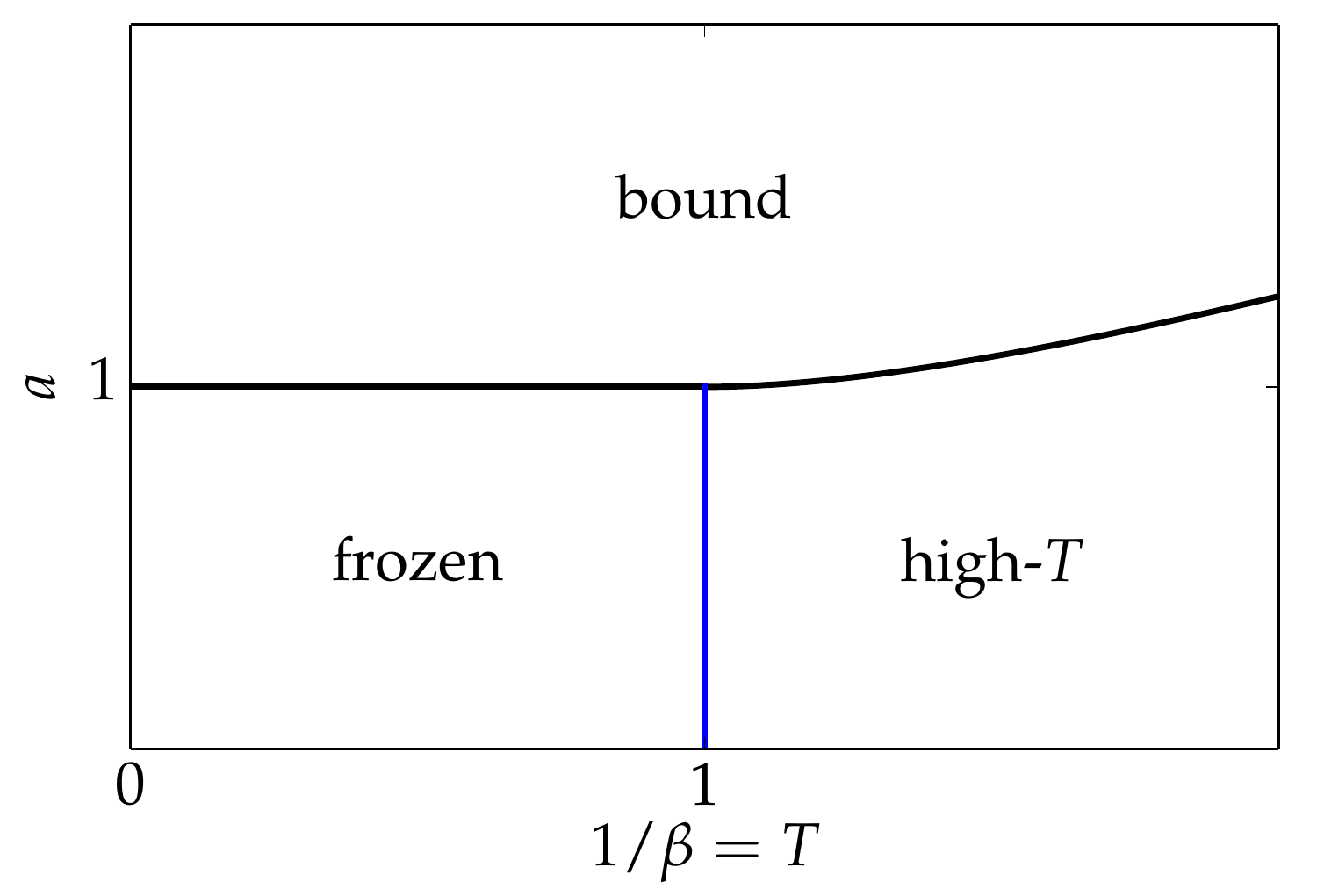}\includegraphics[scale = .5]{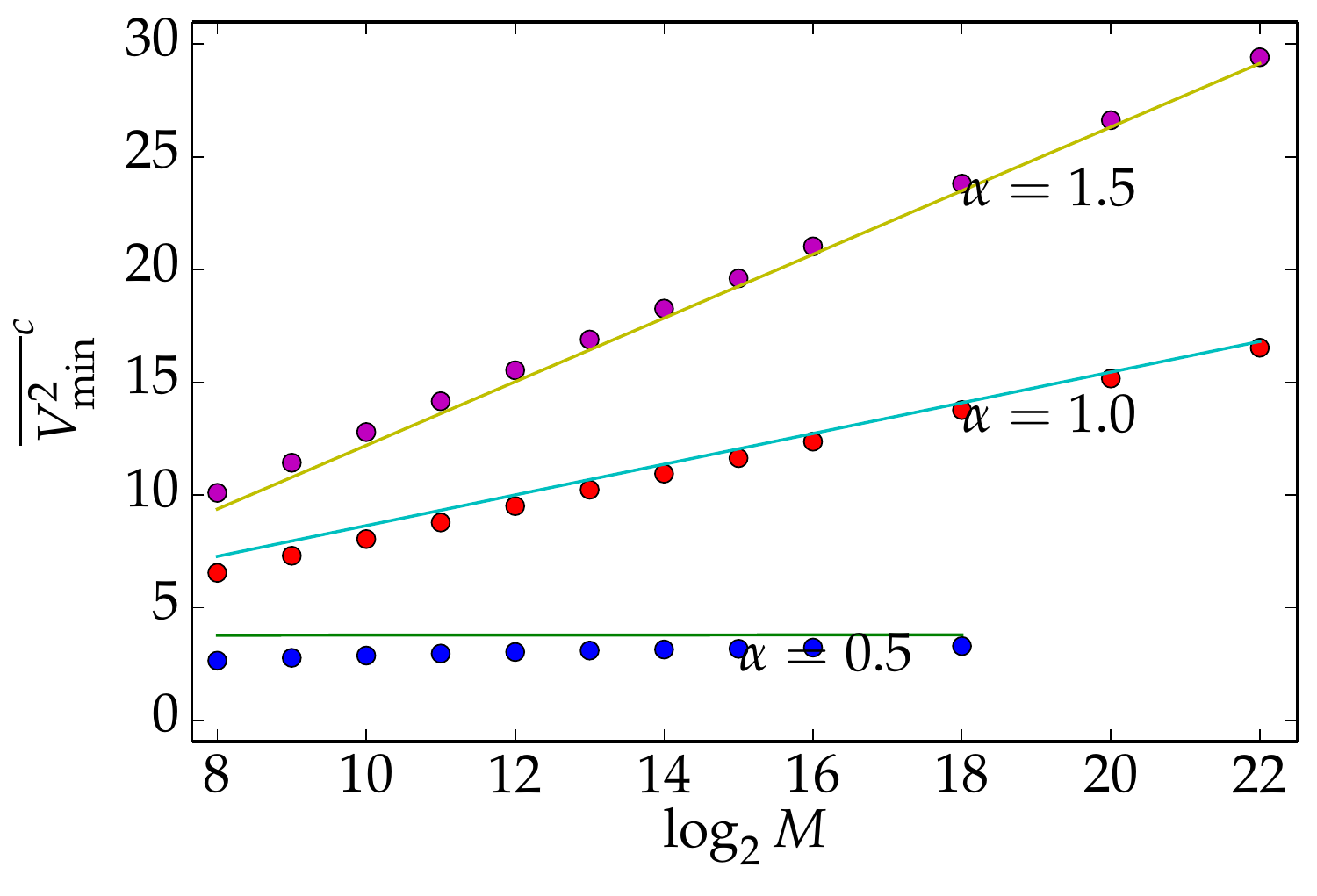}
\caption{(Left) The phase diagram of  \glspl{logrem} with a logarithmic potential, such as the Morris circular model. The blue vertical line is the freezing transition, while the other (near horizontal) curve is the binding transition line, $a = Q / 2$, see eq. \eqref{eq:nobinding}. (Right) The minimum variance of Morris circular model as a function of the log of system size, compared to $2\ln M + c$ ($\alpha = 1.5$, bound phase), $\ln M + c'$ ($\alpha = 1$, zero temperature critical) and $c''$ ($\alpha = .5$, frozen phase). } \label{fig:binding}
\end{figure}

The binding transition happens when $a$ is large enough that the thermal particle is trapped at $z = 1$. The easiest way to establish the phase diagram (in the plane of $(\beta, \alpha)$) is the following argument (already present in \cite{carpentier2001glass}, Appendix D, see also \cite{cao16liouville}). The free energy of the particle sitting at $z = 1$ is $F_1 = V_1 = -\alpha \ln M + O(\sqrt{\ln M})$; on the other hand, the free energy staying away from $z = 1$ is simply that of the circular model, $F_0 \sim Q \ln M + O(1)$, see eq. \eqref{eq:freezedual}. Then, the criterion of \textit{no--binding} is $F_0 \ll F_1$ as $\ln M \to \infty$, which gives
\begin{equation} \alpha < Q/2  \,,\, Q = \begin{cases}
\beta^{-1} + \beta \,,\, & \beta < 1\,, \\  2 \,, & \beta \geq 1 \,.
\end{cases}  \label{eq:nobinding} \end{equation} 
Therefore, the Morris circular model has three phases, the high temperature phase $\beta < 1, \alpha < \beta^{-1} + \beta$, and frozen phase $\beta > 1, \alpha < 2$, and the \textit{bound phase} $\alpha > Q / 2$ with $Q$ defined above. The phase diagram is drawn in Figure \ref{fig:binding} (Left panel). 

In the high temperature and frozen phase, the free energy distribution of the Morris circular model can be studied using the same method as applied to the interval model \cite{fyodorov2009statistical}; in particular, we still have $\mathcal{F} = - Q \ln M + \eta \ln \ln M + O(1)$. In the bound phase, the thermal particle is trapped at $j=1$, so its free energy $\mathcal{F} \approx V_{1}$, thus we predict
\begin{equation}
\overline{\mathcal{F}} = - \alpha \ln M + O(1) \,,\, \overline{\mathcal{F}^2}^c = 2 \ln M + O(1) \,, \alpha > 1 \,.
\end{equation}
So, the variance of the free energy is not $O(1)$ but is extensive. A numerical test (in the zero temperature) of this latter prediction is given in Figure \ref{fig:binding} (Right panel). We observe also that $\overline{\mathcal{F}^2}^c \approx \ln M + O(1)$ in at the (zero--temperature) binding transition $\beta = \infty, \alpha = 1$.

The discrete definition eq. \eqref{eq:Morrisdef}, and the no--binding condition eq. \eqref{eq:nobinding} are not specific to the Morris circular model, but apply to any \gls{logrem}. In the $d$-dimensional Euclidean space, the covariance matrix has a logarithmic decay given by eq. \eqref{eq:logdecay}, so the mean value is
\begin{equation}\label{eq:logpotential}
\overline{V_j} = U(\mathbf{x}_j) \,,\, U(\mathbf{x}) \sim 2d \alpha \ln \abs{\mathbf{x} - \mathbf{x_1}} \,,
\end{equation}
where $(\mathbf{x_j})_{j=1}^M$ is the lattice of the \gls{logrem}. Equation \eqref{eq:nobinding} gives the condition that the thermal particle is not bound near $\mathbf{x_1}$. The consideration generalize naturally to potentials with several logarithmic singularities: they will be used when studying the mapping of \glspl{logrem} to \acrlong{lft} in section \ref{sec:liouville}. 

\section{Application of Jack polynomials}\label{sec:Jack}
This section will review applications of Jack polynomials to the circular model of $1/f$-noise and its variants, based on results reported in \cite{cao15gff} (section \ref{sec:Dirichlet}), and in \cite{cao16maxmin} (sections \ref{sec:EA} and \ref{sec:minmax}).  Although applications of Jack polynomials to \glspl{logrem} are rather recent (in  \cite{fyodorov2015moments} and \textit{op.cit.}), the idea is very natural and straightforward. Indeed, the replica trick applied to \glspl{logrem} produces systematically Coulomb gas integrals of the following form:
\begin{equation}   
\int \dif z_1 \dots \int \dif z_n  \prod_{a<b} \abs{z_a - z_b}^{-2\beta^2} f(z_1, \dots, z_n)  \,,\,
\end{equation}
where $f(z_1, \dots , z_n)$ is a \textit{symmetric function} of $z_1, \dots, z_n$. The Dyson integral eq. \eqref{eq:Dyson} is a simplest case where $f = 1$. In general, such integrals are difficult to evaluate into a form that can then be analytically continued (to $n$ non--integer!). Yet, this task becomes possible when $f$ is a Jack polynomial (or product of two of them). So the strategy will be to use the Jack polynomials as a \textit{basis} to expand $f$, and then integrate term by term. For this, we need some mathematical results, which we review in the following section.

\subsection{Jack polynomials} \label{sec:jackintro}
Jack polynomials \cite{jack1970class} are a family of \textit{symmetric polynomials}. The latter is a classical subject in mathematics, related to representation theory of semi--simple Lie groups and permutation groups, as well as to combinatorics; see \cite{stanleyEC2}, last chapter for a comprehensive introduction and extended references. The classical (19th century) theory is focused on the \textit{Schur polynomials}. In the 20-th century, a few other families of symmetric polynomials have been considered as non-trivial generalizations of the Schur class, \textit{e.g.}, the Jack polynomials and the Macdonald polynomials. A classical treatment is MacDonald's book \cite{macdonald1998symmetric}, which will be our main reference and set the conventions. 

The last two polynomials have important physical applications. The Macdonald polynomials play an important rôle in the development of quantum integrable system related to the \acrfull{kpz} equation in $(1+1)$-dimension \cite{borodin2014mac} (it is impossible to give a fair review of this fast growing field here, however, see \cite{borodin2014maclec} for introduction and \cite{corwin2014macdonald} for summary). The Jack polynomials can be obtained as a degenerate case of Macdonald polynomials, but have arguably wider applications. They are closely related to eigen-functions of the Calegero-Sutheland quantum integrable system (see \cite{pasquier1994CSM,Lesage1995csm} for introduction), and thus related to fractional statistics (see \cite{bernard1994some} for introduction). More recent developments include important applications in fractional quantum Hall effects \cite{bernevig08jack}, and in \acrlong{cft} with extra symmetry \cite{estienne09jack,estienne10jack,estienne12jack}.

\subsubsection{Definition}
The Jack polynomials $P_\lambda^{(\alpha)} (\underline{z}), Q_\lambda^{(\alpha)}(\underline{z})$ are symmetric polynomials in the set of $n$ variables $\underline{z} = (z_1, \dots, z_n)$, \textit{i.e.}, 
$P_\lambda^{(\alpha)} (z_1, \dots, z_n) = P_\lambda^{(\alpha)} (z_{\sigma(1)}, \dots, z_{\sigma(n)})$ for any permutation $\sigma$ of $n$ numbers. They depend on a complex parameter $\alpha$ and are indexed by an \textit{integer partition} $\lambda$. This is an important combinatorial notion for the theory of symmetric polynomials, so we briefly review it here (for a detailed treatment, see the first section of \cite{macdonald1998symmetric}).

\begin{figure}
\center 
\begin{tabular}{|c|c|c|c|}
\hline
 Young diagram & $\lambda$ & $\ell(\lambda)$ & $\abs{\lambda}$  \\ \hline
 \includegraphics[scale=.4,valign=c]{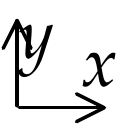}   & $() = \emptyset$   &   $0$ & $0$ \\ \hline
 \includegraphics[scale=.4,valign=c]{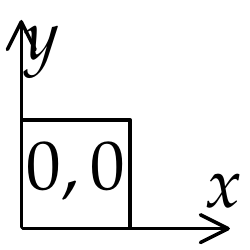}  & $(1) = \square$  &   $1$ & $1$ \\ \hline
  \includegraphics[scale=.4,valign=c]{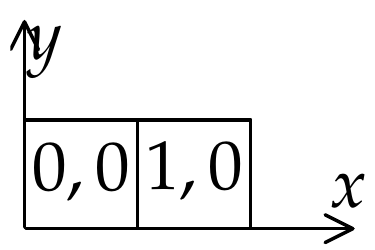}   & $(2)$  &  $1$  & $2$ \\ 
 \includegraphics[scale=.4,valign=c]{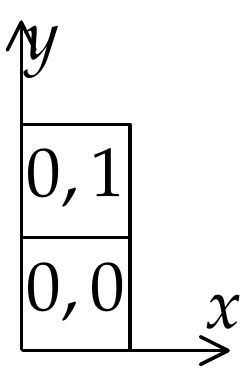}    & $(1, 1)$  & $2$ & $2$ \\  \hline 
 \includegraphics[scale=.4,valign=c]{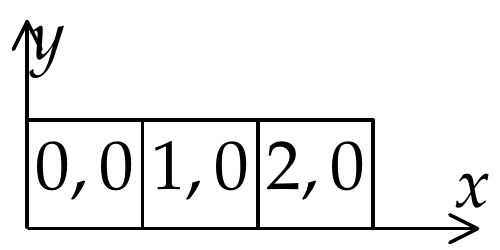}    & $(3)$  & $1$ & $3$ \\
  \includegraphics[scale=.4,valign=c]{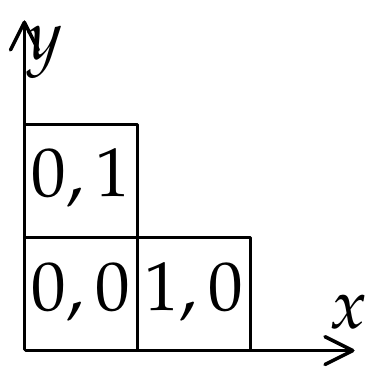}   & $(2, 1)$  & $2$ & $3$ \\  
  \includegraphics[scale=.4,valign=c]{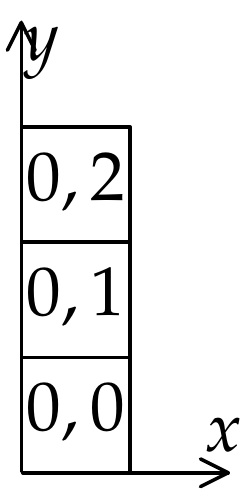}   & $(1, 1, 1)$  & $3$ & $3$ \\   \hline       
\end{tabular}
\caption{Enumeration of all the partitions of size $\leq 3$. } \label{fig:partitions}
\end{figure}

An integer partition $\lambda$ can be defined as a finite sequence of decreasing positive integers $\lambda = (\lambda_0 \geq \lambda_1 \dots \lambda_{\ell(\lambda)} > 0)$, where $\ell(\lambda)$ is called its \textit{length}, while its \textit{size} is defined as $\abs{\lambda} = \lambda_1 + \dots + \lambda_{\ell(\lambda)}$. It is clear that integer partitions of size $n$ and length $\ell$ are in one--to--one correspondence with the ways of decomposing $n$ as sum of $\ell$ non-ro integers, if re-ordering is not considered to give a different sum (this is the reason of the name). 
A common representation of the integer partitions is the \textit{Young diagram}. The Young diagram of a partition $\lambda$ is the set 
$$ \lambda \mapsto \{ (x,y) \in \N^2 \vert x = 0, 1  \dots, \lambda_y - 1 \,,\, y = 0, 1, \dots, \ell(\lambda)-1 \}$$
of integer pairs. In the following, we will identify $\lambda$ with the right hand side. Graphically, each pair is drawn as a box in the 2d plane. To illustrate, we enumerate all the partitions with size $\leq 3$ in Table \ref{fig:partitions}. Note that the \textit{empty} sequence $\emptyset$ is also considered as a partition of si/length $0$. 

The Young diagram representation make it easy to define the \textit{transpose} of a partition, $\lambda^t$, as obtained by applying the $(x,y) \mapsto (y,x)$ transform. In terms of the sequence, it is not hard to see that $\lambda^t_j = \abs{i: \lambda_i \geq j}$. For instance, when $\lambda = (3)$, $\lambda^t = (1,1,1)$. 

\subsubsection{Main properties}
The explicit form of Jack polynomials are not so important for our applications as their properties: 
\begin{enumerate}
\item $P_\lambda^{(\alpha)}(\underline{z})$ and $Q_\lambda^{(\alpha)}(\underline{z})$ are related by a factor 
\begin{equation}
Q_\lambda^{(\alpha)}(\underline{z}) = J_\lambda^{(\alpha)} P_\lambda^{(\alpha)}(\underline{z}) \,.
\end{equation}
The explicit form of $J_\lambda^{(\alpha)}$ is not important here.
\item $P_\lambda^{(\alpha)}(\underline{z})$ is defined for $\underline{z} = (z_1, \dots, z_n)$ with different numbers variables $n$. They are related by 
\begin{equation}
P_\lambda^{(\alpha)}(z_1, \dots, z_{n-1}) = P_\lambda^{(\alpha)}(z_1, \dots, z_{n-1}, z_{n} = 0) \,,
\end{equation} 
and the same is true for $Q_\lambda^{(\alpha)}$. This fact justifies omitting the dependence on $n$ in the notations. In fact, it is sometimes useful to define Jack polynomials as symmetric polynomials of an infinite sequence of variables $z_1, \dots, z_n, \dots$, and obtain the finite-$n$ versions by setting $z_{n+1} = z_{n+2} = \dots = 0$.
\item Both $P_\lambda^{(\alpha)}(\underline{z})$ and $Q_\lambda^{(\alpha)}(\underline{z})$ are \textit{homogeneous} polynomials of order $\abs{\lambda}$, \textit{i.e.}, 
\begin{equation}\label{eq:homogeneity}
(P_\lambda^{(\alpha)}(a z_1, \dots, a z_n) = a^{\abs{\lambda}} P_\lambda^{(\alpha)}(z_1, \dots, z_n) \,,\,
\end{equation}
and the same is true for $Q_\lambda^{(\alpha)}$.
\item \textit{Cauchy identity}:
\begin{equation}\label{eq:cauchy}
\prod_{a=1}^n \prod_{b=1}^m (1 - z_a w_b)^{-1/\alpha} = \sum_\lambda P_\lambda^{(\alpha)}(\underline{z}) Q_\lambda^{(\alpha)}(\underline{w}) \,,
\end{equation}
where $\underline{w} = (w_1, \dots, w_m)$. Note that the sum in the right hand side is over \textit{all} the partitions, with no restriction of size, so it is an infinite series. Technically, the identity holds in a sense of formal power series (\textit{i.e.}, matching of all coefficients). 
\item \text{Orthogonality} with respect to integration on the unit circle:
\begin{subequations} \label{eq:orthogonality}
\begin{align}
&\int_0^{2\pi} \prod_{a=1}^n \frac{\dif \theta_a }{2 \pi }\prod_{a<b} \abs{z_a - z_b}^{\frac2\alpha} P_\lambda^{(\alpha)}(\underline{z}) Q_\mu^{(\alpha)}(\underline{z^*}) = \delta_{\lambda \mu} p_n^\lambda(\alpha) c_n(\alpha) \, , 
\end{align}
where $\delta_{\lambda\mu} = 1$ if and only if $\lambda = \mu$ are the same partition, but $0$ otherwise, 
\begin{align}
&c_n(\alpha) = \int_0^{2\pi} \prod_{a=1}^n \frac{\dif \theta_a }{2 \pi } \prod_{a<b} \abs{z_a - z_b}^{\frac2\alpha}  = \frac{\Gamma(1 + n/\alpha)}{\Gamma(1 + 1/\alpha)^n} \, , \label{eqdyson} 
\end{align}
is the Dyson integral (eq. \eqref{eq:Dyson}), and finally
\begin{align}
& p_n^\lambda(\alpha) =   \prod_{(x,y)\in\lambda} \frac{\alpha x + n - y}{\alpha(x + 1) + n - (y + 1)}\, , \label{eq:pnJack}
\end{align}
is a product over all the boxes in the Young diagram of $\lambda$. Note that the orthogonality relation depends explicitly on the number of variables $n = 1, 2, 3, \dots$. 
\end{subequations}
\end{enumerate} 
Among the above properties, the last two (Cauchy identity and orthogonality) are the most important and will be crucial for calculating the Coulomb gas integrals in the next two sections. Remark that orthogonality property makes the Jack polynomials the multi--variable analogue of the exponential/trigonometric functions $e^{\im k \theta}, k = 0, \pm 1, \pm 2, \dots$ with respect to the integral measure $(2 \pi)^{-1}\int_0^{2\pi} \dif\theta [\dots]$ on the unit circle. Now, another property of the trigonometric functions is that they form a \textit{complete} basis (of the Hilbert space of square--integrable functions on the unit circle). Similarly, the Jack polynomials form a complete basis of the symmetric functions (defined on the torus $\set{(z_1, \dots, z_n): \abs{z_i} = 1}$). Although we will not need this property explicitly in sections \ref{sec:Dirichlet} through \ref{sec:minmax}, we believe that it will be important for future applications, as we discuss in section \ref{sec:finallogrem}. 

\subsection{Dirichlet circular model}\label{sec:Dirichlet}
The Dirichlet circular model was formally introduced in section \ref{sec:IRUV}, see eq. \eqref{eq:mukDirichlet}, as a deformation of the standard circular model with different a \gls{ir} data , eq. \eqref{eq:circularIR}. The motivation of studying this model (whose definition went back to \cite{fyodorov2009statistical}) was the following more ambitious questions, which are all still open: how can we extend the exact calculation of the free energy distribution of the circular and interval model to more general curves or 2d domains? Does the freezing--duality conjecture (see discussion around eq. \eqref{eq:freezedual}) in general?
A more specific motivation (stressed in \cite{cao15gff}) is the \gls{ir} divergence of 2D \gls{gff}, see section \ref{sec:Gaussian} for more discussion. Indeed, 2D \gls{gff} must be defined on a finite geometry, \textit{i.e.}, a 2D sphere or a disk. However, the 2D Coulomb--gas integrals on these geometries are not exactly solved (this is a major open problem in mathematical physics, with applications in 2D one-component plasma \cite{forrester2010log} and in fractional Quantum Hall Effect, see \cite{laskin2015collective,can2014fractional,can2015geometry} for recent developments). So we shall simplify the problem to considering the \glspl{logrem} obtained by restricting a finite-domain 2D \gls{gff} to a circle in that domain. The 2D \gls{gff} on the sphere will be discussed in more detail in section \ref{sec:liouville}. As shown in \cite{cao15gff}, the circular model defined on a sphere is easily related to the standard one. So we will turn to the 2D \gls{gff} on a disk. 

\subsubsection{2D GFF on a disk}
\begin{figure}[h]
\center \includegraphics[scale=.6]{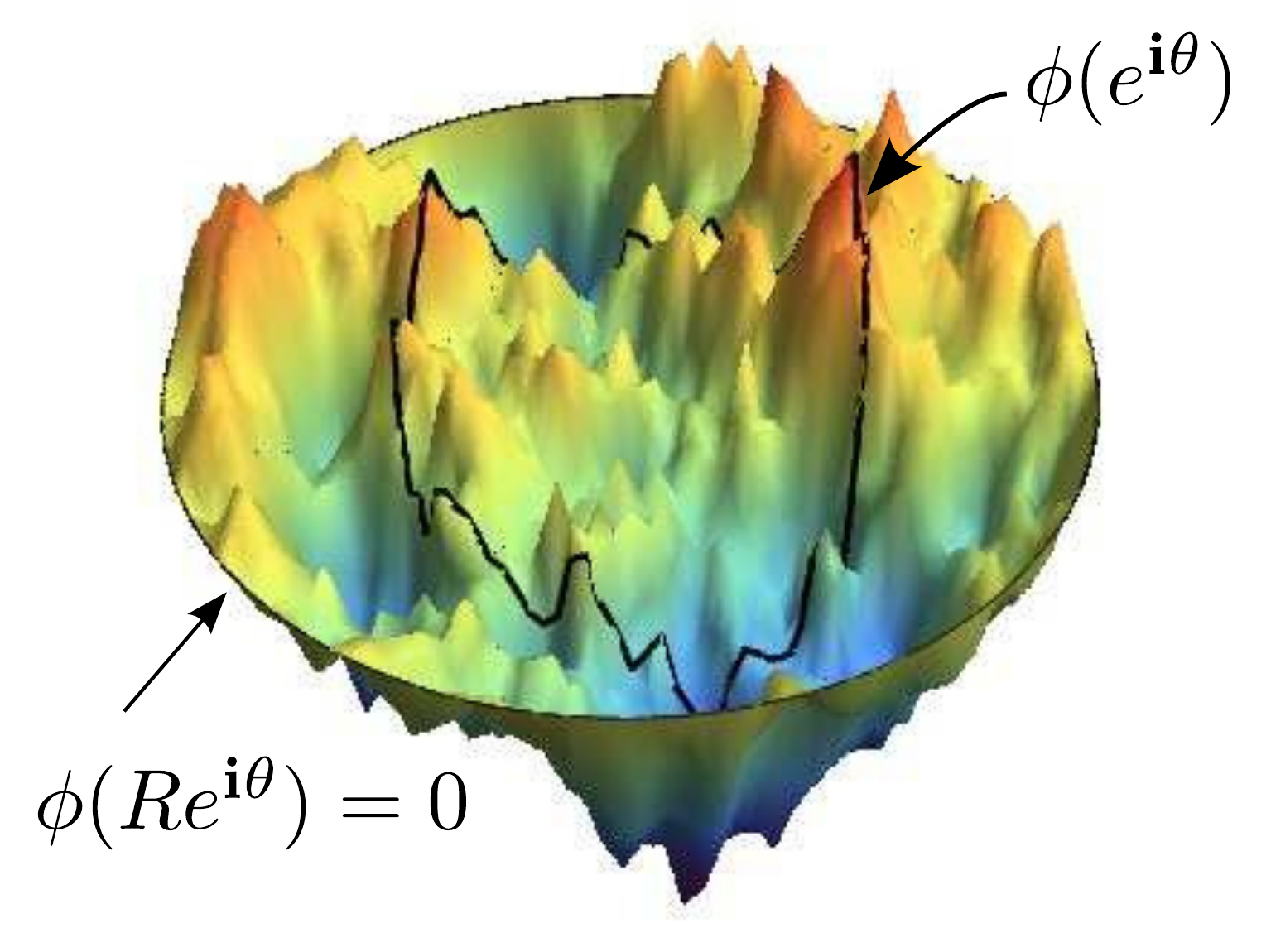}
\caption{A sample of the 2D \gls{gff} in a unit disk with Dirichlet boundary condition, and its restriction on a centred circle. The latter defines the circular Dirichlet model, after removing the zero mode.}
\end{figure} 
Consider a centred disk of radius $R > 1$ in the complex plane, $\set{z : \abs{z} < R}$. The 2D \gls{gff} with Dirichlet boundary condition is defined by the following Green function:
 \begin{equation}
 \overline{\phi(z) \phi(w)} = G_R(z,w) =  -2 \ln \abs{\frac{z - w}{R - z w^*/R}} \,,\, \abs{z}, \abs{w} < R\,.
 \label{eq:GFFDirichlet}
 \end{equation}
$G_R(z,w)$ satisfies the Laplace equation $\Delta_z G_R = 0$ with the Dirichlet boundary condition $G_R(R e^{\im \theta}, w) = 0$ for any $\theta \in \R$ and $w\in\C$. Rewriting eq. \eqref{eq:GFFDirichlet} as follows,
\begin{equation}\label{eq:Dirichlet0}
G_R(z,w)= - 2 \ln\abs{\frac{z - w}{1 - q zw^*}} + 2 \ln R \,,\, q = R^{-2} \in (0,1) \,,
\end{equation}
we can identify it with the Dirichlet circle case in eq. \eqref{eq:circularIR}, if the term $2 \ln R$ is discarded. So if we restrict the Dirichlet 2D \gls{gff} (eq. \eqref{eq:GFFDirichlet}) to the unit circle, we will obtain a 1D log-correlated potential which is that of the Dirichlet circular model plus a \textit{zero-mode}, \textit{i.e.}, we may define $\phi(e^{\im \theta}) = \varphi(e^{\im \theta}) + \phi_0$, where $\phi_0$ is a centred Gaussian distribution of variance $2 \ln R$ and is independent of $\varphi$, while the latter satisfies 
\begin{equation} \label{eq:IRDicichlet}
 \overline{\varphi(z) \varphi(w)} =  - 2 \ln\abs{\frac{z - w}{1 - q zw^*}} \,,\, \abs{z} = \abs{w}  = 1 \,.
\end{equation}
The effect of the zero mode to the free energy distribution is trivial, so we will focus on eq. \eqref{eq:IRDicichlet}. Nevertheless, with respect to the discussions in section \ref{sec:Gaussian} (around eq. \eqref{eq:FkGaussian}), we remark that as $q\to 0$ ($R \to \infty$), the circular model in the Dirichlet disk \textit{with} ro mode (\textit{i.e.}, defined by eq. \eqref{eq:Dirichlet0}), is an \gls{ir} divergent \gls{logrem}, in the sense that its free energy is the convolution of a Gaussian of diverging variance ($2 \ln R$) coming from the  ro--mode, and a correction distribution, which is the free energy of the model with ro--mode removed, eq. \eqref{eq:IRDicichlet}. Here, the correction is a well--defined probability distribution (in contrast to the Gaussian model), thanks to the fact that eq. \eqref{eq:IRDicichlet} is defined on a finite geometry. 

We have seen in section \ref{sec:IRUV} (eq. \eqref{eq:mukDirichlet}) how to define a discrete \gls{logrem} whose \gls{ir} data is eq. \eqref{eq:IRDicichlet}. Note that the limit $q \to 0$  gives back the original circular model corresponding to the infinite plane ($R \to \infty$). The Dirichlet circular model is also defined formally for $q \in [-1, 0)$; a rather artificial physical interpretation exists in terms of anti-symmetrid 2D \gls{gff} on the sphere (\cite{cao15gff}, section 1). 

\subsubsection{Solution by Jack polynomials}
Now let us sketch the solution of the Dirichlet circular model. The approach is identical to that of the circular model \ref{sec:treetoplane}, which starts by considering the continuum partition function $$Z = \int_0^{2\pi}\frac{\dif \theta}{2\pi} \exp(-\varphi(e^{\im \theta})) \,$$
where $\varphi$ satisfies eq. \eqref{eq:IRDicichlet}, and its replicated averages $\overline{Z^n}$. Therefore The essential technical novelty consists in extending the Dyson integral eq. \eqref{eq:Dyson} to the following Coulomb gas integral, 
\begin{align}
\overline{Z^n} = \int_{0}^{2\pi} \prod_{a = 1}^n \frac{\dif \theta_a}{2\pi} 
\prod_{a<b} \abs{\frac{z_a - z_b}{1 - q z_a z_b^*}}^{-2\beta^2},\;
z_a := e^{\im \theta_a} \,. \label{eq:ZnDir} 
\end{align}
This is where Jack polynomials become useful, if the Jack parameter is related to $\beta$ by $\alpha = -\beta^{-2}$. Indeed, we can use the Cauchy identity eq. \eqref{eq:cauchy} (and the homogeneity eq. \eqref{eq:homogeneity}) to expand the denominator above
$$ \prod_{a<b}\abs{\frac{1}{1 - q z_a z_b^*}}^{-2\beta^2} = (1 - q)^{-n\beta^2}
\sum_\lambda q^{|\lambda|}  P_{\lambda}^{(\alpha)}(\underline{z}) Q_{\lambda}^{(\alpha)}(\underline{z}) \,, $$
and then integrate each term obtained by the orthogonality relation eq. \eqref{eq:orthogonality}. The end result is:
\begin{equation}
\overline{Z^n} =  (1 - q)^{-n\beta^2} \frac{\Gamma(1-n\beta^2)}{\Gamma(1-\beta^2)^n}  \sum_\lambda q^{|\lambda|}\prod_{(x,y)\in\lambda} \frac{x + (y-n)\beta^2}{x + 1 + (y+1 - n) \beta^2}  \,. \label{eq:ZnsolveDir}
\end{equation}
Although it is not an closed form formula, it provides an analytical continuation to $n \in \C$ of the Coulomb gas integral eq. \eqref{eq:ZnDir}, generalizing the Dyson integral (which is retrieved at $q = 0$). Therefore we have the free energy distribution in the $\beta < 1$ phase:
\begin{align}
&\overline{\exp(tf)} =  \Gamma(1 + t\beta) \mathbf{s}(t,\beta,q) \,,\, \beta < 1 \,,\, \label{eq:etfDirichlet} \\
&\mathbf{s}(t,\beta,q) = \sum_\lambda q^{|\lambda|} 
\prod_{(x,y)\in\lambda} \frac{x\beta + y\beta^{-1} + t}{(x+1)\beta + (y+1)\beta^{-1} + t}  \,. \label{eq:sumDirich}
\end{align}
Here, $f$ is related to the free energy by a shift, which absorbed the factors $(1 - q)^{-n\beta^2} \Gamma(1-\beta^2)^{-n}$ in eq. \eqref{eq:ZnsolveDir}. 

\begin{figure}
\center
\includegraphics[scale=.4]{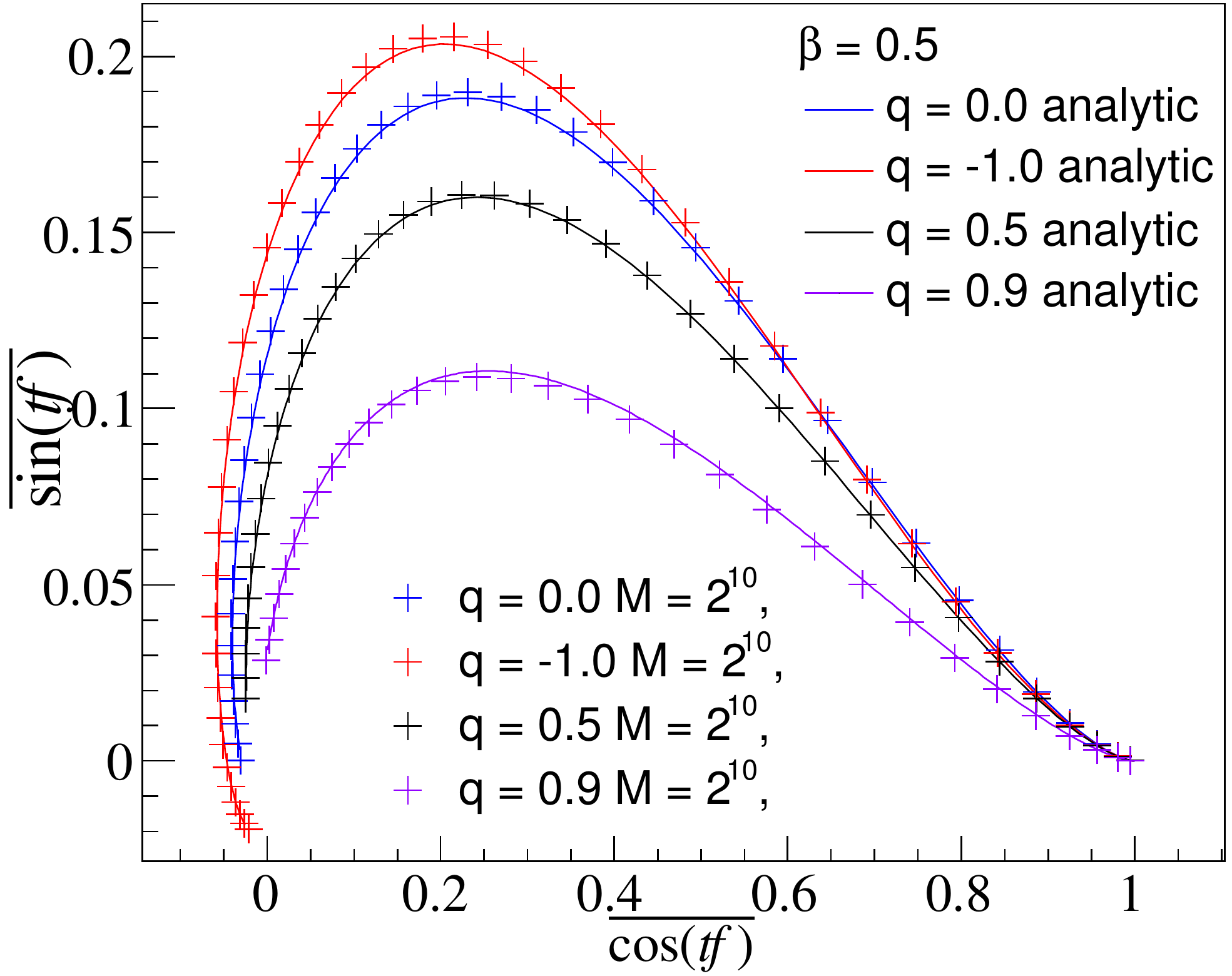}
\caption{Taken from \cite{cao15gff}. Real and imaginary part of $\overline{\exp(\im tf)} = \overline{\cos(tf)} + \im  \overline{\sin(tf)}$ with $t \in \R$. The data points are obtained from numerically simulations, while the curves are the prediction eq. \eqref{eq:sumDirich}. In this Figure, both the predicted and numerical distribution have been shifted so as to have vanishing mean value. However, we have checked that the mean value can be also predicted by taking into account the \gls{uv} data, see eq. \eqref{eq:FshifthighT}. }\label{fig:DirichlethighT}
\end{figure}
Equations \eqref{eq:etfDirichlet} and \eqref{eq:sumDirich} are already non--trivial and non--rigorous predictions that should be checked. In \cite{cao15gff}, two tests were provided. The first (appendix B therein) is a comparison with the analytic high temperature expansion.  The second is against numerical simulations. For this we need to evaluate the infinite sum eq. \eqref{eq:sumDirich} to a sufficient approximation. Fortunately, this can be efficiently done by a transfer matrix method, describe in Appendix A of \cite{cao15gff}. The comparison with numerical simulation confirms the analytic continuation eq. \eqref{eq:etfDirichlet}, see Figure \ref{fig:DirichlethighT} for an example. 

\begin{figure}
\center
\includegraphics[scale=.37]{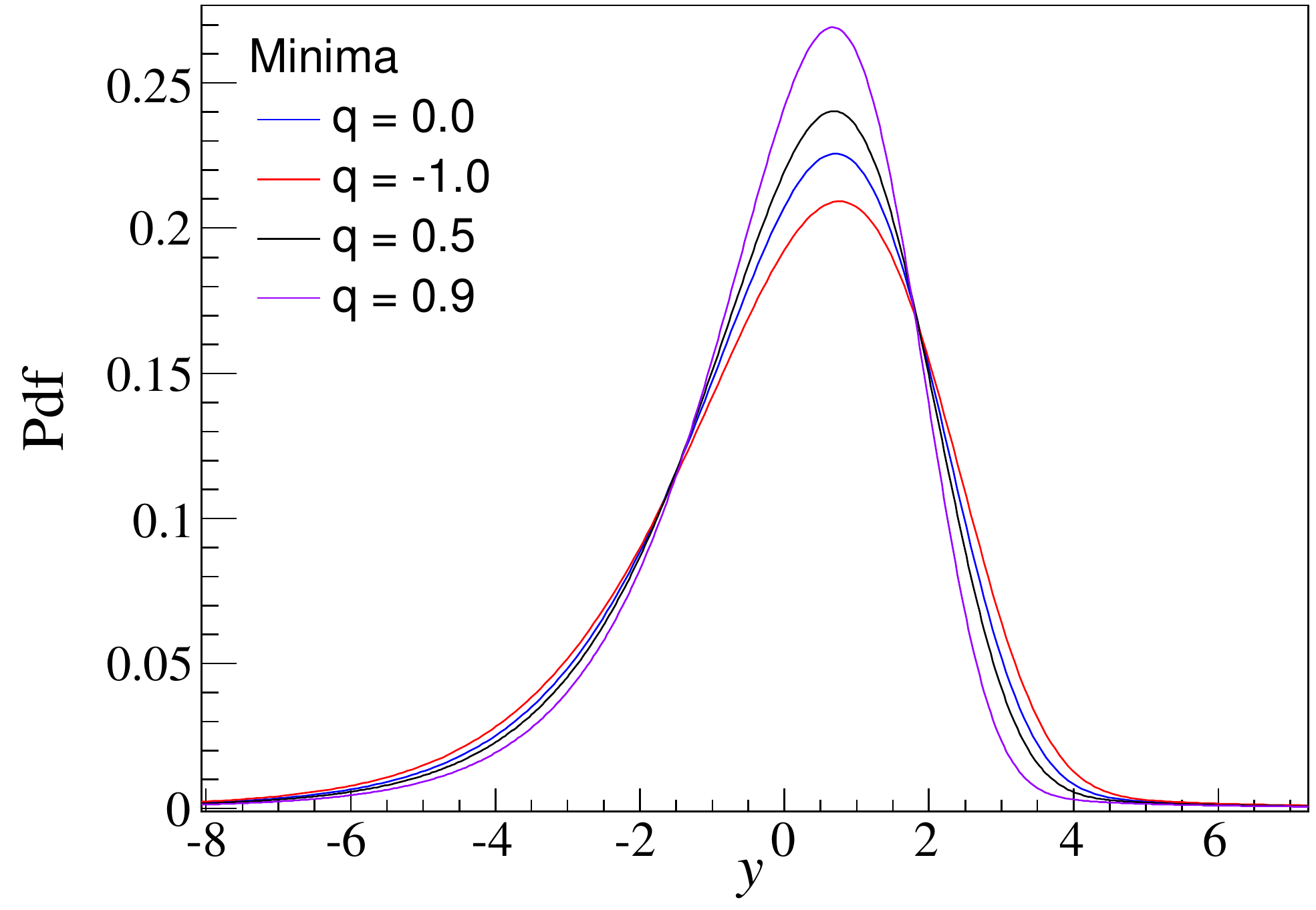}
\includegraphics[scale=.37]{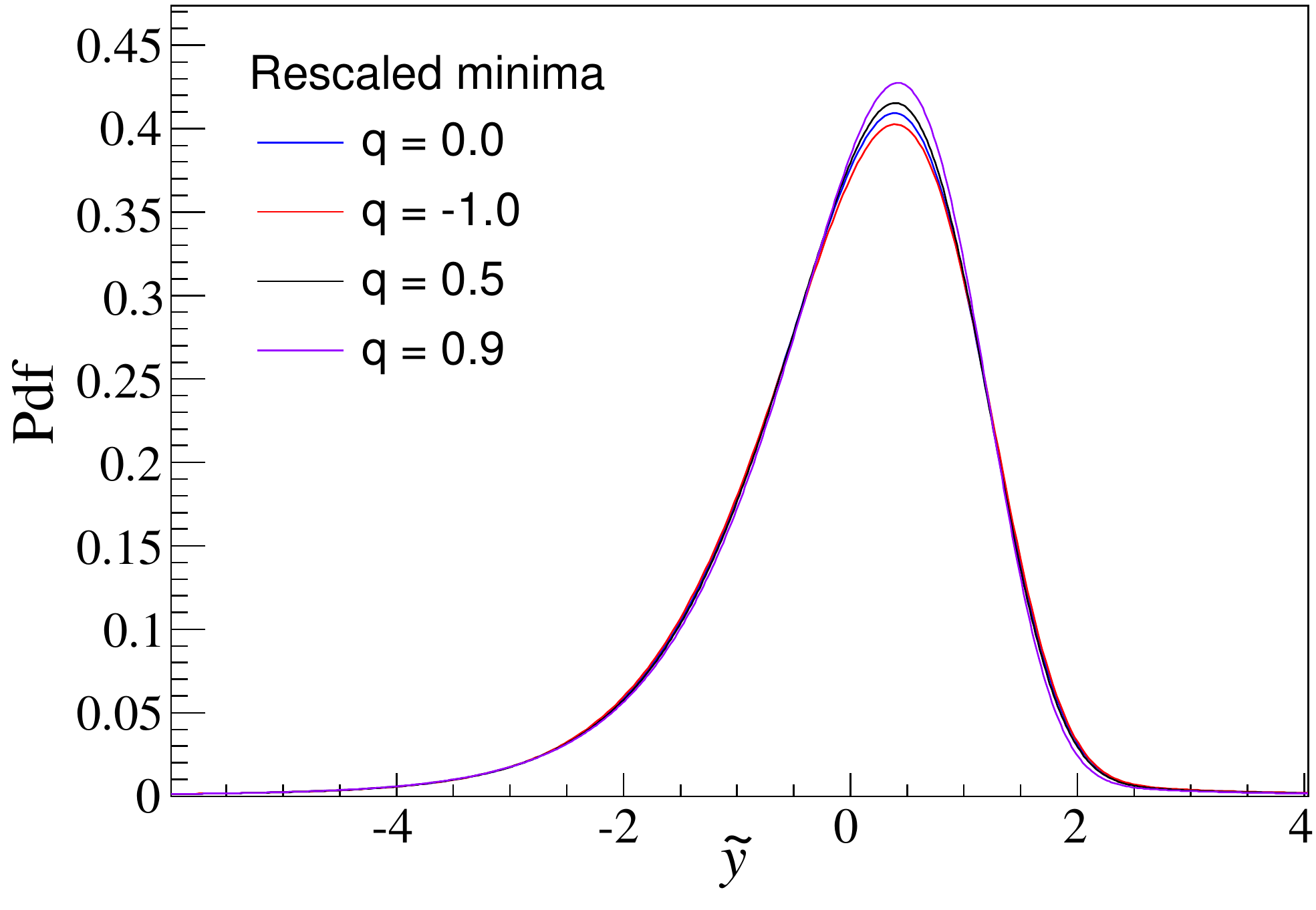}
\caption{Taken from \cite{cao15gff}. \textit{Left panel}: Distribution of the shifted minimum of the Dirichlet circular model for various values of $q$. All the distributions are obtained by numerical inverse Fourier transform applied to eq. \eqref{eq:etyDiri}, whereas the infinite sum in the last equation is efficiently calculated using the methods of \cite{cao15gff}. \textit{Right panel}: Distribution of the shifted minimum, rescaled so as to have unity variance. The rescaled distributions have still a non--trivial, although weak, dependence on the parameter $q$.} \label{fig:minimaDirichlet}
\end{figure}

Given the solution in the $\beta < 1$ phase, we can apply the freezing scenario, in particular the freezing of limit shape in eq. \eqref{eq:GbetaGFFshape} (which is by now a consequence of the \gls{rsb}, see section \ref{sec:freezingRSB}) to the $\beta > 1$ phase. In particular, the distribution of the (shifted) minimum, $y$, is given by the following (via Laplace transform),
\begin{equation}\label{eq:etyDiri}
\overline{\exp(ty)} = \Gamma(1 + t)^2 \mathbf{s}(t,1,q) \,.
\end{equation}
The inverse Laplace transform can be done numerically, and some results are plotted in Figure \ref{fig:minimaDirichlet}. The minimum distribution depends non--trivially on $q$, yet most of the dependence can be absorbed in a rescaling; the variance of the minimum distribution decreases as $q \to 1$, which is expected given the interpretation $q = R^{-2}$: when $R \to 1$, the unit circle approaches the disk boundary where the \gls{gff} is zero by the Dirichlet boundary condition. 

Remark that by eq. \eqref{eq:sumDirich}, $\mathbf{s}(t,\beta = 1, q)$ have poles only at $t = -2, -3, \dots$, so the rightmost pole of $\overline{\exp(ty)}$ is at $t = -1$ and of order $2$. By inverse Fourier transform, this implies $P(y\rightarrow-\infty) \sim A |y| e^{-|y|}$ for any $q$. This is a confirmation of the universality of the left--tail, predicted for general \glspl{logrem}, see eq. \eqref{eq:GbetaGFFtail}.

\subsubsection{Duality}
An interesting feature of the solution eq. \eqref{eq:etfDirichlet} and \eqref{eq:sumDirich} is its duality. Indeed, it is not hard to see from eq. \eqref{eq:sumDirich} that 
\begin{equation}\label{eq:Sdual}
\mathbf{s}(t,\beta^{-1},q) = \mathbf{s}(t,\beta,q) \,,
\end{equation}
because the term corresponding to $\lambda$ is transformed by the change of variable $\beta \mapsto \beta^{-1}$ to that of $\lambda^t$, the transpose of $\lambda$ (see section \ref{sec:jackintro}).
As a consequence, by eq. \eqref{eq:etfDirichlet}, $\overline{\exp(tf)\Gamma(1 + t/\beta)}$ is also invariant under the change of variable $\beta \to \beta^{-1}$. Therefore, the Dirichlet circular model provides a non-trivial check of the freezing-duality conjecture (section \ref{sec:treetoplane}). 

Moreover, the duality invariance is satisfied not only for the whole infinite series $\mathbf{s}(t,\beta^{-1},q)$, but also for every pair of transpose partitions (and every partition that is its own transform). This offers infinite potential opportunities of checking the freezing-duality conjectures. However, a prerequisite is finding the corresponding physical observables. The next section makes a first step in this direction, by interpreting the term corresponding to the first non-trivial term (corresponding to $\lambda = (1) = \square$), in terms of the Edwards-Anderson order parameter.   

\subsection{Edwards-Anderson order parameter} \label{sec:EA}
In this section and the next one, we shall go back to the original circular model ($q = 0$). Up to now, the only observable we have considered is the free energy distribution, related to extreme \textit{value} statistics. What about the extreme \textit{position} statistics, or, in finite temperature, \textit{Gibbs measure statistics}? Because of rotational invariance, the disorder--averaged Gibbs measure is trivially uniform. No higher order correlation function of the Gibbs measure is exactly known for the circular model. 

In this respect, the Edwards--Anderson (EA) order parameter is among the few exact results (besides the prediction of Liouville theory, see \ref{sec:liouville}, and predictions of \cite{fyodorov2015moments}) concerning the Gibbs measure/position of the circular model. Its name is motivated by seeing the position, $\exp(2\pi \im j / M)$ on the unit circle as a $O(1)$ (XY--model) spin. For a given disorder average, one defines its \textit{thermal} average as
\begin{equation}
\left<  \xi \right> =  \frac{1}{\mathcal{Z}}\sum_{j=1}^M  e^{-\beta V_j} \exp(2\pi \im j / M) \,,\, 
\mathcal{Z} = \sum_{j=1}^M  e^{-\beta V_j} \,, \label{eq:EAdef}
\end{equation}
where $V_{j}, j = 1, \dots, M$ are the potential values of the circular model of size $M$. We define the disorder--averaged modulus square of the above as the EA order parameter of the circular model, following the original proposition \cite{edwards1975theory}. 

\begin{figure}
\center
\includegraphics[scale=.5]{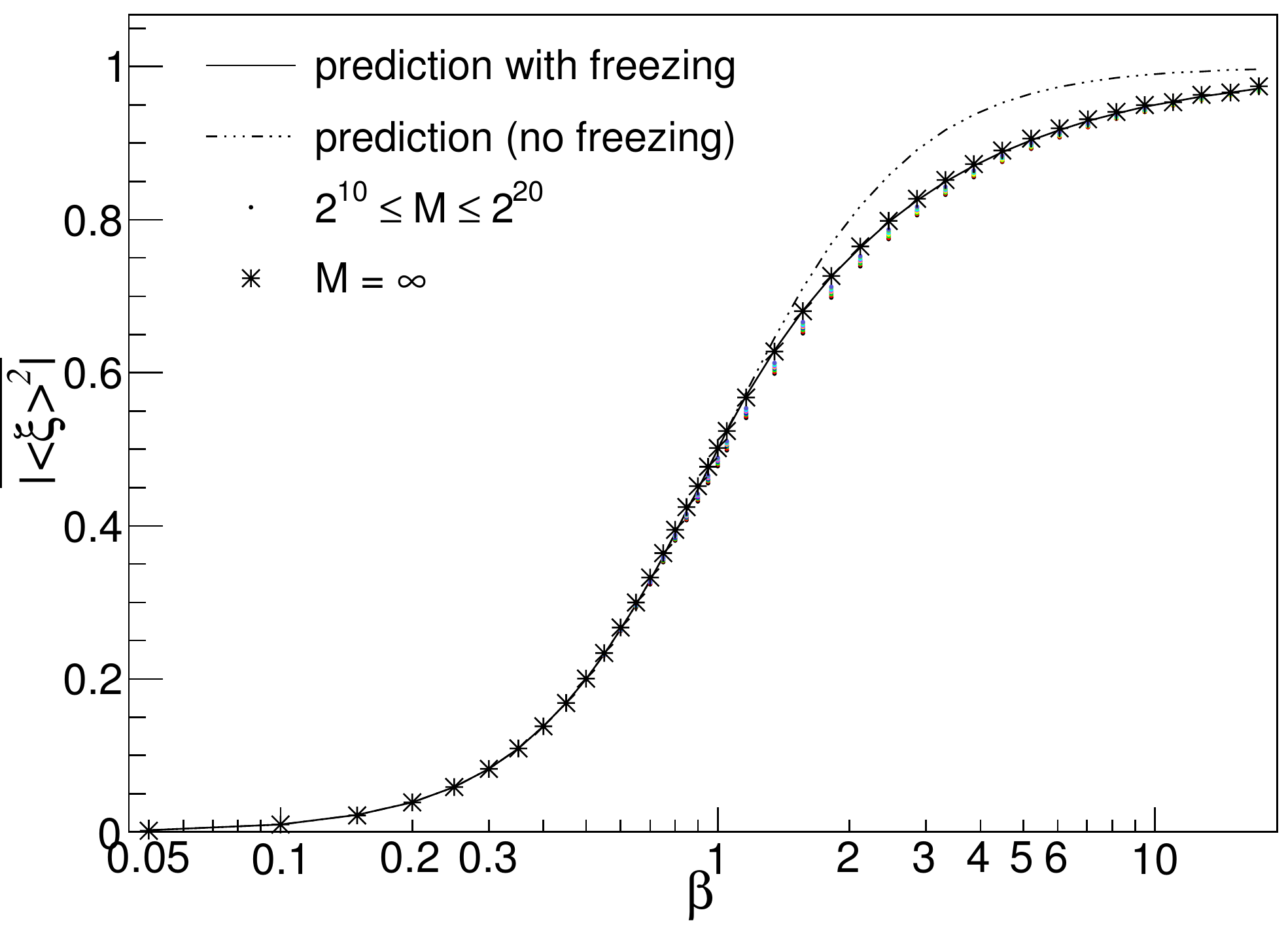}
\caption{Taken from \cite{cao16order}. Numerical check of the Edwards--Anderson order parameter of the circular model. The full curve plots the prediction eq. \eqref{eq:EAresult}; the dashed curves plots the $\beta < 1$ expression continued into the $\beta > 1$ phase. The points come from direct numerical simulation.} \label{fig:EA}
\end{figure}

In \cite{cao16maxmin}, the Edwards--Anderson parameter was exactly calculated to be the following:
\begin{equation}
\overline{\abs{\langle \xi \rangle}^2} \stackrel{M\rightarrow\infty}=
\begin{dcases} 
\frac{\beta^2}{1 + \beta^2} \, , & \beta \leq 1\, , \\
\frac{2\beta - 1}{2\beta}\, , & \beta \geq 1 \, .
\end{dcases} \label{eq:EAresult}
\end{equation}
We can make two remarks. First, the EA parameter is non-zero at any finite temperature. More interestingly, the non-analyticity at the freezing transition $\beta=1$ is very non-trivial. There is no evident reason to rule out the $\beta < 1$ expression as a solution in the $\beta > 1$ phase: the difference seems only quantitative. Nevertheless, the prediction eq. \eqref{eq:EAresult} is very well confirmed by numerical simulation, see figure \eqref{fig:EA}. 

Now let us briefly review two derivations of eq. \eqref{eq:EAresult}, that of \cite{cao16maxmin}, using freezing--duality conjecture, and that of \cite{cao16order} (Appendix B), using \gls{1rsb}.

\subsubsection{Freezing--duality conjecture}
As is usually the case, the two approaches are the same in the $\beta < 1$ phase, where it suffices to work in the continuum, and use the standard replica trick. For this, we consider the following observable $(z_a = e^{\im \theta_a})$
\begin{align}
 \overline{Z^n \left< \xi \right>^2} = \int_0^{2\pi} \prod_{a=1}^n \frac{\dif \theta_a}{2\pi}
 \prod_{a<b} \abs{ z_a - z_b}^{-2\beta^2} z_1 z_2^{*}  \label{eq:ZnEA}
\end{align}
 We can calculate it by relating it to eq. \eqref{eq:ZnsolveDir}. For this, we expand the latter as a power series in $q$, and takes out the $q^1$ term:
 \begin{equation}
 \int_0^{2\pi}  \prod_{a<b} \abs{ z_a - z_b}^{-2\beta^2} \sum_{a,b= 1}^n z_a z_b^{*} = \frac{\Gamma(1-n\beta^2)}{\Gamma(1 - \beta^2)}  \frac{ n }{n - 1 + \alpha} \,,
 \end{equation}
 where the right hand side correspond to the partition $\lambda=(1) = \square$. This above equation is related to eq. \eqref{eq:ZnEA} by symmetrization. Combining them, and expressing in terms of free energy ($n = -t/\beta$), we have
 \begin{equation}
\overline{e^{t F}\left(\beta - (t + \beta) \abs{\langle \xi \rangle}^2 \right)} \stackrel{\beta<1}{=} \frac{\Gamma(1+t\beta)}{\Gamma(1-\beta^2)^{-\frac{t}{\beta}}} \times \frac{1}{t + \beta + \beta^{-1}} \,. 
\label{eq:etfEA}
 \end{equation}
 Note that setting $t = 0$ gives already the $\beta < 1$ phase part of eq. \eqref{eq:EAresult}. 
 
 Now, the last factor of eq. \eqref{eq:etfEA} is the term corresponding to $\lambda  = \square = (1)$ in the infinite sum \eqref{eq:sumDirich}. The Young diagram is symmetric, so the term is a function of the dual variable $\beta + \beta^{-1}$. So it is reasonable to apply the freezing--duality conjecture to that term. The remaining term is equal to $\overline{e^{tF}}$ itself, of which we know the $\beta > 1$ phase behaviour. Defining the shifted free energy $f = F - \beta^{-1} \ln \Gamma(1 - \beta^2)$, the product of $\overline{e^{tf}} = \Gamma(1 + t\beta)$ with $\Gamma(1 + t/\beta)$ is dual and freezes in the $\beta > 1$ phase.
 \begin{equation}
\Gamma(1 + t/\beta) \overline{e^{t f}\left(\beta - (t + \beta) \abs{\langle \xi \rangle}^2 \right)}  \stackrel{\beta>1}{=}
 \Gamma(1+t)^2 \frac{1}{t + 2} \,.
 \end{equation}
 Setting $t = 0$ gives the $\beta > 1$ phase part of eq. \eqref{eq:EAresult}. Such a treatment of the EA order parameter is an example to be generalized to further terms. Doing this would provide infinite series of duality--invariant observables, indexed by (pairs of) partitions, and hopefully a clarification on the origin and generality of the duality invariance.
 
\subsubsection{1RSB insight}
The above application of the freezing--duality conjecture is quite tricky, although the result is confirmed very well by numerics. However, the $\beta > 1$ phase behaviour of the EA order parameter can be much better understood by the \gls{1rsb}. A more detailed account can be found in \cite{cao16order}, appendix B. Here, let us give a simple argument in terms of the overlap distribution. 

For this, note that the EA order parameter can be expressed as the inner product of the position of two independent thermal particles in a circular model: $\overline{\abs{\langle \xi \rangle}^2} = \overline{\langle \xi_1 \rangle \langle \xi_2 \rangle^*}$, where $\xi_1$ and $\xi_2$ are the position of the two independent thermal particles. 

Now, recall from section \ref{sec:RSB1}, discussion around eq. \eqref{eq:PoverlaplogREM} and \eqref{eq:moptimum}, that the two thermal particles have either overlap $\overlap = 1$ or $\overlap = 0$. The probability of $\overlap = 1$ is $\beta^{-1}$ in the $\beta > 1$ phase; by definition, when the overlap is $1$, the two particles are at the same continuum position, so $ \overline{\left<\xi \right>\left<\xi^*\right>} \stackrel{\overlap=1}=1$. The alternative conditional average can be obtained at the critical temperature, at which the overlap is always $0$: $\overline{\left<\xi \right>\left<\xi^*\right>} \stackrel{\overlap=0} = \overline{\left<\xi \right>\left<\xi^*\right>}_{\beta = 1}$. Adding the two possibilities, we obtain
\begin{equation}
\overline{\abs{\langle \xi \rangle}^2}_{\beta > 1} = 1 - \beta^{-1} \left( 1 -\overline{\abs{\langle \xi \rangle}^2}_{\beta = 1}\right) \,. \label{eq:freezingofEA}
\end{equation} 
We can check that this formula correctly reproduces the result eq. \eqref{eq:EAresult}. Moreover, the \gls{1rsb} reasoning reveals that $\overline{\abs{\langle \xi \rangle}^2}_{\beta > 1}$ is \textit{linear} in the $\beta^{-1}$ in the $\beta > 1$ phase; since it must be equal to $1$ at zero temperature, its value at $\beta = 1$ suffices to determine completely its glassy phase behaviour.   

\subsection{Minimum and maximum} \label{sec:minmax}
Another Jack polynomial application to the circular model is the joint min--max distribution. The treatment is quite similar to the Dirichlet circular model and was reported in considerable detail in \cite{cao16maxmin}; so we will sketch the approach and some main results below. Before that let us give a few motivations of the question in a more general context (following \cite{cao16maxmin}): 
\begin{itemize}
\item  The joint min--max distribution contains the distribution of the min--max difference. This quantity is more accessible to experiments, since it is the extremal width of the interface (log--correlated interfaces are studied experimentally \cite{aarts2004direct,devilleneuve2008statistics}), whereas the to define minimum value itself, one need a reference point. 
\item  Properties of opposite extrema are related to the \textit{diffusion dynamics} of an over-damped Langevin particle in the 1d potential. In particular, the max--min difference is the barrier that the particle should surmount to explore the whole system, and is thus related to Arrhenius passage times and to the diffusion coefficient in the periodic potential \cite{derrida1983velocity,ledoussal1995creep,dean14diffusion}. In the log--correlated 1D case, the freezing transition of \gls{logrem}'s manifests itself also in the dynamical exponents \cite{castillo2001freezing}. 
\item  Since the opposite extrema are far apart in space and in value, they are often assumed to the independent. As shall see below, this is a good approximation, but a  min--max correlation of order unity persists in the thermodynamic limit, and we will exactly predict (and test) it for the circular model. 
\end{itemize}

Now let us review the technical part. In order to access the maximum (and minimum) from a thermodynamic approach as a zero temperature limit, one should simply consider a \textit{negative} (and positive) temperatures. So, let us define the continuum partition functions
\begin{equation}
Z_{\pm} = \int_0^{2\pi} \frac{\dif \theta}{2\pi} e^{\mp \beta \phi(z)} \,,\, z = e^{\im \theta} \,,
\end{equation}
where $\phi(z)$ is the planar \gls{gff}: $\overline{\phi(z) \phi(w)} = -2 \ln\abs{z-w}$. Then we consider the replicated averages, which can be written out as Coulomb gas integrals:
\begin{align}
&\overline{Z_+^mZ_-^n} {=}
 \int \mu_n^{\alpha}(\underline{\xi})\mu^{\alpha}_m(\underline{\eta}) \prod_{a,b} \abs{1 - \xi_a^* \eta_b}^{-2/\alpha}, \label{eqznmdef} \\
&1/\alpha = -\beta^2, \, \mu_n^{\alpha}(\underline{\xi}) = \prod_{a=1}^n \frac{\dif \xi_a}{2\pi\im \xi_a} \prod_{a<a'}  \abs{\xi_a-\xi_{a'}}^{2/\alpha}\, .  \label{eq:alphabeta} 
\end{align}
Note that the product $ \abs{1 - \xi_a^* \eta_b}^{-2/\alpha} = (1 - \xi_a^* \eta_b)^{-1/\alpha}
(1 - \xi_a \eta_b^*)^{-1/\alpha}$ can be again written as an infinite sum of Jack polynomials using the Cauchy identity, eq. \eqref{eq:cauchy}. After that, orthogonality relations \eqref{eq:orthogonality} can be used to evaluate the integrals term by term. The result is 
\begin{align}
\overline{Z_+^n Z_-^m} = \frac{\Gamma(1 - n \beta^2)\Gamma(1 - m \beta^2)}{\Gamma(1 - \beta^2)^{m+n}}	 \sum_\lambda p_n^\lambda(\alpha) p_m^\lambda (\alpha) \,, \label{eqsmall:zn}
\end{align}
where $p_n^{\lambda}(\alpha)$ is  defined in \eqref{eq:pnJack}. After switching to the free energy $F_{\pm} = \mp\beta^{-1} \ln Z_{pm}$, and removing the shift as usual by defining $f_\pm = F_{\pm} \mp \beta^{-1} \ln \Gamma(1-\beta^2)$, we obtain the free energy distribution (in terms of Laplace transform) in the $\beta < 1$ phase:
\begin{align}
&\overline{\exp({t_1 f_+  - t_2 f_- })} \stackrel{\beta < 1}{=}  S_{\beta}(t_1,t_2)\prod_{i=1}^{2} \Gamma(1+\beta t_i) \,,\label{eqetfpm}  \\
&S_{\beta}(t_1,t_2) = \sum_\lambda \prod_{\substack{(x,y)\in \lambda \\ i=1,2}} \frac{x \beta^{-1} + y \beta + t_i  }{(x + 1)\beta^{-1} + (y + 1)\beta + t_i}\, ,  \label{eqS2v}
\end{align}
Note that the infinite sum $S_{\beta}(t_1,t_2)$, which is the novel term with respect to two copies of Dyson integrals, is again duality invariant, in the same manner as the sum eq. \eqref{eq:sumDirich} that appeared in the Dirichlet circle model. So we can implement the freezing scenario and the freezing--duality conjecture similarly. The result, in the $\beta \to \infty$ limit, is:
\begin{align} 
\overline{\exp(t_1 v_{+} - t_2 v_{-})} = S_{1}(t_1,t_2) \prod_{i=1}^2 \Gamma^{2}(1+t_i) \,, \label{eqetvv}
\end{align}
where $v_+$ and $v_-$ are the re-shifted minimum and maximum respectively. 

\begin{figure}
\center
\includegraphics[scale=.5]{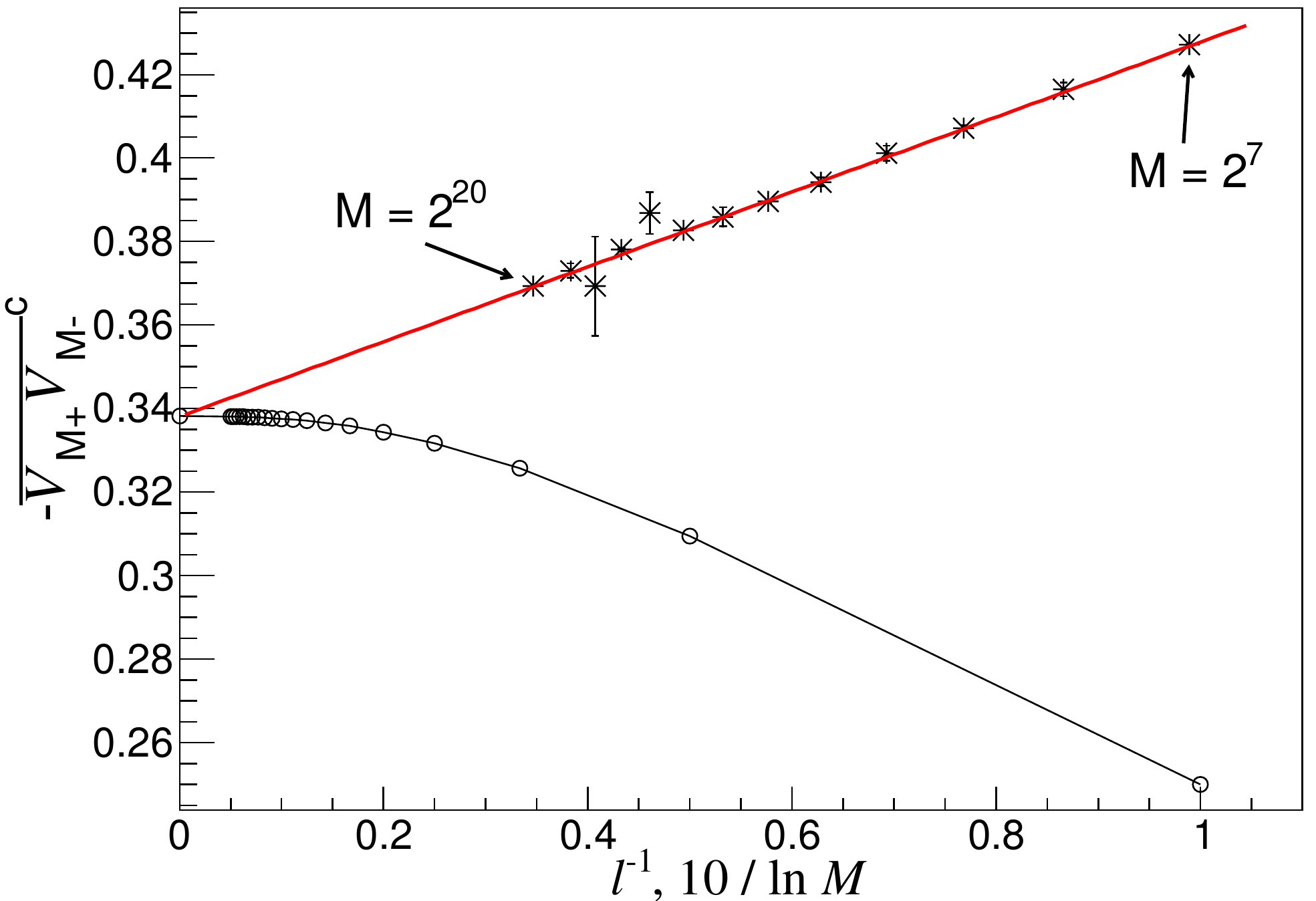}
\caption{Taken from \cite{cao16maxmin}. Numerical check of eq. \eqref{eqsumcov}, taken from \cite{cao16maxmin}. The numerical data ($*$) $2^7 \leq M \leq 2^{20}$ are consistent finite-size scaling $a + b / \ln M$, with $b = 0.89(1)$ and $a = 0.338(1)$, in $3$-digit agreement with \eqref{eqsumcov}. The sums over partitions in this work are all convergent and calculated by the method of \cite{cao15gff}, which involves a truncation size $l$. The sum \eqref{eqsumcov} truncated to $l = 1, \dots, 20$ are ploted ($\circ$) to appreciate convergence; in all cases $l \sim 10^2$ yields sufficient precision. }\label{fig:minmax}
\end{figure}
The main message of the result is that even in the $M\to \infty$ limit, the minimum ($V_{M+}$) and maximum ($V_{M-}$) are correlated, and the correlation is encoded in $ S_{1}(t_1,t_2)$. In particular, the min--max covariance can be calculated as 
\begin{align}
& -\overline{V_{M+} V_{M-}}^{c} \stackrel{M\rightarrow\infty}\longrightarrow  -\overline{v_{+} v_{-}}^{c} = \left. \frac{\partial^2 S_{1}}{\partial t_1 \partial t_2}\right\vert_{t_1,t_2 = 0} \nonumber \\
 = & \sum_{\lambda\neq \emptyset} 
 \frac{1}{4} \prod_{x,y \neq (0,0)} \frac{(x + y)^2}{(x + y + 2)^2}  = 0.338\dots \, , \label{eqsumcov}
\end{align}
This result was checked numerically in \cite{cao16maxmin}, see also Figure \ref{fig:minmax}. This numerical result is worth commenting as it shows clearly the strong finite size correction in the $\beta > 1$ phase of \glspl{logrem}, which should be taken into account to test thermodynamic limit predictions. In most cases, a low--order polynomial in $1/\ln M$ suffices to fit the data (this is also done in Figure \ref{fig:EA}; although the finite size correction is less visible there). This should be quite natural given the fact that $\ln M$ is the large parameter in the saddle point integrals that are behind the \gls{rsb} (see section \ref{sec:RSB}). 

\section{Extreme order statistics}\label{sec:orderstat}
In this section, we use the \gls{1rsb} method to study the statistical properties of the ordered sequence of minima of \glspl{logrem} (and of the \gls{rem}). Since this section will have the maximum technical density of the chapter, let us motivate it by pointing out why the question is interesting. 

Indeed, the ``frozen'' $\beta > 1$ phase of the \glspl{logrem} and the \gls{rem} is characterized the vanishing of extensive entropy, $S = \partial_{1/\beta} \mathcal{F} = o(\ln M)$, see eq. \eqref{eq:freezedual}. Therefore its thermal properties are governed by a few lowest energy levels. In this respect, we could have studied these models starting from the zero--temperature limit, by directly looking at the statistical properties of the lowest energy levels; then we could determine the boundary of the frozen phase, \textit{i.e.}, the temperature at which the lowest energies no longer dominate, and study the high temperature $\beta < 1$ phase by some other means. Such a strategy is common in many contexts with or without disorder. In quantum physics, a \textit{quantum} phase transition is by definition one at the zero (or vanishing) temperature, and is governed by the properties of the ground state (and the first excited ones). In the theory of disordered systems, the Ruelle cascade \cite{ruelle1987mathematical} is also a description of the minimal energy levels of mean--field spin glasses and underlies in fact the \acrlong{rsb}.

Our approach in this chapter is the opposite: we start from the high temperature phase and access the $\beta < 1$ phase and the zero--temperature limit by crossing the freezing transition, by employing the \acrlong{rsb} machinery. Although the latter is not at all rigorous and has many uncontrolled aspects, it has the advantage of providing access to model dependent \gls{ir} and \gls{uv} data, and determine the rôle they played in the statistical properties of the minima. 

For this, we need to develop a replica approach to second, third, \textit{etc}, minima. This is not an obvious task:  up to now, we have been considering the free energy, which tends to the minimum in  the zero--temperature limit. New finite--temperature observables should be designed that give access to the second minimum, \textit{etc}. They will be introduced in section \ref{sec:REMorder}, and applied to the \gls{rem} to retrieve the well--known results.  Then, section \ref{sec:2ndmin} applies the method to \glspl{logrem} (using again the circular model as example). We will give and test predictions on the distribution of second minimum value and of the gap between first and second minima, following \cite{cao16order}. In this work, the approaches were generalized to higher order minima to determine the full minima process. The result will be summarized and explained in section \ref{sec:fullmin}.

\subsection{The method and the REM case}\label{sec:REMorder}
\begin{figure}
\center \includegraphics[scale=.6,valign=c]{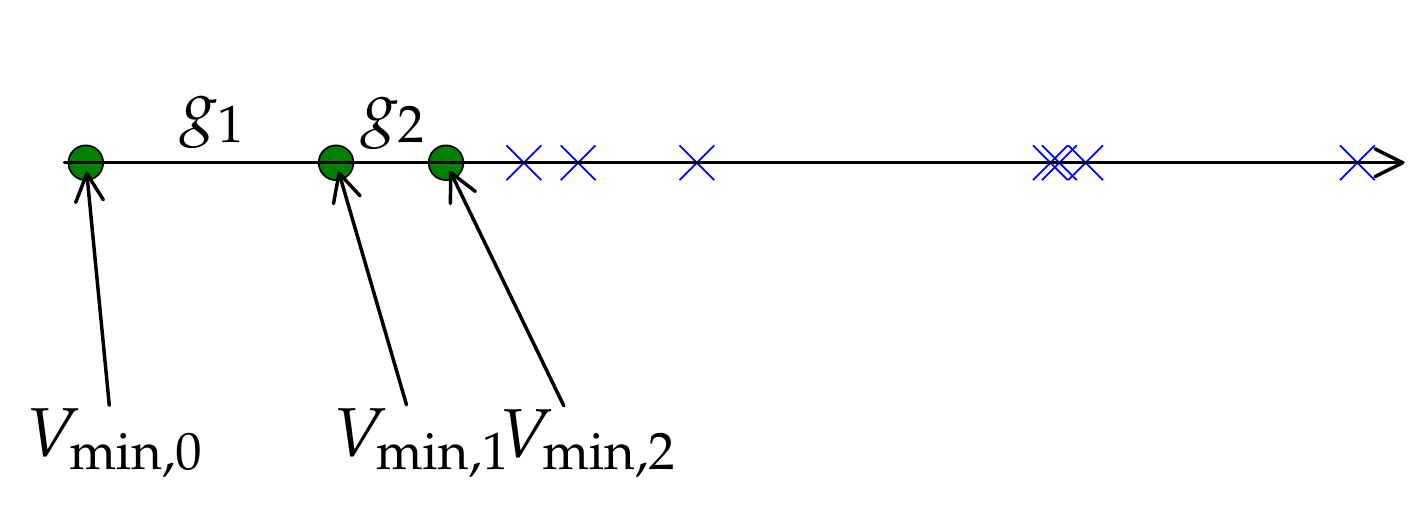}$V_j$
\caption{A sketch of the basic objects of extreme order statistics: the minima value $V_{\min, k}, k = 0, 1, \dots$ and the gaps $g_k = V_{\min,k} - V_{\min, k-1}$.}\label{fig:higherminsketch}
\end{figure}
Extreme order statistics studies the statistical properties of the ordered sequence of the minima
\begin{equation}
V_{\min} = V_{\min,0} < V_{\min, 1}  < V_{\min, 2} < \dots 
\end{equation}
among a set of random variables $V_1, \dots, V_M$, which will be the energy levels/potential levels of the \gls{rem} or some \gls{logrem}. We stress that in our notation, $V_{\min,0} = V_{\min}$ is the absolute minimum value, $V_{\min,1}$ is the \textit{second} minimum value, \dots, $V_{\min,k}$ the $(k+1)$-th minimum value. We wish to calculate the joint distribution of $V_{\min}, V_{\min, 1}, \dots, V_{\min,k}$, for any fixed $k$ in the $M \to \infty$ limit. In more mathematical terms, the goal is the \textit{minima process}. Note that the Ansatz for the extreme value statistics extends to the minima process: all the minima need to be shifted by a leading behaviour $a_M$, so that $(V_{\min,k}-a_M)_k$ has a limit joint distribution. Besides $V_{\min,k}$  themselves, it is also important to consider the \textit{gaps} between minima:
\begin{equation}
g_k = V_{\min,k} - V_{\min,k-1} \,,\, k = 1,2,3,\dots \,.
\end{equation}
The joint distribution of $(g_k)_{k\geq 1}$, often called minima process \textit{seen from the tip}, is defined regardless of the re-shifting by $a_M$. A sketch of the gaps and higher minima is provided in Figure \ref{fig:higherminsketch}. 

\subsubsection{Statistical mechanics approach to higher minima}
Now we discuss the finite--temperature observables that we will calculate and which will give higher extreme order statistics in the zero--temperature limit. For this, recall that for the absolute minima,  we defined the following \eqref{eq:Gasproduct}
\begin{equation}
G_\beta(y) = \overline{\prod_{j=1}^M  \theta_\beta(V_i - y)} \,, \theta_\beta(x) = \exp(-e^{-\beta x}) \,.
\end{equation}
Since $\theta_\beta (x) \to \theta(x) = \begin{cases} 0 & x<0 \\ 1 & x > 0\end{cases}$ as $\beta \to \infty$, $G_\infty(y) = \overline{\prod_{j=1}^M \theta(V_i - y)} = \overline{\theta(V_{\min,0} > y)}$ is the cumulative distribution function of $V_{\min,0}$. To go a step further to the second minimum, we consider the following observable:
\begin{equation}
H_\beta(y_0, y) = \sum_{j=1}^M \overline{ (1 -  \theta_\beta(V_j - y_0)) \prod_{i\neq j} \theta_\beta(V_i - y) } \,,\quad y_0 < y \,.\label{eq:defH}
\end{equation}
Its $\beta \to \infty$ limit can be similarly taken: 
\begin{align}
&H_\infty(y_0, y) = \sum_{j=1}^M \overline{ (1 -  \theta(V_j - y_0)) 
\prod_{i\neq j} \theta(V_i - y) } = \overline{(1 - \theta(V_{\min} - y_0)) \theta(V_{\min,1}-y) } \,. \nonumber
\end{align}
Indeed, it is clear that for $j$ fixed, $\overline{ (1 -  \theta(V_j - y_0)) \prod_{i\neq j} \theta(V_i - y) }$ is the probability that $V_j < y_0$ is the absolute minimum, and that the second minimum is larger than $y$. Summing over $j$ forgets about the minimum position and gives the last expression. So $H_\infty$ is a joint cumulative distribution of $V_{\min}$ and $V_{\min,1}$. Taking derivatives, we obtain
\begin{align}
& - \partial_{y,y_0} H_\infty = \overline{\delta(V_{\min} - y_0) \delta(V_{\min,1} - y)} \,,\, y_0 < y \,. \label{eq:Hinfty}
\end{align}

So calculating $H_\beta(y_0, y)$ for any temperature is more than sufficient to determine the joint distribution of $V_{\min}$ and $V_{\min,1}$, and we will calculate $H_\beta(y_0,y)$ using the replica trick and the \gls{1rsb}. The method is again comparable to that of $G_\beta(y)$, which we recall can be seen as an exponential generating function of the replicated partition sums: $G_\beta(y) = \sum_{n=0}^{\infty} \overline{\mathcal{Z}^n}(-e^{\beta y})^n / n!$, eq. \eqref{eq:Gasseries}. A similar development applies to $H_\beta$. Indeed, eq. \eqref{eq:defH} implies that
\begin{align*}
&H_\beta(y_0, y) \\
 =& \sum_{j=1}^M \overline{\left[\theta_\beta(V_j + \infty - y) 
   \prod_{i\neq j} \theta_\beta(V_i - y)  - \theta_\beta(V_j + (y-y_0) - y) 
 \prod_{i\neq j} \theta_\beta(V_i - y) \right]} \\
 =& -\sum_{j=1}^M \left.\overline{\theta_\beta(V_j + \Delta - y) 
  \prod_{i\neq j} \theta_\beta(V_i - y)} \right\vert^{y-y_0}_{\Delta = + \infty} \,,\, 
\end{align*}
where we used the ``Newton--Leibniz'' notation 
\begin{equation}
f(x)\vert_{x=a}^b = f(b) - f(a) \label{eq:NewtonLeibniz} \,.
\end{equation}
Now, using the definition of $\theta_\beta$,  we have 
\begin{align}
H_\beta(y_0, y)& =  -\sum_{n=0}^{\infty} \frac{(-e^{\beta y})^n}{n!} \sum_{j=1}^M  \left. \overline{\mathcal{W}^n(j, \Delta)} \right\vert^{y-y_0}_{\Delta = + \infty} \,, \label{eq:HGen} \\
\mathcal{W}(j, \Delta) & =  e^{-\beta (V_j + \Delta)} + \sum_{k\neq j} e^{-\beta V_k} \,. \label{eq:Wdef}
\end{align}
In other words, the ``partition fucntion'' $\mathcal{W}(j, \Delta)$ differs from $\mathcal{Z}$ in that the energy level $V_j$ is shifted by $\Delta$: $V_j \leadsto V_{j} + \Delta$. 

Our basic working hypothesis is that the \gls{1rsb} applies to the replica sum $\overline{\mathcal{W}^n(j,\Delta)}$ as well as to $\overline{\mathcal{Z}^n}$, and the grouping of the replicas are still described by  \eqref{eq:moptimumREM} and \eqref{eq:moptimum} for the \gls{rem} and the \glspl{logrem}. This is reasonable because the \gls{rsb} Ansätzse in general are assertions about which configurations of replica positions $j_1, \dots, j_n$ dominate $\overline{\mathcal{W}^n}$ in the thermodynamic $M \rightarrow \infty$ limit. As a consequence, the outcome of the approach depends only on the thermodynamics of the model, and is not altered by a finite (as $M\rightarrow\infty$) energy shift performed in eq. \eqref{eq:Wdef}. 

\subsubsection{The REM case}
Let us illustrate how to proceed in the simple \gls{rem} case. First, we write down the replicated partition sum explicitly
\begin{equation}
\sum_{j=1}^M \left.\overline{\mathcal{W}^n(j, \Delta)}\right\vert_{\Delta=+\infty}^{y-y_0} = 
\sum_{j=1}^M \sum_{j_1, \dots, j_n} \overline{\exp\left( -\beta \sum_{a=1}^n V_{j_a} \right)} 
\left.\left[ \prod_{a: j_a = j} e^{-\beta \Delta} \right]\right\vert_{\Delta=+\infty}^{y-y_0} \,.
\label{eq:Wnsum}
\end{equation}
This formula is true in general and not restricted to the \gls{rem}. An important general observation is that any term for which $j_a \neq j$ for any $a$ vanishes, see eq. \eqref{eq:Wdef}. So the sum over $n+1$ positions $j, j_1, \dots, j_n = 1, \dots, M$ is in fact restricted to that over $j_1, \dots j_n$. In the \gls{1rsb} approach, these $n$ replicas form $n/m$ groups of $m$ replicas, each occupying a distinct position; the group size is given by eq. \eqref{eq:moptimumREM}, copied below:
$$ m = \begin{cases}
1 & \beta \leq 1 \,,\\
1/\beta & \beta >  1 \,.
\end{cases}$$
 Recall also that the sum over $j_1, \dots, j_n$ reduces essentially to the combinatorics of partitioning $n$ replicas into groups of $m$, see discussions around eq. \eqref{eq:REMZnreplica}. The only novelty for $\overline{\mathcal{W}^n(j,\Delta)}$ compared to $\overline{\mathcal{Z}^n}$ is that, one need to choose one group (among $n/m$) to which the ``replica'' $j$ belongs. Note also that extra factor $\prod_{a: j_a = j} e^{-\beta \Delta} = e^{-\beta m \Delta}$ in any case. With these observations in mind, by comparing the \gls{1rsb} solution to $\overline{\mathcal{Z}^n}$ and $\sum_{j=1}^M \left.\overline{\mathcal{W}^n(j, \Delta)}\right\vert_{\Delta=+\infty}^{y-y_0}$, it is not hard to show that 
\begin{equation}
\sum_{j=1}^M \left.\overline{\mathcal{W}^n(j, \Delta)}\right\vert_{\Delta=+\infty}^{y-y_0} = 
 \frac{n}{m} e^{-\beta m \Delta} \overline{\mathcal{Z}^n} = \begin{dcases}
 n e^{-\beta \Delta}  \overline{\mathcal{Z}^n}  & \beta \leq 1 \,, \\ 
 n\beta e^{- \Delta}  \overline{\mathcal{Z}^n}  & \beta >  1 \,, 
 \end{dcases}  \quad \Delta = y - y_0 \,.
\end{equation}
Using the respective generating function formulas eq. \eqref{eq:Gasconvolution} and \eqref{eq:HGen}, the above equation translates into 
\begin{align}
H_\beta(y_0,y) &= -\partial_y G_\beta(y) K_\beta(y-y_0)  \,,\, \label{eq:HGrelationREM} \\
 K_\beta(\Delta) &= K^{\text{REM}}_\beta(\Delta) = \begin{dcases}
 \beta^{-1} e^{- \beta \Delta}  \,, & \beta \leq 1 \,, \\
  e^{- \Delta}  \,, & \beta > 1 \,,
\end{dcases}  \label{eq:KREM}
\end{align} 
Since $G_\beta(y)$ for the \gls{rem} is known (eq. \eqref{eq:GbetaREMhighT} and \eqref{eq:GbetaREMfreeze}), the above two equations complete the \gls{1rsb} calculation of the observable $H_\beta(y_0, y)$, designed to recover the joint distribution of the first and second minima value at the $\beta \to \infty$ limit, by eq. \eqref{eq:Hinfty}. Before applying it, we have a few remarks. First, eq. \eqref{eq:HGrelationREM} and \eqref{eq:KREM} can be checked to be exact in the $M \to \infty$ limit without using \gls{1rsb}. Second, as we will see in section \ref{sec:2ndmin}, eq. \eqref{eq:HGrelationREM} still holds for \glspl{logrem}, upon a non-trivial modification of $K_\beta(\Delta)$ in the $\beta > 1$ phase. 

\subsubsection{Minima and gap distributions}
Now let us come to the $\beta \to \infty$ limit. Applying eq. \eqref{eq:Hinfty} to eq. \eqref{eq:HGrelationREM} gives the joint \gls{pdf} of the first and second minima:
\begin{align}
&\overline{\delta(V_{\min,1} - y) \delta(V_{\min} - y_0)} = -G''_\infty(y) K'_\infty (y - y_0)- G'_\infty(y)  K''_\infty (y - y_0)  \,,  \label{eq:pdfmin1}
\end{align}
where $y_0 < y$. It is not hard to derive some marginal distribution, in which we are more interested. For example, the distribution of the (first) gap $g_1 = V_{\min,1}-V_{\min}$ can be obtained by integrating over $y$ while keeping $y-y_0 = \Delta$ fixed:
\begin{align}
\overline{\delta(g_1 - \Delta)} = K_\infty''(\Delta) \,,\,  \label{eq:gappdf}
\end{align}
of which a useful consequence is obtained by integrating twice while assuming $K_\infty(\Delta)$ (as well its derivative) vanishes at $\Delta\to +\infty$:
\begin{align}
 &\overline{\delta(g_1 - \Delta)} = - K_\infty'(\Delta) \,, \label{eq:gapcdf}\\
 & K_\infty(0) = \overline{g_1}\,. \label{eq:Kroisg1}
\end{align}
On the other hand, by fixing $y$ fixed and integrating eq. \eqref{eq:pdfmin1} over $\int_{-\infty}^{y} \dif y_0$ we obtain the \gls{pdf} of the second minimum 
\begin{align}
&\overline{\delta(V_{\min,1} - y)} =  - G'_\infty(y) + \overline{g_1} G''_\infty(y) \Leftrightarrow
\overline{\theta(V_{\min,1} - y)} =  G_\infty(y) -  \overline{g_1}  G'_\infty(y)  \,, \label{eq:pdf2ndmin}
\end{align}
where we also used eq. \eqref{eq:Kroisg1} and \eqref{eq:gappdf}. Note that equations \eqref{eq:pdfmin1} through \eqref{eq:pdf2ndmin} are all consequences of eq. \eqref{eq:HGrelationREM}, and will be used in section \ref{sec:2ndmin}. 

We now specify to the \gls{rem} case. For the gap distribution, combining eq. \eqref{eq:gappdf} and eq. \eqref{eq:KREM} gives the standard exponential distribution:
\begin{equation}
\overline{\delta(g_1 - \Delta)} = e^{-\Delta} \,,\, \Delta \geq 0 \,. \label{eq:REMgap} 
\end{equation}
In particular $\overline{g_1} = 1$. For the second minimum, let us recall from \eqref{eq:GbetaREMfreeze} that up to a shift $a_M = -2 \ln M + \frac12 \ln \ln M + c$, $G_\infty(y - a_M) = \exp(-e^y)$ (in the $M\to \infty$ limit). Then, eq. \eqref{eq:pdf2ndmin} gives the distribution of the second minimum shifted by the same $a_M$:
\begin{equation}
\overline{\delta(V_{\min,1} - a_M - y)} = \exp\left(2y - e^{y}\right) \,. \label{eq:REM2nd}
\end{equation}

\subsubsection{Poisson point process}
Equations \eqref{eq:REMgap} and \eqref{eq:REM2nd} are well--known results of the \gls{rem}. Indeed, the whole minima process of the \gls{rem}, when shifted by $a_M$, is known to be a \gls{ppp} of density $\rho(y) \dif y = e^{y} \dif y$, called a \textit{Gumbel} \gls{ppp}. In general, a \acrlong{ppp} with density $\rho(y) \dif y$ can be described as follows: we cut the real line into infinitesimal intervals $[y, y + \dif y]$. For each interval, we put a point into it with probability $\rho(y) \dif y$; otherwise we leave the interval empty. This procedure is repeated independently for each interval, and the resulting \textit{random} point set is a realization of the \gls{ppp} with density $\rho(y) \dif y$. The well-known result for the \gls{rem} is that, if $\rho(y) = e^y$, the \gls{ppp} (called the Gumbel \gls{ppp}, for a reason discussed below) have the same statistical properties as the set of minima $\{ (V_{\min,k} - a_M)_{k=0}^{\infty} \}$ of the \gls{rem}, in the $M\to \infty$ limit. 

The simplest argument that allows to understand this result is to recall that the mean number of \gls{rem} energy levels in $[y + a_M, y + a_M + \dif y]$ is (see eq. \eqref{eq:REMentropy}) $$\frac{M}{\sqrt{4\pi \ln M}}  \exp\left(- \frac{(y+a_M)^2}{4 \ln M} \right)  \dif y \to e^{y} \dif y \,,$$
with $a_M = - 2\ln M + \frac12 \ln \ln M + \ln \sqrt{4\pi}$. The independence between the \gls{rem} energy levels leads reasonably (although we do not prove it here) to the independence between intervals $[y, y + \dif y]$. 

Now, given the definition of the \gls{ppp} and its identification to the \gls{rem} minima process, we can calculate the joint distribution of the minima $  V_{\min,0} - a_M, \dots,  V_{\min,k} - a_M$, $k \geq 1$. For example, let us consider $\overline{\delta(V_{\min,0} - a_M - y_0) \theta(V_{\min,1} - a_M - y)}$ ($y_0 < y$). It is not hard to see that this is the probability (density) that there is a particle is put in the interval at $y_0$, and that no particle is put in all other intervals from $-\infty$ to $y$: 
\begin{align}
&\overline{\delta(V_{\min} - a_M - y_0) \theta(V_{\min,1} - a_M - y)} 
= e^{y_0} \prod_{y'=-\infty}^y (1 - e^{y'} \dif y') \nonumber \\
=& e^{y_0} e^{-\int_{-\infty}^{y} e^{y'} \dif y' }=  \exp\left(y_0 - e^y \right) \,. \label{eq:PPP2nd}
\end{align}
It is quite straightforward to check eq. \eqref{eq:REM2nd} and \eqref{eq:REMgap} as consequences of the above equation. The calculation can be easily generalized to the joint distribution of $V_{\min,0}, \dots, V_{\min,k}$ for any $k$ (see for example, Appendix C of \cite{cao16order}). A particular nice consequence is the \gls{pdf} of the $(k+1)$-th minimum $V_{\min,k}$:
\begin{equation}
\overline{\delta(V_{\min,k} - a_M - y)} = \frac{1}{k!} \exp\left((k+1)y - e^{y}\right) \,.
\end{equation}
In particular, for $k = 0$, $V_{\min,k}-a_M$ has a (negative) Gumbel distribution. This justifies the name ``Gumbel \gls{ppp}''.

\subsection{Second minimum in logREMs}\label{sec:2ndmin}
Unlike the \gls{rem}, the full minima process is \textit{explicitly} known for no \glspl{logrem}. We emphasized the adjective ``explicit'', which means that, for instance, the first gap distribution is exactly known in terms of an explicitly enough formula for no \glspl{logrem}; the same can be said about the second minimum. Nevertheless, we emphasize that much is known about the full minima process of \glspl{logrem}, as we will review in section \ref{sec:fullmin}. 

The literature on this topic is very mathematical. In particular, there was no comprehensive account of how to understand the results using the \gls{1rsb} approach. Such an account was provided in  \cite{cao16order}. Working at a physicist's level of rigour, we were able to consider general \gls{logrem} (in the sense discussed in section \ref{sec:logREMdef}). Most importantly, the \gls{1rsb} approach allowed to see how the \gls{ir} and \gls{uv} data determine different parts of the full minima process, as we anticipated in section \ref{sec:IRUV}. The minimum distribution (up to an $O(1)$ shift) depends solely on the \gls{ir} data (this was already seen in section \ref{sec:freezingRSB}), while the minima process \textit{seen from the tip} (\textit{i.e.}, the distribution of the gaps) depends solely on the \gls{uv} data. This claim is shown \cite{cao16order} in full generality. In fact, it is better understood using the notion of decorated \acrlong{ppp}, that we will discuss in section \ref{sec:fullmin}. 

An advantage of the replica approach is that almost all the essential points can be appreciated at the level of first and second minimum. The generalization to higher orders is tedious but straightforward. Therefore, we shall restrict the discussion of this section to first and second minimum, using again the circular model as the example, on which we shall also test our predictions numerically.

\subsubsection{Second minimum by 1RSB}
We shall study the same replicated sum $\sum_{j=1}^M \left.\overline{\mathcal{W}^n(j, \Delta)}\right\vert_{\Delta=+\infty}^{y-y_0}$ given by eq. \eqref{eq:Wnsum}, using the \gls{1rsb} approach. Given the preparation of section \ref{sec:REMorder} ($\mathcal{W}$ in the \gls{rem}) and \ref{sec:freezingRSB} ($\mathcal{Z}$ in \glspl{logrem}), we prefer not to follow in detail the general account, which can be found in \cite{cao16order}, section IV.C/4.3. Rather, let us highlight the differences of \gls{logrem} compared to the \gls{rem}: 
\begin{enumerate}
\item The replicas of different groups still have non--trivial interactions, which are determined by the \gls{ir} data, and rise to a Coulomb gas integral of $n/m$ charges (eq. \eqref{eq:DysonRSB}), each with ``renormalized'' (attractive) charge $\beta \to  m\beta$, where the group size $m$ is still given by eq. \eqref{eq:moptimum}. However, the Coulomb gas integral given by $\sum_{j=1}^M \left.\overline{\mathcal{W}^n(j, \Delta)}\right\vert_{\Delta=+\infty}^{y-y_0}$ is the same as that of $\overline{\mathcal{Z}^n}$. So the \gls{ir} data is \textit{not} the crucial point here.
\item In \glspl{logrem}, The replicas of a same group are not on the same lattice point, but are confined in a block of size $N$, $1 \ll N \ll M$. This gives rise to the non--trivial \textit{intra}--group energies $E_{m,\beta}^{n/m}$, see eq. \eqref{eq:Exi}, when calculating $\overline{\mathcal{Z}^n}$. Now, for $\sum_{j=1}^M \left.\overline{\mathcal{W}^n(j, \Delta)}\right\vert_{\Delta=+\infty}^{y-y_0}$, the ``replica'' $j$ must be in one of the blocks occupied by a replica group $g$. So, the intra--group energy of other groups is still $E_{m,\beta}^{n/m}$ but that of the group $g$ is changed to 
\begin{align}
D_{m,\beta}(\Delta) = & \frac1N \sum_{i=1}^N \sum_{i_1, \dots, i_m = 1}^N
\prod_{l,l' = 1}^m \exp(\beta^2 (e_0 - f(i_l-i_{l'})) / 2) \nonumber \\
\times & \prod_{i: i_{l} = i}
\exp(-\beta \Delta)    \label{eq:Dxi} 
\end{align}
\end{enumerate}
The above explanation, translated into equation, reads simply as follows:
\begin{equation}
\sum_j \left. \overline{\mathcal{W}^n(j, \Delta)}\right\vert_{\Delta = +\infty}^{y- y_0}
\nonumber =  \frac{n}{m} \overline{\mathcal{Z}^n} \left. \frac{ D_{m,\beta}(\Delta) }{E_{m,\beta} } \right\vert_{\Delta = +\infty}^{y - y_0} \,, \label{eq:WnZn1}
\end{equation}
where the fraction $D/E$ ensures that the intra--group interaction of the group $g$ becomes the correct one. Using the generating function formulas eq. \eqref{eq:HGen} and \eqref{eq:Gasseries}, we see that
\begin{equation}
H_\beta(y_0, y) = - \partial_y G_\beta(y) K_\beta(\Delta) \left. \frac{ D_{m,\beta}(\Delta) }{E_{m,\beta} } \right\vert_{\Delta = +\infty}^{y - y_0} \,,\, \beta \geq 1 \,. \label{eq:main1}
\end{equation}
The $\beta < 1$ phase expression is slightly different, but will not interest us here. Eq. \eqref{eq:main1} entails that, eq. \eqref{eq:HGrelationREM} still holds for \glspl{logrem}, with
\begin{equation}
K_\beta(\Delta) = \left.\frac{ D_{m,\beta}(\Delta) }{E_{m,\beta} } \right\vert_{\Delta = +\infty}^{y - y_0}  \,,\, \beta \geq 1 \,, \label{eq:KDE}
\end{equation}
which is different from eq. \eqref{eq:KREM}. Note that $K_\beta(\Delta)$ depends \textit{only} on the \gls{uv} data, as do both $E$ and $D$. This is why we call it the \textit{UV corrector}, following \cite{cao16order}. Because by eq. \eqref{eq:gappdf}, the \gls{pdf} of the first gap is given by $K_\infty''(\Delta)$, it is only affected by the \gls{uv} data, as we claimed in section \ref{sec:IRUV}.

\begin{figure}
\center
a. \includegraphics[scale=.5,valign=t]{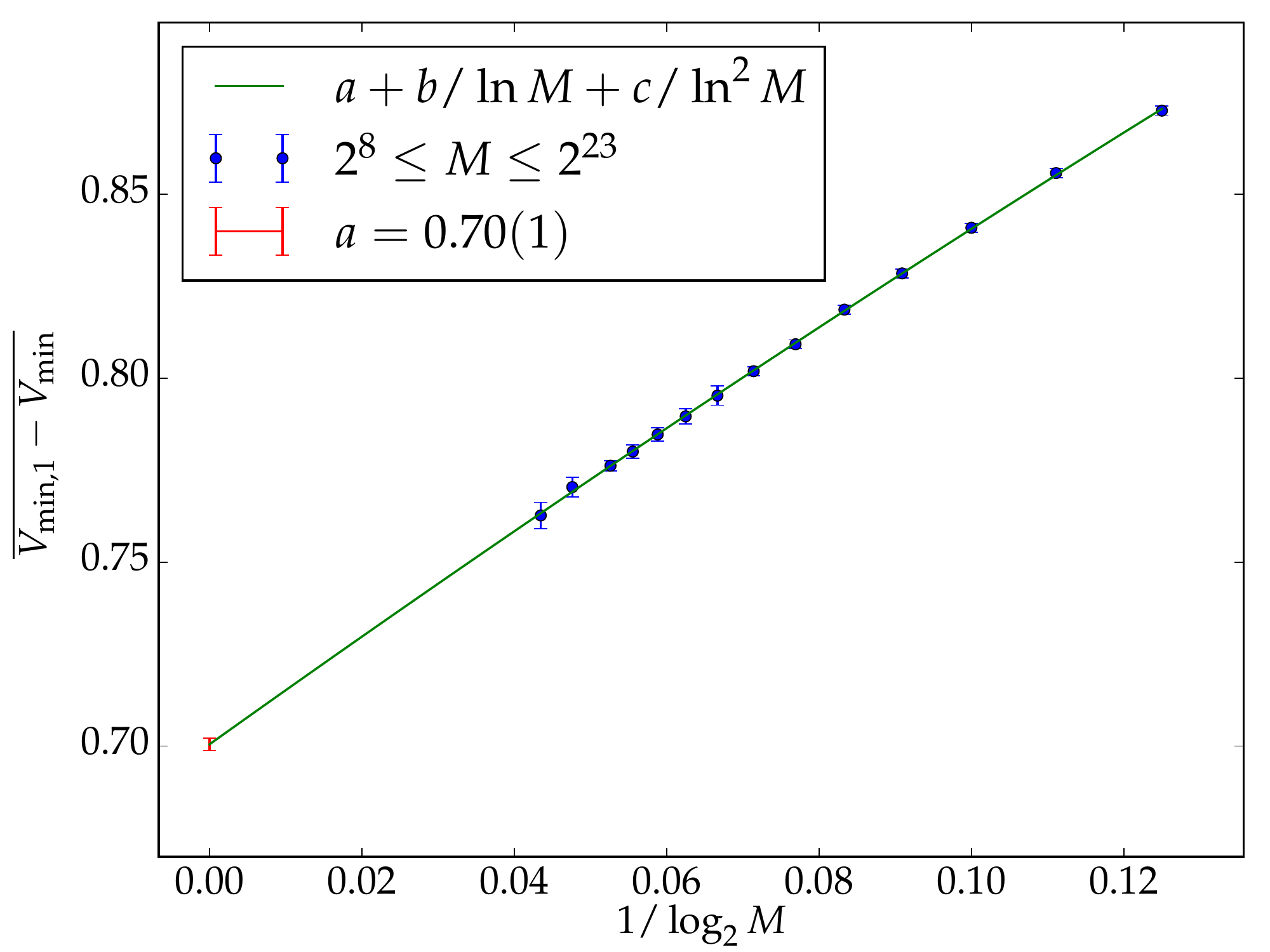}
\center b. \includegraphics[scale=.5,valign=t]{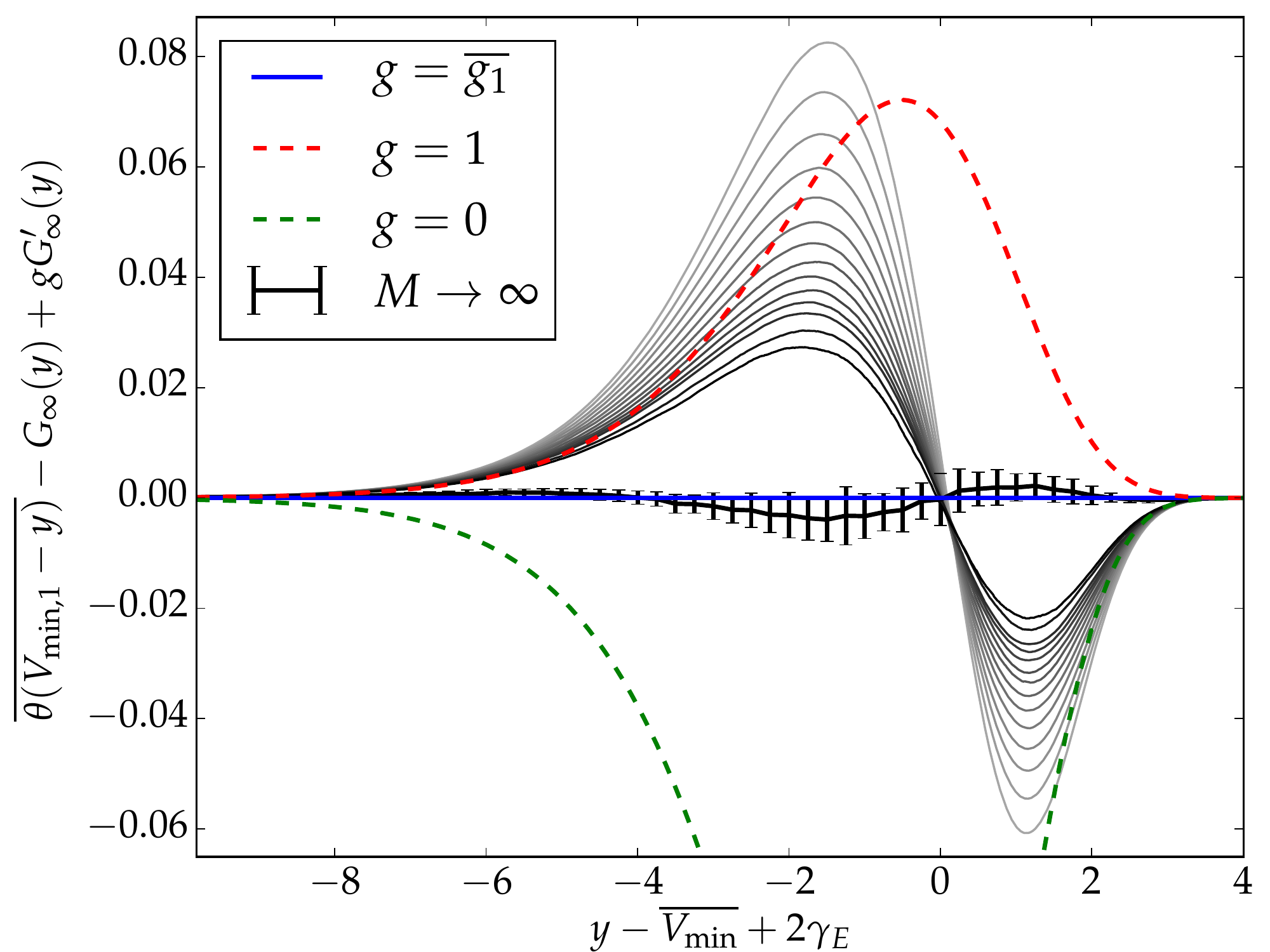}\\
\caption{Taken from \cite{cao16order}. a. The numerical measure of the mean of the first gap, as a function of the system size (points), is well described by a quadratic finite size Ansatz $a + b / \ln M + c / \ln^2 M$. We use it to extract the $M \rightarrow \infty$ value $\overline{g_1} = a = 0.70(1)$. b. The cumulative distribution function of the second minimum $V_{\min,1}$ of the circular $1/f$-noise model, with the theoretical prediction \eqref{eq:pdf2ndmin} subtracted, and the parameter $g = \overline{g_1}$ fed by the previous measurement. Grey curves are numerical data with system sis $2^{8} \leq M \leq 2^{23}$, and the extrapolation to $M\rightarrow \infty$ (thick black curve with error bars) is performed by applying the quadratic Ansatz point--wise. The error bars combine the error in the distribution with that in $\overline{g_1}$. For comparison we plot in dash lines \eqref{eq:pdf2ndmin} with other values of $g$.}
 \label{fig:2ndmin}
\end{figure}

The claim that eq. \eqref{eq:HGrelationREM} holds in \glspl{logrem} for \textit{some} $K_\beta(\Delta)$ is already non--trivial, and can be numerically tested. We did this for the circular model in \cite{cao16order}. Recall that for this model, the minimum distribution is (eq. \eqref{eq:etyFB}):
\begin{equation}
\overline{\theta(V_{\min} > y + a_M)} = G_{\infty}(y + a_M) = 2e^{y/2} K_1(2e^{y/2}) \,.
\end{equation}
Here $K_n$ is the Bessel $K$-function. The unknown (up to an $O(1)$ constant) shift $a_M$ can be fixed by the average value: $a_M = \overline{V_{\min}} + 2 \gamma_E$ ($\gamma_E = -\Gamma'(1)$ is the Euler's constant). Now, eq. \eqref{eq:pdf2ndmin} implies the following \gls{pdf} of the second minimum of the circular model: 
\begin{align}
&\overline{\theta(V_{\min,1} > y)} =  G_{\infty}(y) - g  G'_{\infty}(y) =
2e^{\widetilde{y}/2} K_1(2e^{\widetilde{y}/2}) + 2 g e^{\widetilde{y}}K_0(2e^{\widetilde{y}/2}) \,,\quad \label{eq:prediction1} \\ &\widetilde{y} = y - \overline{V_{\min}} - 2 \gamma_E \,,\, g = \overline{g_1} \,,
\end{align}
In other words, the limit distribution of the second minimum is predicted up to one parameter $g$, which is the mean value of the first gap. So we can measure $g$ numerically, feed the value into the prediction eq. \eqref{eq:prediction1}, and compare it with direct numerical measure of the second minimum distribution, shifted by the \textit{same} $a_M$. The result in shown and discussed in Figure \ref{fig:2ndmin}. We note here that for both numerical measures, finite size effects must be taken into account to yield sound results. 

\subsubsection{Local \glspl{logrem} and gap distribution}
Now let us come back to the \gls{uv} corrector $K_\beta(\Delta)$, defined in terms of $E_{m,\beta}$ eq. \eqref{eq:Exi} and $D_{m,\beta}(\Delta)$ eq. \eqref{eq:Dxi}, copied below:
\begin{align*}
E_{m,\beta} =  &e^{\beta^2 m^2 e_0/2} \frac{1}{N} \sum_{i_1, \dots, i_m = 1}^N  \prod_{l,l' = 1}^m \exp(- \beta^2 f(i_l-i_{l'}) ) \\
D_{m,\beta}(\Delta) = & \frac1N \sum_{i=1}^N \sum_{i_1, \dots, i_m = 1}^N
\prod_{l,l' = 1}^m \exp(\beta^2 (e_0 - f(i_l-i_{l'})) / 2) \nonumber \\
\times & \prod_{i: i_{l} = i}
\exp(-\beta \Delta)  \,.
\end{align*}
It is important now to make sense of them for $m = 1/\beta$, $\beta > 1$, and in particular, in the $\beta \to \infty$ limit. Unfortunately, we are unable to solve these Coulomb gas sums in a form that can be analytically continued to non-integer $m$. 

The approach taken in \cite{cao16order} is to define \textit{local \glspl{logrem}}. It is a sequence of Gaussian variables $u_i, i = 1, \dots, N$, with zero mean value and the following covariance:
 \begin{equation}
 \overline{u_i} \defeq 0 \,,\, \overline{u_i u_j}^c  \defeq C_N + f(i-j) \, . \label{eq:localrem}
 \end{equation}
 Here the over-line $\overline{[\dots]}$ means averaging over the \textit{local} disorder $(u_i)_{i=1}^N$. $C_N$ is a large positive number chosen to make the covariance matrix positive-definite (whose precise value turns out to be irrelevant, see below). Then we can define the local analogues of $\mathcal{Z}$ and $\mathcal{W}(j,\Delta)$:
 \begin{align}
 z_\beta = \sum_{i=1}^N \exp(-\beta u_i) \label{eq:smallw} \,,\, w_\beta(j,\Delta)  \defeq  \exp(-\beta (u_j + \Delta)) +  \sum_{i \neq j} \exp(-\beta u_i) \,.
 \end{align}
 Then it is not hard to write $E_{m,\beta}$ and $D_{m,\beta}(\Delta)$ in terms of moments (replica sums). Their ratio then gives the following expression of the \gls{uv} corrector, by eq. \eqref{eq:KDE}:
 \begin{equation}
 K_\beta(\Delta) =  \lim_{N\rightarrow\infty} \sum_{j=1}^N \frac{ \overline{w_{\beta}^{m}(j,\Delta)} - \overline{w_{\beta}^{m}(j,\infty)}}{\overline{z_\beta^{m}}} \,.\label{eq:Koffd}
 \end{equation}
 This equation is now defined for any $m$, not necessarily integer. This is how to define $E_{m,\beta}$ and $D_{m,\beta}(\Delta)$ for non--integer $m$. We see that a variation of $C_N$ will have the same effects on both the numerator and denominator, and does not affect $K_\beta(\Delta)$. 
 
 We can now take the $\beta\to \infty$ limit of eq. \eqref{eq:Koffd}. As explained in detail in \cite{cao16order}, section II.B.4, the result can be written in terms of the first and second minima $u_{\min}, u_{\min,1}$ of the local \gls{logrem}. Combined with eq. \eqref{eq:gapcdf}, we have:
 \begin{equation}
 \overline{\delta(g_1 - \Delta)} = -K'_\infty(\Delta) =  e^{-\Delta} \frac{\overline{\theta(u_{\min,1} - u_{\min} - \Delta) \exp(-u_{\min})}}{\overline{\exp(-u_{\min})}} \,.\label{eq:gapstat1} 
 \end{equation}
 This is a testable prediction: the left hand side is the cumulative distribution function of the (global) first gap; the right hand side can be measured in local \glspl{logrem}. We did this in \cite{cao16order} for the circular model, and the result is shown in Figure \ref{fig:gapstats} (a.). In that plot, we multiplied the quantities in both sides by $e^{\Delta}$, so as to highlight the difference of the gap distribution in the circular model from the simple standard exponential of the \gls{rem}. The proportion in the $M\to \infty$ limit is not known analytically; however, we find that the proportion can be recovered by $\frac{\overline{\theta(u_{\min,1} - u_{\min} - \Delta) \exp(-u_{\min})}}{\overline{\exp(-u_{\min})}}$ measured in local \glspl{logrem}, as $N \to \infty$, but with much smaller finite size effect. Therefore, the prediction eq. \eqref{eq:gapstat1} provides a way to access the gap distribution more efficiently.
 
 \subsubsection{Remote second minimum: exponential gap}
\begin{figure}
\center
a.\includegraphics[scale=.5,valign=t]{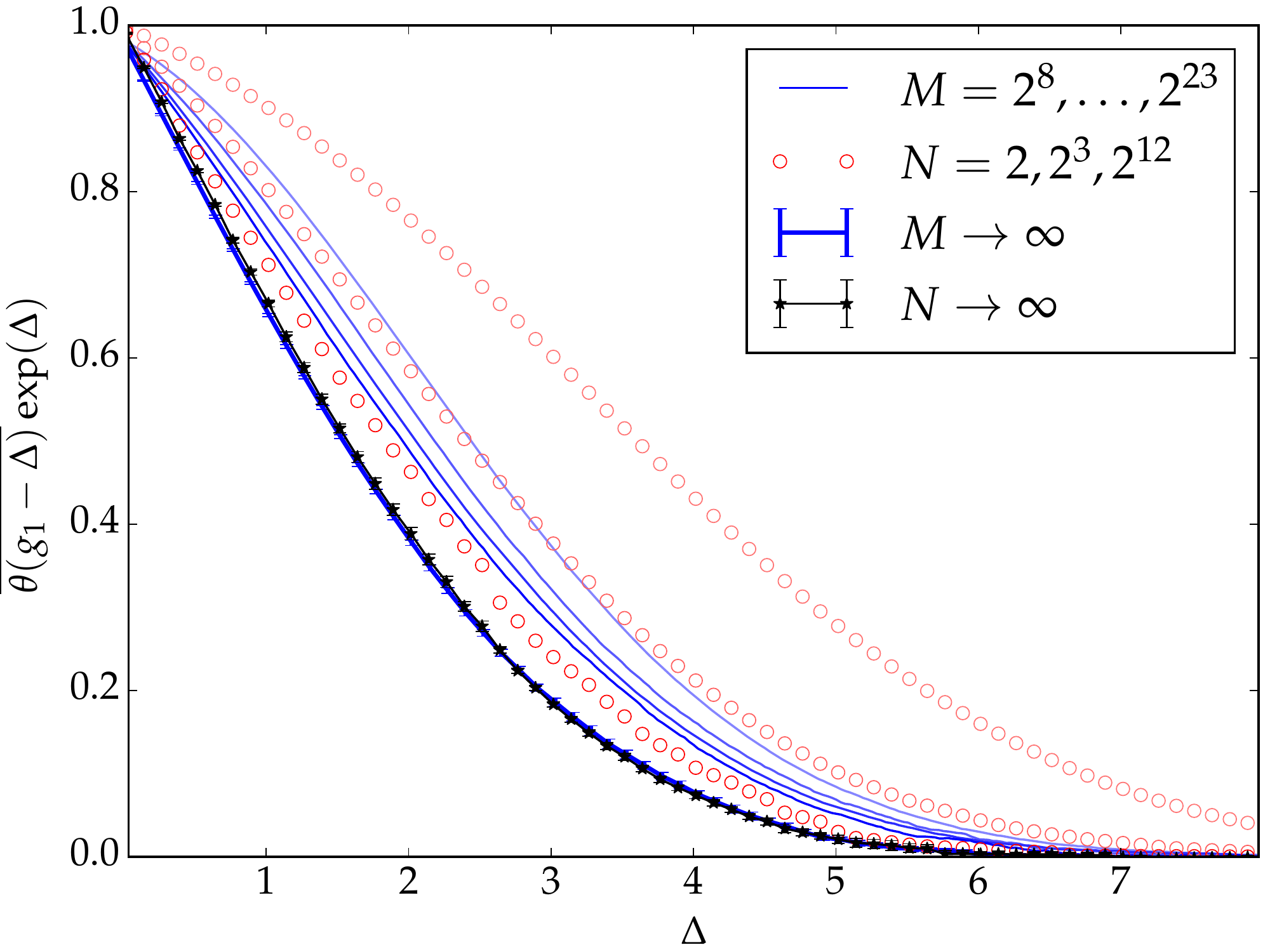}
b.\includegraphics[scale=.5,valign=t]{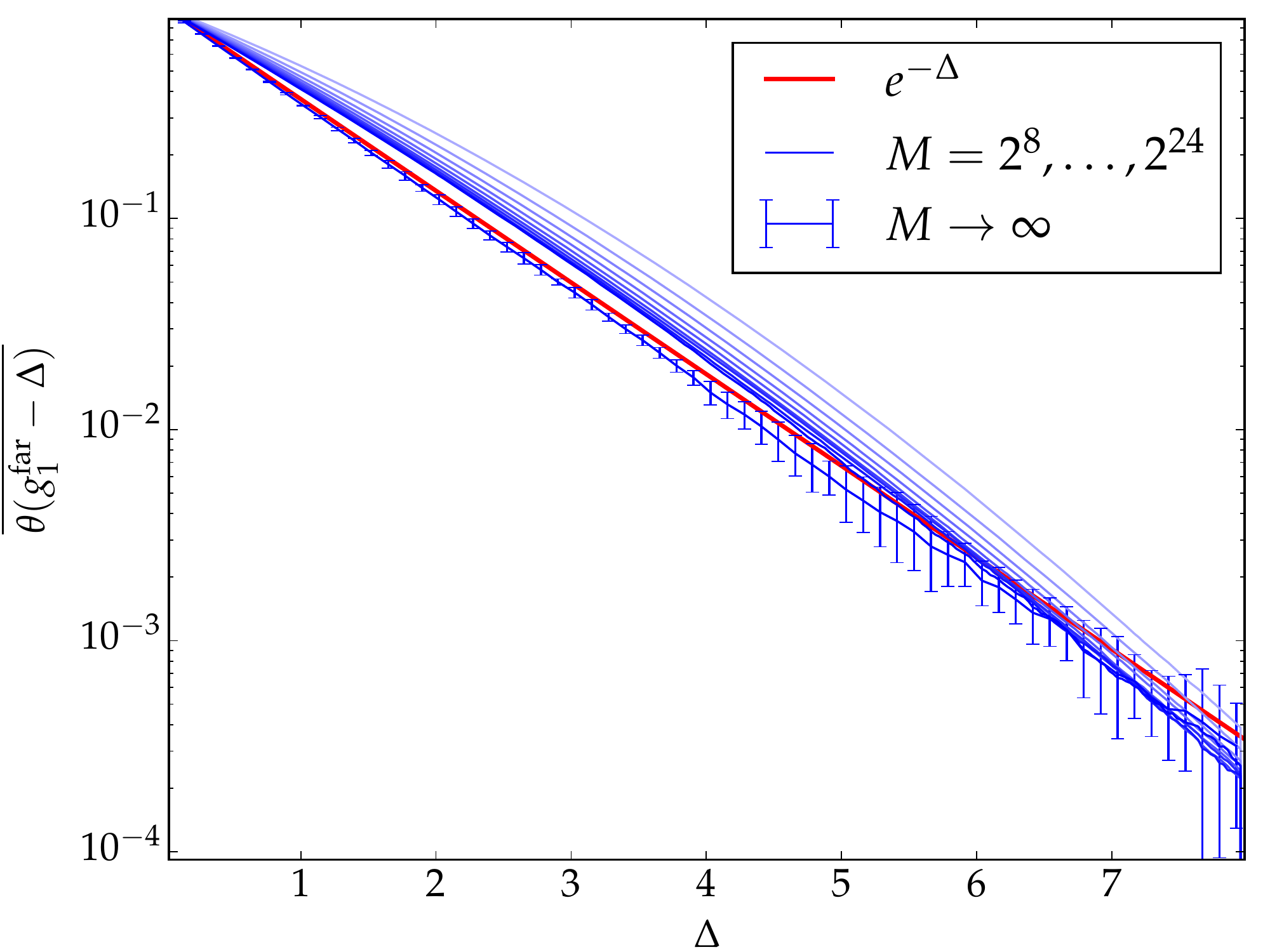}
 \caption{Taken from \cite{cao16order}. a. Global \textit{vs} local first gap distribution in the circular $1/f$-noise model. The global gaps are measured directly from systems of sis $M = 2^{8}, 2^{13}, 2^{18}, 2^{23}$, and extrapolated point wise by the $1/\ln M$-quadratic finite size Ansatz. The results (blue curves) are multiplied by $e^\Delta$ (so that an exponential distribution would give a horizontal line at height $1$). The first gap in local \glspl{logrem} are measured in much smaller sis $2\leq N \leq 2^{12}$ (red circles), and extrapolated using the same Ansatz to $N \rightarrow \infty$ (black circles). The good agreement between the latter curve (\textit{without} multiplying by $\exp{\Delta}$) and the $M \rightarrow\infty$ curve confirms the prediction \eqref{eq:gapstat1}. b. The distribution of the remote gap $g_{1}^\text{far}$, with $N = \sqrt{M}$, compares well to the exponential $e^{-\Delta}$.}\label{fig:gapstats}
\end{figure}

 It is unfortunate that we cannot calculate analytically the contribution of the \gls{uv} data. Nevertheless, there is a way to get rid of it, and recover the exponential gap distribution of the \gls{rem}. For this, we need to consider, instead of the second minimum, the \textit{remote second minimum} and its gap with the absolute minimum. They are defined in the circular model as:
 \begin{equation} V^{\text{far}}_{\min,1} \stackrel{\text{def}}= \min(V_{j,M}, \abs{j - j_{\min}} > N/2)  \,,\, g_{1}^{\text{far}} \stackrel{\text{def}} = V^{\text{far}}_{\min,1} - V_{\min} \,, \label{eq:fargap} \end{equation} 
 where $j_{\min}$ denotes the position of the minimum. We look at the limit in which both $N\to \infty$ and $M\to \infty$ but with $N\ll M$ in the thermodynamic limit; in the numerical simulations, $N = \sqrt{M}$ is used. The idea is that when looking for the second minimum located far from the absolute minimum, the \gls{uv} data trivializes and one retrieves the exponential gap distribution in the thermodynamic limit:
 \begin{equation}
 \overline{\theta(g_{1}^{\text{far}} - \Delta)} \rightarrow \exp(-\Delta)\,, M\rightarrow\infty\,, \label{eq:exponential}
 \end{equation}
 which is the case for the random energy model with uncorrelated potential. This prediction is well verified by the numerical data in the circular model, which are shown in Fig. \ref{fig:gapstats} (b). It is possible to translate the above intuition into a \gls{1rsb} derivation of eq. \eqref{eq:exponential} for general \gls{logrem}, and this is done in \cite{cao16order}, section IV.C.2.  
  
 \subsubsection{Biased minima process}
 The fraction $\frac{\overline{\theta(u_{\min,1} - u_{\min} - \Delta) \exp(-u_{\min})}}{\overline{\exp(-u_{\min})}}$ admits an important interpretation, namely, as the cumulative distribution of the local \gls{logrem} gap $u_{\min,1} - u_{\min}$, but \textit{weighted with} the (minus) exponential of the minimum $\exp(-u_{\min})$. Therefore, we can define a \textit{biased minima process}, $ 0 = v_{\min,0} < v_{\min,1}  <  v_{\min,2} \dots $, by 
 \begin{equation}
 \overline{\delta(v_{\min,1}-v_{\min,0} - \Delta)} = 
 \frac{\overline{\delta(u_{\min,1}-u_{\min,0} - \Delta_1) \exp(-u_{\min})}}{\overline{\exp(-u_{\min})}} \,, \label{eq:biaseddef}
 \end{equation}
 and similarly for the higher orders (as explicitly written in eq. (44) in \cite{cao16order}). Note that the biased minima process is defined in the ``seen from the tip'' fashion, \textit{i.e.}, only the gaps are meaningful observables. Using eq. \eqref{eq:biaseddef}, eq. \eqref{eq:gapstat1} can be written as 
 \begin{equation}
 \overline{\delta(g_1 - \Delta)} =  e^{-\Delta} \, \overline{\theta(v_{\min,1}-v_{\min} - \Delta)} = \overline{ \theta(g_1^{\text{REM}}-\Delta) \theta(v_{\min,1}-v_{\min} - \Delta)} \,,
 \label{eq:gapstatbiased}
 \end{equation}
 where we interpreted the formula further by introducing a standard exponential random variable $g_1^{\text{far}}$ which is independent of the biased minima. In other words, the gap $g_1$ of \gls{logrem} has the same distribution as $\min(g_1^{\text{far}}, v_{\min,1}-v_{\min})$, the minimum of a remote gap and the first gap of the biased minima process, \textit{if} they are independent. From this discussion emerges a picture in which the biased minima process describe the minimal values that are \textit{close}. This idea is systematized by the notion of \textit{decorated} \acrlong{ppp}, that we discuss in the following section. 

\subsection{Full minima process}\label{sec:fullmin}
The notion of decorated \acrlong{ppp} emerged in the study of the full minima process in hierarchical \glspl{logrem}, such as the \acrlong{bbm} (see section \ref{sec:DPCT} for introduction). This field was initiated by the authors of \cite{brunet2011branching} and elaborated in full rigour in \cite{arguin2011genealogy,arguin2012poissonian,aidekon2013branching,arguin2013extremal}. It is now known that in the thermodynamic limit, the minima process tends to a \gls{sdppp} \cite{subag2015freezing}. Such a picture is expected to apply also to non--hierarchical \glspl{logrem}. A first rigorous result in this direction was obtained recently in \cite{biskup2016full} for the discrete 2D \gls{gff}, based on previous work of the same Authors  \cite{biskup2016extreme,biskup2014conformal}. Moreover, the above works covered not only the \textit{values} of the minima, but also their \textit{positions}; in the \gls{bbm} context, the latter is referred to as the \textit{genealogy} of minima \cite{arguin2011genealogy,derrida16kppfinitesize}. So in what follows, the term \textit{full minima process} will refer to the joint statistical properties of minima positions \textit{and} values.

In \cite{cao16order}, working on the physicist's level of rigour, we described the full minima process of general \glspl{logrem}, using entirely the \gls{1rsb} approach, which is a (quite heavy) generalization of the derivations of section \ref{sec:2ndmin} to the higher order statistics. It is not worthwhile to review the technicalities here, because the result is not surprisingly in agreement with the picture established independently by \cite{biskup2016full}. Therefore, we shall provide a light discussion of the results.

\begin{figure}
  \center \includegraphics[scale=.5]{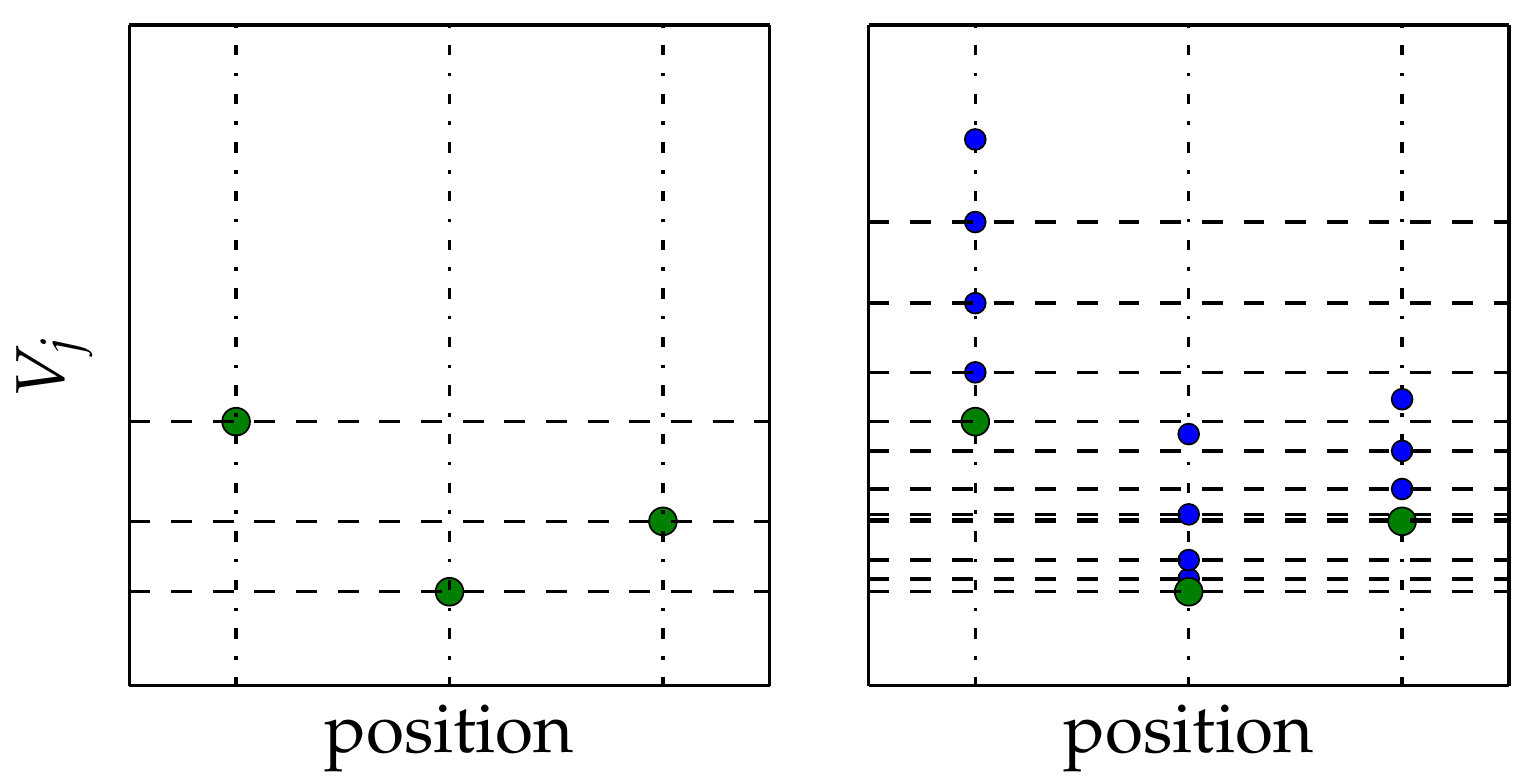}
 \caption{An illustration of the clustering of the minima process and the recipe to generate it. The first step (left panel), generates the local minima (large green dots) by a \acrlong{ppp} with random density, eq. \eqref{eq:PPPdensity} ; the second step (right panel), generates the decorations by (small blue dots), by applying the rule eq. \eqref{eq:decoprocess}.}\label{fig:sdppp}
\end{figure}
Let us start by a qualitative account. The minima of \glspl{logrem} defined on Euclidean spaces display a \textit{clustering structure} in the $M\to \infty$ limit: they form clusters that occupy the same position in the thermodynamic/continuum limit. A realistic illustration is already given in Figure \ref{fig:minimaposlogREM}, while a sketch is provided here in Figure \ref{fig:sdppp}. Each cluster has its own minimum, which is called its \textit{local minimum} (they are draw in larger dots in Figure \ref{fig:sdppp}). The full minima process is generated in two steps. To be concrete, we will again use the example of the circular model. 
\begin{enumerate}
\item First, the process of the local minima (position and value) is generated as a \acrlong{ppp} in $(1+1)$-d with the following density
\begin{equation}
 e^{- \phi(z)} \frac{\dif \theta}{2\pi} \theta e^{y} \dif y \,,\, z = e^{\im \theta} \,, \theta \in [0,2\pi), y \in \R \,, \label{eq:PPPdensity}
\end{equation} 
where $\phi(z)$ is the 2D planar \gls{gff} whose restriction on the unit circle gives random potential of the circular model. Therefore, the density of the \acrlong{ppp} itself is random. We will explain what this means exactly below. In general, this step depends only on the \gls{ir} data of the \gls{logrem}. 
\item Let $(\theta, Y)$ be one of the local minima generated by the previous step. We replace it by the set of points 
\begin{equation}
(\theta,  Y), (\theta, Y + v_{\min,1}), (\theta, Y + v_{\min,2}), \dots \in [0,2\pi) \times \R \,, \label{eq:decoprocess}
\end{equation}
where $0 = v_{\min,0} < v_{\min,1} < \dots$ is the biased minima process in defined in eq. \eqref{eq:biaseddef}. That is, the biased minima process is the \textit{decorating process}. Now we repeat this for each local minima, each with an independent realization of the decorating process. In general, this step depends only on the \gls{uv} data of the \gls{logrem} in question.
\end{enumerate}
The result of the above construction is a random set of points in $[0, 2\pi) \times \R$. The central result of \cite{cao16order,biskup2016full} is that, the statistical properties of that set generated is statistical equivalent to the position and properly shifted values of the \gls{logrem} (the circular model in our case). The shift is applied to all the minimal values $V_{\min,k} \leadsto V_{\min,k} - a_M$, with $a_M =- 2\ln M + \frac32 \ln \ln M + c$ so as to remove the $M$ dependence. 

\subsubsection{\Acrfull{sdppp}}
The best way to explain the above construction, particularly its first step, is to go through examples. To begin with, let us focus on the minimal values. Let $Y_0 < Y_1  < \dots $ be the ordered sequence of the local minima (generated by step 1). As a warm up (and as a consistency check), let us consider the  minimum and calculate $\overline{\theta(Y_0 - y)}$, which is the probability that there is no point in the domain $\{ (z,y'): y' < y \}$. We use a similar method that led to eq. \eqref{eq:PPP2nd} for the density eq. \eqref{eq:PPPdensity}, with an extra average over $\phi$:
\begin{equation}\label{eq:PPPmin}
\overline{\theta(Y_0 - y)} = \overline{\exp\left( -\int_{-\infty}^y  \dif y' \int_{0}^{2\pi} \frac{\dif \theta}{2\pi} e^{y'} e^{-\phi(z)}  \right)}=  \overline{\exp\left( e^y  Z_{1} \right)} 
\end{equation}
where $Z_{b} = \int_{0}^{2\pi}\frac{\dif \theta}{2\pi} e^{-b \phi(z)}$ is the continuum partition function of the circular model. As we discussed in section \ref{sec:freezingRSB}, $Z_{b}$ diverges to $0$ as $b\nearrow 1$ as a precursor of the $\ln \ln M$ corrections when $\beta \geq 1$. However, as we are interested in the minimal values up to a shift in the current discussion, we believe that this problem is not fatal. The right hand side of eq. \eqref{eq:PPPmin} can be written as $\overline{\exp\left( e^{y - F} \right)}$ where $F = -\ln Z_{b \nearrow 1}$ is the re--shifted critical--temperature free energy. So $Y_0 = -g + F$ is a convolution where $g$ is standard Gumbel independent of $F$. Note that by the freezing scenario, $Y_0$'s distribution is just the limit distribution of the minimum in the circular model. This is expected, because after the decoration, the absolute minimum of the whole process is still $Y_0$. 

Let us add the second minimum $Y_1$. The calculation is a combination of eq. \eqref{eq:PPPmin} and \eqref{eq:PPP2nd}, except that one needs to integrate over $\theta$ because we do not restrict the position of the first local minimum. 
\begin{align}
\overline{\delta(Y_0 - y_0) \theta(Y_1 - y)}& =  \int_0^{2\pi} \frac{\dif \theta}{2 \pi} e^{-\phi(z)} e^{y_0} \overline{\exp\left(- e^y   Z_{1} \right)} \nonumber \\
 &=  \exp((y_0 - F) + e^{y-F}) \,.
\end{align} 
Compared to eq. \eqref{eq:PPP2nd}, the above equation means that $Y_0, Y_1$ can be generated by taking the first and second minima of a Gumbel \gls{ppp}, and shift both of them by $F$, which is independent of the \gls{ppp}. In fact, it is in general true that the values of the local minima generated by the \gls{ppp} with random density eq. \eqref{eq:PPPdensity} is a randomly shifted Gumbel \gls{ppp}, where the random shift has the same law as the re--shifted critical--temperature free energy. 

Now, eq. \eqref{eq:decoprocess} implies step 2 projects onto the following decorating operation on the follows: $Y_q \leadsto Y_q , Y_q + v_{\min,1}, Y_q + v_{\min, 2}$, $q = 0, 1, \dots$. The decoration process used for each $Y_q$ is an independent realization. The obtained point process is the \gls{sdppp}, and describes the re--shifted full minima process of \glspl{logrem}. 
In particular,
\begin{itemize}
\item The random shift has the same distribution as the critical temperature free energy, and depends only on the \gls{ir} data. 
\item The decoration process depends only on the \gls{uv} data. 
\end{itemize}
This is the main result of \cite{cao16order} on the extreme order statistics of \glspl{logrem}, and its relation to \gls{ir} and \gls{uv} data. 

\subsubsection{Minima positions and Gibbs measure}
This paragraph discusses the minima positions (forgetting their values whenever possible), as well as the closely related Gibbs measure. A purpose is to prepare for the next section on the \acrlong{lft} mapping. We also enrich and understand better the results in section \ref{sec:2ndmin}.
 
Starting again from the minimum, let us compute $\overline{\delta(\xi_{0} - z)}$ where $\xi_{\min}$ is the minimum position, and the Dirac $\delta$ is with respect to the integral $\int \frac{\dif\theta}{2\pi}$ (in the scope of the present work, the delta is always dual to the integral measure used in the continuum partition function; more general situations are discussed in appendix E of \cite{cao16order}). Since $\xi_{\min}$ is also the minimum position before the decoration step, we can compute its distribution by a \gls{ppp} calculation using eq. \eqref{eq:PPPdensity}:
\begin{align}
\overline{\delta(\xi_{0},z)} = &\int_\R \dif y  \, \overline{e^{-\phi(z)} e^y \exp\left(- \int_{-\infty}^y e^{y'}  \dif y' Z_{1} \right)} = \overline{Z_{1}^{-1} 
e^{-\phi(z)}} \nonumber \\
 = & \overline{p_{1}(z)} \,. \label{eq:PPPposition0} 
\end{align}
 where $p_\beta(z)$ is, by definition, the (continuum) Gibbs measure. To our best knowledge, it does not suffer from the $\beta \nearrow 1$ problem as does the partition function. The above equation is indeed the $\beta \to \infty$ case of the freezing scenario for the Gibbs measure \cite{fyodorov2010freezing,fyodorov2015moments}, which states that 
 \begin{equation}
  \overline{p_{\beta > 1}(z)} =   \overline{p_{1}(z)} \,. \label{eq:freezingpos}
 \end{equation}
 In \textit{op. cit.}, the freezing of Gibbs measure was also observed to be accompanied by the duality invariance property, providing further checks of the freezing--duality conjecture.
 
Digressing a bit, we emphasize that only the one point correlation function of the Gibbs measure freezes. For the two point function, we have 
 \begin{equation}
  \overline{p_\beta(z_1)p_\beta(z_2)} = (1-\beta^{-1}) \delta(z_1,z_2)\, \overline{p_1(z_1)} + \beta^{-1}\overline{p_1(z_1)p_1(z_2)} \,, \beta > 1 \,.\label{eq:pbpblowT} 
 \end{equation}
 This follows immediately from the overlap distribution of \glspl{logrem} eq. \eqref{eq:PoverlaplogREM} (the two terms of the right hand side correspond to $\overlap = 1$ and $\overlap = 0$ respectively): compare also to the $\beta > 1$-phase behaviour of the EA order parameter \eqref{eq:freezingofEA}. Equations eq. \eqref{eq:freezingpos} and \eqref{eq:pbpblowT} will be used in section \ref{sec:liouville}. 
  
  \begin{figure}
  \center \includegraphics[scale=.6]{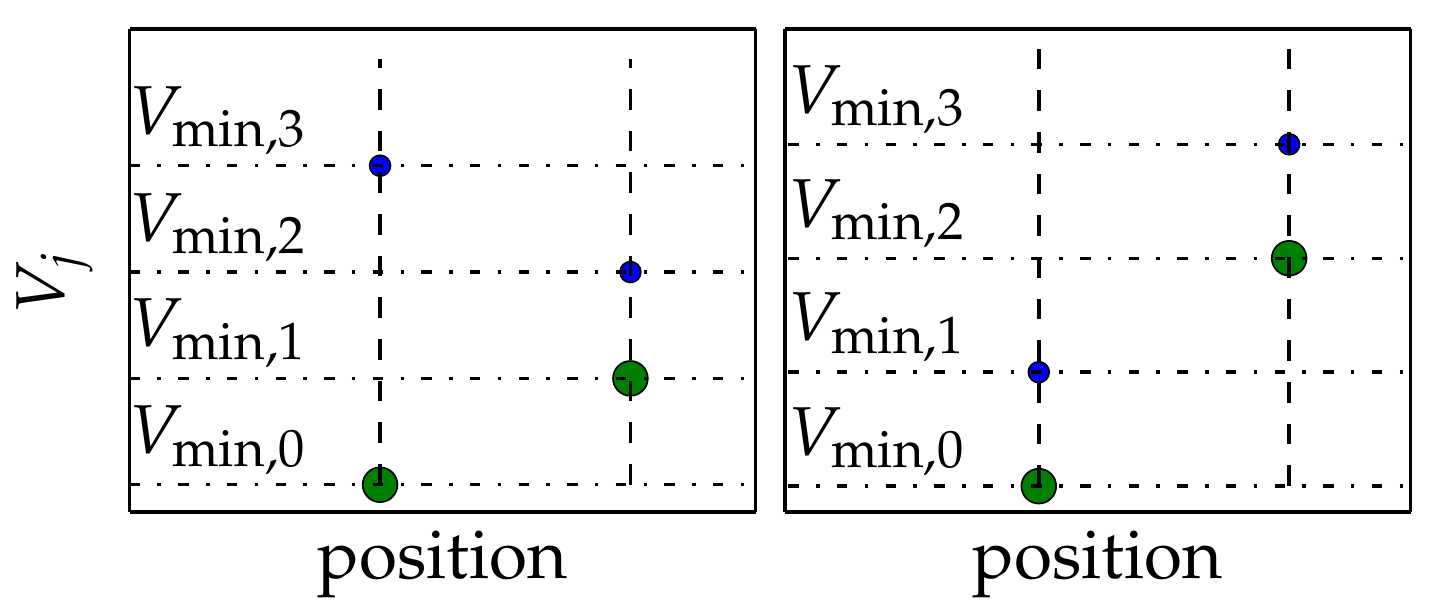}
  \caption{Another sketch of the clustering of the minima in a \gls{logrem}. The two cases that can happen to the first and second minima are illustrated. \textit{Left}: $V_{\min}$ and $V_{\min,1}$ are in two different clusters; \textit{right}: $V_{\min}$ and $V_{\min,1}$ are a same cluster. The larger green dots represent the local minima, which are generated by the \acrlong{ppp}; the smaller blue dots are generated by the decoration process.}\label{fig:sdppp2}
  \end{figure}
 Notice that the $\beta \to \infty$ limit of eq. \eqref{eq:pbpblowT} does \textit{not} give information about the second minimum position. The correct finite temperature observable is a generalization of $H_\beta(y,y_0)$ which we will not discuss in this work (we refer again to \cite{cao16order}, Appendix E). Instead, we apply again the full minima process result. Note that because of the decorating process, the two first minima can be in the same cluster or not (see Figure \ref{fig:sdppp2}), and the \textit{values} will play a rôle in deciding which situation will happen. Therefore, we first calculate the joint law of the two first \textit{local} minima positions $\xi^{\text{far}}_{0}, \xi^{\text{far}}_{1}$ and their value difference $Y_1 - Y_0$,  using eq. \eqref{eq:PPPdensity}: 
 \begin{align}
 &\overline{\delta(Y_1 - Y_0 - \Delta) \delta(\xi^{\text{far}}_{0} , z_0)\delta(\xi^{\text{far}}_{1} , z_1)} 
 =  \int_\R \dif y  \overline{e^{-\phi(z_0)} e^{-\phi(z_1)} \exp(-e^y Z_{1}) } e^{y}e^{y-\Delta} \nonumber \\
 =& \overline{Z^{-2}_1 e^{-\phi(z_0)} e^{-\phi(z_1)} } e^{-\Delta} =
  \overline{p_1(z_0) p_1(z_1) }e^{-\Delta} \,. \label{eq:PPP2pos}
 \end{align} 
 Therefore, the pair of positions $(\xi^{\text{far}}_{0}, \xi^{\text{far}}_{1})$ and $Y_1 - Y_0$ are \textit{independent}; the former's distribution is given by a Gibbs measure two--point correlation at $\beta = 1$, and the latter is exponentially distributed. It is not hard to convince oneself that this is nothing but the remote gap distribution result eq. \eqref{eq:exponential}, because $Y_1 - Y_0$ \textit{is} the remote gap $g_1^{\text{far}}$ eq. \eqref{eq:fargap} according to the clustering picture.
 
 Now, while $Y_0 = V_{\min}$ and $\xi^{\text{far}}_{0}= \xi_0$ for sure, $Y_1$ is not necessarily the second minimum $V_{\min,1}$: it has to compete with the second minimum in the same cluster, whose value is $Y_0 + v_{\min,1}$, where $v_{\min,1} > 0$ is the first gap of the decorating process. The true second minimum value is the smallest among the two candidates: 
 $$ V_{\min,1}= \min(Y_1, Y_0 + v_{\min,1} ) = V_{\min} + g_1 \,,\, g_1 = \min(v_{\min,1}, g_1^{\text{far}} ) \,.$$
 Let $c_0$ be the probability that $v_{\min,1} <  g_1^{\text{far}}$, \textit{i.e.} that the first two minima are in the same cluster. Then, 
  \begin{equation} 
 c_0 = \int_0^{+\infty} \dif v \int_v^{+\infty} \dif \Delta \,\overline{\delta(v_{\min,1}-v)} e^{-\Delta}
     =  \int_0^{+\infty} e^{-v} \overline{\delta(v_{\min,1}-v)}\dif v \,.
  \end{equation}
  It can be checked to be equal to $1- \overline{g_1}$, where the mean value of the gap is:
 \begin{equation}
 \overline{g_1} = \int_0^{+\infty} \dif v \int_0^{+\infty} \dif \Delta \, \overline{\delta(v_{\min,1}-v)} e^{-\Delta} \min(v, \Delta) = 1-c_0 \,. \label{eq:g1isc0}
 \end{equation} 
This is an interesting relation that equates the mean value of the first gap to the probability  that the first two minima are in distinct clusters. 

Combining the last result  with eq. \eqref{eq:PPP2pos} and \eqref{eq:PPPposition0}, we obtain the following joint distribution of the first and second minima positions, $\xi_0, \xi_1$:
\begin{equation}
P(z_0, z_1) \defeq \overline{\delta(\xi_0 , z_0)\delta(\xi_1, z_1)} = (1-\overline{g_1}) \delta(z_0, z_1) + \overline{g_1} \times \overline{p_1(z_0) p_1(z_1) } \,. \label{eq:P2minpos}
\end{equation}
This relation will also be useful in the next section.

\section{Relating logREMs to Liouville field theory (LFT)}\label{sec:liouville}
This section is based on the recent work \cite{cao16liouville}, which relates \acrfull{lft} to \glspl{logrem} generated by 2D \gls{gff}, and then to \glspl{logrem} in general. No preliminary expertise in  \gls{lft}, or in \acrfull{cft} in general, is necessary to follow the exposition. Neither do we claim any new result on the  \gls{lft} \textit{per se}. Throughout this section, we will recall necessary background and specific results of \gls{lft} (with helpful historical and bibliographical comments), and discuss how to apply them to \glspl{logrem}. Nevertheless, given the historical and intellectual stretch of the topic, a more general overview is in order.
 
\subsection{Overview}\label{sec:lftoverview}
 \gls{lft} is one of the most studied 2D \glspl{cft}. It originated in string theory \cite{polyakov1981quantum} and is a key ingredient of the continuum theory of 2D quantum gravity, \textit{i.e.}, the geometry of random surfaces, see \cite{zamolodchikov2007lectures} for an introduction. We will comment briefly on the random--geometric aspects of \gls{lft} when we introduce it formally in eq. \eqref{eq:Sliouville}.
 Another well-developed application of \gls{lft} is its holography correspondence to $(2+1)$-D gravity, see \cite{carlip05gravity} for a review. Although \gls{lft} is an interacting field theory, it is closely related to the 2D \gls{gff}, in the way that we will describe in section \ref{sec:mapping0}. This relation was observed very early and led to important developments of \gls{lft} \cite{goulian1991correlation,dorn1994two,zamolodchikov1996conformal}, and has been the basis of more recent mathematical developments \cite{duplantier2014critical,david2016liouville,kupiainen2015conformal,david2016liouvilletori}, see \cite{duplantier2014log,kupiainen16course} for review and introduction. Section \ref{sec:mapping0} will review the convenient representation of \gls{lft} correlation functions in terms of averages (functional integrals) over the 2D \gls{gff}. 

 This representation in turn  the basis of the \gls{lft}--2D \glspl{logrem} mapping, \textit{i.e.}, between \gls{lft} and \textit{\glspl{logrem}} whose random potential is generated by the 2D \gls{gff}. This mapping is interesting principally because \gls{lft} has been \textit{exactly solved}: \textit{i.e.}, we know the spectrum of operators in \gls{lft} and can compute (in principle) all their correlation functions. This achievement is possible  thanks to the conformal invariance of \gls{lft}, which is fully exploited in the \textit{conformal bootstrap} calculation of the \gls{lft} correlation functions. The results can then be translated by the \gls{lft}--2D \glspl{logrem} mapping to exact prediction on the \glspl{logrem}.
 
 More precisely, correlation functions of \gls{lft} describe naturally the correlation functions of the \textit{Gibbs measure} of these \glspl{logrem}, \textit{i.e.}, \gls{lft} describes the \textit{positions} of thermal particles in 2D \gls{gff}--random potential. This link was observed for the first time by Kogan, Mudry and Tsvelik \cite{kogan96prelocalised}, in the context of 2D Dirac fermions in a random magnetic field. However, at their time, this insight did not lead to precise predictions on \glspl{logrem} by the \gls{lft}, because of three difficulties:
\begin{enumerate}
\item  The thermodynamics of the \glspl{logrem} was not yet developed. By now, this difficulty is sufficiently resolved.
\item  The authors of \cite{kogan96prelocalised} investigated the mapping on the \textit{infinite} plane, which is the most tricky case.  
One of the contributions of our work \cite{cao16liouville} is to solve the difficulty related to the infinite geometry. In fact, one needs to consider an \textit{IR divergent \gls{logrem}} (see section \ref{sec:Gaussian}), made of the planar 2D \gls{gff} plus a logarithmic confining potential.
\item  The \textit{conformal bootstrap} techniques for calculating \gls{lft} correlation functions were just initiated by the Zamolodchikov brothers \cite{zamolodchikov1996conformal}, and their numerical implementation was not widely available. 
\end{enumerate}
 Section \ref{sec:mapping1} will be devoted to the \gls{lft}--2D \gls{logrem} mapping, focusing on the above points. The highlight will be the numerical test of an exact prediction, eq. \eqref{eq:mainhighT}, which equates the disorder--averaged Gibbs measure of a 2D \gls{logrem} and a \gls{lft} \textit{four--point} correlation function. We emphasize ``four--point'' because in general, a \gls{cft} is completely solved if and only if we can calculate its four--point functions. So, remarkably, the simplest \gls{logrem} application involves all the field--theory features of \gls{lft}.
 
The major consequence is that, the {\gls{ope}}, which determines the short--distance behaviour of \gls{lft}, provides new predictions about \textit{universal features} of \glspl{logrem}. In particular, we will unveil important statistical significations of the subtle structure of the Liouville \gls{ope}: the presence/absence of the so--called \textit{discrete terms}. This \gls{lft} feature was sketched already in \cite{zamolodchikov1996conformal}, and elaborated in more recent works \cite{teschner2001liouville,teschner2015supersymmetric,belavin2006integrals,aleshkin2016construction}. Because their importance, we will devote the section \ref{sec:OPE} to discuss them. The applications to \glspl{logrem} will be then presented in section \ref{sec:LFTtree}. There are two stages: the first, straightforward stage, concerns the asymptotic behaviours of Gibbs measure correlations in 2D \glspl{logrem}. The second, more ambitious step, extends the results to \glspl{logrem} in general, \textit{i.e.}, in other dimensions, and on the Cayley tree.

The principle message of these results is that: the presence/absence of discrete terms induces a non-analytical dependence of critical exponents on the parameters, \textit{i.e.}, a transition. This transition is nothing but the termination point transition, discussed in section \ref{sec:multifracintro}. This will be argued in section \ref{sec:LFTmultifrac}, where we will apply Liouville \gls{ope} to the multi--fractal properties of \glspl{logrem} in the \textit{annealed ensemble}. Note that multi--fractality was the initial motivation of the pioneering work \cite{kogan96prelocalised} (and also \cite{chamon1996localization}). Our main new contribution is the \textit{logarithmic corrections} of the inverse participation ratio in the termination point phase. They turn out to be very similar to the $\ln \ln M$ corrections associated with the freezing transition. 

To end this introduction, we mention that, as a by--product of our study, we have investigated the importance of discrete terms for the consistency of \gls{lft}, \textit{i.e.}, we have checked numerically that the \textit{crossing symmetry} is satisfied if and only if the discrete terms are correctly implemented; to our best knowledge, such numerical checks were not done before.

\subsection{LFT and 2D GFF}\label{sec:mapping0} 
The \acrlong{lft} can be defined on any 2D surface, and we will restrict to \textit{closed surfaces}. In general, The surface will be denoted by $\Sigma$, parametrized by a (local) complex coordinate $z = x + \im y$ (and $z^* = x - \im y$), and endowed with the surface element $\dif A$. 
The action for the \acrlong{lft} is given by
 \begin{equation}
\mathcal{S}_b = \int_{\Sigma} \left[ \frac{1}{16\pi} (\nabla \varphi)^2 - \frac1{8\pi}Q \hat{R} \varphi + \mu e^{-b\varphi}  \right] \dif A  \,.\label{eq:Sliouville} 
\end{equation}
Here, $\varphi$ is the Liouville field, $\mu > 0$ is the coupling constant (also called the ``cosmological constant''), $\hat{R} = \hat{R}(z)$ is the Ricci curvature of the surface, $b > 0$ is the parameter defining the \gls{lft}, and 
\begin{equation}
Q = b + b^{-1} \,. \label{eq:defQ}
\end{equation}
This coefficient of the coupling of the Liouville field to the curvature is the most important ingredient of \gls{lft}, and guarantees its conformal invariance among other properties, see \cite{zamolodchikov2007lectures} for detailed demonstration. We mention that the \textit{central charge} of \gls{lft} is 
\begin{equation}
c = 1 + 6 Q^2 \geq 25 \,,\, \text{ if } b \in \R \,,\label{eq:centralcharge}
\end{equation}
although we will not need this notion in our technical treatments.

Referring to \cite{zamolodchikov2007lectures} for details, we recall briefly geometric/physical meaning of eq. \eqref{eq:Sliouville}. Indeed, regarding it as a classical action, the Euler--Lagrange equation will be the Liouville equation
\begin{equation}\label{eq:liouville}
\frac1{16 \pi} \Delta \varphi =  -b \mu e^{-b \varphi} \,.
\end{equation}
Note that the term $\propto \hat R$ has no classical contribution because of the Gauss--Bonnet theorem, see eq. \eqref{eq:gaussbonnet} (this term is topological). It is well--known in differential geometry that, if $\varphi(z)$ satisfies eq. \eqref{eq:liouville}, a surface with metric (line element) $\abs{\dif s} =  e^{-b\varphi/2} \abs{\dif z}$ has a constant \textit{negative} curvature (if $\mu > 0$). Such surfaces are the classical solutions to the Einstein equation in 2D. So, \gls{lft} is a quantization of eq. \eqref{eq:liouville}, and accounts for the \textit{fluctuation} of the metric determined by $e^{-b\varphi}$.

In a quantum theory, the fundamental objects calculated are the correlation functions. The ones considered in \gls{lft} are those of \textit{vertex operators}, defined as exponential fields
\begin{equation} \label{eq:defvertex}
\vertex_a (w) = e^{-a \varphi(w)} \,,
\end{equation} 
 where $a$ is called the \textit{charge} of the vertex operator. The $n$-point correlation functions are defined by the usual functional integral 
 \begin{equation}
 \left< \prod_{i=1}^n \vertex_{a_i} (w_i) \right>_b = \int \mathcal{D}\varphi \, e^{-\mathcal{S}_b} 
 \prod_{i=1}^n e^{-a_i \varphi(w_i)} \,. \label{eq:LFTcorrelation}
 \end{equation}
 In \gls{lft}, it is customary not to normalize the correlation function. That is, there is no factor $1 / \left< 1 \right>_b$ involving the zero--point function in the right hand side. This is due to a particularity of \gls{lft}: its zero--point function is not well--defined in general. Indeed, the Seiberg bound (see eq. \eqref{eq:seiberg2} below) cannot satisfied with no vertex operators, \textit{i.e.}, $\sum_i a_i = 0$, when $\chi \geq 0$, \textit{i.e.}, on the sphere and on the torus. From a classical geometry point of view, the reason is that the Liouville equation eq. \eqref{eq:liouville} describes a surface with constant negative curvature. However, such surfaces cannot be a sphere or a torus, by the Gauss--Bonnet theorem, eq. \eqref{eq:gaussbonnet}, unless we insert some \textit{conical singularities}, \textit{i.e.}, Dirac $\delta$'s of curvature.  In the quantum theory, this amounts to inserting enough vertex operators in the correlation functions. 
 
 \subsubsection{Integration of zero--mode}
 The connection between eq. \eqref{eq:LFTcorrelation} and 2D \gls{gff} was established by Goulian and Li \cite{goulian1991correlation} and is the basis of the recent rigorous developments \cite{david2016liouville}. Note that although the correlation functions of any quantum field theory can be written in terms of \gls{gff} (this is the basis of perturbation theory), for \gls{lft}, thanks to its conformal invariance, this strategy turns out particularly fruitful and has led (in \cite{dorn1994two,zamolodchikov1996conformal}) to its complete solution (beyond perturbation theory).
 
 The starting point is the decomposition of the Liouville field into a \textit{zero--mode} $\zmode$ and a fluctuating part $\tilde{\varphi}$, and a similar factorization of the functional integral:
 \begin{equation}\label{eq:goulianlidecomp}
 \varphi = \zmode + \tilde{\varphi} \,,\, \text{such that} \int_{\Sigma} \tilde{\varphi} \dif A = 0\,,\, 
\int \mathcal{D} \varphi = \int_{\R}  \dif \zmode  \times \int \mathcal{D} \tilde\varphi \,.
 \end{equation}
 The action eq. \eqref{eq:Sliouville} can be then written as 
 \begin{align}
 \mathcal{S}_b &= \int_{\Sigma} \left[ \frac{1}{16\pi} (\nabla \tilde\varphi)^2 - \frac1{8\pi}Q \hat{R} \tilde\varphi \right] \dif A  + \mu e^{-b \zmode} \int_\Sigma e^{-b\tilde\varphi} \dif A+  \zmode \int_{\Sigma} \frac1{8\pi}Q  \dif A \nonumber \\
& = \int_{\Sigma} \left[ \frac{1}{16\pi} (\nabla \tilde\varphi)^2 - \frac1{8\pi}Q \hat{R} \tilde\varphi \right] \dif A  + \mu e^{-b \zmode} Z_0 + \varphi_0 Q \chi / 2 \,, \label{eq:Sbzero}
 \end{align}
 where in the second line we defined the continuum partition function
 \begin{equation}
 Z_0 \defeq \int_\Sigma e^{-b\tilde\varphi} \dif A \,, \label{eq:defZ0}
 \end{equation}
 and used the Gauss-Bonnet theorem
\begin{equation}
  \int_{\Sigma}  \hat{R}   \dif A   =  4 \pi \chi \,,  \label{eq:gaussbonnet}
 \end{equation}
where $\chi$ is the \textit{Euler characteristic} of the surface $\Sigma$. It depends only on the topology of the surface: for a sphere, $\chi = 2$, for a torus, $\chi=0$, and $\chi = 2 - 2 g$ for general closed surface with genus $g$ handles. We can decompose similarly the vertex operator eq. \eqref{eq:defvertex} $\vertex_a(z) = e^{-a\zmode} e^{-a\tilde\varphi}$. Plugging this and eq. \eqref{eq:Sbzero} and into eq. \eqref{eq:LFTcorrelation}, we can integrate out the dependences on the zero mode:
\begin{align}
\left< \prod_{i=1}^n \vertex_{a_i} (w_i) \right>_b = &\int \mathcal{D} \tilde\varphi  \, e^{-\int_{\Sigma} \left[ \frac{1}{16\pi} (\nabla \tilde\varphi)^2 -\frac1{8\pi}Q \hat{R} \tilde\varphi \right] \dif A +\sum_{i=1}^n a_i \tilde{\varphi}(w_i) } \times \nonumber\\
& \int_\R \dif \zmode 
\exp(-\mu e^{b\zmode} Z_0) e^{\left(Q\chi/2 - \sum_{i=1}^n a_i \right) \zmode}  \nonumber 
\end{align}
The zero mode integral (the second line) is convergent if and only if
\begin{equation}
s \defeq \sum_{i=1}^n a_i - Q\chi/2 > 0 \,, \label{eq:seiberg2}
\end{equation}
which is called a Seiberg bound \cite{seiberg1990notes}. As we have remarked above after eq. \eqref{eq:LFTcorrelation}, this is why the zero--point correlation function is not well--defined on the sphere or on the torus. When eq. \eqref{eq:seiberg2} is satisfied, the $\zmode$ integral can be evaluated using a Gamma function:
\begin{align} \label{eq:GoulianLi}
\left< \prod_{i=1}^n \vertex_{a_i} (w_i) \right>_b = \frac{\Gamma\left[\frac{s}{b}\right]}{b\mu^{\frac{s}{b}}} \int \mathcal{D} \tilde\varphi  \, e^{-\int_{\Sigma} \left[ \frac{1}{16\pi} (\nabla \tilde\varphi)^2 -\frac1{8\pi}Q \hat{R} \tilde\varphi \right] \dif A +\sum_{i=1}^n a_i \tilde{\varphi}(w_i) } Z_0^{-s/b} 
\end{align}
 We see now that the terms in the exponential become at most quadratic, and can be interpreted in terms of a \textit{free} field. In order to be completely clear, let us denote by $\phi$ the 2D \gls{gff} on $\Sigma$. Its Green function is the solution to Poisson equation
  \begin{equation}
  \overline{\phi(z) \phi(w)} = K(z,w) \,,\, \Delta_z K(z,w) = 8 \pi \left(V^{-1} - \delta_{z,w}\right)  \,, \int_\Sigma K(z,w)  \dif A =0 \,, \label{eq:generalgff}\end{equation}
  where $V = \int_{\Sigma} \dif A$ is the total area of the surface and $\Delta$ is the Laplace--Beltrami operator on $\Sigma$.  Since $\tilde{\varphi}$ has no zero mode by eq. \eqref{eq:goulianlidecomp}, the functional integral over it is proportional to an average over the 2D \gls{gff} defined on the surface $\Sigma$:
 \begin{equation}
 \int \mathcal{D} \tilde\varphi  \, e^{-\int_{\Sigma} \left[ \frac{1}{16\pi} (\nabla \tilde\varphi)^2 \right]} \mathcal{O}[\tilde\varphi] = N_{\Sigma} \overline{\mathcal{O}[\phi]} \,. \label{eq:Dphioverline}
 \end{equation}
 Here, $\mathcal{O}[\dots]$ denotes any observable, and $N_{\Sigma}$ is a normalization factor (the partition function of free boson on $\Sigma$), and $\overline{[\dots]}$ in the right hand side denotes the average over $\phi$.
 
 From now on, we will identify $\phi$ and $\tilde\varphi$, \textit{e.g.} in \eqref{eq:defZ0}. Combined with \eqref{eq:GoulianLi} we have the following representation of \gls{lft} correlation functions in terms of 2D \gls{gff}:
 \begin{equation}\label{eq:GoulianLi2}
 \left< \prod_{i=1}^n \vertex_{a_i} (w_i) \right>_b = C \overline{\exp\left( \int_\Sigma Q\hat{R} \phi / 2  - \sum_{i=1}^n a_i \phi(w_i) \right)  Z_0^{-s/b} } \,,
 \end{equation}
 where $C = \Gamma(s/b) b^{-1} \mu^{-\frac{s}{b}} N_{\Sigma}$. Note that the dependence on the coupling constant $\mu$ is included in this global factor.
 
 We remark that, if we neglected the Seiberg bound requirement, eq. \eqref{eq:seiberg2}, and supposed that $-s / b = n$ was an integer in eq. \eqref{eq:GoulianLi2}, its right hand side would be a Coulomb--gas integral over $n$ moving charges on the 2D surface. This is reminiscent of the Coulomb--gas representation of correlation functions in minimal \glspl{cft} \cite{dotsenko1984conformal}, which describes critical statistical models in 2D, like the Ising model. Now, by observing the constant below eq. \eqref{eq:GoulianLi2}, we see that the residue at $-s / b = n$ of the correlation function in the left--hand side of eq. \eqref{eq:GoulianLi2} is given by a Coulomb--gas integral; so, a crucial step of the bootstrap solution of \gls{lft} can be seen as the \textit{analytical continuation} of the Coulomb gas integrals to non--integer values of $n$. In the literature, the equation $-s/b = n, n = 1, 2, 3, \dots$ is also called a ``screening condition''. We stress that in our applications, this will never be satisfied, because eq. \eqref{eq:seiberg2} will always hold: $s > 0$. 
  
\subsection{LFT and 2D logREMs}\label{sec:mapping1}
The right hand side of eq. \eqref{eq:GoulianLi2} is an involved observable of the 2D \gls{gff} on $\Sigma$, with \textit{a priori} no clear physical interpretation. Our next goal is to show that with a suitable choice of the charges $a_i$, and a  ``complete--the--square'' trick, we can interpret eq. \eqref{eq:GoulianLi2} as a correlation function of a Gibbs measure. We state here below the exact connection between 2D \glspl{logrem} and \gls{lft}:\vspace{.2cm}

\fbox{
\begin{minipage}{.99\textwidth}
Consider, on the one hand the \gls{lft} correlation function 
$\left< \prod_{j=1}^\ell \vertex_{a_j} (w_j) \prod_{i=1}^k \vertex_{q_i b} (z_i) \right>_b$
 with the following constraints on the parameters $a_1, \dots, a_\ell, q_1 b, \dots, q_k b,$
\begin{equation}
\sum_{i=1}^\ell a_i = Q \chi / 2 \,,\, b  \leq 1 \,, \forall a_j < \frac{Q}2 \,,\, \forall q_i < \frac{Q}{2b} \,.   \label{eq:sumcharges0}
\end{equation} 
On the other hand, let $U(z)$ be a deterministic logarithmic potential defined  by the Laplace equation, eq. \eqref{eq:Ugeneral} below:
\begin{equation}
\Delta_z U(z) = 8\pi( a_1 \delta_{z,w_1} + \dots + a_{\ell} \delta_{z,w_\ell})  -Q\hat{R} \,. \label{eq:Ugeneral0}
\end{equation}
We defined the Gibbs measure $p_b(z)$ (at inverse temperature $\beta = b \leq 1$) of a \gls{logrem} made of the 2D \gls{gff} plus $U(z)$: \eqref{eq:Gibbsgeneral}:
  \begin{equation}
  p_b(z)\defeq \frac1Z e^{-b(\phi(z)+ U(z))}\,,\, Z \defeq  \int e^{-b(\phi + U)} \dif A \,,
  \end{equation}
Then, we have the exact correspondence:
 \begin{equation}\label{eq:maingeneral}
 \left< \prod_{j=1}^\ell \vertex_{a_j} (w_j) \prod_{i=1}^k \vertex_{q_i b} (z_i) \right>_b = C' \overline{\prod_{i=1}^m p_b^{q_i}(z_i)} \,,
 \end{equation}
 where $C'$ is a constant independent of $(z_i)$. 
\end{minipage} }\vspace{.3cm}

The 2D \gls{logrem}--\gls{lft} dictionary is also summarized in Table \ref{tab:Dict} below. This section goes through its derivation, and discusses further conditions that appeared in the above mapping, \textit{i.e.} the full set of Seiberg bounds, see eq. \eqref{eq:seiberg1}. 

\subsubsection{Complete--the--square trick}
To show the above quoted result, we need to separate the vertex operators into two groups to which we associate different meaning. Let $\ell < n$ and consider the first $\ell$ vertex operators. For the mapping to be correctly established, we shall require that their sum 
\begin{equation}
\sum_{i=1}^\ell a_i = Q \chi / 2 \,,\label{eq:sumcharges}
\end{equation}
where $\chi$ is the Euler characteristics. This is indeed the \textit{charge neutrality condition} (which is \textit{not} the screening condition mentioned above) of the following Poisson equation for $U(z)$: 
\begin{equation}
\Delta_z U(z) = 8\pi( a_1 \delta_{z,w_1} + \dots + a_{\ell} \delta_{z,w_\ell})  -Q\hat{R} \,, \label{eq:Ugeneral}
\end{equation} 
where the Dirac deltas are respect to the surface integral $\int \dif A$. The necessary and sufficient condition for eq. \eqref{eq:Ugeneral} to have a solution is that the integral right hand side over the surface vanishes. But this is precisely guaranteed by eq. \eqref{eq:sumcharges} and the Gauss-Bonnet theorem eq. \eqref{eq:gaussbonnet}:
$$ \int \dif A \left[ 8\pi( a_1 \delta_{z,w_1} + \dots + a_{\ell} \delta_{z,w_\ell})  -Q\hat{R}\right] = 8 \pi Q\chi/2 - 4 \pi Q\chi = 0 \,.$$
Provided eq. \eqref{eq:sumcharges}, the solution $U(z)$ to eq. \eqref{eq:Ugeneral} is unique up to a constant, and will be the deterministic background potential of the 2D \gls{logrem} that we define now.

For this, we denote $u(z)$ the right hand side of eq. \eqref{eq:Ugeneral}, and perform the  \textit{complete--the--square} trick, which is a standard exercise of Gaussian integral (we shall recall it for completeness below). It is also known as Girsanov transform \cite{girsanov1960transforming}, see \cite{david2016liouville} for rigorous treatment in similar context. Indeed, denoting $\mathcal{O}[\phi] $ a general observable of the \gls{gff} $\phi$, and using definition of its average, eq. \eqref{eq:Dphioverline}, we have 
\begin{align} 
 & \overline{  \exp\left[ - \int u(z) \phi(z) \dif A \right] \mathcal{O}[\phi]  } = N_{\Sigma}^{-1} \int \mathcal{D}\phi  
 \exp\left[ \int \left( \frac{1}{16\pi} \phi\Delta\phi - u \phi \right) \dif A   \right] \mathcal{O}[\phi]  \nonumber \\
   = &  N_{\Sigma}^{-1} C_U \int \mathcal{D}\phi   \exp\left[ \int \left( \frac{1}{16\pi} (\phi-U) \Delta (\phi-U) \right) \dif A   \right] \mathcal{O}[\phi] 
   \nonumber \\
  = & C_U N_{\Sigma}^{-1}   \int \mathcal{D}\phi   \exp\left[ \int \left( \frac{1}{16\pi} \phi \Delta \phi \right) \dif A   \right] \mathcal{O}[\phi + U]
   = C_U \overline{\mathcal{O}[\phi + U]} \,,\label{eq:complete} \\
 \text{where }&  C =  \exp\left( - \frac{1}{16 \pi} \int U \Delta U \dif A  \right) \,. 
 \end{align}
 Note that the constant $C$ is formally infinite in the continuum. Remark that, to resolve this problem, one may carry out the same trick directly for the discrete \gls{logrem}, and obtain a constant $C$ which is finite but diverges as one removes the \gls{uv} cut--off. We will not do this here since fixing the normalization constant is not our goal here.
 
 Now, let us apply this to eq. \eqref{eq:GoulianLi2}. For this, let us  rename the second group of vertex operators
  \begin{equation}
  a_{\ell+i} = q_i b \,,\, w_{i + \ell} =  z_i\,,\, i = 1, \dots, m = n - \ell \,. 
  \end{equation} 
  The above trick will transform the partition function \eqref{eq:defZ0} into 
 \begin{equation}\label{eq:ZZ0}
 Z \defeq Z_0[\phi\leadsto \phi + U] = \int e^{-b(\phi + U)} \dif A \,.
 \end{equation}
  Note also that eq. \eqref{eq:sumcharges} and \eqref{eq:seiberg2} imply 
  \begin{equation} s = \sum_{i=1}^m q_i b  \label{eq:seiberg22} \,,\end{equation}
  so we have nicely 
 \begin{align}
&\left< \prod_{j=1}^\ell \vertex_{a_j} (w_j) \prod_{i=1}^k \vertex_{q_i b} (z_i) \right>_b = C'
 \overline{Z^{-s/b} \prod_{i=1}^m e^{-b q_i( \phi(z_i) + U(z_i) )}}  \nonumber \\
 = & C' \overline{\prod_{i=1}^m \left[e^{-b q_i( \phi(z_i) + U(z_i) )} Z^{-q_i}\right]} = C' \overline{\prod_{i=1}^m p_b^{q_i}(z_i)}  \,, \label{eq:mainLFTcontinuum}
 \end{align}
 where $C' = C_U \Gamma(s/b) b^{-1} \mu^{-\frac{s}{b}} N_{\Sigma}$ is another constant, and 
 \begin{equation}\label{eq:Gibbsgeneral}
 p_b(z)\defeq \frac1Z e^{-b(\phi(z)+ U(z))}
 \end{equation}
 is the Gibbs measure of a thermal particle in the potential $\phi(z) + U(z)$ made of a random 2D \gls{gff} and a deterministic logarithmic potential. This equation is valid with further conditions (Seiberg bounds), as we discuss below.

 Equation \eqref{eq:mainLFTcontinuum} indicates, at a continuum level, that \gls{lft} correlation functions correspond to the \textit{Gibbs measure statistics} (multi-point correlations of powers of the Gibbs measure) of a \gls{logrem} with a composite potential $\phi(z) + U(z)$. This is the core result of the exact mapping established in \cite{cao16liouville}. We would like to underline the \textit{naturalness} of the link: eq. \eqref{eq:sumcharges} is required by the ``charge neutrality condition'' imposed by eq. \eqref{eq:Ugeneral}, and guarantees \textit{at the same time}, \textit{via} eq. \eqref{eq:seiberg22}, that the $Z^{-s/b}$ factor has exactly the correct power to allow an interpretation as Gibbs measure. 
 
 In this respect, we make an important remark. In order to take into account the  free energy distribution, we would need to add another factor $Z^{-t/b} = \exp(t F)$ in the right hand side of eq. \eqref{eq:mainLFTcontinuum}; then, its left hand side would be the correlation function of a field theory defined by the  action eq. \eqref{eq:Sliouville} but with $Q \neq b + b^{-1}$. Unfortunately, such a theory is \textit{not} conformal invariant and not exactly solved. Therefore, we are unable to extend fruitfully the current mapping to the include any information on the free energy of the \gls{logrem}. Therefore, the exact calculation of the free energy distribution in 2D \gls{logrem} remains an open question.
 
 \subsubsection{Seiberg bounds}  
 Eq. \eqref{eq:mainLFTcontinuum} is obtained by continuum manipulations, so it describes correctly the Gibbs measure of the \textit{discrete} \gls{logrem} in the thermodynamic/continuum limit only in the high-temperature phase. We have taken care of the normalization of the 2D \gls{gff} in eq. \eqref{eq:generalgff} such that $\overline{\phi(z) \phi(w)} \sim -4 \ln \abs{z-w}$ as $z \to w$, in agreement with eq. \eqref{eq:logdecay} ($d=2$), so the critical temperature is $\beta_c = 1$. When $\beta < 1$, the temperature of the \gls{logrem} corresponds simply to the parameter $b$ in \gls{lft}.  When $\beta > 1$, we must combine the \gls{1rsb}/freezing results and \gls{lft} predictions at $b = 1$ to describe the Gibbs measure of the discrete \gls{logrem}, see eq. \eqref{eq:freezingpos} and \eqref{eq:PPP2pos}. In summary, the following notation will be convenient:
 \begin{equation}\label{eq:bandbeta}
 b = \begin{cases}
 \beta \,, & \beta < 1 \,, \\  1\,,  & \beta \geq 1 \,.
 \end{cases}
 \end{equation}
 Curiously, by eq. \eqref{eq:defQ}, $Q = b + b^{-1} = -\frac{\mathcal{F}}{\ln M}$ is the (minus) free energy density of \glspl{logrem}, eq. \eqref{eq:freezedual}. 
 
The condition $\beta = b \leq 1$ is not the only condition for eq. \eqref{eq:mainLFTcontinuum} to hold. The others come from the \textit{Seiberg bounds} \cite{seiberg1990notes,david2016liouville}. These bounds, together with $b \leq 1$ (known as the $b=1$ ``barrier'' in the \gls{lft} language) are conditions for \gls{lft} correlations to be represented by 2D \gls{gff} in a probabilist sense. The Seiberg bounds require that 
\begin{enumerate}
\item $\sum_{i} a_i > Q \chi/2$, where $a_i$ are the charges in the \gls{lft} correlation. We have already seen it in \eqref{eq:seiberg2} as the condition of convergence of the zero mode integral. In our applications, this is always true, by eq. \eqref{eq:sumcharges}, as long as $q_i > 0$ (we will not consider negative powers of Gibbs measure). 
\item $a_i < Q / 2$ for any of charges, \textit{i.e.}, in eq. \eqref{eq:mainLFTcontinuum},
\begin{subequations}\label{eq:seiberg1}
\begin{align}
 & a_i < Q/2 \,,\, i = 1, \dots, \ell \,, \label{eq:seibeig11} \\
 & q_i < Q/(2b) \,,\, i = 1, \dots, m \,.  \label{eq:seibeig12} 
\end{align}
\end{subequations}
For each $a = a_i$, the first condition coincides with the no--binding condition, eq. \eqref{eq:nobinding}. This is the correct interpretation since eq. \eqref{eq:Ugeneral} implies that $U(z) \sim -4\ln \abs{z-w_i}$ near the charge, in agreement with eq. \eqref{eq:logpotential} ($d=2$). Therefore eq. \eqref{eq:seibeig11} is the condition under which none of the singularities of $U(z)$ is too strong to trap the thermal particle at its bottom. Note that eq. \eqref{eq:seibeig11} and eq. \eqref{eq:sumcharges} imply that for sphere like surfaces, $\chi=2$, we need a potential $U(z)$ with $\ell \geq 3$ singularities; on the other hand, on the torus, $\chi = 0$, $\ell = 0$ ($U(z) = 0$) is permitted. 

For each $q = q_i$, and when $\beta = b \leq 1$, eq. \eqref{eq:seibeig12} coincides with the phase boundary of the \acrfull{ipr} exponent in the annealed ensemble, eq. \eqref{eq:IPRann}. This is not a coincidence, as we will discuss in section \ref{sec:LFTmultifrac}. Before that section, we will only consider cases where $\forall i, q_i = 1$, so eq. \eqref{eq:seibeig12} is always true when $b < 1$.
\end{enumerate}
Summarizing the discussions so far, we arrive at the main result eq. \eqref{eq:maingeneral}. The Table \ref{tab:Dict} is also a good summary of the mapping established above.

\begin{table}
\center
\begin{tabular}{|c|c|c|}
\hline
\gls{lft} &  \textit{via}, see also & 2D \gls{logrem} \\ \hline
$\mu$ cosmological constant        &  &  no meaning \\ \hline
$b$ in eq. \eqref{eq:Sliouville}      &   $b = \min(1,\beta)$, \eqref{eq:bandbeta} & inverse temperature $\beta$ \\ \hline
Vertex operator $\vertex_a(w)$ & \eqref{eq:Ugeneral0} & Background potential $U(z)$ \\ \hline 
Seiberg bound $a < Q / 2$  & \eqref{eq:seibeig11} & no binding/escaping \\ \hline
Vertex operator $\vertex_{qb}(z)$ &  $b = \min(1, \beta)$ & Gibbs measure $p_\beta^{q}(z)$  \\ \hline
Seiberg bound $q b < Q / 2$ &  \eqref{eq:seibeig12}  & no termination point transition \\ \hline
Correlation function $ \left< \dots \right>_b $ &  $\stackrel{\beta<1}\propto$, \eqref{eq:maingeneral}   & average over disorder  $\overline{[\dots]}$  \\ \hline
Seiberg bound $\sum \text{charges} > Q$ & \eqref{eq:seiberg2} & always satisfied \\ \hline
\end{tabular}
\caption{The summary of the \gls{lft}--2D \glspl{logrem} mapping as a dictionary.} \label{tab:Dict}
\end{table}

 \subsubsection{Infinite plane case, numerical test}
We illustrate and check numerically the general result in the special case of infinite plane, which is the one considered in \cite{cao16liouville} (main text). It is well--known (see for example \cite{zamolodchikov2007lectures}) that the \gls{lft} can be considered on the infinite plane \textit{plus a point}, $\C \cup \set{\infty}$. This surface is closed, topologically identical to the round sphere ($\chi=2$), but its geometry resembles more the flat Euclidean plane. Its surface element $\dif A = \dif^2 z$; its curvature $\hat{R} = 8 \pi \delta(z-\infty)$ vanishes everywhere but is concentrated at infinity, so that $\int_{\C \cap \infty} \hat{R} \dif A = 4 \pi \chi$ in agreement with the Gauss--Bonnet theorem eq. \eqref{eq:gaussbonnet}. 

The application of the \gls{lft} on such a surface to statistical models is problematic. Indeed, the infinite plane case is not yet covered by the rigorous treatments in \cite{david2016liouville}, so presents a new technical challenge and an interesting subject of numerical study. The reason is that the 2D \gls{gff} on $\C \cup \set{\infty}$ is ill-defined, and should be considered as the $R \to \infty$ limit of the 2D \gls{gff} on a flat domain of linear size $R$. In practice, we use the periodic boundary condition and the fast Fourier transform, eq. \eqref{eq:FFT2D}, to generate it. Note that the resulting covariance is $\overline{\phi(z) \phi(w)} = -4\ln \abs{z-w}$, see eq. \eqref{eq:2DGFFdis}, and satisfies  eq. \eqref{eq:generalgff} where $V^{-1}= 0$ because the area is infinite. So the 2D \gls{gff} is in fact on a \textit{torus} of size $R$ and lattice spacing $\epsilon$. The obvious objection is then: how can the planar/spherical \gls{lft} make prediction about this setting? 

\begin{figure}
\includegraphics[scale=.7,valign=t]{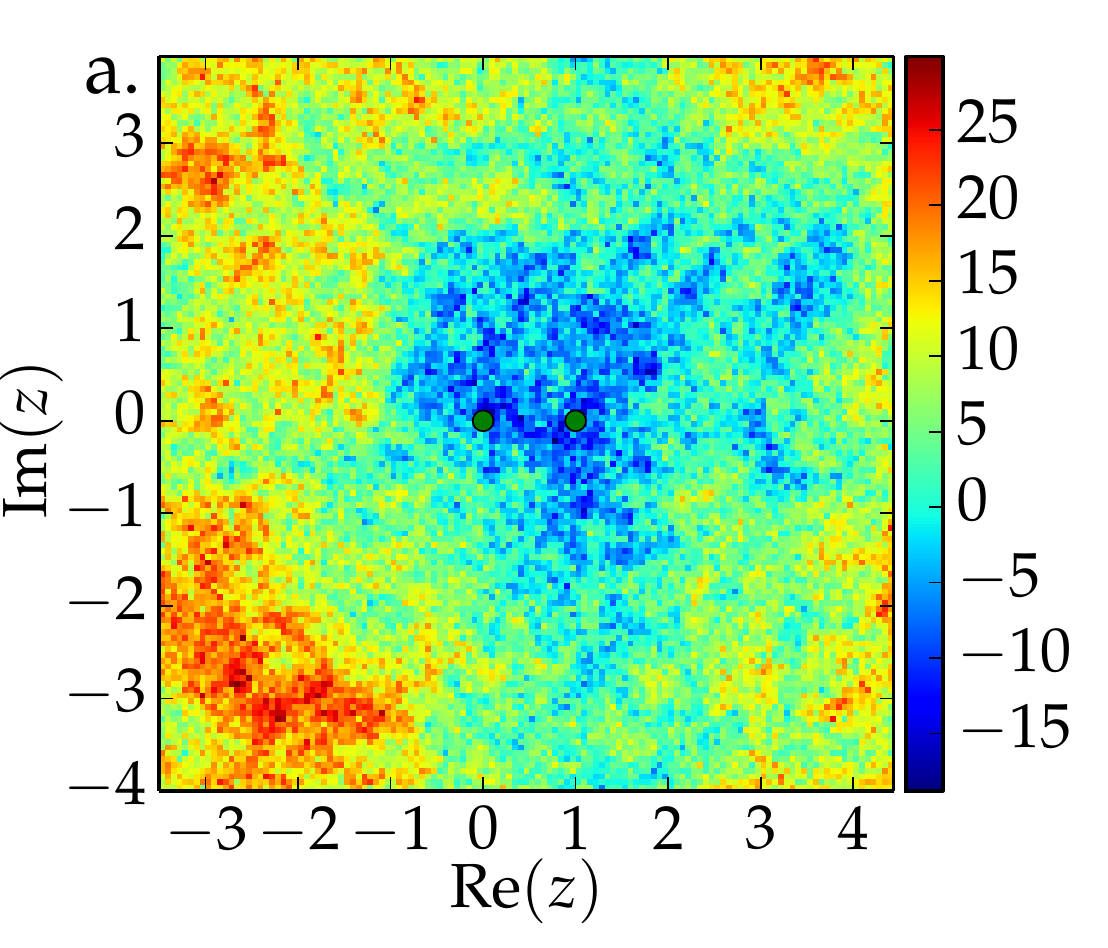} 
\includegraphics[scale= .8,valign=t]{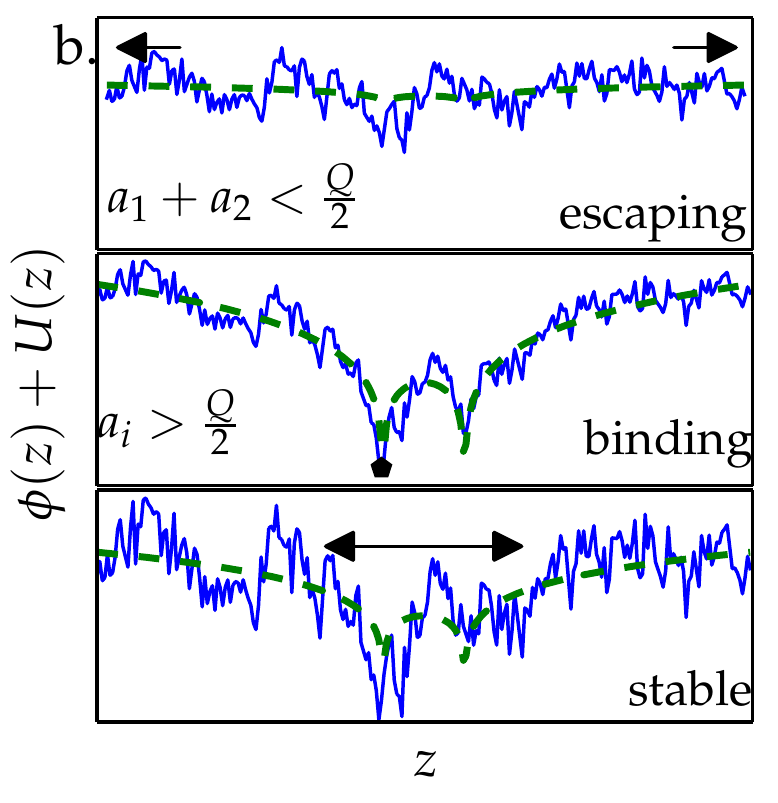}
\caption{Taken from \cite{cao16liouville}. a. Colour plot of a sample of 2D \gls{gff} plus the log confining potential $U(z)$ \eqref{eq:defUsimple} with $a_{1,2} = .8, .6$. The two singularities $z=0,1$ are indicated by green dots. The domain has lattice spacing $\epsilon = 2^{-5}$ and size $R = 8$, with periodic boundary condition.  b. Top: When the potential is too shallow the particle escapes to $\infty$, and the Gibbs measure vanishes as $R \to \infty$. Middle: When the potential is too deep, the Gibbs measure becomes a $\delta$ peak as $\epsilon \to 0$. Bottom: When all the Seiberg bounds are satisfied, the extent of the central region is stable as $R \to \infty, \epsilon \to0$ and the limiting Gibbs measure can be compared to planar \gls{lft}.} \label{fig:photo}
\end{figure}
The reason is the existence of the potential $U(z)$. Recall that it is defined by a set of charges $(a_i, w_i), i = 1, \dots, \ell$, such that $\sum_i a_i = Q$ \textit{via} eq. \eqref{eq:Ugeneral}. The Seiberg bounds eq. \eqref{eq:seibeig11} imply $\ell \geq 3$. We take the minimum $\ell = 3$, and set $(w_1, w_2, w_3) = (0, 1, \infty)$. Then eq. \eqref{eq:Ugeneral} is solved by 
\begin{equation}\label{eq:defUsimple}
U(z) = 4 a_1 \ln \abs{z} + 4 a_2 \ln \abs{z-1} \,, 
\end{equation}
up to a constant. The Seiberg bound eq. \eqref{eq:seibeig11} for the charge at infinity, $a_3 < Q / 2$, becomes equivalent to $a_1 + a_2 > Q / 2$ because $a_3 = Q-a_1 -a_2$, by eq. \eqref{eq:sumcharges}. Since the point is at infinite, the Seiberg bound acquires another interpretation: it guarantees that thermal particle does not \textit{escape to} $w_3 = \infty$. 

A simple way to understand this is to proceed by analogy with the analysis of the Gaussian model in section \ref{sec:Gaussian} (see the paragraph before eq. \eqref{eq:ZnGaussian}). Recall that the free energy of the \gls{logrem} with potential $\phi(z)$ alone (without $U(z)$) has extensive free energy $\mathcal{F} = -Q \ln M = - 2Q \ln R+ 2 \ln \epsilon $ ($M = (R/\epsilon)^2$ is the size of the \gls{logrem}). Now, when $U(z)$ is added, the free energy cost for a particle to escape to $R$ is $F(R) = \mathcal{F} + U(R) \approx - 2 Q \ln (R) + 4 (a_1 + a_2) \ln R + c = 4 (a_1 + a_2 - Q/2) \ln R + c$ where $c$ is $R$-independent. So $a_1 + a_2 > Q/2$ is equivalent to $F(R) \to + \infty$  as $R \to \infty$, \textit{i.e.}, escaping is unfavourable. Thus, as we anticipated in section \ref{sec:Gaussian}, the \gls{logrem} with potential $\phi(z) + U(z)$ considered here is a \textit{IR divergent \gls{logrem}}. Its free energy distribution suffers from the same problems as the Gaussian model; yet, its Gibbs measure can be still studied. 
 
Thus, when the $a_1 < Q/2, a_2 < Q/ 2$, but $a_1 + a_2  > Q/2$, the particle in the potential $ \phi(z) + U(z)$ is neither trapped at $0$ or $1$ nor escaping to infinity, and its Gibbs measure is expected to be stable in the limit $R \to \infty$  (as well as $\epsilon \to 0$). Since the scale $R$ becomes irrelevant, the detail of \gls{ir} regularization (\textit{i.e.} the periodic boundary condition) is irrelevant, and the \gls{lft} on $\C \cup \set{\infty}$ is suited to describe it. The simplest prediction made by eq. \eqref{eq:maingeneral} relates the disorder--averaged Gibbs measure to a 4 point function of \gls{lft} on $\C \cup \set{\infty}$:
\begin{equation}
\overline{p_\beta(z)} \propto \left\langle  \vertex_{a_1} (0)  \vertex_{a_2} (1) \vertex_b(z) \vertex_{a_3} (\infty)  \right\rangle_{b} \,,\, a_3 = Q - a_1 - a_2 \,. \label{eq:mainhighT}
\end{equation}
Although eq. \eqref{eq:maingeneral} is limited to the $\beta < 1$ phase, we have seen in eq. \eqref{eq:freezingpos} that $\overline{p_\beta(z)}$ becomes temperature--independent in the $\beta > 1$ phase. So eq. \eqref{eq:mainhighT} still holds in that phase, thanks to the notation eq. \eqref{eq:bandbeta}. 

\begin{figure}
\center
\includegraphics[scale=.65]{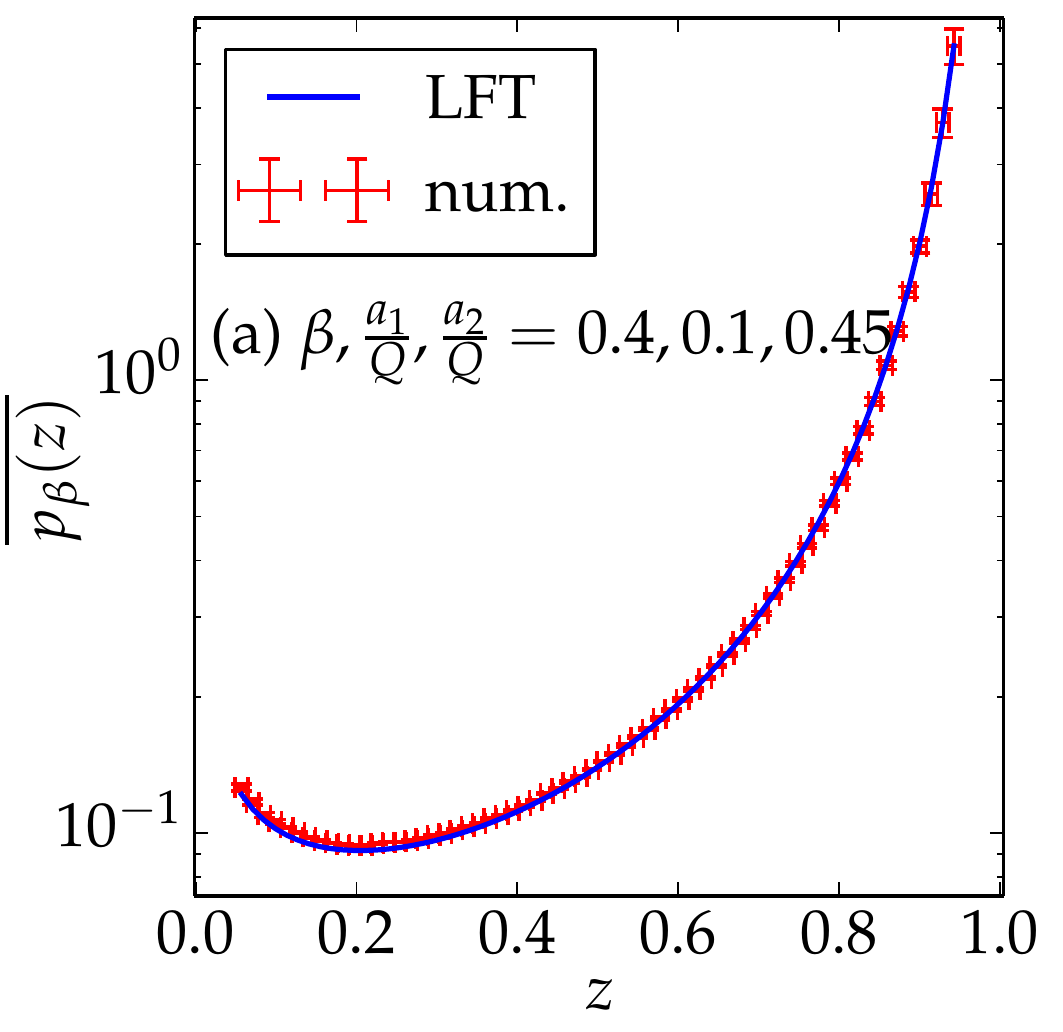}
\includegraphics[scale=.65]{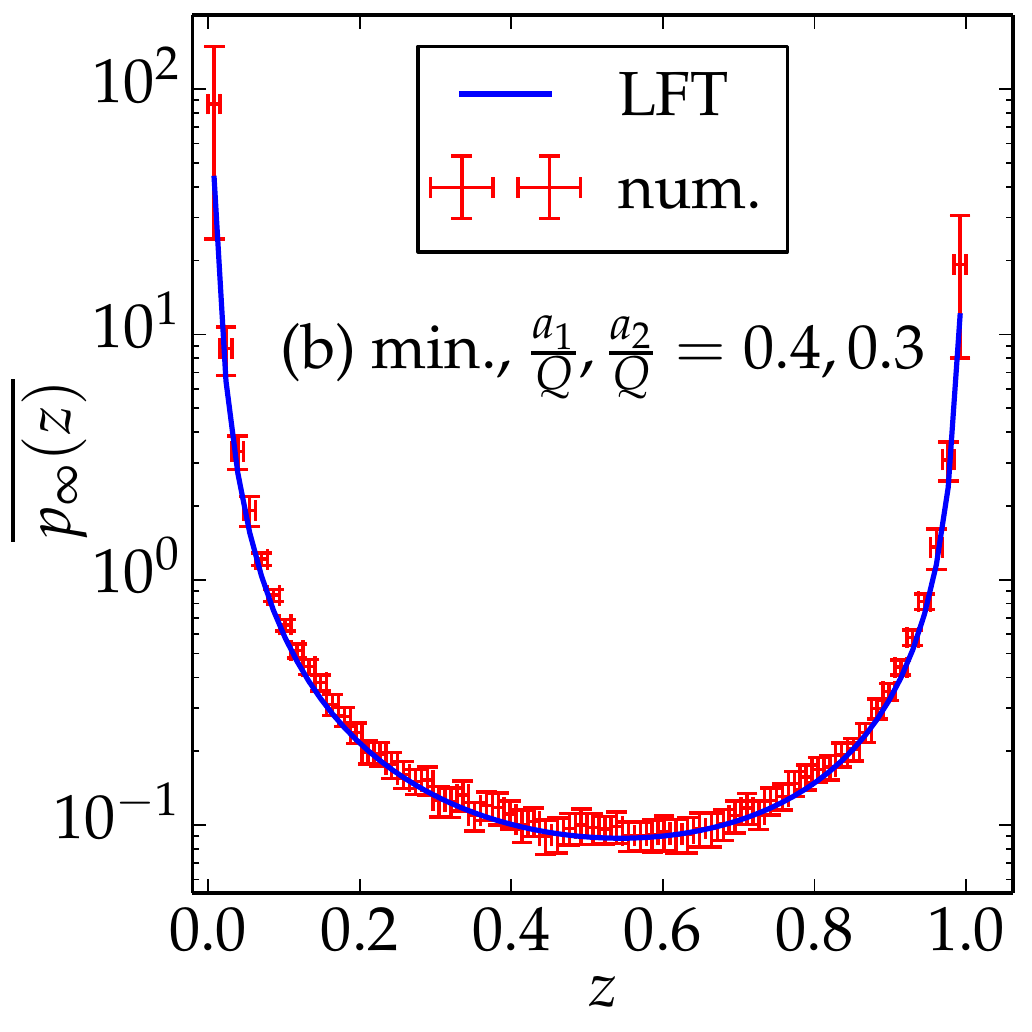}
\caption{Taken from \cite{cao16liouville}. Test of \eqref{eq:mainhighT} on the segment $z\in [0,1]$. (a) High-$T$ regime ($\beta=.4, a_1 / Q = .1, a_2 / Q = .45$). (b) Minimum position distribution versus \gls{lft} with $b=1$ ($a_1 / Q = .4, a_2 /  Q = .3$). The 2D \gls{gff} is generated on a square lattice with size $R = 2^{3}$ and lattice spacing $\epsilon = 2^{-9}$, with periodic boundary condition, using eq. \eqref{eq:FFT2D}. There are $5 \times 10^6$ independent samples for each measure. }\label{fig:LFTnum0}
\end{figure}
Equation \eqref{eq:mainhighT} was tested numerically in \cite{cao16liouville}. Its left hand side was measured on large scale simulations of 2D \gls{gff} (see Figure \ref{fig:LFTnum0} for parameters). The right hand side can be calculated using the conformal bootstrap solution of \gls{lft}, in terms of an involved analytical expression, as will be described in the next section. Fortunately, the code-base \cite{codebootstrap} is a powerful and accessible toolbox for numerical computation of \gls{lft} four--point functions. For the present application, we extend the code to take into account discrete fusions between vertex operators of type $\vertex_a$, $a\in (0, Q/2)$ (this point will be discussed in more detail in the next section \ref{sec:OPE}), and calculates easily the right hand side of \eqref{eq:mainhighT} with $10^{-5}$ precision, which is enough for the present application	. The left hand side of  \eqref{eq:mainhighT} is measured on extensive simulations of discrete 2D \gls{gff}. The results are reported in Figure \ref{fig:LFTnum0}. In each test, the values of both sides of \eqref{eq:mainhighT} for $x = z \in (0,1)$ are considered; the only unknown parameter is the global normalisation factor, which is fixed by matching the empirical mean of the logarithms. The results confirm well the prediction eq. \eqref{eq:mainhighT}. 

In summary, in this section we established the \gls{lft}--2D \glspl{logrem} mapping, and tested it numerically in its simplest setting on the infinite plane. The latter turns out also a tricky case, and is yet not covered by the rigorous treatments. Given the nice agreements obtained, we may be confident about the mapping and explore its consequences.

\subsection{Conformal Bootstrap and OPE}\label{sec:OPE}
All applications  of the \gls{lft}--\glspl{logrem} that we will discuss in this thesis will rely on the analytical results available about the \gls{lft} correlation functions. These results come from the conformal bootstrap solution of \gls{lft}. As for now, this is a very well understood subject; a systematic introduction can be found in \cite{ribault2014conformal} and in \cite{ribault2015liouville}, to which we will refer to. 

In the conformal bootstrap solution of \gls{lft}, we have the following expression for the 4--point function of \gls{lft} 
 \begin{align}
 \label{intcont}
 &\left< \vertex_{a_1}(0)\vertex_{a_4}(z)\vertex_{a_2}(1)\vertex_{a_3}(\infty)\right>_b 
 \nonumber \\
= &\int_{\mathcal{A}_L} C^{\text{DOZZ}}(a_1,a_4,a)C^{\text{DOZZ}}(Q-a,a_2,a_3)|\mathcal{F}_{\Delta_{a}}(\{a_i\},z)|^2 \abs{\dif a}  \,, \forall a_{i} \in \mathcal{A}_L
\end{align}
Let us explain the notations of the right hand side:
\begin{itemize}
\item The integral is over the \textit{spectrum} of \gls{lft}:
\begin{equation}
\label{LFTspect}
\mathcal{A}_L = \frac{Q}{2} + \im \R =  \left\lbrace \frac{Q}{2} + \im P: P \in \R \right\rbrace \,.
\end{equation}
\item $C^{\text{DOZZ}}$ is the \gls{dozz} \textit{structure constants} of \acrlong{lft} (the two groups of authors found them independently \cite{dorn1994two,zamolodchikov1996conformal}). We will discuss them in more detail below. 
\item $\Delta_a$ is the \textit{conformal dimension} of the field $\vertex_a(z)$. In \gls{lft}, 
\begin{equation}\label{eq:LFTDeltaLFT}
\Delta_a = a (Q - a) \,.
\end{equation}
We note that a general \gls{cft} is determined completely by its spectrum, the conformal dimensions, and its structure constants. 
\item $\mathcal{F}_{\Delta_{a}}(\{a_i\},z)$ is the \textit{conformal block}. It is a function of $z$, and the five conformal dimensions $\Delta_a, \Delta_{a_1}, \dots, \Delta_{a_4}$. It is known analytically, and efficiently calculated by the code--base \cite{codebootstrap}. It is universal to all 2D \glspl{cft}, \textit{i.e.}, the same function would appear if we consider eq. \eqref{intcont} for another \gls{cft}.
\end{itemize}

\subsubsection{The DOZZ structure constant}
The \gls{dozz} structure constants are known to be:
\begin{align}
&\label{DOZZ}
C^{\text{DOZZ}}(a_1,a_2,a_3)= \nonumber \\
& \frac{\left[b^{\frac{2}{b}-2b} \mu\right]^{Q-a_1-a_2-a_3}\prod_{i=1}^{3}\Upsilon_b(2 a_i)}{\Upsilon_b(\sum_{i=1}^3 a_i-Q)\Upsilon_b( a_1+a_2-a_3)\Upsilon_b(a_1-a_2+a_3)\Upsilon_b(-a_1+a_2+a_3)} \,.
\end{align}
The function $\Upsilon_b(x)$ is related to the $a$-Barnes function by (\cite{fateev2000boundary}, eq. 3.16)
\begin{equation} \Upsilon_b(z) = \Barnes_b(z) \Barnes_b(Q - z) \,, \label{eq:UppBarne} \end{equation} 
see eq. \eqref{eq:Barnes}. A few useful analytical properties of $\Upsilon_b(z)$ are recorded in \cite{ribault2014conformal} and \cite{cao16liouville} (Supplmental Material C.2). Here, the only analytical property we need is that: $\Upsilon_b(z)$ is analytic on $\mathbb{C}$, with infinitely many simple zeros:  
\begin{align}
 \Upsilon_b(z) = 0  \Leftrightarrow  z \in &\left\{ -  b m - b^{-1} n : m,n= 0,1,2 \dots  \right\}  \nonumber \\
 \bigcup &\left\{ Q +  b m +  b^{-1} n : m,n= 0,1,2 \dots  \right\}  \label{eq:Uppzero}
\end{align}
Observe they are organized into two \textit{lattices}: one ranging from $0$ to $-\infty$, the other from $Q$ to $+\infty$, and the two related by the symmetry $a \mapsto Q - a$. 

An important consequence of eq. \eqref{eq:Uppzero} is that, the \gls{dozz} formula eq. \eqref{DOZZ} has a simple zero at $a_3 = Q / 2$ (coming from $\Upsilon_b(2 a_3)$ in the numerator) 
\begin{equation}
C^{\text{DOZZ}}(a_1,a_2, Q / 2 + p) = c_1 p + O(p^2)  \,,\, \abs{p} \ll 1 \,. \label{eq:DOZZzero}
\end{equation}
where $c_1$ is some factor, for \textit{generic} $a_1$ and $a_2$, \textit{i.e.}, when all the other \gls{dozz} functions do not vanish. When they do, the zero might be cancelled by one in the denominator, or become of higher order. 

\subsubsection{Discrete terms}
It is important to note that eq. \eqref{intcont} holds if $\Re(a_1), \dots, \Re(a_4) = Q / 2$, so we cannot \textit{a priori} apply it to eq. \eqref{eq:mainhighT}. To obtain the correct formula, we need to perform an analytic continuation of the correlation functions to $\Re(a_i) \in (0, Q/2)$. A detailed account was provided in \cite{cao16liouville}, Supplemental Material C.3. For general discussion of this procedure, one may refer to \cite{aleshkin2016construction} or \cite{ribault2014conformal}. Here, let us explain the basic idea. 

\begin{figure}
\center \includegraphics[scale=.75]{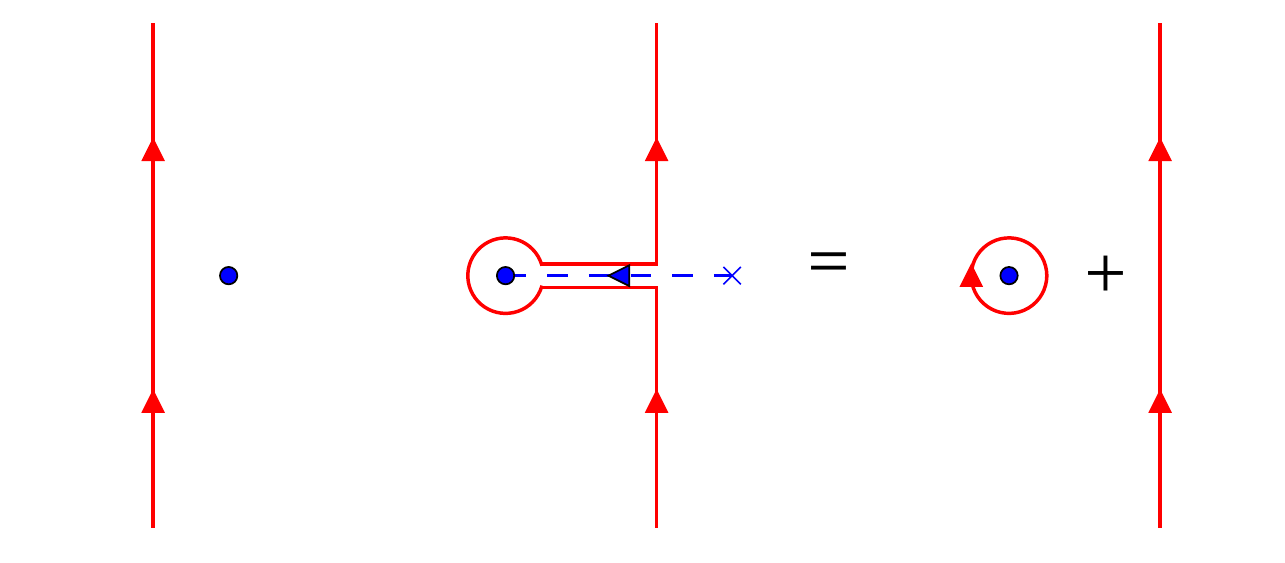}
\caption{An illustration of the deformation of a contour integral. \textit{Left}: an integral contour is drawn in red line (with arrow) and a pole at its initial condition is drawn as a blue dot. \textit{Middle}: the pole moves smoothly to the left of the contour. In order to prevent crossing, the contour is deformed. \textit{Right}: Applying Cauchy's theorem to the previous contour reduces it into the sum of the contribution of a residue (\textit{discrete term}) and that of a continuous integral.}\label{fig:contour}
\end{figure}
The integral eq. \eqref{intcont} is a contour integral of a meromorphic function of $a$. It has many poles which come from the zeros of the $\Upsilon$'s in the denominator of the \gls{dozz} formula eq. \eqref{DOZZ}, so their positions depend on $a_i$. Now, when $a_i$ moves smoothly away from $\mathcal{A}_L$ to their positions in eq. \eqref{eq:mainhighT}, some poles may cross the integral contour $\mathcal{A}_L$. To analytically continue the contour integral, the contour needs to be deformed to prevent the pole from crossing it, in the way indicated by Figure \ref{fig:contour}. Then, by Cauchy's theorem, the new contour integral can be evaluated as the integral along the old contour, plus a \textit{discrete term}: 
\begin{equation} \label{eq:discreteterm}
D_p = \pm 2\pi \;\text{Res}_{a \to p} \left[C^{\text{DOZZ}}(a,a_1,a_4) C^{\text{DOZZ}}(Q - a,a_2,a_3)\right]  \abs{\mathcal{F}_{\Delta_{a}}(\{a_i\},z)}^2 \,,
\end{equation}
where the sign $\pm$ depends on whether the pole is left--crossing or right--crossing (the example of Figure \ref{fig:contour} has a minus sign). This needs to be done for all the crossing poles. The list of them is analysed in \cite{cao16liouville}, Supplemental Material C.3, using the zeros of the $\Upsilon$ function eq. \eqref{eq:Uppzero}. In particular, the list of left--crossing poles of $C^{\text{DOZZ}}(a,a_1,a_4)$ is 
\begin{equation}
P_-^{14} = \left\{ x \in a_1 + a_4 + m b + n b^{-1}: \Re(x) \in  (0, Q / 2), m, n = 0,1,2, \dots \right\}  \,.
\end{equation}
It turns out that the right--crossing poles give the same discrete term contributions as the left--crossing ones. The analysis for the other $C^{\text{DOZZ}}$ constant in eq. \eqref{intcont} is similar. Therefore, the correctly generalized version of eq. \eqref{intcont} is 
 \begin{align}
 &\left< \vertex_{a_1}(0)\vertex_{a_4}(z)\vertex_{a_2}(1)\vertex_{a_3}(\infty)\right> = 
 2 \sum_{p \in P_-^{14}} D_p + (14 \leadsto 23) +   \nonumber \\
 & \int_{\mathcal{A}_L} C^{\text{DOZZ}}(a_1,a_4,a)C^{\text{DOZZ}}(Q-a,a_2,a_3)|\mathcal{F}_{\Delta_{a}}(\{a_i\},z)|^2 \abs{\dif a}  \,.  \label{intpdis}  
\end{align}
One contribution of the work \cite{cao16liouville} is adding the implementation of eq. \eqref{intpdis} to the code--base \cite{codebootstrap}, which we then use for the numerical simulation reported in Figure \ref{fig:LFTnum0}. 

\subsubsection{Asymptotic behaviour (\gls{ope})}
We consider the asymptotic behaviour of eq. \eqref{intpdis} as $z \to 0$, assuming $a_i \in (0, Q / 2)$. For this, note that the $z$-dependence of the $4$-point function \eqref{intpdis} comes from the conformal blocks. The $z\to 0$ series expansion of the latter is well-known:
\begin{equation}
\label{cbexp}
\mathcal{F}_{\Delta_a}(\{\Delta_{a_i}\},z) = z^{-\Delta_{a_1}-\Delta_{a_4}+\Delta_a} \left(1+  O(z)\right) \,,
\end{equation}
where higher order terms can also be explicitly written but this is unnecessary for our purposes here. Therefore, to compute the dominant asymptotic behaviour of eq. \eqref{intpdis} as $z\to0$, we need to consider the internal charges $a \in P_- \cup (Q/2 + \im \R)$ involved, and find the \textit{smallest} scaling dimension $\Delta_a = a(Q - a)$. We have to distinguish three cases, illustrated in Figure \ref{fig:fusionrule}:
\begin{figure}[h]
\center
a.\includegraphics[scale=.42,valign=t]{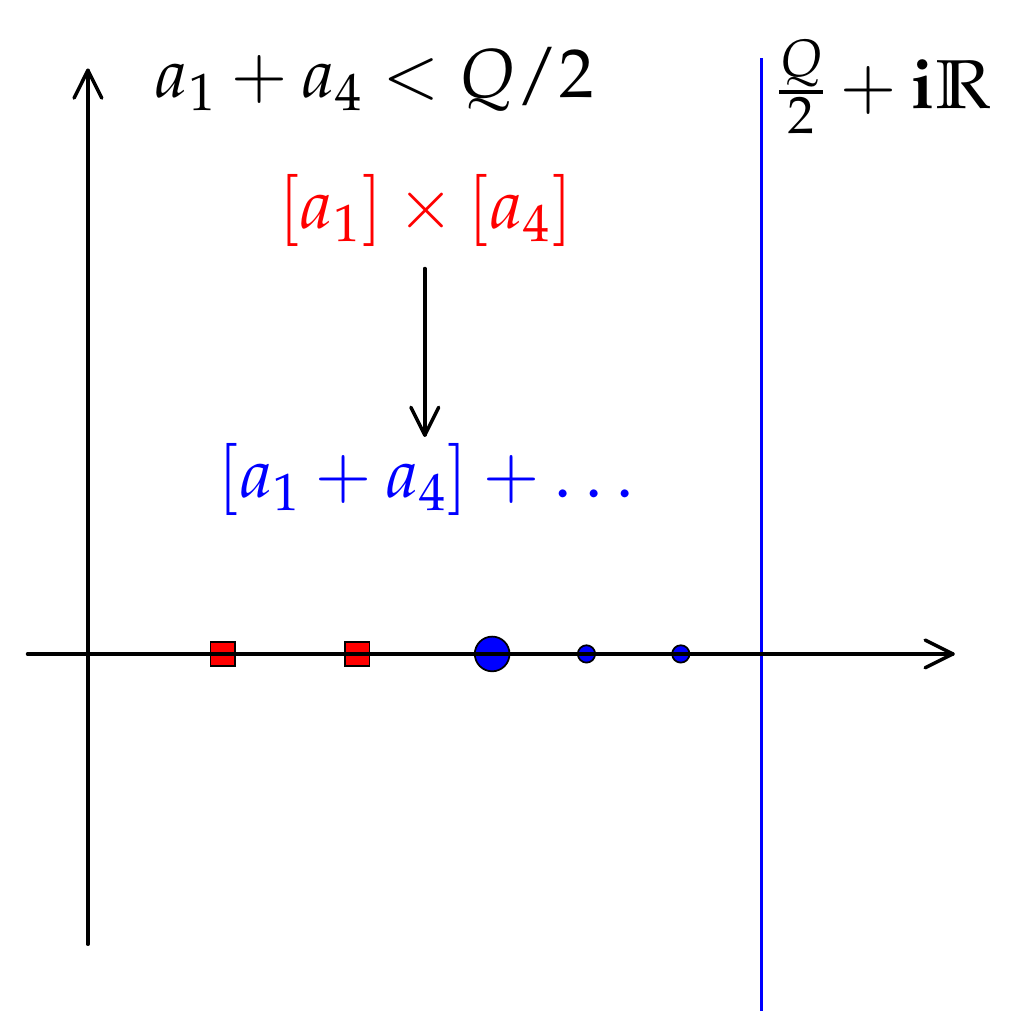}
b.\includegraphics[scale=.42,valign=t]{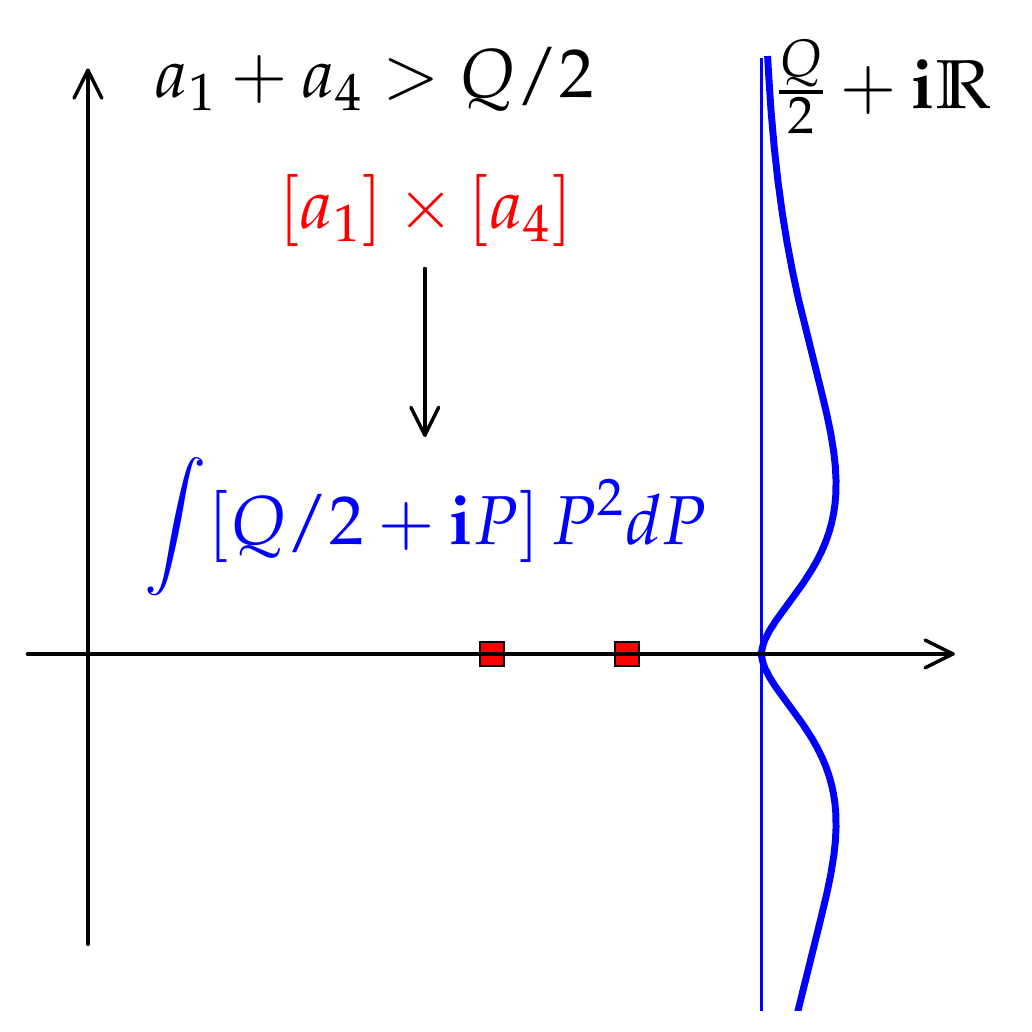}
c.\includegraphics[scale=.42,valign=t]{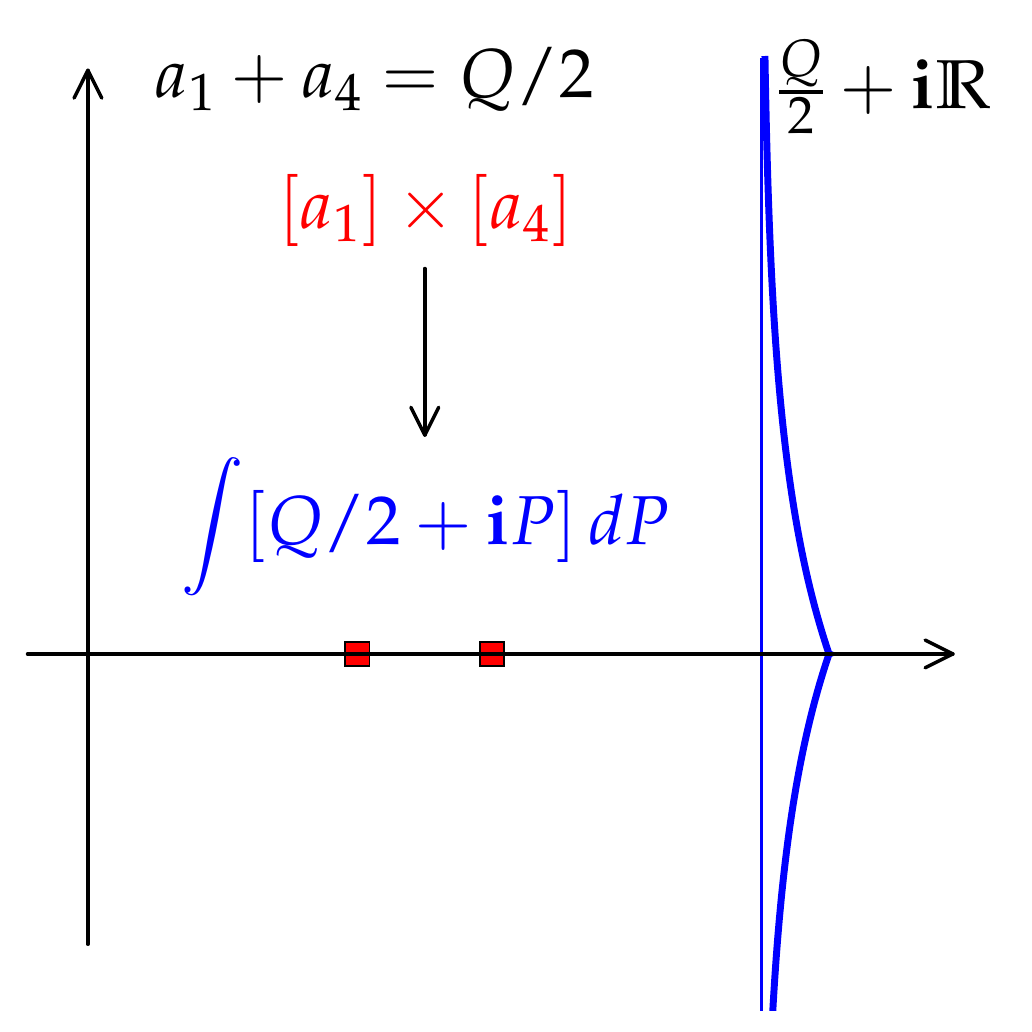}
\caption{Illustrations of the three cases of \gls{ope}. In each case the (upper-half) complex plane of charge values are drawn. Red squares indicate the charges $a_1$ and $a_4$. The charge(s) dominating the \gls{ope} is written in blue. The Blue dots indicate the positions of poles included in \eqref{intpdis}; the largest one dominates the \gls{ope}. The blue straight line is the \gls{lft} spectrum. The blue thick curves are cartoons of the value of the integrand in eq. \eqref{intcont}, highlighting their behaviour near $a = Q / 2$.}
\label{fig:fusionrule}
\end{figure}
\begin{itemize}
\item[(a)] $a_1+a_4< \frac{Q}{2}$ (pole crossing).
The smallest scaling dimension is given by the discrete term $a = a_1 + a_4 \in P_-$. So \eqref{cbexp} implies
\begin{align}
\label{disbehav}
 \left< \vertex_{a_1}(0)\vertex_{a_4}(z)\vertex_{a_2}(1)\vertex_{a_3}(\infty)\right> 
\underset{z\to 0}{\sim} |z|^{-2 \delta_0} \,,  \\ 
\delta_0 = \Delta_{a_1} + \Delta_{a_4} - \Delta_{a_1 + a_4} = 2 a_1 a_4 \,. \nonumber
\end{align}
\item[(b)] $a_1+a_4> \frac{Q}{2}$ (no pole crossing).
There are no discrete terms, and the smallest dimension is given by $a= Q / 2$ in the continuous integral. Now, recalling that the $Q/2$ belongs to the continuous spectrum $(Q/2 + \im \R)$, we need to consider the vicinity of $Q / 2$. Moreover, $C^{\text{DOZZ}}(a, a_1, a_4)$ and $C^{\text{DOZZ}}(Q-a, a_2, a_3)$ have a simple zero at $a = Q / 2$ (eq. \eqref{eq:DOZZzero}), so eq. \eqref{cbexp} and \eqref{intpdis} imply
\begin{align} 
&\left< \vertex_{a_1}(0)\vertex_{a_4}(z)\vertex_{a_2}(1)\vertex_{a_3}(\infty)\right> 
\underset{z\to 0}{\sim}  \int_{\mathbb{R}} |z|^{-2\delta_1 - 2P^2} P^2 \, \dif P  \,, \\ \sim & |z|^{-2\delta_1} \ln^{-\frac32}|1/z| \,,\,  \delta_1 = \Delta_{a_1} + \Delta_{a_4} - \Delta_{Q/2}  \,. \label{intbehav}
\end{align}
\item[(c)] $a_1+a_4 = \frac{Q}{2}$ (marginal case). 
This case is similar to the above one, except that $C^{DOZZ}(a, a_1, a_4)C^{DOZZ}(Q-a, a_2, a_3)$ does not vanish at $a = Q/2$, so we have 
\begin{equation}
\label{marginal}
 \left< \vertex_{a_1}(0)\vertex_{a_4}(z)\vertex_{a_2}(1)\vertex_{a_3}(\infty)\right> 
\underset{z\to 0}{\sim}  \int_{\R} |z|^{-2\delta_0 - 2P^2} \dif P  \sim |z|^{-2\delta_0} \ln^{-\frac12}|1/z|\,.
\end{equation}
This case can also be understood as the pair of dominant poles in case (a) merging at $Q / 2$ and compensating the double zero of case (b).
\end{itemize}
The results \eqref{disbehav}, \eqref{intbehav} and \eqref{marginal} can extend to general \gls{lft} $n$-point functions on \textit{any} closed surface (except for correlation functions of $n\leq3$ points on sphere like surfaces). In summary, the results of this section can be summarized in the following:
\begin{align}
&\langle \vertex_{a_1}(z) \vertex_{a_2}(0) \dots \rangle_b  
\stackrel{z\to 0}\sim \begin{dcases}
 \abs{z}^{-2\eta} \ln^{-\frac{3}{2}}|1/z|  & a_1 + a_2 > Q / 2 \,, \\
  \abs{z}^{-4a_1a_2} \ln^{-\frac{1}{2}}|1/z|  & a_1 + a_2 = Q / 2 \,, \\
 \abs{z}^{-4a_1a_2} & a_1 + a_2 < Q / 2 \,.
\end{dcases} \label{eq:OPE} \\
& \eta \defeq \Delta_{a_1} + \Delta_{a_2} - \Delta_{\frac{Q}{2}} \,, \label{eq:exponent}\, \Delta_a  = a(Q - a) \,.
\end{align}
Therefore, in \gls{lft}, the exponents of the asymptotic behaviour have a non-analytic dependence along the ling $a_1 + a_2 = Q/2$. This comes from the abrupt transition between the presence and absence of discrete terms. Moreover, the \textit{absence} discrete terms make the continuous integral dominate the \gls{ope} and lead to the log corrections. Their exponents $\frac{1}{2}$ and $\frac{3}{2}$ come from the vanishing order the \gls{dozz} structure constants at $Q/2$. For the Reader familiar with \gls{cft}, we stress that \textit{no} logarithmic \gls{cft} is involved here \cite{ribault2015liouville}.

\subsection{Application to logREMs} \label{sec:LFTtree}
Thanks to the mapping eq. \eqref{eq:maingeneral}, the asymptotic behaviour of \gls{lft} eq. \eqref{eq:OPE} and eq. \eqref{eq:exponent} translate easily into predictions concerning the \glspl{logrem} defined by a 2D \gls{gff} plus a logarithmic potential. More precisely, this is true in the $\beta < 1$ phase; in the $\beta > 1$ phase, the \gls{1rsb} results eq. \eqref{eq:freezingpos}, \eqref{eq:pbpblowT} and \eqref{eq:P2minpos} will be used in addition. 

\subsubsection{Near singularity behaviour}
As can be seen in Fig. \ref{fig:LFTnum0}, $\overline{p_\beta(z)}$ diverges as $z$ comes near a log singularity of the potential $U(z)$, say as $z\to 0$ where $U(z) \approx 4 a_1 \ln \abs{z}$. The asymptotic behaviour can be calculated by combining eq. \eqref{eq:mainhighT} (which is true in all temperature) and eq. \eqref{eq:OPE}, with $a_1, a_4 = a_1, b$ (note $b = \min(\beta, 1)$, eq. \eqref{eq:bandbeta}). The results read as:
\begin{align}
\overline{p_\beta(z)} \stackrel{z\to0}{\sim} \begin{cases}   
  \abs{z}^{-4 a_1 b}  & a_1+ b <  Q/2 \\
  \abs{z}^{2b^2 - 2} \ln^{-\frac{1}{2}}|1/z|  & a_1 + b = Q/2 \\
\abs{z}^{\frac{(Q - 2a_1)^2}{2}-2} \ln^{-\frac{3}{2}}|1/z|  & a_1 + b > Q/2 
 \end{cases} \label{eq:logsingularity} \,.
\end{align}
Of course, this asymptotic behaviour depends only on $a$ and the charge of the singularity $a_1$, and is independent of the other details of the \gls{logrem}, \textit{i.e.}, the surface and the global form of the potential. 

\subsubsection{Two thermal particles}
\begin{figure}
\center
\includegraphics[scale=.45]{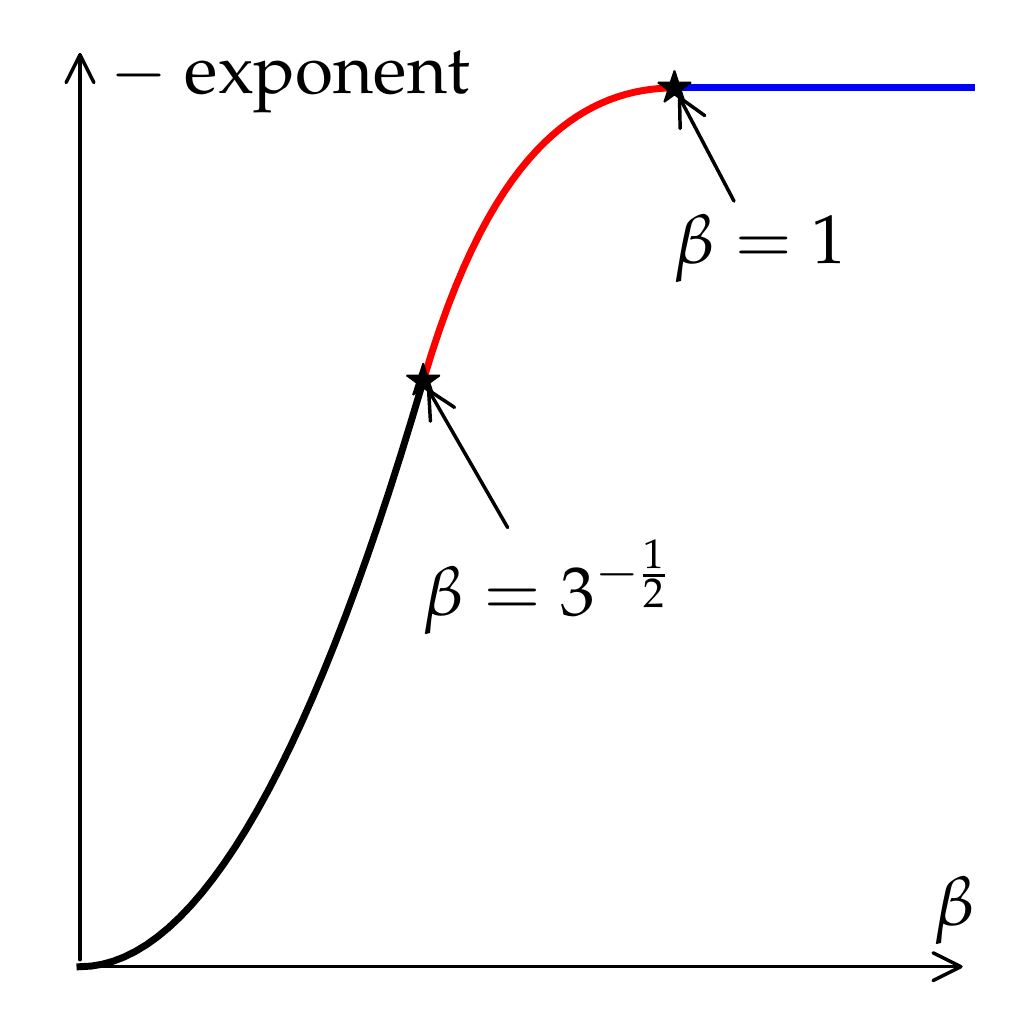}
\caption{A plot of the exponent of $\abs{z}$ in eq. \eqref{eq:pbpb}, as a function of the temperature.}
\label{fig:exponentsbeta}
\end{figure}

Let us now consider two independent thermal particles in a same random potential. This induces an effective attractive interaction between them, since they tend to fall in the same favourable regions of the potential. To characterize their attraction, we can consider the joint probability distribution of their positions, given by the two point correlation function of the Gibbs measure $\overline{p_\beta(w) p_\beta(w + z)}$, and focus on its asymptotic behaviour as $z \to 0$ (the corresponding full exact \gls{lft} correlation function is too complicated to actually compute, even using the conformal bootstrap solution). Provided $w$ is not a log--singularity of the potential $U(z)$, the asymptotic behaviour depends only on the temperature $\beta$,  but not on $U(z)$ or the surface, and is given by eq. \eqref{eq:OPE} with $a_1 =  a_2 = \beta$ in the $\beta < 1$ phase. In the $\beta > 1$ phase, we need to put $a_1 = a_2 = 1$ and apply also eq. \eqref{eq:pbpblowT}. The results read as follows:
\begin{equation}
\overline{p_\beta(w) p_\beta(z + w)} \stackrel{z\to0}{\sim}
\begin{dcases}
\abs{z}^{-4 \beta^2} &  \beta < 3^{-\frac{1}{2}}  \\
\abs{z}^{-4/3} \ln^{-\frac{1}{2}}\abs{1/z} & \beta = 3^{-\frac{1}{2}} \\
\abs{z}^{-3 + \frac{\beta^2 + \beta^{-2}}{2}} \ln^{-\frac{3}{2}}\abs{1/z} &  \beta \in (3^{-\frac{1}{2}},1] \\
 c' \beta^{-1}\abs{z}^{-2} \ln^{-\frac{3}{2}}\abs{1/z} + (1-\beta^{-1})\delta(z) & \beta > 1  \,.
\end{dcases} \label{eq:pbpb} 
\end{equation}
where $c'$ is an unknown constant (which will depend on other details of the \gls{logrem}). Note that the $\beta$ dependence of the exponent has two non-analyticities (see Figure \ref{fig:exponentsbeta}). The one at $\beta = 1$ comes from the freezing transition. The other one is inside the in the $\beta < 1$ phase, but is associated with the logarithmic corrections reminiscent of the freezing transition. Its value is also quite mysterious: we shall interpret in terms of multi--fractality in section \ref{sec:LFTmultifrac}.

We can also combine eq. \eqref{eq:pbpb} and eq. \eqref{eq:P2minpos} to obtain an asymptotic behaviour of the joint distribution $P(z_1, z_2)$ of the first and second minima positions:
\begin{equation}
P(z_1, z_2) \sim  \abs{z_1 - z_2}^{-2} \ln^{-\frac{3}{2}}\abs{1/(z_1 - z_2)}  \,. \label{eq:P2minasymp}
\end{equation} 
This holds for $1 \gg \abs{z_1 - z_2} \gg \epsilon$ ($\epsilon$ is the lattice spacing), while the $\delta$ in \eqref{eq:pbpb} takes over as $\abs{z}\sim \epsilon$). 

\subsubsection{Finite--size correction to overlap distribution}
The previous result eq. \eqref{eq:pbpb} can be applied to a translation invariant setting, \textit{i.e.}, a torus of size $R = 1$ (the thermodynamic limit is obtained by letting the lattice spacing $\epsilon \to 0$), and with $U(z) = 0$. Then, eq. \eqref{eq:pbpb} is related to the distribution of the (squared) distance $r^2 = \abs{z}^2$ between the two thermal particles. Denoting $P(r^2)$ the \gls{pdf}, we have $\dif P = P(r^2) \dif r^2 = \overline{p_\beta(w) p_\beta(r + w)} 2 \pi r \dif r$, so 
\begin{equation} P(r^2) = \pi \overline{p_\beta(w) p_\beta(r + w)}  \,, \label{eq:Pr2LFT} \end{equation}
 where $P(r^2)$ is the \gls{pdf} of $r^2$. 
 
 In \glspl{logrem}, there is another notion of the distance, the overlap $\overlap$. Its relation with the Euclidean distance is given by eq. \eqref{eq:overlapdis}. With $d = 2$ and $R = 1$, it implies  
 \begin{equation} \overlap =  - \ln (r^2) / t  \,,\,  t = \ln M = \ln (1/\epsilon^2) \,. \label{eq:overlapdis2d} \end{equation}
  Recall also that in the thermodynamic $t \to \infty$ limit, the limit distribution of $\overlap$ is known to be  (see eq. \eqref{eq:PoverlaplogREM})
 \begin{equation} 
 P(\overlap) \to \begin{cases} \delta(\overlap) & \beta < 1 \\ \beta^{-1} \delta(\overlap) + (1 - \beta^{-1})\delta(1 - \overlap) & \beta > 1 \end{cases}  \label{eq:PoverlaplogREM1}
  \end{equation}  
 
Now, equations eq. \eqref{eq:overlapdis2d} and eq. \eqref{eq:Pr2LFT} applied to \gls{lft} prediction eq. \eqref{eq:pbpb}, give us \textit{finite--size corrections} to eq. \eqref{eq:PoverlaplogREM1}:
 \begin{align}
  P(\overlap) &= \frac{\dif r^2}{\dif \overlap} \pi \overline{p_\beta(w) p_\beta(r + w)} = e^{\overlap t} \pi\overline{p_\beta(w) p_\beta(r + w)}  \nonumber \\
 & \sim 
 \begin{dcases}
 e^{(2 \beta^2 - 1)t \overlap} \, t &  \beta < 3^{-\frac{1}{2}}  \\
 e^{- \overlap t /3} \, \overlap^{-\frac{1}{2}}  t^{\frac{1}{2}} & \beta = 3^{-\frac{1}{2}} \\
 e^{- (\beta - \beta^{-1})^2 t\overlap / 4} \, t^{-\frac{1}{2}} \overlap^{-\frac{3}{2}} &  \beta \in (3^{-\frac{1}{2}},1) \\
 t^{-\frac{1}{2}}\overlap^{-\frac{3}{2}}  & \beta \geq 1, q \ll 1 \,.
 \end{dcases} \label{eq:overlapfinite}
 \end{align}
 In the last case $\beta > 1$, we do not know how to transform the Dirac peak $\delta(z)$ in eq. \eqref{eq:pbpb} to the variable $\overlap$; qualitatively, a probability mass of $1 - \beta^{-1}$ must be attributed to the regime $\overlap \sim 1$ in the $t \to \infty$ limit. So the formula given in eq. \eqref{eq:overlapfinite} can be only right for $\overlap \ll 1$. In the other cases, the expressions in eq. \eqref{eq:overlapfinite} are consistent with eq. \eqref{eq:PoverlaplogREM1} at $\overlap \sim 1$, so we believe that they are correct for all $\overlap$.
 
 Note that, while eq. \eqref{eq:pbpb} is a prediction for a \gls{logrem} in 2D, eq. \eqref{eq:overlapfinite} makes sense for \textit{any \glspl{logrem}}. Motivated by the universality of \glspl{logrem}, we conjecture that  eq. \eqref{eq:overlapfinite} holds for general \glspl{logrem}. We do not know yet a complete argument supporting this conjecture, say, using the replica approach. More precisely, we know how to recover the leading behaviours $e^{-c\overlap}$ in eq. \eqref{eq:overlapfinite}, using \gls{1rsb}, by generalizing the arguments of \cite{fyodorov2009pre}; the difficulty lies in recovering the log--corrections. 
 
 However, a confirmation of our conjecture comes from a recent result \cite{derrida16kppfinitesize}, which studies the same quantity on the directed polymer on the Cayley tree model. Their results  concern only the $\beta > 1$ phase and the critical point $\beta=1$, but is valid for all $\overlap \in (0,1)$. We quote them below (eq. 6 and 7, \textit{op. cit.}, with $\beta_c = 1$, $v(\beta) =- \beta - \beta^{-1}$, $v''(\beta_c) = 2$):
 \begin{equation}
 P(\overlap) \sim t^{-\frac{1}{2}}\overlap^{-\frac{3}{2}} \frac{1}{\beta\sqrt{4\pi}} (1- \overlap)^\eta \,,\, 
 \eta = \begin{cases}
 \frac12 & \beta = 1 \\ \frac32 & \beta > 1 \,. \label{eq:PoverlapDerr}
 \end{cases} 
 \end{equation}
 So it agrees with eq. \eqref{eq:overlapfinite} while giving more precision on the pre--factor.
 
 \subsubsection{General logREMs}
 Assuming the validity of eq. \eqref{eq:overlapfinite} for general \glspl{logrem}, we propose two applications, one on the Cayley tree, and the other on \glspl{logrem} of general dimension.
 
\textit{\Acrfull{dpct}}. For the \gls{dpct} model of branching number $\kappa$, recall that $t = \ln M =n \ln \kappa$, where $n$ is number of generations, and that the overlap is given by $\overlap = \hat{\overlap} /  n = \hat{\overlap} \ln \kappa / t$, see eq. \eqref{eq:defoverlap}. Applying this change of variable to eq. \eqref{eq:overlapfinite}, we obtain
\begin{equation}
P(\hat{\overlap})  \sim 
\begin{cases}
\kappa^{(2 \beta^2 - 1)\hat{\overlap}} \,,&  \beta < 3^{-\frac{1}{2}},  \\
\kappa^{- \hat{\overlap} /3} \hat{\overlap}^{-\frac{1}{2}}\,, & \beta = 3^{-\frac{1}{2}}, \\
\kappa^{- (\beta - \beta^{-1})^2 \hat{\overlap}/4} \, \hat{\overlap}^{-\frac{3}{2}} \,,&  \beta \in (3^{-\frac{1}{2}},1) \, \\
 \hat{\overlap}^{-\frac{3}{2}} \beta^{-1},  & \beta \geq 1 \,, \hat\overlap \ll n \,.
\end{cases}  \label{eq:overlapDPCT}
\end{equation}
Note that on the Cayley tree, $\hat{\overlap}$ is an integer, and eq. \eqref{eq:overlapDPCT} holds only for $\hat{\overlap} \ll 1$; indeed, $\hat{\overlap} \sim O(1)$ is the \gls{ir} limit of the model, and we cannot hope the asymptotic prediction of \gls{lft} to hold there. For the \acrfull{bbm} model, eq. \eqref{eq:overlapDPCT} holds with $\kappa = e$.

Observe that, nicely, changing the variable from $\overlap$ to $\hat{\overlap}$ absorbs the system--size dependence of eq. \eqref{eq:overlapfinite}, and provides the correct way of resolving the $\delta$--peak of eq. \eqref{eq:PoverlaplogREM1} at $\overlap = 0$. 

\textit{\Glspl{logrem} on $d$--dimensions.}-- Equations \eqref{eq:overlapdis2d} and \eqref{eq:Pr2LFT} have analogues in $d$--dimension, obtained by replacing $r^2$ by $r_d^d$. Applying them to eq. \eqref{eq:overlapfinite}, we obtain:
 \begin{equation}
 \overline{p_\beta(0) p_\beta(\mathbf{x})}^d \stackrel{\abs{\mathbf{x}}\to0}{\sim}
 \begin{dcases}
 \abs{\mathbf{x}}^{-2d \beta^2} &  \beta < 3^{-\frac{1}{2}}  \\
 \abs{\mathbf{x}}^{-2d/3} \ln^{-\frac{1}{2}}\abs{1/\mathbf{x}} & \beta = 3^{-\frac{1}{2}} \\
 \abs{\mathbf{x}}^{-3d/2 + \frac{(\beta^2 + \beta^{-2})d}{4}} \ln^{-\frac{3}{2}}\abs{1/\mathbf{x}} &  \beta \in (3^{-\frac{1}{2}},1] \\
  c' T \abs{\mathbf{x}}^{-d} \ln^{-\frac{3}{2}}\abs{1/\mathbf{x}} + (1-T) \delta(\mathbf{x}) & \beta > 1 \,.
 \end{dcases} \label{eq:pbpb1}
 \end{equation}
 In particular, these results apply to 1D \glspl{logrem} such as the circular model. Unfortunately, even for the latter model, the Coulomb gas integral needed to check it using the replica trick is not exactly solvable. However, it is hopeful to use results in \cite{fyodorov2015moments} to provide an analytical check.

\subsection{Multi--fractality revisited} \label{sec:LFTmultifrac}
 We observed that eq. \eqref{eq:pbpb} has a non--analyticity at $\beta = 1/\sqrt3$. What transition does it correspond to? In \cite{cao16liouville}, we claimed that the answer is the termination point transition, discussed in section \ref{sec:multifracintro}. Here we will explain this claim, and give an argument for the last equation in the main text of \cite{cao16liouville}. Throughout this section, we will assume $\beta = b < 1$.  
 
 Recall that if $p_{\beta,j},j=1,\dots,M$ is the Gibbs measure of a discrete \gls{logrem} of size $M$, the \acrfullpl{ipr} in the \textit{annealed ensemble} is given by eq. \eqref{eq:IPRann}:
 \begin{equation}
\overline{P_q} = M \overline{p_{\beta,j}^q} \sim M^{-\tau_q} \,,\, \tau_q = \begin{dcases}
q \beta \left( Q - q \beta \right) - 1 \,,   &  q < Q/(2\beta) \,, \\
Q^2 / 4  - 1 \,, & q \geq Q/(2\beta) \,,
\end{dcases} \label{eq:IPRann2}
 \end{equation}
 where in the first equation, we assume that the \gls{logrem} is homogeneous.  Our goal is to give arguments for the \textit{logarithmic correction} to the leading behaviour, announced in \cite{cao16liouville}: 
 \begin{align}  \overline{P_q} \stackrel{\beta<1}\sim  \begin{cases}  
  M^{-\tau_q} & q \beta  < \frac{Q}{2}  \\
 M^{-\tau_q} \ln^{\frac12} M  & q \beta  =  \frac{Q}{2} \\
 M^{-\tau_q} \ln^{\frac32} M  & q \beta  > \frac{Q}{2}  \\
 \end{cases} \,. \label{eq:Pq}
  \end{align}
We will use \gls{lft} to argue for this in the 2D context; extension to general \glspl{logrem} can be argued again based on the their universality. Before proceeding, we note that the analogue of eq. \eqref{eq:Pq} for the uncorrelated \gls{rem} is known \cite{fyodorov2009pre}. The exponents $\tau_q$ is the same, but the corrections are different: there is only a $\frac{1}{2}\ln\ln M$ correction when $q \beta > Q/2$. For the \glspl{logrem}, the logarithmic corrections in the {\it typical} ensemble (see section \ref{sec:multifracintro}) were calculated much earlier in Sec. VI.B in \cite{carpentier2001glass}, without using \gls{lft}. 

When $q \beta < Q / 2$, the vertex operator $\vertex_{q\beta}$ satisfies the Seiberg bound eq. \eqref{eq:seibeig12}, and can be used to describe the power of the Gibbs measure $p_{\beta,j}^q$. Since the latter is on the discrete lattice, the corresponding operator is the \textit{bare} one, and its one point function is (in general \glspl{cft}) given by 
\begin{equation} \label{eq:vertexbare} \left<\vertex_{a}\right>_b \sim M^{-\Delta_a}  \,,\end{equation}  where $\Delta_a = a(Q -a)$ (eq. \eqref{eq:LFTDeltaLFT}) is the conformal dimension. Therefore, 
  \begin{equation} 
  \overline{P_q}= M \overline{p_{j,\beta}} = M \left<\vertex_{q \beta}\right>_b \sim M^{-\tau_q} \,,\, \tau_q  = \Delta_{q\beta}-1 \,,
  \end{equation}  
giving the first case. Note that the matching  between $\Delta_a$ and the \gls{ipr} exponent in \gls{logrem} was observed at least as early as from the pioneering work \cite{kogan96prelocalised}.
 
When $q \beta \geq Q/2$, $\vertex_{q\beta}$ can no longer represent $p_{\beta,j}^q$. To proceed, we split $q = q_1 + q_2 + \dots + q_n$ such that $q_i \beta < Q/2$, and represent $p_{\beta,j}^q$ by a product of vertex operators $ \vertex_{q_1\beta} \dots \vertex_{q_n\beta}$, evaluated at close but not identical points. Then we apply the \gls{lft} fusion rules (\cite{ribault2014conformal}, Exercise 3.3) to them. Let us explain with a case where $n = 2$ suffices. The relevant \gls{lft} fusion rule reads (see also Figure \ref{fig:fusionrule}): 
  \begin{equation} \label{eq:fusion}
  \vertex_{a_1} \vertex_{a_2} \sim \begin{dcases}
   \vertex_{a_1 + a_2} + \dots   &  a_1 + a_2 < Q / 2 \,, \\
  \int \vertex_{\frac{Q}2 + \im P} \dif P  & a_1 + a_2 =Q / 2\,, \\ 
  \int \vertex_{\frac{Q}2 + \im P} P^2 \dif P  & a_1 + a_2 > Q / 2 \,, \\
  \end{dcases}  a_1, a_2 \in (0, Q/2) \,. 
  \end{equation}
Both integrals are performed on $(-\epsilon, \epsilon)$, a small interval around $P = 0$. Combined with eq. \eqref{eq:vertexbare}, we have, when $q \beta > Q / 2$,
\begin{align}
& \overline{P_q} = M \overline{p_{j,a}} \sim M \left<\vertex_{q_1\beta} \vertex_{q_2\beta}\right> \sim M \int \left< \vertex_{\frac{Q}2 + \im P} P^2 \dif P \right> \nonumber \\ \sim  &  M \int M^{-\Delta_{\frac{Q}2 + \im P}}  P^2\dif P =  M^{1 - Q^2/4 } \int M^{-P^2} P^2\dif P 
=  M^{1-Q^2/4} \ln^{-\frac{3}{2}} M  \,, \label{eq:Pq32}
\end{align}
which is the third case of eq. \eqref{eq:Pq}. The second case is completely similar: the $\ln^{-\frac{1}{2}} M$ correction comes from $ \int M^{-P^2} \dif P$. Remark also that if $a_1 + a_2 < Q / 2$, the first fusion rule would lead to no logarithmic correction, in agreement with eq. \eqref{eq:Pq}. Note that the end result does not depend on how we split $qa$, because of eq. \eqref{eq:fusion}. For general $n > 2$, the above procedure can be repeated (by using other fusion rules of \textit{op. cit.}); we can still show that the end result is eq. \eqref{eq:Pq}, regardless of how we split $qa$.

 \begin{table}
 \center
 \begin{tabular}{c|c|c|c}
 \hline
  $a_1 + a_2$ & $\vertex_{a_1} \times \vertex_{a_2}$ & $\left\langle \vertex_{\alpha_1}(z) \vertex_{\alpha_2}(z'\to z) \times \dots \right\rangle_{b}$ & exponent $\delta$ \\ \hline 
  $<\frac{Q}{2}$ & $\vertex_{a_1 + a_2} + \dots$ & $\abs{z-z'}^{-2 \delta }$ & $\Delta_{a_1 + a_2}-\Delta_{a_1} - \Delta_{a_2}$  \\
  $= \frac{Q}{2}$ & $ \int \vertex_{\frac{Q}{2} + \im P} \dif P$ &$\abs{z-z'}^{-2\delta} \ln^{-\frac{1}{2}}(1/\abs{z-z'})$ & $\Delta_{\frac{Q}2}-\Delta_{a_1} - \Delta_{a_2}$ \\
  $> \frac{Q}{2}$ & $ \int \vertex_{\frac{Q}{2} + \im P} P^2 \dif P $ & $ \abs{z-z'}^{-2\delta} \ln^{-\frac{3}{2}}(1/\abs{z-z'}) $ & $\Delta_{\frac{Q}2}-\Delta_{a_1} - \Delta_{a_2} $\\ \hline
 \end{tabular}
 \caption{Comparing the fusion rules eq. \eqref{eq:fusion} and the \gls{ope}, eq. \eqref{eq:OPE}. $\Delta_a = a (Q-1)$ is the conformal dimension of the operator $\vertex_a$. In all cases, it is assumed $\Re(a_1), \Re(a_2) \in (0, Q / 2).$} \label{table:fusionvsope}
 \end{table}
Although we will not formally define the notion of fusion and discuss its relation with operator product expansion in section \ref{sec:OPE}, let us compare eq. \eqref{eq:fusion} to eq. \eqref{disbehav}, \eqref{marginal} and \eqref{intbehav} (see also Table \ref{table:fusionvsope}). The vertex operators appearing at the right hand side of \eqref{eq:fusion} have the charge $a$ in the respective cases. The exponents in the \gls{ope} formulas are given by the difference of scaling dimensions in the way shown in Table \ref{table:fusionvsope}. This is standard in any critical field theory. Finally, the integrals, when they exist, are all over the \gls{lft} spectrum (eq. \eqref{LFTspect}), near $Q/2$ where the conformal dimension is the smallest; the ``spectral density'' ($\dif P$ and $P^2 \dif P$ respectively) coming from the \gls{dozz} formula is also present. In light of this, the transition in eq. \eqref{eq:pbpb} at $\beta = 1/\sqrt{3}$ corresponds to the termination point transition of $q=2$. In hindsight,  this interpretation is natural, because $\overline{p_\beta(z) p_\beta(z')}$, when $z' \to z$, is approximately $\overline{p^2_\beta(z)}$ in the large scale (compared to $\abs{z-z'}$).

\begin{figure}
\center 
\includegraphics[scale=.6]{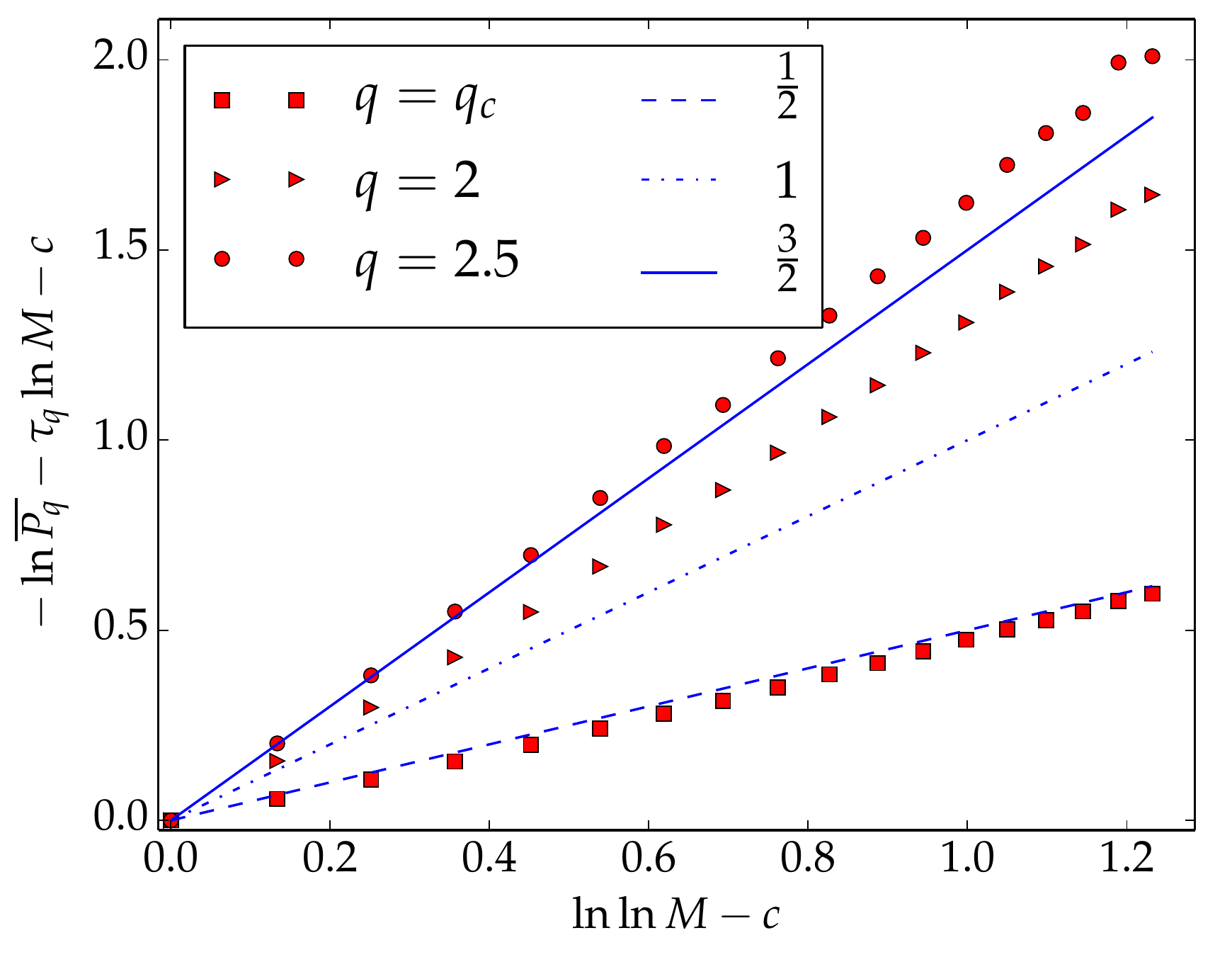}
\caption{Numerical measure the annealed \gls{ipr} with $\beta = .75$, with $q=q_c = (Q/2\beta), 2$ and $q = 2.5$, in the circular model, with $M = 2^7, \dots, 2^{24}$. We take the logarithm, remove the leading order in eq. \eqref{eq:Pq} and compare the resulting correction to $\frac32\ln\ln M$, $\ln\ln M$, $ \frac12 \ln \ln M$ (which are the slopes of the straight lines through origin). For better comparison, we translate each data set so that its first point is at the origin. } \label{fig:iprcir}
\end{figure}

We note that the multi--fractal exponents can be obtained in the \gls{1rsb} framework; such an analysis was carried out in \cite{fyodorov2009pre}. The same method can apply to find the \textit{leading exponents} of the asymptotic behaviours we predicted in section \ref{sec:LFTtree}. So, the novelty of the \gls{lft} mapping is essentially the log--corrections. As we have seen in section \ref{sec:RSB}, predicting log--corrections is a major difficulty of the \gls{rsb} approach. On the other hand, the derivation of this section is also very heuristic. To further support the claim eq. \eqref{eq:Pq}, we have performed numerical measures of eq. \eqref{eq:Pq} on the circular model, which is a \textit{1D} \gls{logrem}. So strictly speaking, the derivation of this section does not directly apply. However, our preliminary result, as shown in Figure \ref{fig:iprcir}, gives encouraging support to eq. \eqref{eq:Pq} as a \textit{general} \gls{logrem} prediction.

To conclude this section, we emphasize that the above discussion is limited to the $\beta < 1$ phase. Indeed, it is clear that eq. \eqref{eq:Pq} cannot hold in the $\beta > 1$ phase, since for $q = 1$, we would be in the termination point phase, but the log--correction cannot be present because $\overline{P_1} = 1$ by definition. 

\section{Summary and Perspectives}\label{sec:finallogrem}
In this chapter, we reviewed the historical developments that led to the field of \acrlongpl{logrem} (section \ref{sec:logremintro}): the \gls{rem}, hierarchical models (\gls{bbm}/\gls{dpct}), and the Euclidean space ones. They are known to have common thermodynamic properties, in particular the freezing transition. Moreover, in the frozen phase the sub--leading behaviour and fluctuation distribution of the free energy enjoy universal features, known as the freezing scenario (eq. \eqref{eq:GbetaGFF}). This is the genesis of the \textit{universality class} of \glspl{logrem}. 

In section \ref{sec:logREMwhat}, we characterized quantitatively the common feature (``log--decaying correlation'') shared by the seemingly different members in the \gls{logrem} class, and proposed the notion of \gls{ir} and \gls{uv} data to organize the different models of the \gls{logrem} class (more precisely, those defined on the Euclidean spaces). 

As we showed in \ref{sec:RSB}, thanks to \textit{ultrametricity} (eq. \eqref{eq:ultrametcity}),  \glspl{logrem} and their multi--scale generalizations can be analysed by the \acrfull{rsb}, although they are finite--dimensional systems. The result of the analysis, specified to \glspl{logrem}, is that they can be solved by the \acrfull{1rsb} Ansatz very similar to that of the \gls{rem}, despite important differences. We then implemented a \gls{1rsb} formalism on Euclidean \glspl{logrem} (using the circular model as example) such that the model--dependent \gls{ir} and \gls{uv} data are taken into account. This is done for the free energy/minimum distribution (section \ref{sec:freezingRSB}), leading to a (partial) understanding of the freezing scenario.

With more work, the approach is extended to (finite--temperature extensions of) higher extreme order statistics and the full minima process (section \ref{sec:orderstat}), which describes the minima positions and values in terms of a decorated Poisson point process, in which the minima form clusters that occupy lattice--spacing scale regions and are separated by system--size scale distances. Intuitively, the \gls{uv} data determines the \textit{decoration process}, \textit{i.e.}, the structure of the minima in a single cluster. As a consequence of the general structure of decorated Poisson point process, the \gls{uv} data determines the minima process ``seen from the minimum'' (\textit{i.e.}, the gaps). On the other hand, the \gls{ir} data determines the minimum/free energy distribution and that of the  positions of the clusters. Within this big picture, two types of questions can be posed for finer characterisation (even exact solution) of specific models: the \gls{ir} ones and the \gls{uv} ones. 

Technically speaking, the \gls{ir} questions can be reduced to the solution and analytical continuation of Coulomb gas integrals, the simplest being the Dyson integral (eq. \eqref{eq:Dyson}) used for the circular model. In section \ref{sec:Jack}, we see how Jack polynomials appear as natural and useful tool to tackle (1D) cases where Dyson--type closed form formulas are not available. Indeed, Jack polynomials are the suitable  basis to expand symmetric functions arising in the Coulomb gas integrals, for two reasons. First, because the resulting terms can be then integrated and analytically continued term by term (and some infinite sums can be calculated efficiently using the method of \cite{cao15gff}). Second, the terms generated enjoy nice duality properties (see for example eq. \eqref{eq:Sdual}), suggesting their individual physical interpretation as \textit{freezing} quantities (in view of the freezing--duality conjecture, see around eq. \eqref{eq:freezedual}). We reported the first step of this project, giving rise to the Edwards--Anderson order parameter (section \ref{sec:EA}), whose (linear) glassy  phase behaviour can be simply understood by \gls{1rsb} (in fact, by the combinatorics of replica grouping) in general. It is interesting to carry out more steps to see the general pattern with the perspective of giving a replica--combinatorics understanding of the Jack polynomials. Another exciting direction is to use Jack polynomials to study \glspl{logrem} defined on closed curves, \textit{i.e.}, geometric deformations of the circular model, of which the Dirichlet circular model (section \ref{sec:Dirichlet}) is the first explicit example. Such an endeavour will probably lead to a general understanding of the origin of duality.  

The other analytical tool to study the \gls{ir} question is the \acrfull{lft}. In a sense that was clearly present in the pioneering works \cite{zamolodchikov1996conformal,goulian1991correlation,dorn1994two,dotsenko1984conformal}, the solution of \gls{lft} can be reduced to the analytical continuation of 2D Coulomb gas integrals: indeed, every \gls{lft} correlation function can be seen as a 2D Coulomb gas integral, \textit{with a non--integral number of charges}. In this respect, our contribution has been to clearly identify the  (continued) Coulomb gas integrals with their \gls{logrem} interpretation. For this, one should realize that \gls{lft} describes \textit{only} the Gibbs measure of \glspl{logrem}  (this was known by the authors of \cite{kogan96prelocalised,carpentier2001glass}), and take care of the coupling to curvature/infinity boundary condition (this is our contribution). 

The future directions that can emerge from this work are as numerous as the ramifications and connections of \gls{lft} itself. To start with, one may extend the mapping to boundary \glspl{lft} and connect them to 1D \glspl{logrem} like the circular model. A more ambitious project could concern \glspl{logrem} with \textit{imaginary temperature} (defined by $Z = \int e^{-\im \beta \phi}$). A particle in a energy landscape with complex temperature can be seen as the \gls{rem} for modelling quantum interference \cite{dobrinevski11complex}. In the log--correlated case, the basic phase diagram was also established \cite{lacoin15complex}, yet the mapping to some $c \leq 1$ \gls{cft} is not yet found. This task can be considerably harder than the one undertaken here, because the recently constructed $c\leq 1$--\gls{lft} \cite{ribault2015liouville} is not the only candidate \gls{cft} (there are also the minimal models \cite{belavin19843cft}!). However, the corresponding Coulomb gas integrals are the usual (repulsive) ones that are used in $\beta$--random matrix theory \cite{dumitriua2002matrix} and in fractional quantum Hall effect \cite{can2014fractional}, so we may expect more exciting connections. 

However, the most exciting aspect of the \gls{lft}--\gls{logrem} mapping is the ability of the former to predict the termination point transition \textit{and the log--corrections}. In other words, it has over--performed the \gls{rsb} approach, which is designed to study transitions, but has difficulty in giving sub--leading corrections. This suggests that \gls{lft} may be far from being merely a solution to an \gls{ir} problem of some 2D \glspl{logrem}. Some structures of the \gls{lft} may be relevant for the \gls{logrem} \textit{universality class}. A subtle hint supporting this point of view is the striking similarity between the discrete/continuous dichotomy in \gls{lft} which governs the termination point transition (see Fig. \ref{fig:fusionrule}) and the one that corresponds to the \textit{freezing} transition of the \gls{rem} (see Fig. \ref{fig:REM}). To the extent indicated by eq. \eqref{eq:rhoxtpertube} of the \gls{kpp}--\gls{bbm} analysis, the same can be said for the freezing of \glspl{logrem}. Moreover, the argument leading to the $\frac32$ log--correction in the \gls{kpp} equation (eq. \eqref{eq:threehalf}) is also an integral of type $\int_{\R} M^{-P^2} P^2 \dif P$, like the one in eq. \eqref{intbehav} and \eqref{eq:Pq32}. Could \gls{lft} be relevant for the \gls{logrem} freezing transition and describe the associated universal behaviours?

Note that this is \textit{not} a new question, but goes back at least to \cite{carpentier2001glass}, which noticed and investigated the relation between the $c = 25$ ($b = 1$, eq. \eqref{eq:centralcharge}) ``barrier'' and the freezing transition. Reopening the question with the new results obtained, there are still serious obstacles, which can be summarized as: the complete mapping is essentially limited to the \textit{Gibbs measure} of \textit{2D} \glspl{logrem} in the \textit{high--temperature phase}. Indeed, even in 2D, if we extended the mapping beyond Gibbs measure to include free energy distribution, the corresponding field theory would be \eqref{eq:Sliouville} with $Q \neq b + b^{-1}$: the conformal invariance will be lost, and the massive theories have \textit{non--conformal} response to geometry that is a subject of current research, see \textit{e.g.}, \cite{can2014fractional,can2015geometry,wiegmann16cft,laskin2015collective}. \textit{A fortiori}, it is completely unclear which properties of \gls{lft} will be preserved in the (missing) field theory description of \textit{general} \gls{logrem} freezing transition. This being said, remark that \gls{lft} is a key ingredient of the continuous theories of 2D random surfaces with statistical models defined on them (Liouville quantum gravity). We know now that these \textit{discrete 2D gravity models} can be described either by the graphical expansion of matrix models \cite{gross902DQG,zamolodchikov2007lectures}, or purely combinatorially \cite{schaeffer1997bijective,bouttier2004planar,legall2013}. To simplify enormously, the main object of the latter approach is essentially a \gls{dpct} model defined on a \textit{random} tree (note that the relation between trees and Liouville quantum gravity has been further unveiled in \cite{duplantier2014liouville}). Therefore, it cannot be excluded that \gls{lft} could be relevant for statistical models beyond 2D \glspl{logrem}.

So far we have not evoked the \gls{uv} questions. They are  much less studied, because the lack of analytical/integrability methods in this \textit{discrete} context. One may also question the importance of such questions, other than, say, calculating exactly the gap distribution of the \gls{bbm} model or the circular model, which is dependent on the \gls{uv} data. We believe that, regardless of the solvability of specific cases, the understanding of \glspl{logrem} from the \gls{uv} point of view is not sufficient and may be an obstacle to the full understanding of \glspl{logrem} from the \gls{rsb} perspective. Indeed, Derrida and Mottishaw's \cite{derrida16kppfinitesize} finite--size correction result on the overlap distribution, eq. \eqref{eq:PoverlapDerr} suggests strongly that the freezing transition is of \gls{uv}--nature, because it is the exponent at $\overlap = 1$ that changes at $\beta = 1$.

Last but not lease, there is an important and rapidly growing application of \glspl{logrem} where the discrete structure is inherent: the characteristic polynomials $\chi$ of random matrices and the closely related Riemann zeta function  on the critical line $\zeta(\frac12 + \im t)$ (for an introduction, see \cite{Bourgade2013}). Indeed, the logarithm of $\abs{\chi}$ is shown to be asymptotically Gaussian and log--correlated \cite{Keating2000,keating03rev}, making predictions on \glspl{logrem} are relevant \cite{FHK12,FyoKeat14,FyoSim15} for these central objects of mathematics and mathematical physics, and driving a new stream of activity \cite{ABB2015,arguin2015maxima,PaqZeit2016,ChMadNaj2016,lambert2016subcritical} (note that there is also a random polynomial approach to random geometry in 2D and beyond, see \textit{e.g.}, \cite{zelditch12bergman,Ferrari2011QG}). In these applications, the discreteness between eigenvalues or $\zeta$ zeros is naturally defined, and for random matrices, the level spacing scale is more important because it corresponds to the long time scale in quantum mechanics. However, these functions are not \textit{a priori} Gaussian in these scales, so the application of current framework will require considerable future work.

\chapter{Anderson localization with long-range hopping}\label{ch:anderson}
\section{Localization with long-range hopping} \label{sec:loc-intro}
The main subject of this chapter is the \textit{localization of eigenvectors of random matrices}. This has become a classic and vast topic, with wide applications in both classical and quantum physics. Although the most famous one is the Anderson localization \cite{anderson1958absence} which concerns the quantum transport of electrons (and matter waves in general \cite{lamarie08anderson}) in disordered media, the earliest one may be Dyson's model of random elastic network \cite{dyson53elastic}. A recent rapidly developing application of the problem of eigenvector localization in quantum physics is the Many--Body Localization \cite{Basko2006mbl,huse15mblreview}, from a Fock--space localization point of view \cite{altshuler97mbl,deluca14bethe,altshuler2016rrg}. For recent applications in classical physics, one may cite the numerous investigations on the vibration modes in disordered systems; they can be hard/soft sphere models of glass forming liquids \cite{manning11softspots,manning15boson,schirmacher98bosonpeak,charbonneau16debye,Berthier16boson} and electron glasses \cite{amir10glass,amir13localisation}. In the latter case, the ``sparse'' long--range random matrix studied in \textit{op. cit.} is quite close in flavour to the ones that we will study in this chapter. 

From a former point of view, the object of study is always a large random matrix $H_{nm}, n, m = 1, \dots, N$, drawn from some simple statistical distribution. We will always assume that $H_{nm}$ is real symmetric, $H_{nm} = H_{mn} \in \R$, and view it as a quantum Hamiltonian $\hat{H} = \sum_{nm} H_{nm} \vert n \rangle \langle m \vert$ acting on the Hilbert space spanned by the basis states $\vert n \rangle, n = 1, \dots, N$ (throughout this chapter, we will use interchangeably the Dirac bracket notation and the usual linear algebra one). Then we focus on the eigenstates (eigenvectors) of $H$, 
\begin{equation}
\hat{H} \vert \phi_i \rangle = E_i \vert \phi_i \rangle \,,\, E_1 < \dots < E_N
\end{equation}
being the spectrum of energies (eigenvalues). The eigenstates will be normalized by default: $\sum_n \abs{\langle n \vert \phi_i \rangle^2} = 1$.  Crudely speaking, the eigenstate $\phi = \phi_i$ is \textit{localized} if $\abs{\langle n \vert \phi \rangle^2}$ is vanishingly small for all but a few sites $n$. On the contrary, it is \textit{extended} (delocalized) if the coefficients' magnitudes are uniform $\abs{\langle n \vert \phi \rangle^2} \sim 1/N$. A most important phenomenon is the \textit{localization transition}, \textit{i.e.}, a sharp change from one behaviour to the other as one varies smoothly the eigenvalue $E$ or other parameters of the random matrix.

\subsection{Anderson model}\label{sec:Anderson}
The Anderson model describes in general the wave mechanics in the presence of disorder. Its discrete versions is defined on any graph with vertices labelled by $n = 1, \dots, N$, in terms of the following $N \times N$ random matrix (Hamiltonian) 
\begin{equation} \label{eq:Anderson}
\hat{H} = \sum_{n=1}^N V_n \vert n  \rangle \langle n \vert  + \sum_{<nm>}  t (\vert n \rangle \langle m \vert  + \vert m \rangle \langle n \vert)\,,
\end{equation}
where the diagonal elements $V_n, n = 1, \dots, N$ of the matrix are \gls{iid} random variables, usually drawn from a uniform distribution $[-W, W]$. Clearly, the eigenvectors of $\hat{H}$ depends only on the ratio $\beta = W / t$ between the disorder width and the hopping amplitude.
Usually, Anderson model is considered on a $d$-dimensional periodic lattice, say, a square lattice (although recent interest in this model is focused on tree--like lattices \cite{biroli2012difference,deluca14bethe,altshuler2016rrg}). 

When $\beta = 0$ ($W = 0$), eq. \eqref{eq:Anderson} reduces to the tight-binding model of basic solid state physics. The eigenvectors are plane waves, thus are all extended. On the other hand, when $ \beta = +\infty$ ($t = 0$), eq. \eqref{eq:Anderson} becomes diagonal, so any eigenvector is localized on a single site: $ \hat{H} \vert n \rangle = V_n \vert n \rangle.$ What happens in between? Is there a sharp transition at some critical $\beta_c$ (when $N \to \infty$)?

\begin{figure}
\center  \includegraphics[scale=.6]{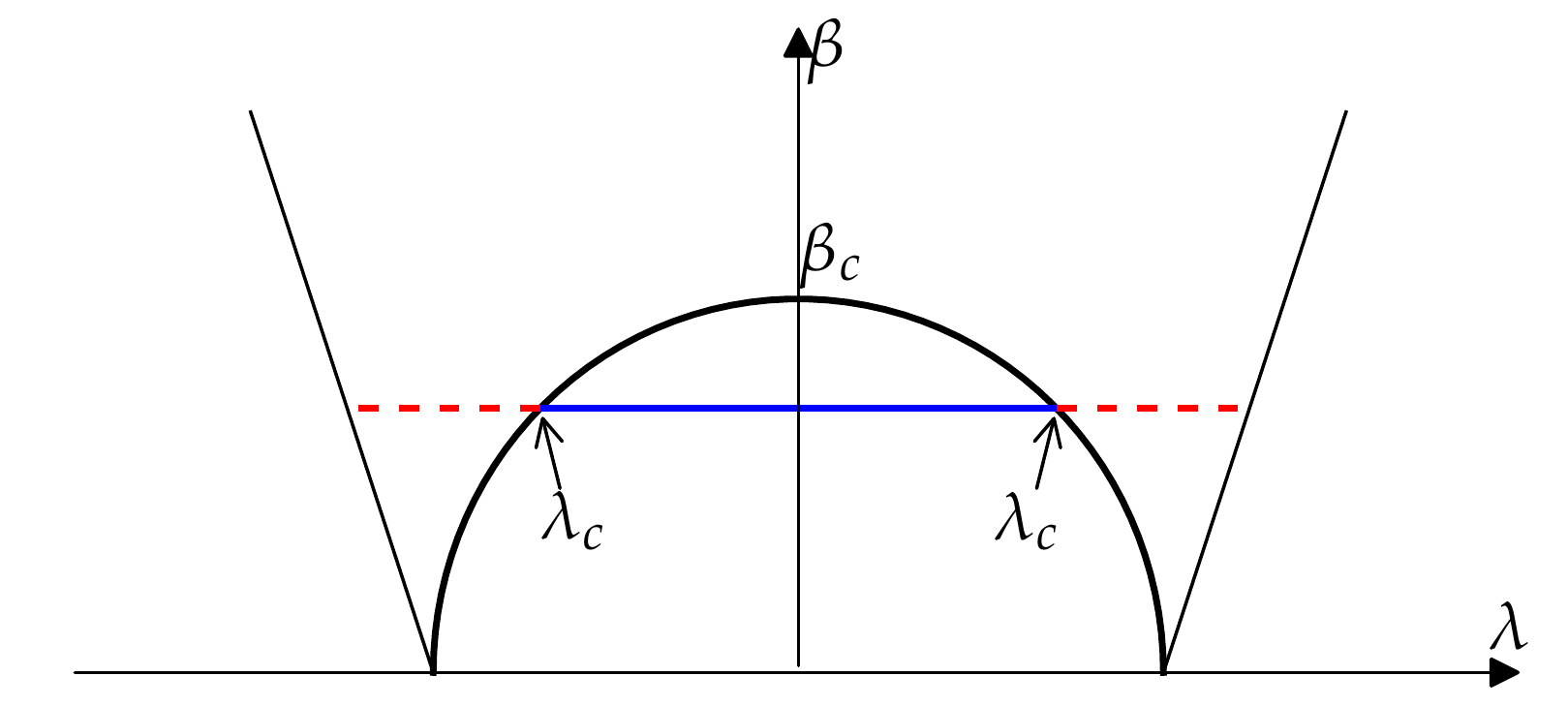}
\caption{A cartoon of the phase diagram of the Anderson model in $d \geq 3$ dimensions. The half--circle like curve separates the localized states (outside) and extended ones (inside). A matrix with mobility edge is represented by the dashed (localized state) and continuous (extended) horizontal lines. The other two straight lines indicate the limits of the spectra. } \label{fig:andersonphase}
\end{figure}

The complete answer to this question in general dimension took quite a long time \cite{abraham79scaling}. In 1d and 2d, for any finite $\beta > 0$, \textit{all} the eigenvectors are localized. For $d \geq 3$, the phase diagram is non-trivial. There exists a $\beta_c > 0$ such that, for $\beta > \beta_c$, all the eigenstates are localized. When $0 < \beta < \beta_c$, the spectrum is divided: there are two \textit{mobility edges} $\pm E_{c}$ (depending on $\beta$), such that eigenstate with $ -E_c < E < E_c$ are extended, while the others are localized. A good summary of the phase diagram is Figure \ref{fig:andersonphase}.  

There are thus two ways of probing the transition: fixing the energy (usually, at $E = 0$, in the middle of the spectrum), and varying $\beta$; or fixing $\beta$, and varying the energy, \textit{i.e.}, crossing the mobility edges. As far as critical behaviours are concerned, the two ways are equivalent; in theoretical and numerical studies, the first way may be preferred. However, what makes Anderson model truly remarkable is the \textit{existence} of mobility edges in \textit{one} matrix. Indeed, if we consider a model of free fermions whose one-particle Hamiltonian has mobility edges, then tuning the chemical potential/Fermi level across a mobility edge can induce a metal--insulator transition. In this work, we shall distinguish localization transitions in the sense of mobility edge as \textit{genuine} localization transitions.

The fact that there is a localization transition only in $d > 2$ is a major difficulty for its analytical and numerical study. This motivated theorists to consider 1D Anderson models with \textit{long--range hopping}. As is common in statistical physics, one hopes to mimic short--range models in higher dimensions by long--range ones in 1D.  

\subsection{Power-law banded random matrices}\label{sec:pbrm}
 The best--known such model is the \gls{pbrm} ensemble \cite{mirlin96pbrm,mirlin00pbrm}. Its elements $H_{nm}$ are independent \footnote{In this chapter, the random matrices we considered are all real symmetric. So when we say that their elements are independent, we mean that the set of variables $\set{H_{mn}, m \leq n}$ are independent, while the other half of the matrix is fixed by symmetry.} centred Gaussian random variables with a variance that decays algebraically with the distance (our notation is related to that in \cite{mirlin96pbrm} by $\mu = \sigma = 2 \alpha - 1 > 0$ and $\beta = b^{-\mu-1} > 0$):
 \begin{eqnarray}
 \overline{H_{nm}^2}^c = \frac{1}{1 + g_{nm}}\, ;\quad g_{nm} = \beta \abs{n-m}^{1+\mu} \,,\, \label{eq:pbrmasymp}  
 \end{eqnarray}
 where $\beta, \mu > 0$ are the parameters of the model \footnote{In this chapter, $\abs{n-m}$ denotes the distance between two lattice points in 1D; in numerical simulations, periodic boundary condition is always assumed and the distance is suitably modified.}. $\mu$ controls how long--range is the model, so $1/\mu$ plays a rôle of effective dimension. $\beta$ controls the ratio between diagonal elements and hopping ones, so it is the parameter of disorder strength.
 
The phase diagram of PBRM was established in \cite{mirlin96pbrm}, and further confirmed numerically in \cite{cuevas2001anomalously}, and re--confirmed recently \cite{quito2016anderson} using the Wegner flow approach: when $\mu < 1$ ($\mu > 1$), all the eigenstates are extended (localized, respectively). We will give a simple argument to (partially) support this statement below. Remarkably, when $\mu = 1$, there is a family of critical models (parametrised by $\beta$) with \textit{multi--fractal} eigenstates, which were studied numerically \cite{cuevas2001anomalously}, and analytically, using super--symmetry methods \cite{mirlin00pbrm,mirlin06exactmulti} (based on earlier studies on banded matrices \cite{mirlin91band}) or strong--disorder renormalization group \cite{levitov99pbrm,mirlin00pbrm}. We refer to \cite{evers2008anderson} for a comprehensive review. These critical models are an important laboratory for testing theoretical predictions of multi--fractal properties of eigenstates \textit{at the critical point}. 

This being said, there is no genuine localization transition with mobility edge in \gls{pbrm}. Worse, even tuning the disorder strength  $\beta$ does not induce a transition; only tuning the ``effective dimension'' $1/\mu$ does so. In this respect, the \gls{pbrm} is not a satisfactory proxy for studying localization transitions in higher dimensions. A main objective of this chapter is to study new long--range 1D random matrix models with a better--behaving localization transition. However, our approach has been inspired by a seemingly unrelated topic, to which we turn in the next section \ref{sec:fisher}. In the remainder of this section, let us review a simple but important argument that allows to understand the above phase diagram.

The  \gls{pbrm} phase diagram can be heuristically understood by a simple resonance argument, which is essentially based on first--order perturbation theory. The small parameter is $T = \beta^{-1}$; when $T = 0$, $\beta = +\infty$, $H_{mn}$ is diagonal (so this is a \textit{strong disorder} argument).  Let us take an unperturbed eigenstate $\vert \phi_0 \rangle = \vert n \rangle$. When $T$ is turned on, it becomes at first order
 \begin{equation}\label{eq:1storderperturb}
  \vert \phi \rangle \propto \vert n \rangle + \sum_{m \neq n}  \frac{H_{mn}}{H_{nn} - H_{mm}} \vert m \rangle + \dots \,.
 \end{equation}
For a typical site, $\abs{m-n} \sim N$ ($N$ is size of the matrix), so the \textit{hopping element} $\abs{H_{mn}} \sim t = N^{-\frac{1 + \mu}{2}}$. On the denominator, the \textit{energy mismatches} $H_{nn} - H_{mm}, m \neq n$ are independent centred Gaussian with variance 2. Since there are $\sim N$ of them, the \textit{level spacing} $\delta E$, \textit{i.e.}, the difference between adjacent energy levels scales as $\delta E \sim N^{-1}$. 

When $\mu < 1$, $t / \delta E \sim N^{(1-\mu)/2} \gg 1$. This means any site $n$ resonates strongly with some site $m$ far away from it. This excludes the existence of any localized states. When $\mu > 1$, $t / \delta E \ll 1$, system--size resonance is impossible. This is consistent with the localization of all the eigenstates. 

\section{Epidemic dynamics with long-range dispersion} \label{sec:fisher}
This section reviews the key results of the long--range epidemic dynamics and \acrfull{fpp} model \cite{hallatschek2014acceleration,chatterjee2016multiple} that is another inspiration of this chapter. 

\subsection{The model and main results}
\begin{figure}
\center 
\includegraphics[scale=1]{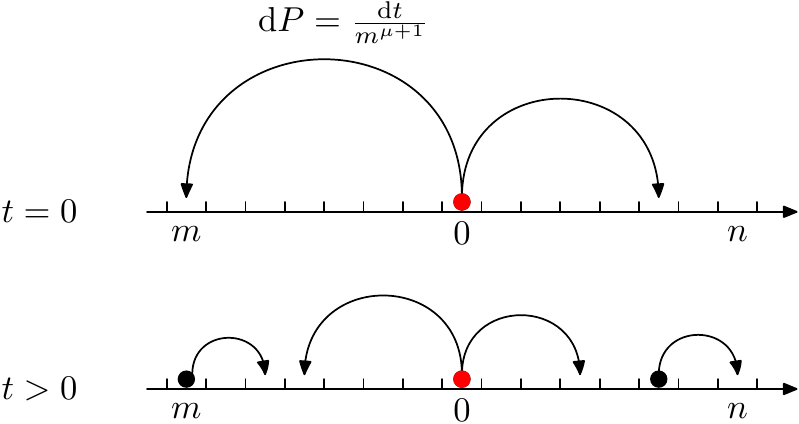}
\caption{An illustration of the dynamics of the epidemics dynamics model of \cite{hallatschek2014acceleration} in $d=1$. Initially ($t = 0$), the only occupied site at $n = 0$ infects other sites at a rate that decays as eq. \eqref{eq:infectionrate}.  Later, $t > 0$, infected sites go on to infect remaining sites.}\label{fig:fishermodel}
\end{figure}
Fisher and Hallatschek studied in \cite{hallatschek2014acceleration} an epidemic dynamics model, which can be viewed as the long--range version of the \textit{Eden model} introduced in section \ref{sec:introkpz}. It can be defined on a lattice in any dimension, and we shall restrict to $d = 1$. For this, consider a linear lattice made of sites $n = -N/2, \dots, N/2-1$. Like the Eden model, every site can be either empty or occupied (infected). The dynamics is as follows, see Figure \ref{fig:fishermodel} for an illustration: 
\begin{itemize}
\item Initially, at $t = 0$, only the site $0$ is occupied, and all the others are empty.  
\item During each infinitesimal time interval $\dif t$, and for any pair of sites $m$ and $n$, such that $m$ is occupied and $n$ is empty, $n$ becomes occupied with probability 
\begin{equation}\label{eq:infectionrate}
\dif P_{mn} = \abs{m-n}^{-\mu - 1} \dif t  \,, 
\end{equation} 
That is, occupied sites ``infect'' all other sites with a rate that decays algebraically as function of the distance. All these events are uncorrelated.
\item Once a site is occupied, it remains so forever. 
\end{itemize}
The work \cite{hallatschek2014acceleration} required $\mu > 0$ in eq. \eqref{eq:infectionrate} so that $\sum_{m} P_{mn}$ does not diverge in the infinite system $N \to \infty$ limit. In the epidemic dynamics context, the divergence would be unrealistic because the total rate at which a site send offspring elsewhere must be finite.

The main question treated in \cite{hallatschek2014acceleration} is: how does the size of the colony (the set of occupied sites),  $L(t)$, grows as a function of time (for $t \to \infty$)? The answer turned out to be quite non--trivial, and enjoys a rich dependence on the value of $\mu$. To get a feeling, let us consider the extreme cases: 
\begin{itemize}
\item As $\mu \to \infty$, $ \dif P_{mn} = \dif t $ if and only if $\abs{m-n} = 1$, and vanishes otherwise. So the model becomes a \textit{short--range} one: the 1d Eden model. Trivially, the colony is always a full interval whose size grows linear in time: $L(t) = 2t$, because only the sites at the boundary can infect its nearest neighbour: all other attempts are suppressed.
\item When $\mu \to 0$,  $\sum_{m} \dif P_{mn}$ becomes divergent at $\abs{m-n}$ large. The model is then approximatively ``infinite--range'', in which all attempts increase actually $L(t)$. Since the number of attempts is proportional to $L(t)$, $L(t) = e^{ct}$ grows exponentially. 
\end{itemize}
Now, for intermediate values of $\mu$, \cite{hallatschek2014acceleration} showed that 
\begin{equation}\label{eq:fisherresult}
L(t) \stackrel{t\to\infty}{\propto} \begin{dcases}
 t\,,  & \mu > 2\,,  \\
 t^{\frac1{\mu - 1}}\,, & \mu \in (1,2)\,, \\
 \exp(c t^{\eta})\,, & \mu \in (0,1)\,, \\
\end{dcases} \quad \text{ where } \eta(\mu) = \log_2 \left(\frac{2}{\mu + 1}\right) \,.
\end{equation}
Observe that the short--range behaviour extends to all values $\mu > 2$. This condition equivalent to the condition that $\sum_{m} m^2 \dif P / \dif t = \sum_m m^{-\mu+2}$ converges at $m \to \infty$. This means that the dispersion of offspring has a finite diffusion constant \cite{hallatschek2014acceleration}. When $\mu < 2$, the diffusion becomes anomalous (Lévy flights), and it is expected the colony size grows supra-linearly. In particular, the algebraic regime $\mu \in (1,2)$ can be understood by dimensional analysis \cite{hallatschek2014acceleration}. For this, we notice that if the model were defined in the continuum, then eq. \eqref{eq:infectionrate} would read 
\begin{equation}\label{eq:dimensionanalysis}
 \dif P = \abs{x - y}^{-\mu-1} \dif t \dif x \dif y \,,
\end{equation}
where $x,y$ are continuous spatial coordinates, with the same dimension as $L$, while the left hand side is dimensionless. Therefore eq. \eqref{eq:dimensionanalysis} implies $[1] = [L]^{-\mu + 1} [t]$, which is the same power law as the $\mu \in (1,2)$ case of \eqref{eq:fisherresult} \footnote{for $\mu > 2$, the solution should be discarded because the growth cannot be sub--linear. In this respect, we note that for a rigorous analysis, this regime turns out to be the most demanding \cite{chatterjee2016multiple}}.
However, there is another non--trivial qualitative change at $\mu = 1$, and an exponential regime at $\mu (0,1)$, which cannot be explained by dimensional analysis (which fails at $\mu = 1$). As we shall discuss in section \ref{sec:stretch}, the colony grows in a fundamentally different way in this regime.

\subsection{First--passage percolation}\label{sec:fpp}
\begin{figure}
\center
\includegraphics[scale=1]{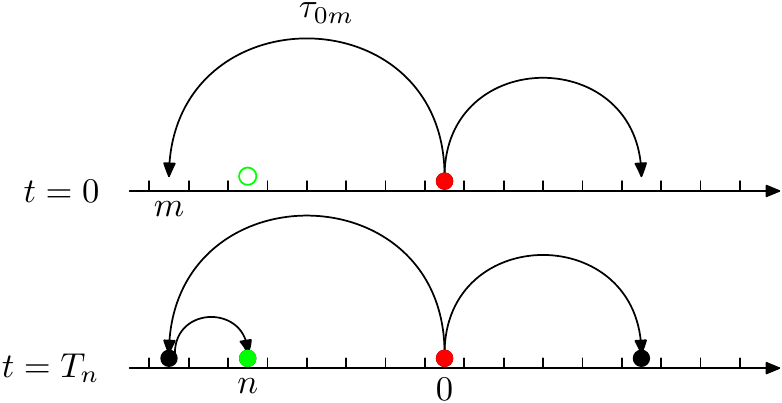}
\caption{An illustration of the long--range \acrlong{fpp} model and its correspondence to the epidemics model, see also Figure \ref{fig:fishermodel}. For each pair of vertices (like $0$ and $m$ in the top panel), one assigns a waitiing time $\tau_{0m}$ which is an exponential random variable of mean value $\abs{m}^{\mu+1}$, see eq. \eqref{eq:Pexy}. Then one studies the first passage time $T_n$, defined as the first moment when $n$ is infected (bottom panel).}\label{fig:FPP}
\end{figure}
As we have seen in section \ref{sec:introkpz}, the Eden growth model is equivalent to a \gls{fpp} model. The same can be said about the epidemic dynamics model just discussed, which gives rise to a \textit{long--range \acrfull{fpp}} model. This model is defined by a symmetric matrix of independent random waiting times  $\tau_{mn}$ having exponential distribution:
  \begin{equation}\label{eq:Pexy}
  \tau_{nn}\equiv +\infty \,,\, \mathbb{P}(\tau_{m\neq n} > t \geq 0) = \exp \left(-\frac{t}{\abs{m-n}^{\mu+1}} \right) \,.
  \end{equation}  
The relation with the epidemic model is ensured by the fact that the mean waiting time $\overline{\tau_{mn}}$ is the inverse of the rate $\dif P_{mn} / \dif t$ in eq. \eqref{eq:infectionrate}. The key observable is the \textit{first passage time}, $T_n$, defined as the moment when the site $n$ becomes occupied. To define the $T_n$ in terms of $\tau_{mn}$, let us call a \textit{path} (polymer) between two points $0$ and $m$ any finite sequence of sites $\mathfrak{p}=(0 = m_0, m_1,\dots, m_s = n)$, $s$ being its length, $s = s[\mathfrak{p}]$. Note that the paths are not directed and the sites can be visited multiple times. The total waiting time of a path is defined as the sum over the edges
\begin{equation}
T[\mathfrak{p}] = \sum_{i=1}^s \tau_{m_i, m_{i-1}} \,. \label{eq:Esupp}
\end{equation}
Then, the following analogue of eq. \eqref{eq:Txypolymer}  holds
\begin{equation}\label{eq:fptsupp}
T_{n} = \min_{\mathfrak{p}:0\rightarrow n} T[\mathfrak{p}] \,. 
\end{equation}
Because of the long--range dispersion, the minimum is over any paths. The fundamental question is to determine how $T_n$ grows as a function of $n$. 

This question is answered rigorously by Chatterjee and Dey \cite{chatterjee2016multiple} (the work was independent to Fisher--Hallatscheck's \cite{hallatschek2014acceleration}), and the answer reads as 
\begin{equation}\label{eq:TnFPP}
T_n \stackrel{n\to\infty}\propto \begin{dcases}
\abs{n}\,, & \mu > 2 \,, \\
\abs{n}^{\mu - 1} & \mu \in (1, 2)   \,, \\
\exp\left(c\sqrt{\ln \abs{n}} \right) &\mu  = 1 \,, \\
\left(\ln \abs{n}\right)^{\kappa(\mu)} \,, & \mu \in (0,1) \\
O(1) \,,\,  & \mu \leq 0 \,,
\end{dcases} \quad \text{ where } \kappa(\mu) =  \frac{\ln 2}{\ln \frac{2}{\mu + 1}} = \eta(\mu)^{-1} \,,
\end{equation}
where $\eta(\mu)$ is defined in \eqref{eq:fisherresult}. One can check that $T_n$ and $L(t)$ are inverse functions of each other in all cases when they can be compared, which is expected. The $\mu=1$ case is a marginal case that is also obtained in \cite{hallatschek2014acceleration}. The $\mu \leq 0$ result justifies excluding this regime from the epidemics point of view: the growth of $L(t)$ is \textit{not} exponential, but \textit{spontaneous} (an exponential growth requires carefully logarithmic fine--tuning of eq. \eqref{eq:Pexy}, as shown in \cite{chatterjee2016multiple}). 

\subsection{Stretch-exponential regime}\label{sec:stretch}
Let us explain the results \eqref{eq:TnFPP} or \eqref{eq:fisherresult} in the stretch exponential regime $\mu \in (0,1)$, following \cite{hallatschek2014acceleration}. This regime is especially important because it is the inspiration of our original arguments in section \ref{sec:RG}. 

\begin{figure}
\center
\includegraphics[scale=1]{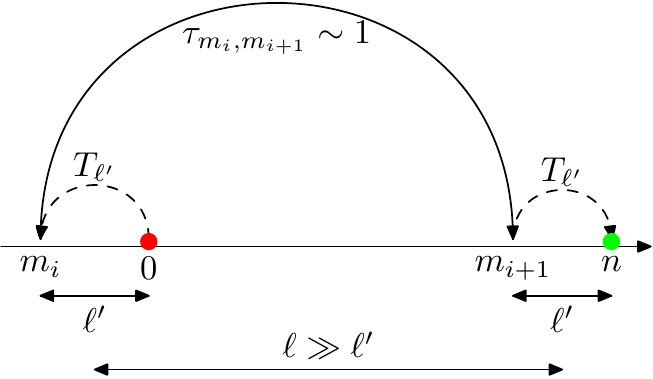}
\caption{An illustration of the typical minimizing path in the $\mu\in(0,1)$ regime. It is made of a long bond $m_i, m_{i+1}$ of length $\ell_k \sim \abs{n}$, and two sub--polymers (drawn in dashed curves) connecting $0\to m_{i}$ and $m_{i+1} \to n$, each of which repeating the structure. }\label{fig:stretch}
\end{figure}
The key characteristics of the stretch exponential regime is the \textit{structure} of the minimizing path of eq. \eqref{eq:fptsupp}, denoted $\mathfrak{p} = (0 = m_0, \dots, m_s = n)$, see Figure \ref{fig:stretch} for illustration. It has typically a bond $m_{i}, m_{i+1}$ of length $\abs{m_i - m_{i+1}} \sim \ell = \abs{n}$ comparable to the distance between two end--points, and such that $\tau_{m_i, m_{i+1}} \sim O(1)$. According to eq. \eqref{eq:Pexy}, the probability that this happens for \textit{one} bond length $\ell$ is $\ell^{-\mu - 1} \ll 1$. However, if one defines the following scale
\begin{equation}\label{eq:recursion0FPP}
(\ell')^2  \ell^{-\mu - 1} = 1 \Leftrightarrow \ell =  (\ell')^{\frac{2}{\mu+1}} \ll \ell 
\end{equation}
when $\mu<1$ and $\ell \gg 1$, then the minimum of the set of $4 \ell'^2$ bonds, $\tau_{mm'}$ for $\abs{m} \leq \ell'$ and $\abs{m-n} \leq \ell'$,  will be an exponential random variable of mean value $1/4$. So there must be some $m, m'$ that can play the rôle of $m_{i}, m_{i+1}$. However, we have no control over their distance from $0$ (and $n$, respectively) other than $\abs{m-0}, \abs{m'-n} \sim \ell'$. Thus, the waiting time two halves of the polymers $0 \to m$ and $m'\to n$ can be estimated by $T_{\ell'}$. Summarizing, we have the following recursion relation for $T_\ell$
\begin{equation}\label{eq:recursion1FPP}
T_{\ell} = T[\mathfrak{p}] = \underbrace{\tau_{m_i, m_{i+1}}}_{O(1)}+  \underbrace{\sum_{j=1}^{i} \tau_{m_{j}, m_{j-1}}}_{T_{\ell'/2}} + 
\underbrace{\sum_{j=i+1}^{s-1} \tau_{m_{j}, m_{j+1}}}_{T_{\ell'/2}} \sim  2 T_{\ell'} \,.
\end{equation}
Now, both eq. \eqref{eq:recursion0FPP} and \eqref{eq:recursion1FPP} can be repeated. When $\mu < 1$, eq. \eqref{eq:recursion0FPP} defines a sequence of lengths 
\begin{equation}\label{eq:recursion3FPP}
\ell_k = \ell_{k-1}^{\frac{2}{\mu+1}} \Rightarrow \ln \ell_k = \left(\frac{2}{\mu+1}\right)^k \ln \ell_0  \Rightarrow \ln \ln \ell_k = \ln \ln \ell_0 +  k \ln \frac{2}{\mu+1}
\end{equation}
where $\ell_0 > 1$ can be fixed arbitrary. On the other hand, eq. \eqref{eq:recursion1FPP} implies 
\begin{equation}\label{eq:recursion4FPP}
T_{\ell_k} = 2^k T_{\ell_0} \Rightarrow  \ln T_{\ell_k} = \ln T_{\ell_0} + k \ln 2 \,.
\end{equation}
Combining eq. \eqref{eq:recursion3FPP} and \eqref{eq:recursion4FPP} to eliminate $k$, we recover the $\mu\in (0,1)$ case of eq. \eqref{eq:TnFPP}, namely:
$$  \ln  T_{\ell} =  \frac{\ln 2}{\ln \frac{2}{\mu+1}}  \ln \ln \ell + c \Rightarrow 
T_{\ell} =  \left(\ln \ell \right)^{\kappa(\mu)} \,.$$
Note that such a strategy is not possible when $\mu \geq 1$ because eq. \eqref{eq:recursion0FPP} would imply $\ell' \gg \ell$. 

In terms of the epidemic dynamics, the above discussion entails that when $\mu < 1$, the colony grows in an explosive manner: from time to time, the colony sends a offspring to a distance which is far greater than the present colony size. The offspring becomes the seed of a new satellite. The colony growth is dominated by the multiplication of satellites. In contrast, when $\mu > 1$, the growth is incremental, in the sense that the dominating dispersion length is much smaller than then the colony size.

\section{Beta Banded Random Matrices (BBRM)} \label{sec:bbrm}
\subsection{Motivation, definition and main results}\label{sec:BBRMoverview}
In this section, we study the \glspl{bbrm}, following \cite{cao16loc}. Before defining it, we summarize the two motivations behind it.

\begin{itemize}
\item The first motivation is to find better 1D long--range proxies than  \gls{pbrm} for the Anderson localization transitions in $d > 2$ dimensions. As we have seen in section \ref{sec:pbrm}, the phase diagram of  \glspl{pbrm} is oversimplified because \textit{direct} hopping dominates the transport in the space that  \glspl{pbrm} describe. Moreover, the argument that leads to the phase diagram is quite general. Indeed, the \gls{pbrm}'s phase diagram would not be changed if we changed the distribution of its matrix elements to 
\begin{equation} 
P(H_{mn}) = \sqrt{g} P_0(\sqrt{g} H) \,,\,  g = \beta \abs{n-m}^{1+\mu} \,,\, \abs{m-n} \gg 1 \,, \end{equation} 
where $P_0$ is any narrow enough distribution (\textit{e.g.}, uniform in $[-1, 1]$) such that the moments obey 
\begin{equation} \label{eq:PBRMmomentdecay}
 \overline{H_{mn}^p} \propto g^{-p/2} \,,\, \abs{m-n} \gg 1  \,,\, p = 1, 2, 3 \dots. \end{equation}
This means that all matrix elements corresponding to a certain (large) distance $|n-m|$ have the same order of magnitude $g_{mn}^{-1/2}$. So, in order to change the phase diagram, a natural direction is to look at banded random matrices whose elements' distribution is radically different. 
\item The second motivation is that the long--range epidemics/\gls{fpp} model gives rise naturally to  \gls{bbrm} via a mapping, and the results and methods of \cite{hallatschek2014acceleration,chatterjee2016multiple} (discussed in section \ref{sec:fisher}) lead to predictions on  \gls{bbrm}. This mapping will be explained in section \ref{sec:mappingAL}. 
\end{itemize}

Now let us define the  \gls{bbrm}. It is inspired directly by the long range FFP model in section \ref{sec:fpp}. Formally, it can be defined as 
\begin{equation}\label{eq:defBBRM1}
Q_{nm} = \exp(-\beta \tau_{nm})  \,,\, n,m = 1, \dots, N \,,
\end{equation}
where $\tau_{nm}$ is the matrix of waiting times eq. \eqref{eq:Pexy}, and $\beta$ is the extra parameter of the matrix model. Equivalently, the  \gls{bbrm} ensemble is defined as a real symmetric $N \times N$ matrix whose diagonal entries are all ro ($Q_{nn} \equiv 0$) and all off-diagonal entries $Q_{nm}$ are independent random variables in $(0,1)$, with a Beta distribution,
 \begin{equation}
 \mathbb{P}(Q_{mn} < q) = q^{1/g} \,,\,  g = \beta \abs{m-n}^{\mu + 1} \,,\, q \in (0,1) \,. \label{eq:Qdefsupp}
 \end{equation}
 So the  \gls{bbrm} model has the same parameters $\mu,\beta$ as  \gls{pbrm}. 
 
  \begin{figure}
  \center\includegraphics[scale=.7]{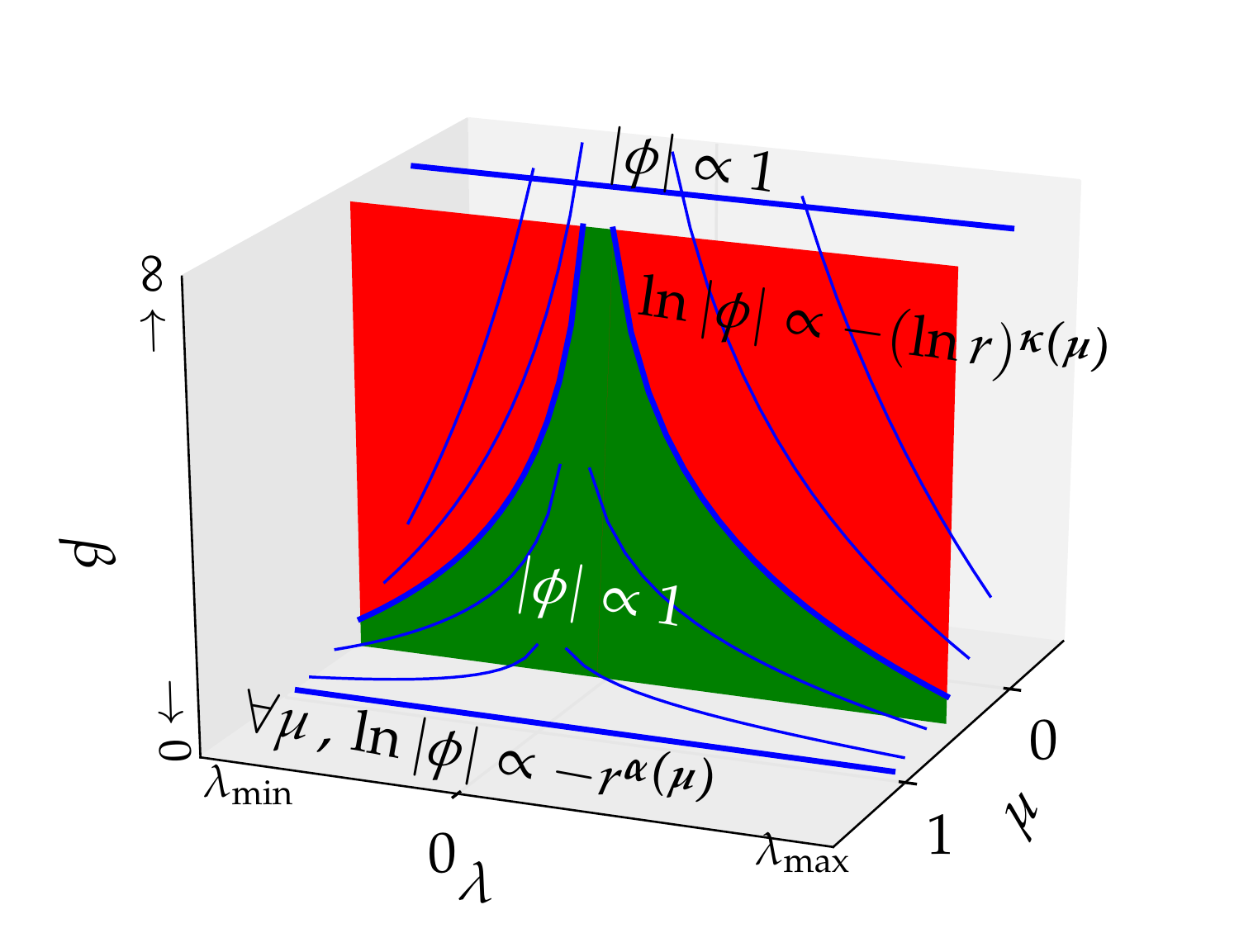}
  \caption{Taken from \cite{cao16loc}. Schematic phase diagram of  \gls{bbrm}s. The decay of the eigenstates $\phi$ in different regimes are indicated. In the regime $\mu \in (0,1)$, there is a mobility edge $\lambda(\beta;\mu)$ (blue curves) separating localized eigenstates (red region) and extended eigenstates (green region). The localized states have a very peculiar decay, Eq. (\ref{eq:logpowerdecay}). When $\mu > 1$, all the eigenstates are localized, but with a stretched exponential decay for $\mu < 2$, Eq. (\ref{eq:expdecay}).}\label{fig:bbrmphase}
  \end{figure}
 We now summarize the main results of \cite{cao16loc} on this model (see Figure \ref{fig:bbrmphase}):
\begin{itemize}
\item  When $\mu \in (0,1)$, the model enjoys a localization transition with mobility edges, separating extended states in the middle of the spectrum from localized states elsewhere. This is the most important result of \cite{cao16loc}. We will discuss it in section \ref{sec:LT}.
\item  When $\mu \in (0,1)$, the localized states have a peculiar decay. Indeed, if $\phi_m$ is a states localized around $m$, we have   
\begin{equation}
   \abs{\langle n \vert \phi_m \rangle} = \exp\left(- C \ln^{\kappa(\mu)} (\abs{n-m} / \xi) \right) \,,\, \text{ where } \kappa(\mu) =  \frac{\ln 2}{\ln (2/(\mu + 1))} \label{eq:logpowerdecay} 
  \end{equation} 
 is the same exponent as in  \eqref{eq:TnFPP}.
\item  When $\mu > 1$, all eigenstates are localized, with a stretch exponential decay: 
   \begin{equation} 
   \label{eq:expdecay}  \abs{\langle n \vert \phi_0 \rangle}  \sim \exp\left(- C \abs{n}^{\min(\mu,2)-1} \right) \,,\, \mu > 1 \,. 
   \end{equation} 
   The predictions eq. \eqref{eq:expdecay} and \eqref{eq:logpowerdecay} come from the mapping to  \gls{fpp} model, as will discuss in section \eqref{sec:mappingAL}.
\end{itemize}

\subsection{Basics: broad distribution, density of states}\label{sec:basics}
 This section discusses two elementary aspects of  \gls{bbrm} that are preliminary to further investigations. 
 
 \subsubsection{Broad distribution of matrix elements}
 First, let us recall some basic statistical properties of its matrix elements:
 \begin{equation}
  \mathbb{P}(Q_{mn} < q) = q^{1/g} \,,\,  g = \beta \abs{m-n}^{\mu + 1} \,,\, q \in (0,1) \,. \label{eq:Qdefsupp1}
  \end{equation}
 Note that the dependence on $m,n$ and $\beta$ is through $g$. When $g = 1$, $Q_{mn}$ is uniformly distributed, when $g \rightarrow 0$, $Q_{mn}\rightarrow 1$ becomes non-random. This justifies calling $\beta\rightarrow0$ the \textit{weak disorder} limit, and $\beta\rightarrow\infty$ the \textit{strong disorder} limit consequently.

    \begin{figure}
    \center \includegraphics[scale=.5]{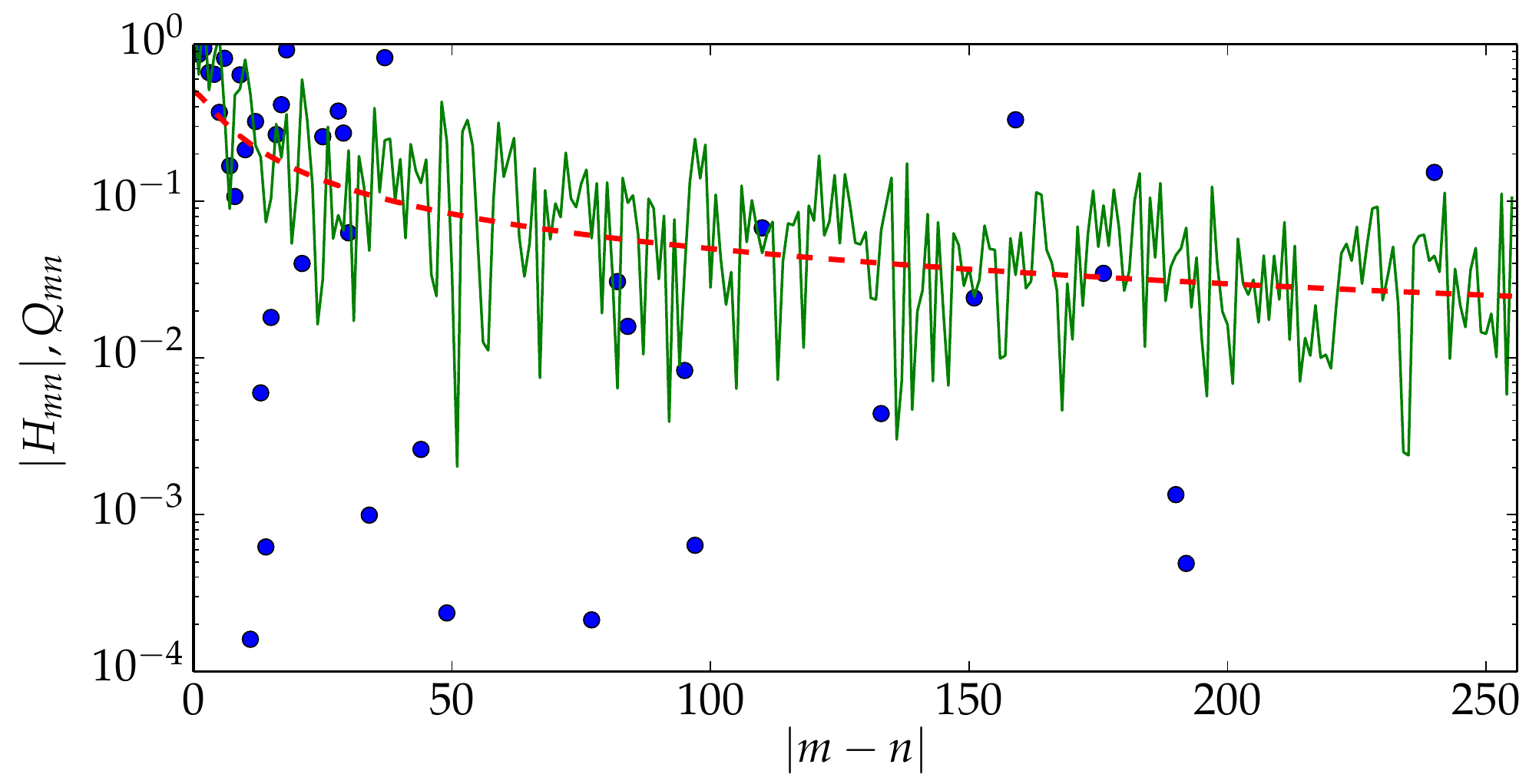}
    \caption{Taken from \cite{cao16loc}. The magnitudes of matrix elements of  \gls{pbrm} (line) and  \gls{bbrm} (dots) on one line of the matrix,as a function of the distance from the diagonal. For both matrices, we used the parameters $\mu = .5, \beta = 0.1$. The red dashed line depicts the standard deviation common to both. Many elements of the  \gls{bbrm} are too small to be drawn on the plot.}\label{fig:matrixele}
    \end{figure}
   It is the behaviours of $Q_{mn}$ as $g\to \infty$ that are crucially different from that of  \gls{pbrm}.  Indeed, it is an example of what we shall call \textit{broad distributions} in section \ref{sec:bbrmgen}: the typical value $Q_{mn}$ is very small compared to its integer moments, which are contributed by occurrences of \textit{black swans} of magnitude $O(1)$ with probability $1/g$ (see Figure \ref{fig:matrixele}). More quantitatively,
   \begin{itemize}
   \item  Its moments are
      \begin{align}
      & \overline{Q_{mn}^k} =  \frac{1}{1 + k g}  \,, \label{eq:moments} \\
      &\overline{\exp(- t Q_{mn})}  = 
      \sum_{k=0}^{\infty} \frac{(-t)^k}{k!} \frac{1}{1 + k g}  = g^{-1} t^{-1/g} (\Gamma (1/g)-\Gamma (1/g ,t))\, . \label{eq:laplace}
      \end{align}
      where $\Gamma(x,z) = \int_{z}^{+\infty} \dif y e^{-y} y^{x-1}$ is the incomplete Gamma function. As a consequence, the second moment of  \gls{bbrm} matrix elements have the same decay as  \gls{pbrm}, eq. \eqref{eq:pbrmasymp}. However, \textit{all} its integer moments decay as $1/g$, which is different from eq. \eqref{eq:PBRMmomentdecay}.
    \item 
       $Q_{mn}$ can be characterised by the property that $-\ln Q_{mn}$ is exponentially distributed with mean value $g$ (see also eq. \eqref{eq:defBBRM1}):
       \begin{equation} \mathbb{P}(-\ln Q_{mn} > l) = \exp(-l/g)  \, ,\, l \geq 0 \,.\label{eq:exp} \end{equation} 
      Therefore, its typical value, defined as the exponential of the mean of log, is
       \begin{equation} Q^{\text{typ}}_{mn} \stackrel{\text{def}}= \exp(\overline{\ln Q_{mn}}) = \exp(-g) \,. \end{equation}
       As $g\rightarrow\infty$,  $Q^{\text{typ}}_{mn} \ll \overline{Q^k_{mn}}$. The same can be said about its median, which is $m(Q_{mn}) = \exp (-g \ln 2)$.
     \item Eq. \eqref{eq:exp} implies also that the occurence probability of black swans
        \begin{equation}
        \mathbb{P}(Q_{mn} > a) \sim \frac{-\ln a}{g} \,,\, g \rightarrow \infty \,, 0 < a < 1 \text{ fixed.}  \label{eq:blackswan}
        \end{equation}
   \end{itemize}
   In summary (see also Figure \ref{fig:matrixele}), when the parameters $\mu$ and $\beta$ are the same,  \gls{bbrm}'s \textit{typical} matrix elements decay in fact faster than  \gls{pbrm}, yet the latter does not have the atypically large elements of \gls{bbrm}.  
   
  \subsubsection{\Gls{dos}}
   It is always helpful to have an idea of the \acrfull{dos} of the random matrix model at hand. The general definition of the \gls{dos} for any random matrix is:
   \begin{equation}
   \rho(\lambda) = \frac1N \sum_{i=1}^N \overline{\delta(\lambda - \lambda_i)} \,,
   \end{equation}
    where $\lambda_1, \dots, \lambda_N$ are the eigenvalues of the matrix. When averaged over disorder, $\rho(\lambda)$ is usually a continuous function that vanishes outside some finite interval $[\lambda_{-}, \lambda_{+}]$ (a notable exception is the Lévy matrix \cite{cizeau1994levy,tarquini16levy}, for which $\lambda_{+} = + \infty$). 
    For example, we recall \cite{mirlin96pbrm} that the \gls{dos} of the \gls{pbrm} is always given by Wigner's semi--circle law (upon a rescaling) $$\rho(\lambda) = \frac{2}{\lambda_{+}\pi} \sqrt{1 - (\lambda/\lambda_{+})^2} \,.$$ 
     \begin{figure}
     \center \includegraphics[scale=.4]{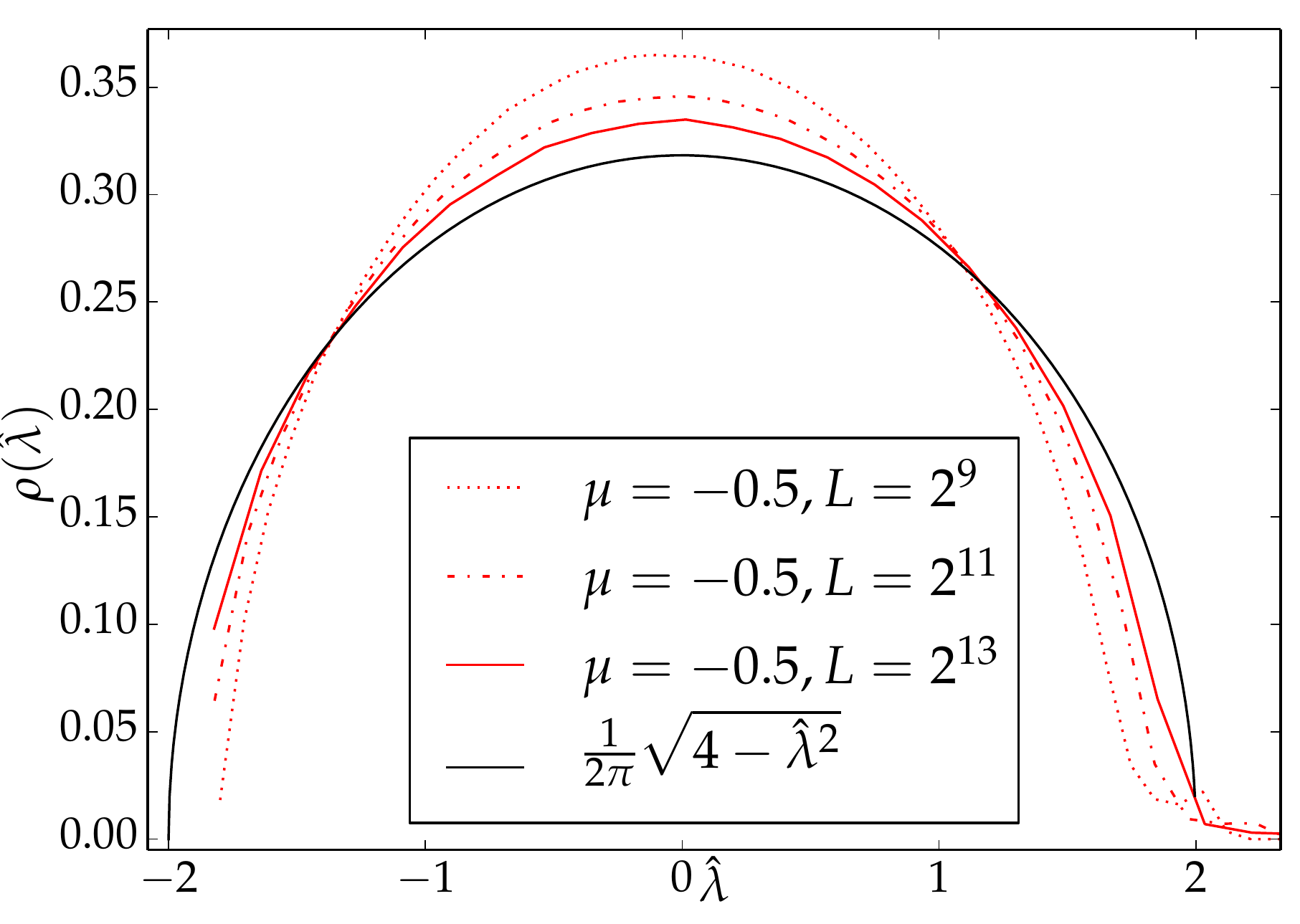}\includegraphics[scale=.5]{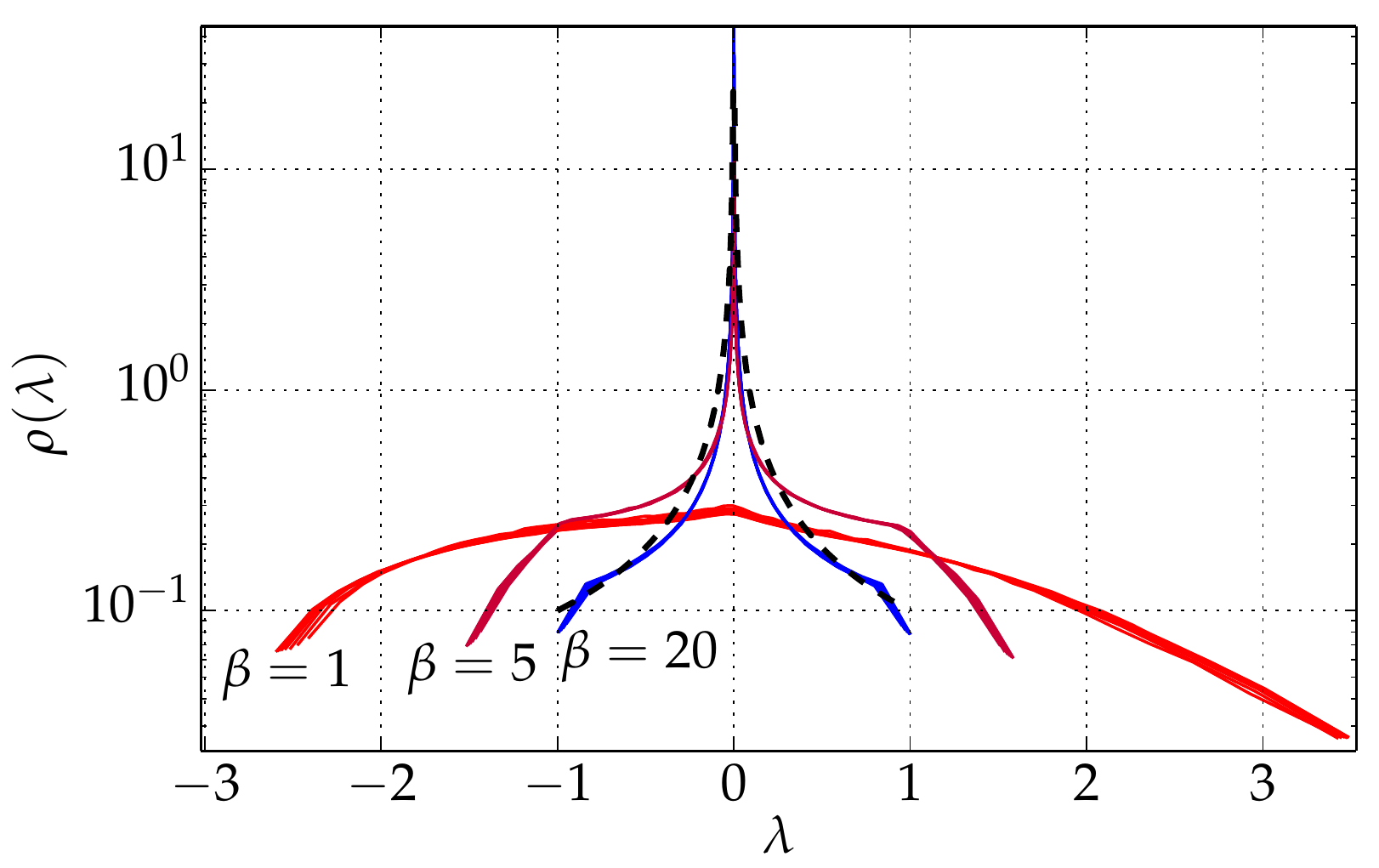}
     \caption{Taken from \cite{cao16loc}. \textit{Left}: Rescaled \gls{dos}  $\rho(\hat{\lambda})$ of the  \gls{bbrm} for $\mu = -.5$ and $\beta = 1$, where $\hat{\lambda} = a + b \lambda$ is the rescaled eigenvalue so that the mean and variance are that of the semicircle law. \textit{Right}: \gls{dos} of the  \gls{bbrm} for $\mu = .5$ and various $\beta$. The curve $\beta = 20$ is compared to the prediction eq. \eqref{eq:rhobigb}, plotted in dashed line. }\label{fig:DoS}
     \end{figure}
   In general, the width of the support of the \gls{dos}  can be estimated by 
    $$ \overline{\lambda^2} \defeq \int \rho(\lambda) \lambda^2 \dif \lambda = \frac1N \sum_{i} \overline{\lambda_i^2} =  \frac1N\overline{\mathrm{Tr}\hat{H}^2} =  \frac1N \sum_{nm} \overline{H_{mn}^2}$$
    For both  \gls{pbrm} and  \gls{bbrm}, the last quantity $\sim \sum_{n=1}^N \beta^{-1} \abs{n}^{-\mu - 1}$ (by eq. \eqref{eq:moments} and eq. \eqref{eq:PBRMmomentdecay}). So, when $\mu < 0$, $\overline{\lambda^2} \sim N^{-\mu}$ diverges as $N \to \infty$. In such cases, according to a theorem of \cite{kus1991density},  \gls{bbrm}'s density of state must also be a semicircle law. This is corroborated by the numerical measure, which we show in Figure \ref{fig:DoS} (left panel). 
    
    However, when $\mu > 0$, due to the absence of diagonal disorder ($Q_{nn} = 0$), its \gls{dos} $\rho(\lambda)$ is in fact quite peculiar. It develops a divergence at $\lambda = 0$, and a non-analyticity at $\lambda = \pm 1$ as $\beta$ increases. For $\beta \gg 1$, most elements are vanishing, and a crude estimate of \gls{dos}  is given by the distribution of the eigenvalues $\pm Q_{01} $ of the $2\times2$ sub-matrix $\begin{pmatrix} 0 & Q_{01}  \\ Q_{01} & 0 \end{pmatrix}$. This leads to:
   \begin{equation} 
   \rho(\lambda) \approx \abs{\lambda}^{-1 + 1/\beta} / (2\beta) \,,\,\text{ if } \abs{\lambda} < 1 \label{eq:rhobigb}
   \end{equation} 
   and $\rho(\lambda) \approx 0$ for $\abs{\lambda}> 1$. In Figure \ref{fig:DoS} (right panel), we show a numerical check of this claim. We observe also that even for $\beta \sim 1$, the \gls{dos} of  \gls{bbrm} differs significantly from the semi--circle law. 
   
\subsection{Mapping to the epidemic dynamics} \label{sec:mappingAL}
\begin{figure}
\center\includegraphics[scale=1]{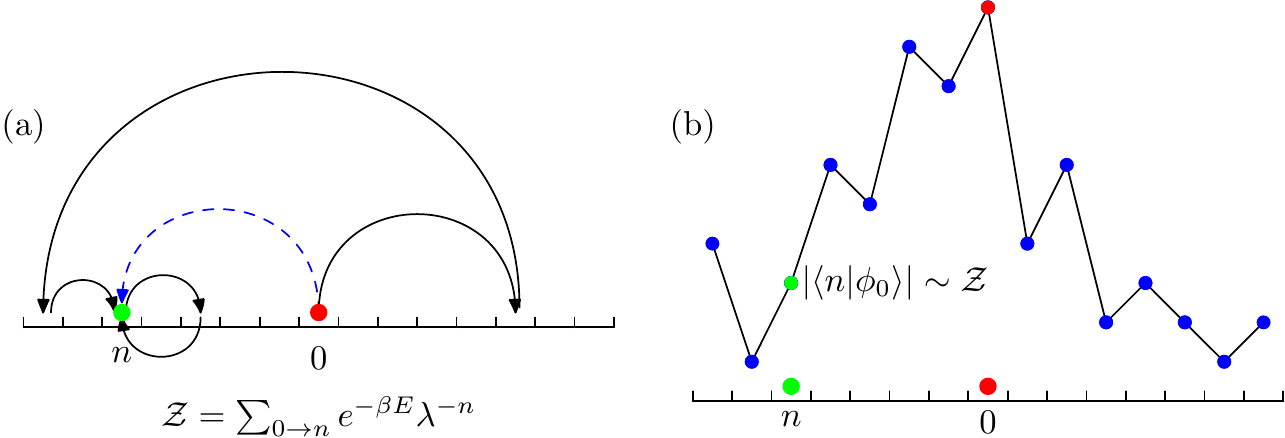}
 \caption{The relation between the polymer model and Anderson localisation. (a) In the polymer model, which is a finite temperature extension of the growth model,  we sum over all paths connecting two points $0$ and $n$, ranging from the direct path (blue, dashed) and detoured paths with loops, for example, the one in black. (b) The amplitude at site $n$ of (strong-disorder)  \gls{bbrm} eigenstates localized around $0$ is in turn related to the polymer partition function by \eqref{eq:roTsccarsup}. } \label{fig:mappingAL}
\end{figure}
This section, based on \cite{cao16loc} (Supplemental Material, section D), will explain the relation between  the \gls{bbrm} and the long--range  \gls{fpp} model, and then use it to derive the decay rate of the localized states of  \gls{bbrm}, eq. \eqref{eq:expdecay} and \eqref{eq:logpowerdecay}. 

The relation between  \gls{bbrm} and the long--range  \gls{fpp} follows directly from the relation of the long--range  \gls{fpp} and a statistical model of long--range ``polymers'' in random media, analogous to the one discussed in section \ref{sec:introkpz}. The latter polymer model is defined by a grand canonical partition function $\mathcal{Z}$ which is a sum over \textit{all} the paths from $0$ to $n$, with energy given by eq. \eqref{eq:Esupp}. The inverse temperature is $\beta$, and the chemical potential associated to the length of the polymer will be denoted $\tau$ (to avoid clash with $\mu$). Namely, we have (see Figure \ref{fig:mappingAL}, (a) for an illustration)
\begin{align}
  \mathcal{Z} & = \sum_{s=0}^{\infty} \sum_{\substack{\mathfrak{p}:(m_0, \dots, m_s) \\ m_0 = 0, m_s = n}}
  \exp(- \beta T[\mathfrak{p}] + s \beta \tau ) \nonumber \\
  & = \sum_{s=0}^{\infty} \sum_{m_1} \dots \sum_{m_{s-1}} \exp\left(- \beta \sum_{i=1}^s (\tau_{m_i m_{i-1}} - \tau) \right)    \,. \label{eq:ZofBBRM}
\end{align}
This is the long--range version of the polymer model, defined in eq. \eqref{eq:polymerZgen} of section \ref{sec:introkpz}. Now, using eq. \eqref{eq:defBBRM1} and the definition of matrix multiplication, eq. \eqref{eq:ZofBBRM} can be written as a resolvent of the  \gls{bbrm} 
  \begin{align}
 \mathcal{Z}  &=  \sum_{s=0}^{\infty} \sum_{m_1} \dots \sum_{m_{s-1}} \prod_{i=1}^s  \left(Q_{m_i m_{i-1}} \lambda^{-1} \right) = \langle n \vert ( 1 - \hat{Q} / \lambda )^{-1} \vert 0 \rangle \,,\, \lambda = e^{-\beta\tau} \,, \label{eq:ZofBBRM2}
  \end{align} 
On the other hand,  The zero--temperature, zero--chemical potential limit of the free energy is just the first passage time $T_n$ defined in eq. \eqref{eq:fptsupp} (this is a long--range analogue of eq. \eqref{eq:TxyZgen}):
 \begin{equation}
 T_n =  \left[ - \beta^{-1} \ln \mathcal{Z}  \right]_{\beta \to \infty, \tau\to 0 } =  \left[ - \beta^{-1} \ln \abs{\langle n \vert ( 1 - \hat{Q} / \lambda )^{-1} \vert 0 \rangle}  \right]_{\beta \to \infty, \tau\to 0 }  \,, \label{eq:Tnisresolvant}
 \end{equation}
 where we applied eq \eqref{eq:ZofBBRM2}. This equation allows to translate the known asymptotic growth of $T_n$ (eq. \eqref{eq:TnFPP}) to the decay of eigenstates in the strong disorder $\beta \to \infty$ limit. To this end, let us fix some $\lambda \neq 0$ while let $\beta \to \infty$, so $\tau = -\beta^{-1} \ln \lambda \to 0$ by eq. \eqref{eq:ZofBBRM2}. So eq. \eqref{eq:Tnisresolvant} can be rewritten as 
 \begin{equation}
\ln \abs{\langle n \vert ( 1 - \hat{Q} / \lambda )^{-1} \vert 0 \rangle} \stackrel{\beta\gg1}{\approx} 
-\beta T_n \,,\, \lambda \neq 0 \,. \label{eq:logresolvant}
 \end{equation}
The decay of the left hand side is related that of eigenstates in a quite standard way, as we briefly review. For this, we write $\hat{Q}$ in the basis of its eigenstates $\vert \lambda' \rangle$ of $\mathbf{Q}$ (with energy $\lambda'$): 
  \begin{equation} \langle n \vert (1 - \hat{Q} / \lambda)^{-1} \vert 0 \rangle= \sum_{\lambda'} \frac{\langle n \vert \lambda' \rangle\langle \lambda' \vert 0 \rangle }{1 - \lambda'/\lambda} \,.  \label{eq:resolvantexpand} \end{equation} 
  Because of the denominator, the sum is dominated by eigenstates with energy close to $\lambda$, and the decay of those eigenstates determines the behaviour $\langle n \vert \lambda' \rangle\langle \lambda' \vert 0 \rangle$ as a function of $n$. Indeed, when $\vert\lambda'\rangle$ is localized around some site $m$ with decay $\ln \abs{\langle n \vert \lambda' \rangle} \propto -T_{\abs{m - n}}$, we have $ \ln \abs{\langle n \vert \lambda' \rangle\langle \lambda' \vert 0 \rangle }= - T_{\abs{m}} - T_{\abs{n-m}} \leq -T_{\abs{n}}$ (the last convexity inequality can be checked for all cases of eq. \eqref{eq:TnFPP}), so the eigenstates localized at $0$ or $n$ will dominate \eqref{eq:resolvantexpand} and give the same contribution:
  \begin{equation}
  \ln \abs{\langle n \vert (1 - \hat{Q} / \lambda)^{-1} \vert 0 \rangle} \approx \ln \abs{\langle n \vert \phi_0\rangle }  \,,
  \end{equation}
  where $\vert \phi_0\rangle$ is an eigenstate localized around $0$ having energy $\approx \lambda$. Combined with \eqref{eq:logresolvant}, we have
  \begin{equation}
  - \ln \abs{ \langle n\vert\phi_0 \rangle} \stackrel{\beta \gg 1}\approx \beta  T_{n}  \,,\, \label{eq:roTsccarsup}
  \end{equation} In particular, if $T_{n}$ does not increase with $\abs{n}$, the state $\vert \phi_0 \rangle$ in fact extended (this is the case when $\mu < 0$). As we have noted, the above equation is valid for $\lambda \neq 0$. More precisely, the requirement is that $\tau = \abs{\beta^{-1} \ln \lambda} \ll 1$. So for a given large $\beta$, eq. \eqref{eq:roTsccarsup} covers the whole spectrum except an exponentially small interval $(-e^{-c\beta}, e^{-c\beta})$ around $0$. 
  
      \begin{table}
      \center	\begin{tabular}{c|c|c|c}
      		$\mu \in$ & $T_{n} \propto$ &  \gls{bbrm} $\abs{\langle n \vert \phi_0 \rangle}\sim$ &   \gls{pbrm} $\abs{\langle n \vert \phi_0 \rangle}\sim$  \\ \hline
      		$(-1,0]$ & $ \stackrel{L\rightarrow\infty}\longrightarrow 0$ & extended &  extended \\  
      		$(0,1)$ &  $\left(\ln \abs{n}\right)^{\kappa(\mu)}$  & $ \exp\left(- c \ln^{\kappa(\mu)}\abs{n}\right)$/extended & extended  \\ 
      		$\{1\}$ & $e^{c\sqrt{\ln \abs{n}}}$ & $\exp\left(- e^{c\sqrt{\ln \abs{n}}}\right)$ / extended & critical \\
      		$(1,2)$ & $\abs{n}^{\mu - 1}$   & $ \exp\left(- c \abs{n}^{\mu-1)}\right)$ &   $ \abs{n}^{-(\mu+1)/2}$ \\
      		$(2, \infty)$ & $\abs{n}$  &  $\exp\left(- c \abs{n}\right)$  & $ \abs{n}^{-(\mu+1)/2}$\\ \hline
      	\end{tabular}
      	\caption{Summary of asymptotic first-passage time \cite{hallatschek2014acceleration,chatterjee2016multiple} of the first-passage percolation model and their implication on the eigenstate decay of  \gls{bbrm}. $\kappa(\mu) = \ln 2/\ln \frac{2}{\mu + 1}$. The results on  \gls{pbrm} \cite{mirlin96pbrm} are also shown to compare. In the $\mu \in (0,1)$ regime,  \gls{bbrm} has localization transitions and extended eigenstates, see section \ref{sec:LT}. In the $\mu < 0$ regime, $T_{n} \rightarrow 0$ in the $L\rightarrow \infty$ limit. By \eqref{eq:roTsccarsup}, this means the eigenstates are extended even in the $\beta \gg 1$ limit. }\label{table:summary}
      \end{table} 
  In Table \ref{table:summary}, we list the results eq. \eqref{eq:TnFPP} and their translation by \eqref{eq:roTsccarsup}. In particular, eq. \eqref{eq:expdecay} and \eqref{eq:logpowerdecay} are contained in the cases $\mu > 1$ and $\mu \in (0,1)$, respectively.   
  We also quoted the analogous results for the  \gls{pbrm}, known in \cite{mirlin96pbrm}. Note that  \gls{bbrm} eigenstates decay always faster than  \gls{pbrm} ones.  In particular  both  \gls{bbrm} and  \gls{pbrm} eigenstates are all localized when $\mu > 1$, the decay are qualitatively different. The  \gls{pbrm} eigenstates are always algebraically decaying, and the decay rate is simply by the typical magnitude of the hopping element $\abs{H_{0n}} \sim \abs{n}^{-(\mu+1)/2}$. On the other hand, the  \gls{bbrm} eigenstates have an exponential decay when $\mu \geq 2$, recovering the short--range Anderson model behaviour. For $\mu< 1$, as we show in \cite{cao16loc} (Supplemental Material, section E), the non--trivial decay rates of  \gls{bbrm} can be interpreted in terms of Anderson models in higher dimensions.  

\begin{figure}
  	\center
  	\includegraphics[scale=0.55]{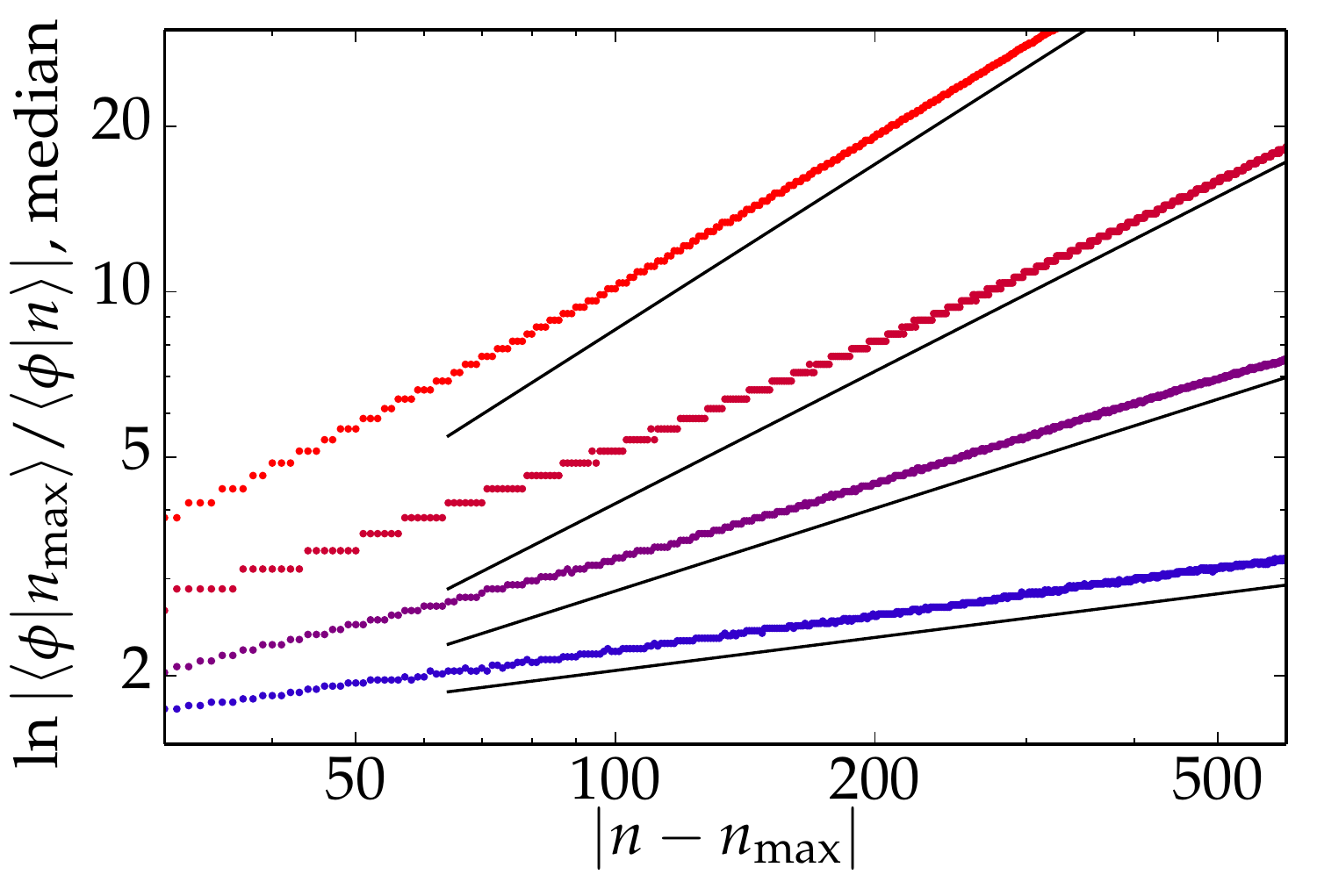}
  	\caption{Taken from \cite{cao16loc}. Numerical test of the  (stretched) exponential decay of the localized states of  \gls{bbrm} for . The prediction eq. \eqref{eq:expdecay} is plotted as lines for $\mu = 2.5, 1.8, 1.5, 1.2$ (from top to bottom). To obtain the numerical data, we diagonalize  \gls{bbrm}s of size $N \times N$, $N = 2^{11}$, with $\beta=.25$, and retain the $N/4$-th to $3N/4$-th eigenstates (ordered by eigenvalue) of each realisation. For each eigenstate $\vert \phi \rangle$, we define its localization centre  $n_{\max}$ as the site with maximum amplitude $\abs{\langle \phi \vert {n}\rangle}$.
  	We then obtain the distribution of $\ln \abs{\langle \phi \vert n_{\max} \rangle / \langle \phi \vert n \rangle}$ for each $\abs{n - n_{\max}}$, and plot the median as data points.  } \label{fig:localised}
  \end{figure}

 The alert Reader may have noticed that eq. \eqref{eq:roTsccarsup} makes only predictions in the $\beta \gg 1$ strong disorder limit, and we have tacitly extended them to \textit{all} localized eigenstates. The usual argument supporting this is a scaling/renormalization group one like that in \cite{abraham79scaling}: as long as one is in the localized phase, even when the disorder is weak ($\beta$ not large), its behaviour at sufficient large scale flows to the $\beta \to \infty$ fixed point. Moreover, in \cite{cao16loc}, we measured numerically the decay of eigenstates of  \gls{bbrm} with various values of $\mu > 1$, with $\beta = .25$, and compared to eq. \eqref{eq:expdecay} (see Figure \ref{fig:localised} for numerics details and data). The results confirm the assertion that the  predictions in the strong disorder limit do extend to small $\beta$. The same test is much less conclusive in the $\mu \in (0,1)$ regime, because not all eigenstates are localized: there is a localization transition.  We claimed this in section \ref{sec:BBRMoverview} (see Figure \ref{fig:bbrmphase}), and will start discussing it in the next section \ref{sec:LT}. 

\subsubsection{Contextual remark}
We remark that the mapping above is a new instance of the well--known interplay between polymer in random media and localization \cite{nss85hops,muller13magneto,pietracaprina16forward,tarquini16levy}. A highlight of this connection has been the relation between the conductance fluctuations in the short-range 2D Anderson model to the $(1+1)$-d \gls{kpz} \cite{kardar1986dynamic} universality class \cite{medina89anderson,somoza2007universal,somoza15anderson}. As we have seen in \ref{sec:introkpz}, the $(1+1)$-d \gls{kpz} class is believed of govern the \textit{fluctuations} of the first--passage time $T(x,y)$ in the short range Eden growth model in 2D. More precisely, denoting $x,y = r\cos\theta, r \sin\theta$, we have 
\begin{equation} T(x,y) = a(\theta) r + b r^{1/3} \chi_{TW} \,, r \to \infty \,. \label{eq:TracyWidom} \end{equation}
So the leading behaviour is linear $\propto r$,  while $a(\theta)$ is some non-universal function controlling the limit shape of the colony. The fluctuation around $a(\theta)$ has the $(1+1)$-d KPZ universal scaling $r^{1/3}$ (but $b$ is a non--universal pre--factor), and $\chi_{TW}$ is a universal Tracy--Widom distribution (see \cite{Halpin-Healy2015} for review of the \gls{kpz} universality class in $(1+1)$-d and beyond). Note that eq. \eqref{eq:TracyWidom} is another instance of the statistics Ansatz eq. \eqref{eq:evsansatz}.  In \cite{somoza2007universal}, it is shown that the log--conductance of the 2D Anderson's model in the insulator phase, as a function of the length $r$ of the sample, satisfies also eq. \eqref{eq:TracyWidom} (further observation of other universal \gls{kpz} predictions was made in \cite{somoza15anderson}). Assuming that the conductance is essentially given by the resolvent eq. \eqref{eq:resolvantexpand}, these results are to be expected in light of the mapping of this section.

For the long--range model considered here, it is the leading behaviour that becomes non--trivial functions of $n$, as given by $T_n$ in Table \ref{table:summary}. The fluctuation laws are not known yet.
  
\subsection{Localization transition}\label{sec:LT}
The existence of localization transition and extended state in \glspl{bbrm} with $\mu \in (0,1)$ is the main claim of \cite{cao16loc}. In this section, we shall support the claim with a first, heuristic, argument based on long--range percolation, and with numerical simulations. In section \ref{sec:bbrmgen}, another more general argument will be presented. 

\subsubsection{Long--range percolation argument}
This argument, presented in \cite{cao16loc} (main text and Supplemental Material, section D), concerns the \gls{bbrm} in the strong disorder $\beta \gg 1$ regime. It starts with the observation that the grand canonical partition function eq. \eqref{eq:ZofBBRM2} is an infinite series that can diverge. Indeed, for any pair $m,m'$, the sub-series made up back--and--forth paths $(0, m,m',m,m',\dots,m,m',n)$, $\ell = 1,2,\dots$, which is
$Q_{0m} Q_{m,m'}Q_{m'n}\lambda^{-3} \sum_{k=0}^{\infty} (Q_{mm'} / \lambda)^{2k}$, diverges when   \begin{equation} Q_{mn} > \lambda =  e^{-\beta \tau} \Leftrightarrow \tau_{mn} < \tau \,,\label{eq:resonancecondition} \end{equation}
see eq. \eqref{eq:ZofBBRM}. Such a divergence is associated with condensation phenomena in statistical physics. At the onset of divergence $Q_{mn} = \abs{\lambda} -\varepsilon$, the ``polymer'' tends to occupy infinite number of times on the link $m \leftrightarrow m'$.  Beyond that point, the statistical mechanics model is non--physical. However, from the quantum mechanics (matrix eigenstates) point of view, such divergence should be re-summed and interpreted as a \textit{resonance} between the sites $m$ and $m'$. From \eqref{eq:Pexy} and \eqref{eq:resonancecondition}, one sees that the probability of such resonance is 
\begin{equation} p_{nm} = 1 - e^{-\tau / \abs{n-m}^{\mu + 1}} \sim \frac{\tau}{\abs{n-m}^{\mu + 1}} \,,\, \abs{m-n} \to \infty \,. \end{equation}
The random graph made of resonating edges defines thus a long--range percolation problem \cite{schulman83lrp}. 

When $\tau = 0$ ($\lambda = 1$), the graph is completely disconnected; when $\tau = \infty$, the graph is completely connected. In the limit of large systems, a sharp percolation transition is expected at some $\tau_c$. When $\tau < \tau_c$, we have the ``insulating'' phase, where connected clusters of the graph are of finite size. When $\tau > \tau_c$, we have the ``percolating'' phase, where infinite clusters appear. A necessary criterion for the existence of the percolating phase at finite $\tau$ is the presence of a resonance crossing any site $i$, of the system. This probability is given by 
$$ p_i =1 - \prod_{n < 0} \prod_{m > 0} (1 - p_{nm}) = 1 - \exp(-\tau \sum_{\ell > 0} \ell^{-\mu}) $$ where $\ell = n-m$. So when $\mu > 1$, the sum over $\ell$ is convergent and $p_i < 1$, which makes percolation impossible (because one needs to accomplish \textit{infinitely} many conditions with probability $p_i < 1$). When $\mu < 1$, on the other hand, $p_i = 1$, and a percolation transition can occur; its existence was indeed proven in \cite{aizenman1986lrp,newman1986lrp}. 

Now, the percolation of the graph of resonance bonds is generally associated with the de-localization of the eigenstate. Therefore, the above results indicate that extended states cannot exist when $\mu > 1$. When $\mu < 1$, there are extended states with exponentially small eigenvalues $\abs{\lambda} < \lambda_c \approx e^{-\beta \tau_c}$ as $\beta\rightarrow\infty$. Since we know from the mapping to  \gls{fpp} that the eigenstates with $\lambda \sim O(1)$ (as $\beta \to \infty$) are localized, we expect a localization transition with mobility edges, at position 
\begin{equation} \lambda_c \approx e^{-\beta \tau_c} \,, \beta \gg 1 \,. \label{eq:mobilityedge}\end{equation} 

This argument is quite heuristic, because percolation does not necessarily imply de--localisation. For example, in 2D short--range lattices, percolation is possible, but the Anderson model does not have an extended phase \footnote{This is true for the usual Anderson model written in section \ref{sec:Anderson}. However, Anderson transitions in 2D are possible in a wider sense, see \cite{evers2008anderson} for a review.}. However, in our case, the estimate eq. \eqref{eq:mobilityedge} captures qualitatively the phase diagram seen in the numerics (see below): the extended phase is indeed in the middle of the spectrum, and its size shrinks (rapidly) as $\beta$ increases. 
 
In the hindsight, we remark that a relation between  \gls{bbrm} and long--range percolation is not a surprise, because the latter is known to be closely related to the long--range  \gls{fpp} model that inspired our definition of the \gls{bbrm}. Indeed, it has been shown \cite{biskup2004,biskup11diameter} that, if $0$ and $n$ are two lattice points that are connected in the long--range percolation model, then the formulas of $T_n$ in Table \ref{table:summary} describe also the asymptotic behaviour of the shortest path connecting them on the random graph. This observation will also inspire the generalization of the \gls{bbrm} results to more general random matrices, in section \ref{sec:bbrmgen}.

\subsubsection{Numerical study of the localization transition}
There are two ways of numerically studying the localization transition: \texttt{i.} studying the eigenstates themselves, and \texttt{ii.} studying the \textit{level statistics} of the eigenvalues. 
 
   The most common observable of the eigenstates is the (generalized) \acrfullpl{ipr}. Recall that for a normalised state $\phi$, it is defined as
   \begin{equation}
   P_q =  \sum_{n=1}^N \abs{\langle \phi\vert n \rangle}^{2q} \,.
   \end{equation}  
 It is the same definition as in eq. \eqref{eq:defIPR} in section \ref{sec:multifracintro}, if we identify the Gibbs measure $p_{\beta,n}$ with the occupying probability $\abs{\langle \phi\vert n \rangle}^2$.. The asymptotic behaviours (as $N \to \infty$) of $P_q$ the extended and localized phase are
 \begin{equation}\label{eq:Pqphases}
  P_q \sim \begin{dcases}
   N^{0} & \text{localized phase} \,, \\ N^{1 - q}  & \text{extended phase} \,.
  \end{dcases}
 \end{equation}
 One can see this by studying the extreme cases. The most localized state is a single site state $\vert \phi \rangle = \vert n_0\rangle$, for which $P_q = 1$ for any $q$. The most de-localized state is uniformly distributed, $ \vert \phi \rangle = N^{-1/2} \sum_{n=1}^N e^{\im \theta_n} \vert n \rangle$, for which $P_{q} = N^{1 - q}$ exactly. At the localization transition, $P_q \sim N^{-\tau_q}$ for some non-linear exponent function $\tau_q$ characterizing the multi--fractal properties of the critical eigenstate \footnote{See section \ref{sec:multifracintro} for more information on this. In this respect, the Gibbs measure of \glspl{logrem} is ``critical'' at all temperature!}. In what follows, the term \gls{ipr} refers to the case $q = 2$, on which we focus exclusively.
  
  \begin{figure}
   \center	\includegraphics[scale=.5]{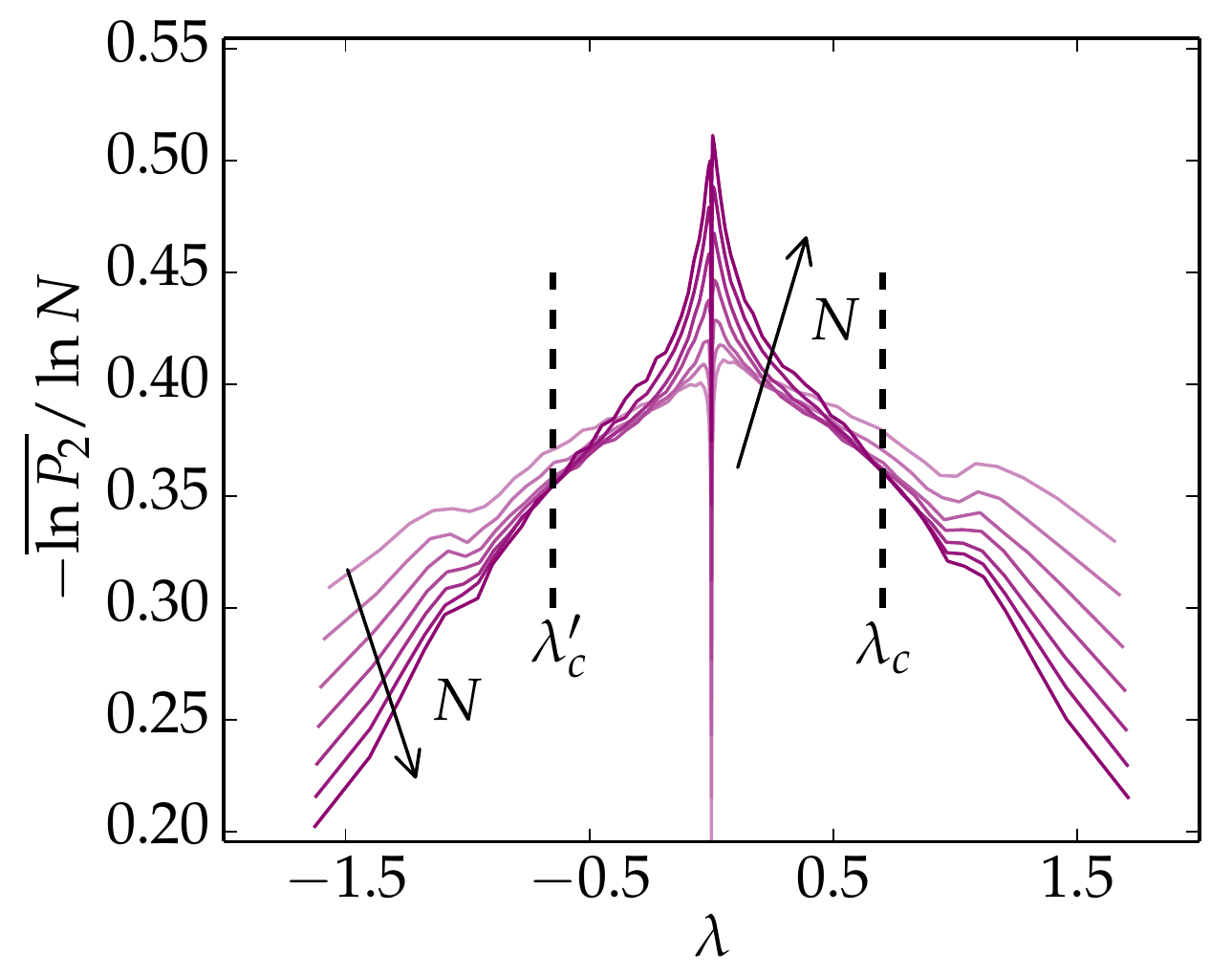} 
   \includegraphics[scale=.5]{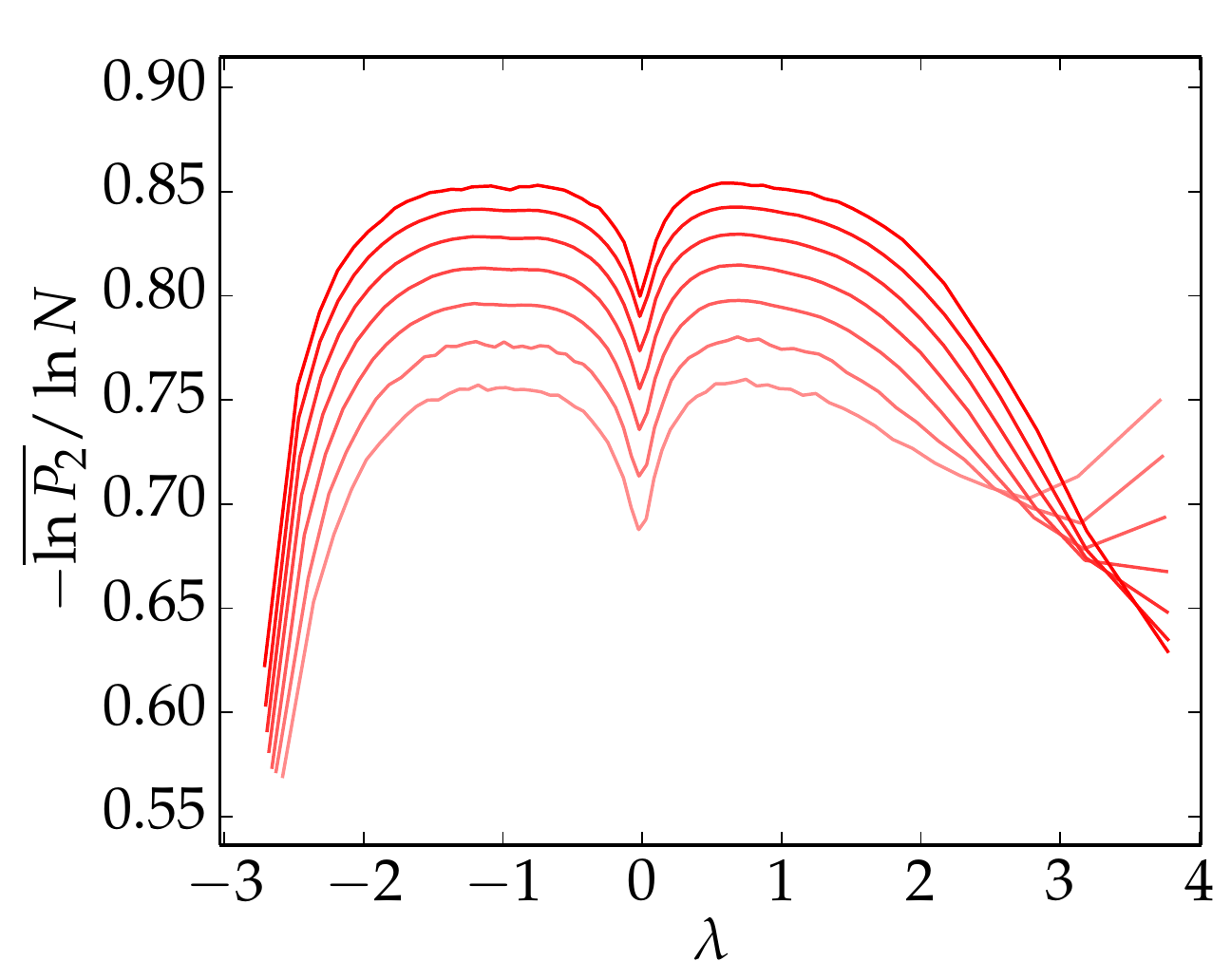}
   \caption{Taken from \cite{cao16loc}. Measure of the \gls{ipr} of  \gls{bbrm} eigenstates as function of eigenvalue $\lambda$, at $\mu = 0.5$ and with various size of matrices, $N = 2^7, \dots, 2^{14}$ (darker colours indicate larger systems). According to eq. \eqref{eq:Pqphases}, extended (localised) states would have $y$ coordinate $-\overline{\ln P_2}/\ln N \to 1$ ($\to 0$, respectively) when $N \to \infty$. \textit{Left panel}: $\beta = 5$. Dashed lines indicate the mobility edges, estimated at $\lambda_c \approx -0.65(5)$ and $\lambda_c \approx 0.70(5)$; they are not accurate estimates, because of the pronounced correction to scaling effect: the crossing of curves move to the outside when $N$ increases. Taking into this effect, a crude estimate of the critical \gls{ipr} is $\tau_c = \left[ \overline{\ln P_2} / \ln N \right]_{\lambda_c} \approx 0.36(5)$. \textit{Right panel}: The same measure for $\mu = 0.5$, $\beta = 1$. As $N$ increases (darker colour), the eigenstates with $\lambda \in (-3,3)$ becomes more and more de-localised. } \label{fig:IPR}
  \end{figure}
\begin{figure}
 	\center
 	\includegraphics[scale=.6]{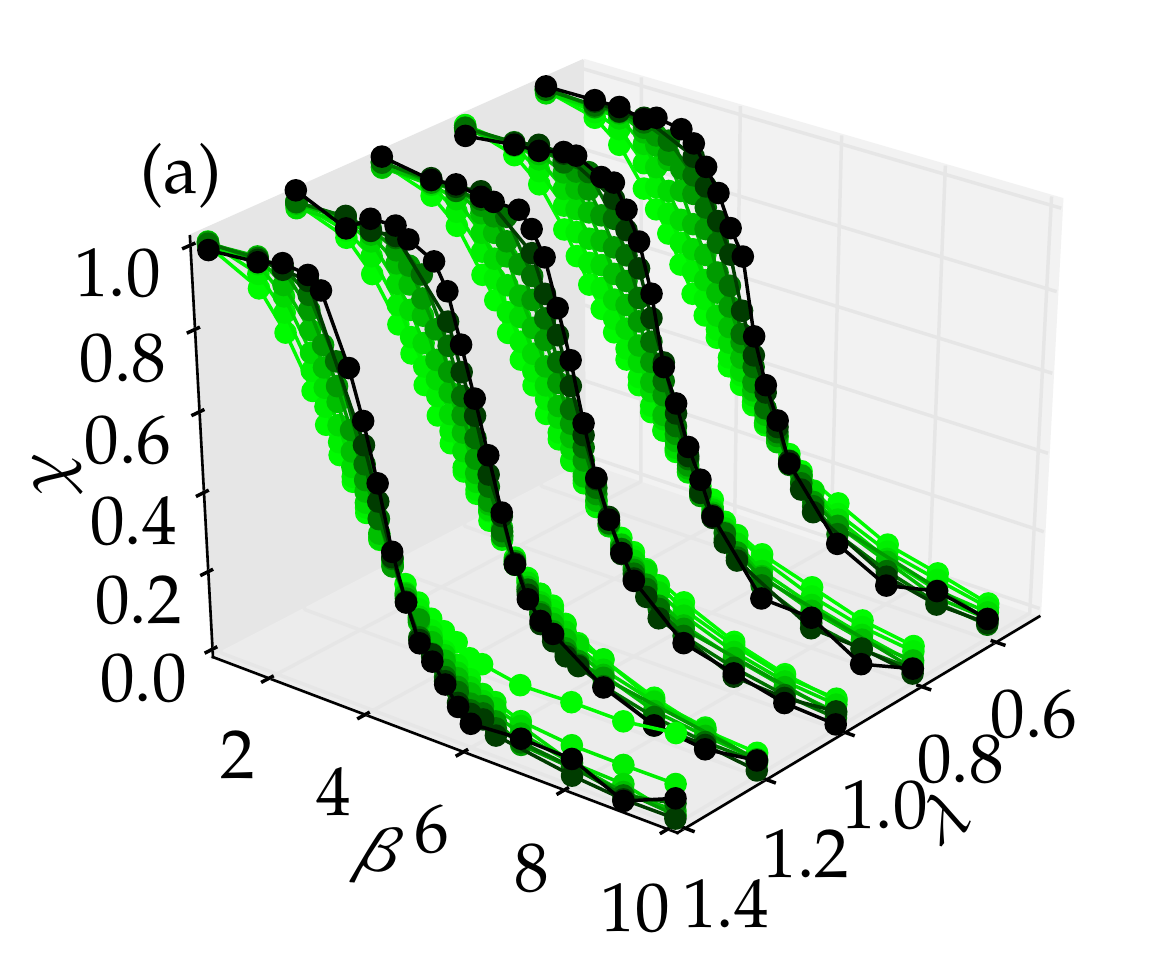}
    \includegraphics[scale=.5]{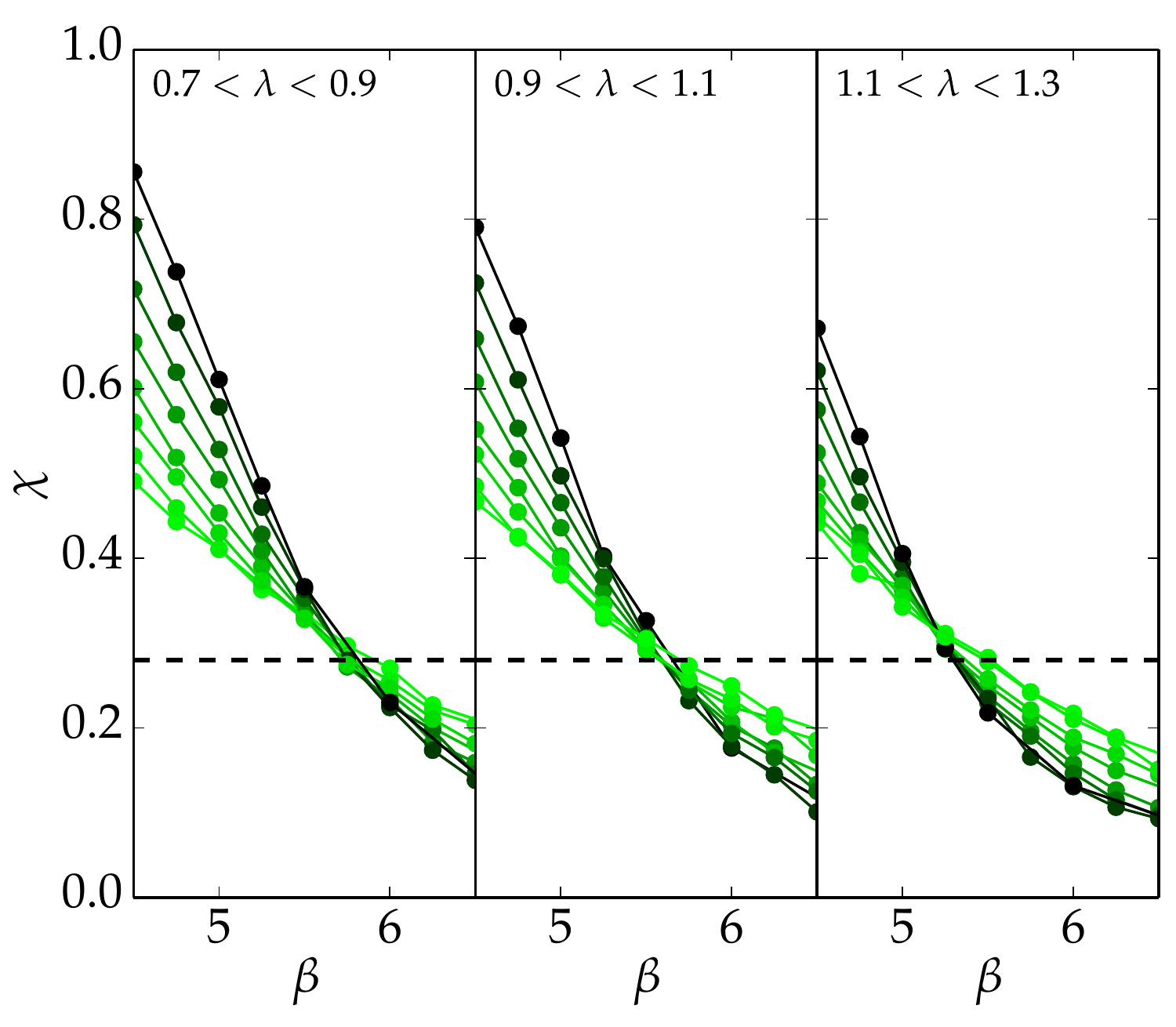}
 	\caption{Taken from \cite{cao16loc}. \textit{Left Panel}. Numerical measure of ratio $\chi$, Eq. (\ref{eq:defchi}) for the  \gls{bbrm} model with $\mu = 0.5$, $\beta \in (1,10)$ and of sis $N = 2^7$ (light color) to $2^{14}$ (dark color). Eigenvalues in $(0.5, 1.5)$ are binned into $5$ bins of equal width. More than $10^5$ different gaps are averaged over for each data point.  \textit{Right panel}. A zoom-in to the critical regime. We locate the critical value of the rescaled gap ratio observable $\chi_c \approx .28(2)$.}\label{fig:transition}
 \end{figure}

In Figure \ref{fig:transition} (Left panel), we show the numerical measure of the \gls{ipr} $P_2$ for the  \gls{bbrm} ensemble with $\mu = 0.5$, $\beta = 5$. For this we generate samples of  \gls{bbrm}, and numerically diagonalize them (using \textsf{LAPACK} wrapped in the \textsf{numpy} package).  We observe clear mobility edges  separating extended states in the middle of the spectrum (containing $\lambda = 0$) and localized states near the edges. 
We considered other values of $\beta$ than $5.0$. For larger values, the results are qualitatively the same, but the extended phase shrinks to the point $\lambda = 0$. For $\beta \approx 1$ (see Figure \ref{fig:transition}, Right panel, the extended phase expands rapidly, and we cannot decide whether the mobility edges disappear or are located near the edge of the spectrum. Thus, we cannot exclude the existence of a value $\beta_{\inf}(\mu)$ such that all eigenstates are extended when $\beta < \beta_{\inf}$.

Now we turn to the level statistics. Because of the non-trivial shape of the \gls{dos} (see Figure \ref{fig:DoS}), the most suitable observable is the \textit{gap ratio} proposed in \cite{oganesyan07huse}. Denoting $\lambda_1 \leq \dots  \leq \lambda_N$ the ordered eigenvalues of a matrix, and $\delta_i = \lambda_{i+1} - \lambda_i$ the level spacings (gaps), we consider the following ratio between successive gaps 
\begin{equation} r_i = \min(\delta_i,\delta_{i+1})/\max(\delta_i,\delta_{i+1}) \,. \end{equation}
The advantage of this observable is that it cancels out the dimension of energy, and does not depend on the DoS. Its mean value is universal in localized and extended phases \cite{atas13ratio}
\begin{equation}
\overline{r} = \begin{dcases}
 r_P  = 2 \ln - 1 \approx 0.39   & \text{localized phase} \,, \\
 r_{GOE} \approx 4 - 2\sqrt3 \approx 0.53 \,,  & \text{extended phase}.
\end{dcases}
\end{equation} 
The localized value $r_P$ comes from Poisson level statistics, while in the extended phase, $r_{GOE}$ is the value of the GOE ensemble; its approximate and numerical values were studied in \cite{atas13ratio}. It is then convenient to define the rescaled gap ratio
 \begin{equation} \chi \defeq \frac{\overline{r} - r_{P}}{r_{GOE} - r_P} \Rightarrow \chi \rightarrow \begin{dcases} 0 &  \text{localized phase / Poisson}, \\ 1 & \text{extended phase / GOE}. \end{dcases} 
 \label{eq:defchi}\end{equation} 
 
We measured this quantity for  \gls{bbrm} with $\mu = 0.5$ for several values of $\beta \in [1,10]$, and for a few windows of $\lambda$ of width $0.2$. As we can see in Figure \ref{fig:transition} (left panel), $\chi$ goes from the GOE (extended) value to the Poisson (localised) value when $\beta$ increases from $1$ to $10$, and the change becomes sharper as the system sis increases. In the right panel of Figure \ref{fig:transition}, we look more closely at $\beta \sim 5.5$ and three windows of eigenvalues to examine the critical region. We observe that the critical value $\beta_c$ depends clearly on energy $\lambda$, going from $\beta_c(.7 < \lambda < .9) \approx 5.8$ to $\beta_c(1.1 < \lambda < 1.3) \approx 5.2$. This means that the function $\beta_c(\lambda)$ is non--trivially decreasing; inverting this function, we conclude that for $ \beta \approx 5.2$ ($5.8$), the  \gls{bbrm} model has a mobility edges at $\lambda_c(\beta) \approx 1.3$ ($0.8$, respectively). Moreover, we remark that the critical value of the rescaled gap ratio $\chi_c \approx .28(2)$ is independent of $\lambda$; this indicates that for a given $\mu \in (0,1)$, there is one unique critical point of localization transition. 

However, notice that this estimate does \textit{not} agree quantitatively with the \gls{ipr} estimate, see Figure \ref{fig:IPR}. In fact, we observe that $\abs{\lambda_c^{\text{IPR}}} < \abs{\lambda_c^{\text{ratio}}}$, \textit{i.e.}, there is a critical regime which seems to be localized according to \gls{ipr} but extended according to level statistics. Such a discrepancy has been observed in other matrix models, like the Lévy random matrix \cite{cizeau1994levy,tarquini16levy} and the Anderson model on the Bethe lattice \cite{biroli2012difference}. In the latter case, this turns out to be a signature of a ``critical'' phase \cite{deluca14bethe,altshuler2016rrg}, which is neither extended nor localised, and in which the eigenstates are multi--fractal. In the Lévy matrix case, such discrepancy is due to large finite size effect that was carefully taken into account very recently in \cite{tarquini16levy}, in which the previously (\cite{cizeau1994levy}) conjectured critical phase is convincingly ruled out. In light of these lessons, we shall refrain from advancing any statement concerning critical phase in \glspl{bbrm}.

To conclude this section, we comment on another banded random matrix model in the literature that exhibits mobility edge: the one with \textit{non--random} hopping, studied in \cite{rodriguez2003anderson}. By definition, its off diagonal elements are deterministic, positive, and depend only on the distance to diagonal, in an algebraic manner: in 1D,  $H_{mn} = \abs{m-n}^{-\mu - 1}$, for $\abs{m-n} \geq  1$. On the diagonal, one has \gls{iid} elements, uniformly distributed in $[-\beta, \beta]$, where $\beta$ is the strength of the disorder. It is shown in \cite{rodriguez2003anderson} that when $\mu < \frac12$, and when $\beta$ is small enough, the matrix ensemble has a mobility edge. In this sense, this model can be seen as the first long--range banded random matrix ensemble to overcome the pathology of \gls{pbrm}. Nevertheless, the de--localization phase in this model is quite peculiar: the extended states are situated near the lower edge of the spectrum (rather than in the middle). The reason for this (explained in \cite{rodriguez2003anderson}) is that, for the pure ($\beta = 0$) system, the level spacing near the low edge of the spectrum scales as $\delta \lambda \sim N^{-\mu}$ (this can be seen by solving the pure system by plane waves). At weak disorder ($\beta \ll 1$), the matrix elements of the their perturbation scale as $N^{-\frac12}$. When $\mu < \frac12$, as $N \to \infty$, the plane--waves with longest wave--lengths are thus too far away from the rest of the spectrum to be perturbed. Therefore, these extended states are localized in the momentum (Fourier) space, and are distinct from the \textit{ergodic} extended states (which are extended in every basis) in the Anderson model. The level statistics in the extended phase is also different from the GOE ensemble. In summary, the localization transition in the model of \cite{rodriguez2003anderson} should not be confused with the one unveiled in this thesis.

\section{Broadly distributed banded matrices} \label{sec:bbrmgen}
In the previous section we have considered the  \gls{bbrm} model, whose matrix elements have a specific distribution (eq. \eqref{eq:Qdefsupp1}), which is inspired by the epidemics model. Its most interesting property is the existence of localization transition in the regime $\mu \in (0,1)$. This begs an important question: can we generalize this result to more general banded matrix models? This section proposes an answer to this question, advanced in \cite{cao16loc}: \vspace{.3cm}

\fbox{\begin{minipage}{.9\textwidth}
\begin{center}
A localization transition prevails in the $\mu \in (0,1)$ regime provided $Q_{nm}$ is a banded random matrix with \textit{broadly distributed} elements. 
\end{center}
\end{minipage}}\vspace{.3cm}

In section \ref{sec:broad}, we will define the term ``broadly distributed'', and discuss the crucial \textit{self--averaging property}. Then, section \ref{sec:RG} will discuss the arguments of \cite{cao16loc} supporting the above claim. By construction, these arguments apply to any ensemble of banded random matrices with broadly distributed elements; the class of these matrix models will be called the \textit{broadly distributed} class. 

\subsection{Broad distributions and self--averaging}\label{sec:broad}
  The elements of a random banded matrix \footnote{Throughout, we still assume the matrix is real symmetric, and the elements are uncorrelated except $Q_{mn} = Q_{nm}$.} is called \textit{broadly distributed} if
  \begin{subequations}\label{eq:broaddefall}
  \begin{equation}
  \overline{\abs{Q_{mn}}^k} \stackrel{g\gg1}\sim  \frac{c(k)}{g} \Leftrightarrow  \ln \overline{e^{-t \abs{Q_{mn}}}} \stackrel{g\gg1}=  - \frac1g f(t) + O(1/g^2)  \,,\,  g = \beta \abs{m-n}^{\mu+1} . \label{eq:broaddef}
  \end{equation}
  where $f(t) = \sum_{k>0} (-1)^{k-1} c(k) t^k / k!$ is some function \textit{independent} of $g$ that grows at most logarithmically 
  \begin{equation} f(t \rightarrow  +\infty) = O(\ln t) \,. \label{eq:ftlog}\end{equation}
    \end{subequations}
  The first condition eq. \eqref{eq:broaddef} is essential, while eq. \eqref{eq:ftlog} is a technical criterion. As we shall see, it is not at all restrictive, and allows some arguments to be technically easier.

  Let us check that the \gls{bbrm} model fits into the definition. Indeed, by eq. \eqref{eq:moments}, we have more explicitly
  \begin{equation}
  f_{\textbf{B}} (t)= \log (t)+\Gamma (0,t)+\gamma_E \,,\, c(k) = 1 / k \,, \label{eq:ftBBRM}
  \end{equation}
  where $\gamma_E = 0.57\dots$ is the Euler constant and $\Gamma(0,t) = \int_{t}^{+\infty} \dif y e^{-y} / y$ is exponentially decaying at $t$. So eq. \eqref{eq:ftlog} is also satisfied. Since the criterion eq. \eqref{eq:broaddefall} concerns only the absolute value $\abs{Q_{mn}}$, it is also satisfied by the  \gls{bbrm} with \textit{random signs} $(\epsilon_{mn} Q_{mn})$, where $\epsilon_{mn} \in \{\pm1\}$ are independent signs.
  
  Another important example is the \textit{randomised sparse matrices} associated to long range percolation \cite{ayadi2009semicircle,ayadi2009asymptotic}. In this case, the \gls{pdf} of $Q_{mn}$ is 
  \begin{equation}\label{eq:sparse}
  P(Q_{mn}) = g^{-1} P_0(Q_{mn}) + \left(1 - g^{-1}\right) \delta(Q_{mn}) \,,\,
  \end{equation}
  where $P_0(x)$ is the \gls{pdf} of a standard Gaussian or a uniform distribution. In other words, $Q_{mn} = 0$ with probability $1-1/g$, and is an $O(1)$ random variable with fixed distribution otherwise. Then, eq. \eqref{eq:broaddefall} is satisfied with 
  \begin{equation} f(t) =  \int \dif v (1-e^{-t \abs{v}}) P_0(v)  \leq 1 \ll \ln t \,. \end{equation}
   Moreover, eq. \eqref{eq:broaddef} is satisfied without $O(1/g^2)$ correction. In this respect, the definition eq. \eqref{eq:broaddefall} is essentially that of a relaxed \textit{sparseness}, which does not require the majority of the elements to be exactly zero, but sufficiently small so that the moments come from the sparse minority.
  
 Now we proceed to show that the other properties of  \gls{bbrm} matrix elements discussed in section \ref{sec:basics} can be reproduced by the criterion eq. \eqref{eq:broaddefall}.
  For this we denote $Q = \abs{Q_{mn}} \geq 0$. Then the Laplace transform of $Q$ can be written as the cumulative distribution of $\ln Q$ convoluted with an random variable $\mathsf{Gum}$ drawn from the standard Gumbel distribution, independent of $Q$ (compare to eq. \eqref{eq:Gasconvolution}): 
  \begin{equation}
  \overline{\exp{(-t Q)}} = \overline{\exp(-\exp(\ln Q + \ln t))} = \mathbb{P}(-\ln Q - \mathsf{Gum} > \ln t)
  \end{equation} 
  where $\theta(x)$ is the Heaviside step function. Now eq. \eqref{eq:broaddef} with $t = 1$ implies 
  \begin{equation}
  \mathbb{P}(-\ln Q - \mathsf{Gum} < \ln t) = f(t) / g + O(1/g^2) \,. \label{eq:cdf}
  \end{equation}
  Since $\mathsf{Gum} \sim O(1)$, this implies $Q \sim O(1)$ with probability $\propto 1/g$. So the ``black swan'' property, eq. \eqref{eq:blackswan}, is common to all broadly distributed banded matrices. 
  
  Now we look at the typical magnitude of $Q^{\text{typ}}$. It can be estimated by the value $\ln t$ for which the cumulative $\mathbb{P}(-\ln Q - G  < \ln t) = 1/2$, namely when $f(t) = g/2$ by eq. \eqref{eq:cdf}. This gives $-\ln Q \sim y \sim \ln f^{\text{inv}}(g)$ and 
  \begin{equation} Q^{\text{typ}} \sim 1 / f^{\text{inv}}(g) < \exp(-cg) \,. \label{eq:Qtype}\end{equation} 
  where $f^{\text{inv}}$ is the inverse function of $f$ which grows at least exponentially according to eq. \eqref{eq:ftlog}. So the only purpose of this technical condition is to guarantee that the typical elements are at least exponentially small.
  
  \subsubsection{Self--averaging property}
  For broadly distributed $Q_{mn}$, the typical value is very small compared to its moments when $g = \beta \abs{m-n}^{\mu+1}$ is large. On the other hand, since all moments exist, for any fixed $g$, the central limit theorem applies. That is, if $Q_i, \dots,  Q_M$ are $M$ independent copies of $\abs{Q_{mn}}$, the sum
     \begin{equation}
     S \stackrel{\text{def}} = \sum_{i=1}^M Q_i  \label{eq:defS}
     \end{equation}
  tends to a Gaussian as $M\rightarrow\infty$ (after proper rescaling), whose moments and typical value are the same. So, when does the crossover happens? 

   The answer is $M \sim g$. In light of eq. \eqref{eq:blackswan}, this is intuitive since this is when the rare event $Q_{mn} \sim O(1)$ begins to occur for one of $Q_1, \dots, Q_M$. Indeed, the distribution of $S$ has a well-defined limit $S_T$ when $M,g\rightarrow\infty$ with $M/g = T$ kept constant. For this, recall the expansion $\overline{\exp(- t Q_{mn})} = 1 - f(t) / g + O(1/g^2)$, Eq. \eqref{eq:broaddef}, which implies that:
   \begin{equation}
   \overline{\exp(-t S)} =  \left[\overline{\exp(- t Q_{mn})}\right]^{M} \stackrel{}\longrightarrow \exp(- T f(t))\,,\, M=Tg\rightarrow\infty\,.\label{eq:limitdis}
   \end{equation}
   So, the distribution of $S$ has a limit $S_{T}$ depending on $T$, given in terms of Laplace transform
   $\overline{ \exp(-t S_{T}) }=  \exp(- T f(t))  \,. $
   For  \gls{bbrm}, by \eqref{eq:ftBBRM}, the cumulants of $S_T$ have a simple form:
   \begin{equation} \overline{S_{T}^k}^c = (-1)^{k-1} T \frac{\dif^k f}{\dif t^k}(0) = T / k \quad \text{(BBRM)}. \end{equation}
   From \eqref{eq:limitdis} we conclude that when $M \sim g$, the distribution of the sum becomes $g$-independent in the $g\rightarrow\infty$ limit. Therefore, when $M \gg g$, we enter the central limit theorem regime, in which $S / \sqrt{M / g}$ tends to a Gaussian. On the other hand, when $M \ll g$, $T \ll 1$, eq. \eqref{eq:limitdis} implies $\overline{ \exp(-t S_{T}) } \sim 1 - T f(t) + O(T^2)$, so the sum $S_T$ becomes itself broadly distributed, with $1 / T$ playing the rôle of $g$.
   
 The reason for which we consider this problem is that in the following section, we will encounter more complicated sums of type  $\sum_{i=1}^M Q_i c_i$ where $c_1, \dots, c_M$ are the pondering coefficients; in this respect, the above considerations are a useful warm--up in which $c_i = 1$.
   
\subsection{General argument for localization transition} \label{sec:RG}
 In this section, we consider a general banded matrix $Q_{mn}$ with broadly distributed elements in the sense of eq. \eqref{eq:broaddefall}, for some $\mu \in (0,1)$; in particular, our analysis applies to  \gls{bbrm} and randomised sparse banded matrices (eq. \eqref{eq:sparse}). Our main goal is to argue that there is a localization transition. The method is directly inspired by that used in the epidemics/\gls{fpp} model \cite{hallatschek2014acceleration,chatterjee2016multiple} considered in section \ref{sec:stretch}.
 
 \begin{figure}[h]
 \center \includegraphics[scale=.6]{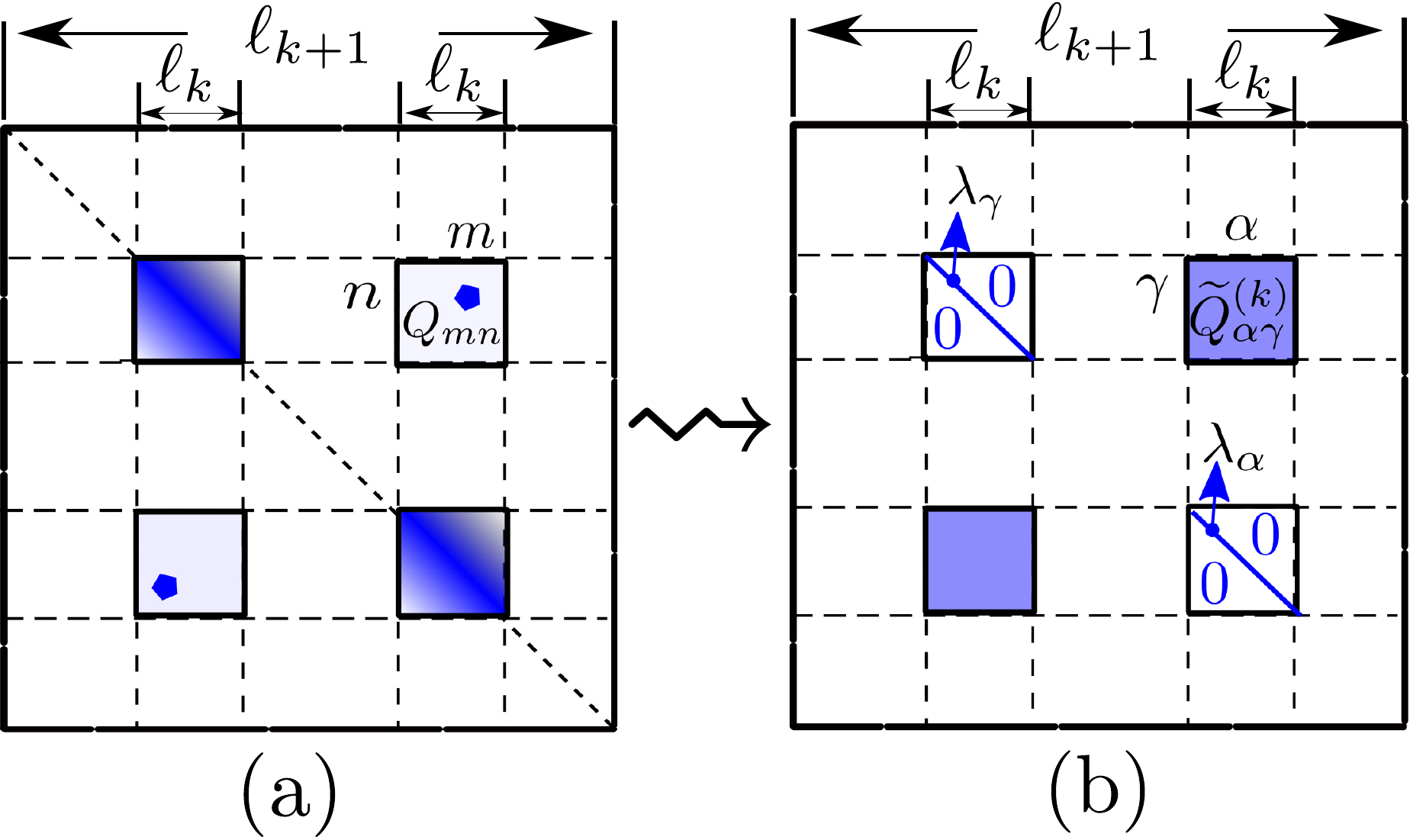}
  \caption{An illustration of the block diagonalization procedure. (a) depicts a diagonal block of size $\ell_{k+1} \times \ell_{k+1}$, which is divided in into block--matrices of si $\ell_k \times \ell_k$. We pick two blocks of sites labelled by $m$ and $n$ respectively. The diagonal $\ell_k$--blocks are banded matrices (represented as squares with gradient--colour). The choice of the scales (eq. \eqref{eq:lk}) ensures that the off--diagonal block has a few large elements $Q_{mn} \sim O(1)$ (represented as a dot) while the typical elements are negligibly small. (b) depicts the matrix $ \widetilde Q^{(k)}_{\alpha \gamma}$ (eq. \eqref{eq:Qtransfred}) after diagonalizing the $\ell_k$--diagonal blocks, so the latter become diagonal. Off--diagonal, the unitary transformations smear out the sparseness of $Q_{mn}$, so that the transformed elements $\widetilde Q^{(k)}_{\alpha \gamma}$ are no longer broadly distributed. }\label{fig:block}
 \end{figure}
Referring to Figure \ref{fig:block} for an illustration, the basic idea is to take a sequence of lengths $\ell_0 < \ell_1 < \dots < \ell_k \ll \ell_{k+1} \ll \dots $, which will be determined later. At step $k$, we divide the matrix $Q$ into blocks of size $\ell_k \times \ell_k$ and let 
$\vert \phi_{\alpha}^{(k)} \rangle$ be the eigenstates of the matrix of diagonal blocks. In that basis, the transformed matrix elements are
\begin{equation}\label{eq:Qtransfred}
\widetilde Q^{(k)}_{\alpha \gamma} =
\sum_{(n,m)} \langle\phi_\alpha^{(k)} \vert n\rangle  Q_{nm} \langle  m \vert   \phi_{\gamma}^{(k)}\rangle \,,
  \end{equation}
 where the sum over all pairs $(n,m)$ where $n$ ($m$) is in the block of $\alpha$  ($\gamma$, respectively). Thus, eq. \eqref{eq:Qtransfred} is a sum of $M = \ell_k^2$ terms, and can be re--written as:
  \begin{align}
 & S[c_i] = \sum_{i=1}^{M} Q_i c_i  \text{ where } M = \ell_{k}^2  \nonumber \\
 & \left\{ c_1, \dots, c_M \right\} =
  \left\lbrace\langle\phi_\alpha^{(k)} \vert n\rangle  \langle  m \vert   \phi_{\gamma}^{(k)}\rangle \right\rbrace 
  \label{eq:defQtilde}
  \end{align}
  are the $M$ coefficients in eq. \eqref{eq:Qtransfred}, and $ Q_i, i = 1, \dots, M $ are the elements $Q_{mn}$ in eq. \eqref{eq:Qtransfred}. For simplicity, we approximate all the distances by the maximal value  $\abs{m-n} = \ell_{k+1}$, so that $Q_i$'s have identical distribution. Observe that the coefficients $c_i$ depend on the diagonal blocks, so are independent from the $Q_i$ in an off--diagonal block, see Fig \ref{fig:block}. 
  
 The length scales $\ell_k$ are chosen to ensure that in each off--diagonal block, there is at least one black swan $\abs{Q_{mn}} \sim  O(1)$. This amounts to requiring 
  \begin{equation} \ell_k^2 = M = \beta \ell_{k+1}^{\mu+1} \,,\label{eq:Misg} \end{equation} which gives a sequence of length scales
 \begin{equation}
 \ln (\ell_k / \xi) = \left(\frac{2}{1+\mu}\right)^{k} \ln (\ell_0 / \xi), \quad \xi \defeq \beta^{\frac{1}{1-\mu}} \,.  \label{eq:lk}
 \end{equation} 
 Note that $\ell_k$ is an increasing series if and only if $\mu < 1$ and $\ell_0 > \xi$. We shall fix $\ell_0$ that is a few times $\xi_0$. Now, for $\beta\ll 1$, all elements in the $\ell_0$ blocks are of $O(1)$, and the iteration starts with $\vert \phi_{\alpha}^{(0)} \rangle$ which are extended. On the other hand, if $\beta \gg 1$, the block eigenstates will be localised at step $0$. Our goal is to carry each of the starting situations through the iteration, by considering them separately.
 
  \subsubsection{Extended case}
  Let us assume that at step $k$, the block eigenstates $ \phi_\alpha^{(k)}$ in eq. \eqref{eq:defQtilde} are extended. Then, the amplitudes $\abs{\langle n \vert \phi_\alpha^{(k)}} \rangle \sim  \sqrt{1/\ell_{k}}$ independently of $n$. Therefore, it is reasonable to assume $c_1, \dots, c_M$ in eq. \eqref{eq:defQtilde} as identically distributed independent random variables, whose distribution is as $v / \ell_k$, where $v$ is some fixed distribution. Then, using eq. \eqref{eq:broaddef} and then eq. \eqref{eq:Misg}, we have
    \begin{align}
    &\overline{\exp(-t S[c_i])} = \left(\overline{\exp(-v t Q / \ell_k )}\right)^M = 
    \left( 1 - g_{k+1}^{-1} \int  f(tv / \ell_k) P(v) \dif v  + O(g_{k+1}^{-2}) \right)^M \nonumber \\ 
  &  \stackrel{M=g_{k+1}}{\longrightarrow} \exp(- f_1(t / \ell_k)) \,,\, f_1(s) =  \int f(t v)  P(v)\dif v  \,.  \label{eq:etSextended}
    \end{align}
  This means that $S[c_i] = S_0 / \ell_k$ where $S_0$ is some fixed random variable (with Laplace transform $\overline{\exp(-tS_0)} = \exp(-f_1(t))$). Therefore, all transformed matrix elements have typical magnitude $\widetilde Q^{(k)}_{\alpha\gamma} \sim 1/\ell_k$: the matrix $\widetilde Q^{(k)}$ is no longer sparse or broadly distributed but is rather like the \gls{pbrm}. This allows us to employ the resonance argument in section \ref{sec:pbrm}, and compare $\widetilde Q^{(k)}_{\alpha\gamma} \sim 1 / \ell_k$ to the typical level spacing on scale $k+1$, which is $\delta_{k+1} \propto 1/\ell_{k+1}$ (because the eigenvalues are of order unity and there are $\ell_{k+1}$ of them). Since $\ell_k \gg \ell_{k+1}$, we conclude that $\ell_{k+1}$ are still extended. Repeating this procedure \textit{ad infinitum} we conclude the eigenstates are extended when $\beta \ll 1$. 
  
 We stress that the block diagonalization procedure is essential for showing the existence of extended phase, because recall from eq. \eqref{eq:Qtype} that the \textit{typical} original matrix elements for $\abs{n-m} \sim 1/\ell_{k+1}$ are exponentially small $Q_{nm} = O(\exp(-c \ell_{k+1})) \ll 1/\ell_{k+1}$, although there are a few ``black swans'' of order unity dominating the second moment $\overline{Q_{nm}^2} \approxeq \overline{H_{nm}^2}$ ($H$ is the  \gls{pbrm} matrix); it is precisely the diagonalization at smaller scale that spreads the magnitudes evenly and reduces the situation to  \gls{pbrm} like. 
  
 The analysis above also sheds light on following numerical observations in section \ref{sec:LT} . Indeed, the de--localization of eigenstates of broadly distributed matrices depends on two ingredients: a rapidly increasing sequence of scales (eq. \eqref{eq:lk} means that $\ell_k = \exp(c' e^{ck})$ is double exponential), and the occurrence of \textit{rare} events. The former should be relevant for the strong finite size effects. The latter is also the case for the Lévy matrix \cite{cizeau1994levy,tarquini16levy}, in which one observes the similar discrepancy between eigenstate and level statistics properties in finite systems. However, the Lévy matrix is a fully--connected model whose mobility edge can be exactly calculated \cite{tarquini16levy}, so it is not obvious to how its methods and results could be adapted to the current models, which have a non--trivial spatial structure.
 
  \subsubsection{Localized case}
  Now we turn to the localized case. Again, we put ourselves at step $k$, and try to estimate the sum eq. \eqref{eq:defQtilde}, by assuming that the eigenstates $\vert \phi_\alpha^{(k)} \rangle$ are localized, with a decay rate $\abs{\langle \phi_\alpha \vert n \rangle} \sim \pm \exp(- a(\abs{n-n_\alpha}))$, where $n_\alpha$ is the localized centre (similarly for $\gamma$).
  
  Since $\abs{n-n_\alpha}$ ranges from $1$ to $\sim \ell_k$, so the coefficients $c_i$ in eq. \eqref{eq:defQtilde} have very different magnitudes, from $1$ (which occurs for few $i$'s) to $e^{-2a(\ell_k)}$ (which occurs for typical $i$'s). Since $(c_i)$ is uncorrelated from $(Q_i)$, and $\abs{Q_i} \sim O(1)$ also for a few $i$'s, the event $Q_i c_i \sim 1$ happens with vanishing probability $1/M$. So we are left with two (extreme) types of contributions to the sum \eqref{eq:defQtilde}:
  \begin{itemize}
  \item[(i)] From the terms with $c_i \sim 1$; since there are only $O(1)$ such terms, black swans in $Q_{nm}$ cannot occur (except in rare events of probability $\sim 1/M$), so we have $Q_{nm} \sim Q^{\text{typ}}$ and the total contribution of these terms is $ \mathrm{I} \sim  Q^{\text{typ}}.$
  \item[(ii)]  From the terms with typical $c_i \sim e^{-2 a(\ell_k)}$; since there are $\sim M$ of them, there will be $O(1)$ black swans $Q_{nm} \sim 1$, so the total contribution is $ \mathrm{II}\sim e^{-2 a(\ell_k)}$.
  \end{itemize} 
  Now, since our model is long--range, it is safe to assume that $a(\ell) < c \ell$ grows at most linearly. Then eq. \eqref{eq:Qtype} ($Q^{\text{typ}} < \exp(-c g_{k+1})$) implies 
  $$ \mathrm{I} < \exp(-c g_{k+1}) = \exp(-c  \ell_{k}^2) \ll \exp(-c \ell_k)  < \mathrm{II} \,. $$
   This indicates that  the sum eq. \eqref{eq:defQtilde} is dominated by the latter case
   \begin{equation}\label{eq:Qtildelocalised}
   \widetilde Q^{(k)}_{\alpha\gamma} \sim \exp(-2 a(\ell_k)) \,.
   \end{equation}
   This conclusion is further supported by more technical arguments and  a numerical check in \cite{cao16loc} (Supplemental Material, C.2, and Appendix 1). 
  
   Now we apply eq. \eqref{eq:Qtildelocalised} to estimate the decay of eigenstates of generation $k+1$, at first order in perturbation theory. Indeed, eq. \eqref{eq:1storderperturb} gives 
   $$ e^{-a(\ell_{k+1})} = \langle m  \vert \phi^{(k+1)}_\alpha \rangle = \sum_{\gamma} \frac{\widetilde Q^{(k)}_{\alpha\gamma}}{\lambda_{\alpha}- \lambda_{\gamma}} \langle m \vert \phi^{(k)}_\gamma \rangle + \dots  $$ 
   where the site $m$ belongs to the $\ell_{k}$ block whose eigenstates do not contain $ \vert \phi^{(k)}_\alpha \rangle$ but the states $ \vert \phi^{(k)}_\gamma \rangle$.
   We note that since the states $\phi^{(k)}_\gamma$ are localized, so $\langle m \vert \phi^{(k)}_\gamma \rangle$ is small except for a few $\gamma$'s localized around $m$. For these $\gamma$'s, $\abs{\lambda_{\alpha}- \lambda_{\gamma}} \sim O(1)$ (resonance comes with vanishing probability), giving a contribution of magnitude $\sim \widetilde Q^{(k)}_{\alpha\gamma}$. On the other hand, the energy mismatch can be as small as $\propto 1/ \ell_{k}$ for a few $\gamma$'s (note that there are $\ell_k$ $\gamma$'s, not $\ell_{k+1}$), for which the magnitude of $ \langle m \vert \phi^{(k)}_\gamma \rangle \sim e^{ -a(\ell_{k}) }$, so such contributions have magnitude $\sim \widetilde  Q^{(k)}_{\alpha\gamma} \ell_{k} e^{-a(\ell_{k})}$. Now if we assume that $a(\ell) \gg \ln \ell$, \textit{i.e.}, the eigenstates decay faster than algebraically, the first kind of contribution dominates, giving 
   $$ e^{-a(\ell_{k+1})} = \langle m  \vert \phi^{(k+1)}_\alpha \rangle  = \widetilde Q^{(k)}_{\alpha\gamma} \sim \exp(-2 a(\ell_k)) \Rightarrow a(\ell_{k+1}) = 2 a(\ell_k) \,.$$
 This is the same recursion as eq. \eqref{eq:recursion1FPP} for $T_k$, so the solution is also the same:
 \begin{align}   
 a(\ell) \propto \ln^{\kappa(\mu)} (\ell /\xi) \,,\,\label{eq:kappamu} \kappa(\mu) = \frac{\ln 2}{\ln \frac{2}{1 + \mu}} > 1 \,,
 \end{align} 
 justifying the assumption $a(\ell) \gg \ln \ell$.
 
Not surprisingly, this is the same decay rate as predicted by eq. \eqref{eq:logpowerdecay}, for the   \gls{bbrm} model in the strong disorder limit; for that specific model, the predictions is obtained by the mapping to the  \gls{fpp} model. Here, adapting the \textit{methods} in section \ref{sec:stretch} to the ``quantum'' setting, we argued that the decay rate applies to general broadly distributed banded matrices.

\subsubsection{Discussions}
Summarizing the two cases, we have shown that the class of broadly distributed banded random matrices, defined by the criteria eq. \eqref{eq:broaddefall}, captures the key characteristics of the \gls{bbrm} model in the $\mu \in (0,1)$ regime: the existence of localization transition and the peculiar decay rate of the localized states. In fact, our arguments can be seen as a renormalization group that runs through the scales defined by eq. \eqref{eq:lk}. In renormalization--group terms, we have shown that the $\beta \to \infty$ and $\beta \to 0$ limits are attractive \textit{fixed points}.
 
We emphasize that not all the properties of the  \gls{bbrm} in this regime extend to the broadly distributed class. For example, the existence of mobility edges for large $\beta$ depends crucially on the absence of diagonal disorder. In this respect, the arguments and evidences in section \ref{sec:LT} supporting the existence of mobility edges in the \gls{bbrm} ensemble are important, despite being model--specific. 
 
More generally, it is more subtle to argue for the existence or the absence of \textit{mobility edges}, \textit{i.e.}, localization transition by varying the $\lambda$ (not $\beta$). Indeed, in the above arguments, the dependence on $\lambda$ is implicitly present when we make comparisons to the level spacing. The latter depends on the \gls{dos}, which in turn depends on $\lambda$. Therefore, it is reasonable to argue that the critical disorder strength $\beta_c$, which depends on the \textit{microscopic} (close to the diagonal) properties of the matrix ensemble, has also a non--trivial $\lambda$ dependence: $\beta_c(\lambda)$. As a consequence, for at least the values $\set{\beta \vert \beta = \beta_c(\lambda)}$, the matrix ensemble with disorder parameter $\beta$ prepossess mobility edges. By this argument, we expect that the existence of mobility edges is \textit{generic} in the broadly distributed class.
 
\subsubsection{Numerics: randomised sparse matrices}
 \begin{figure}
 \center (a)\includegraphics[scale=.6,valign=t]{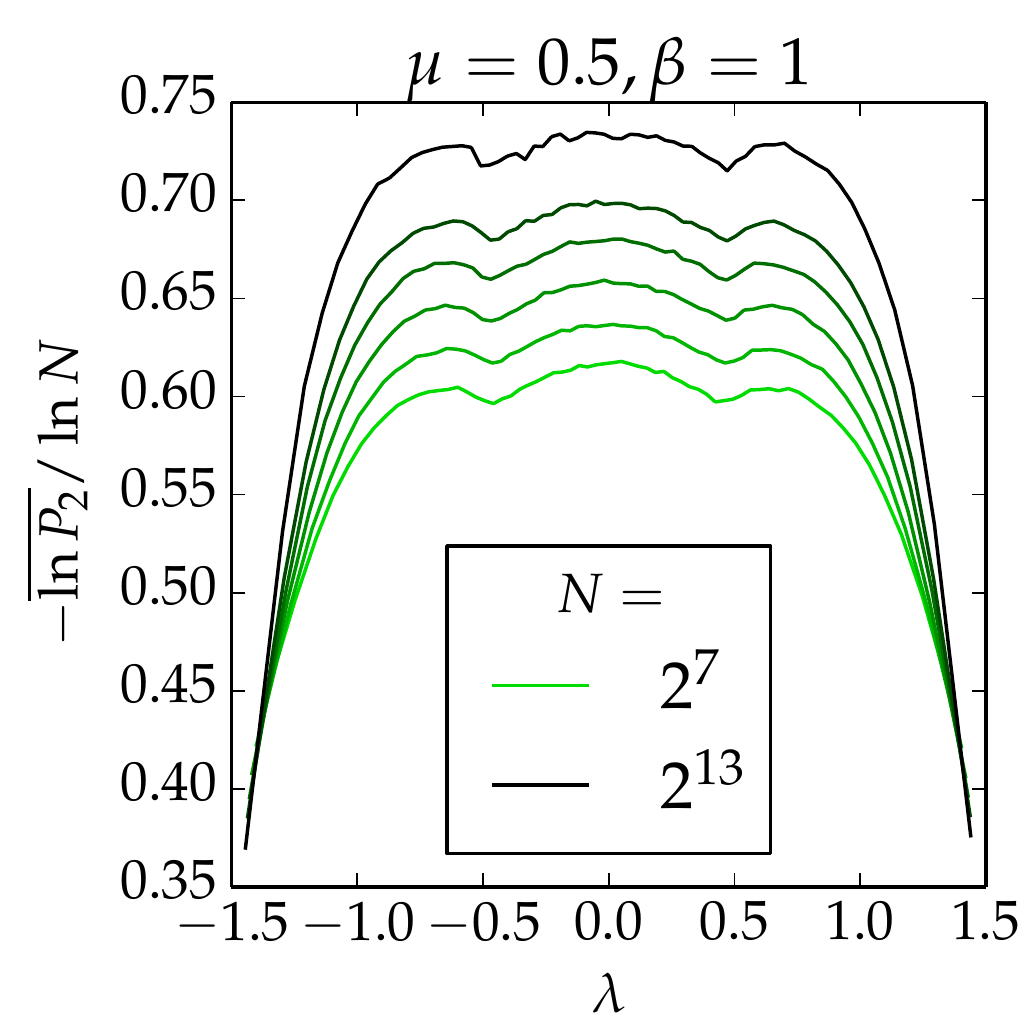}
(b) \includegraphics[scale=.6,valign=t]{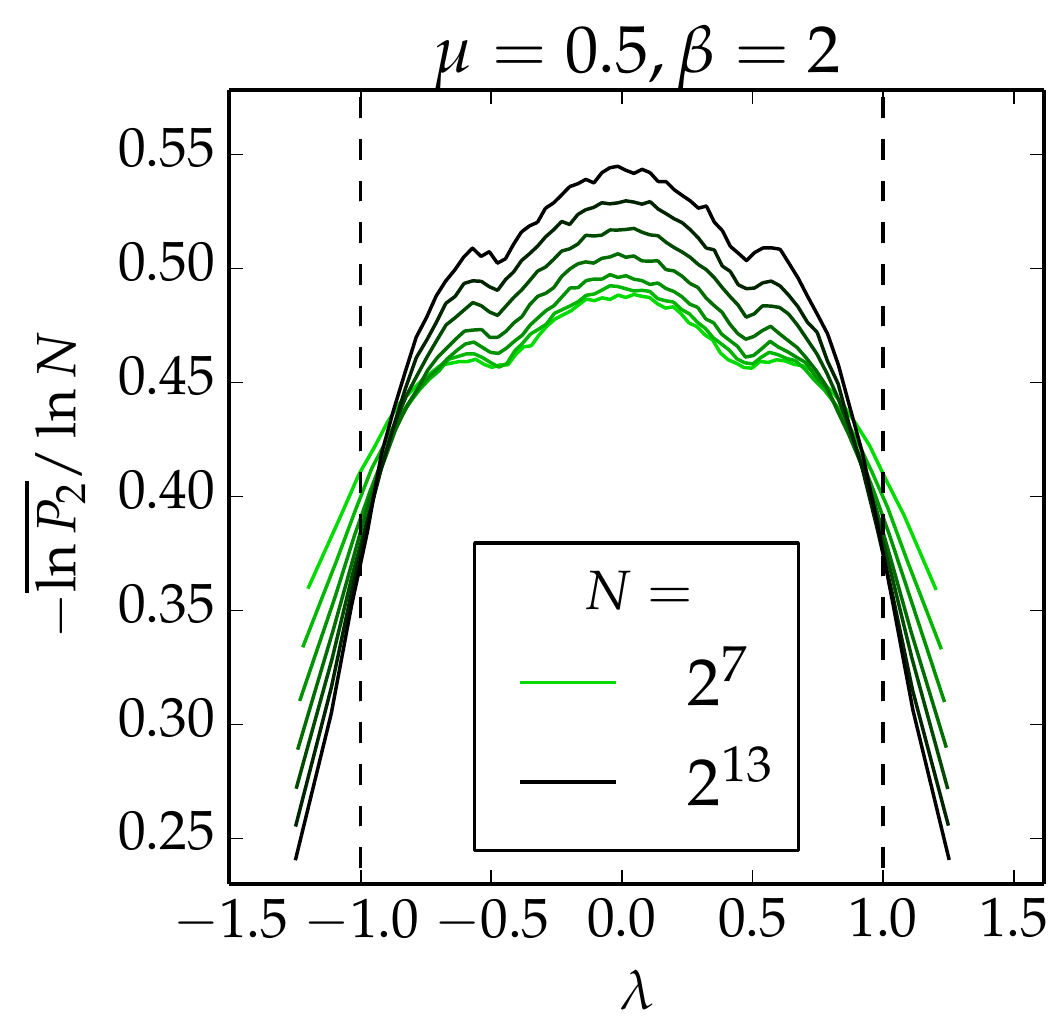}\\
(c)\includegraphics[scale=.6,valign=t]{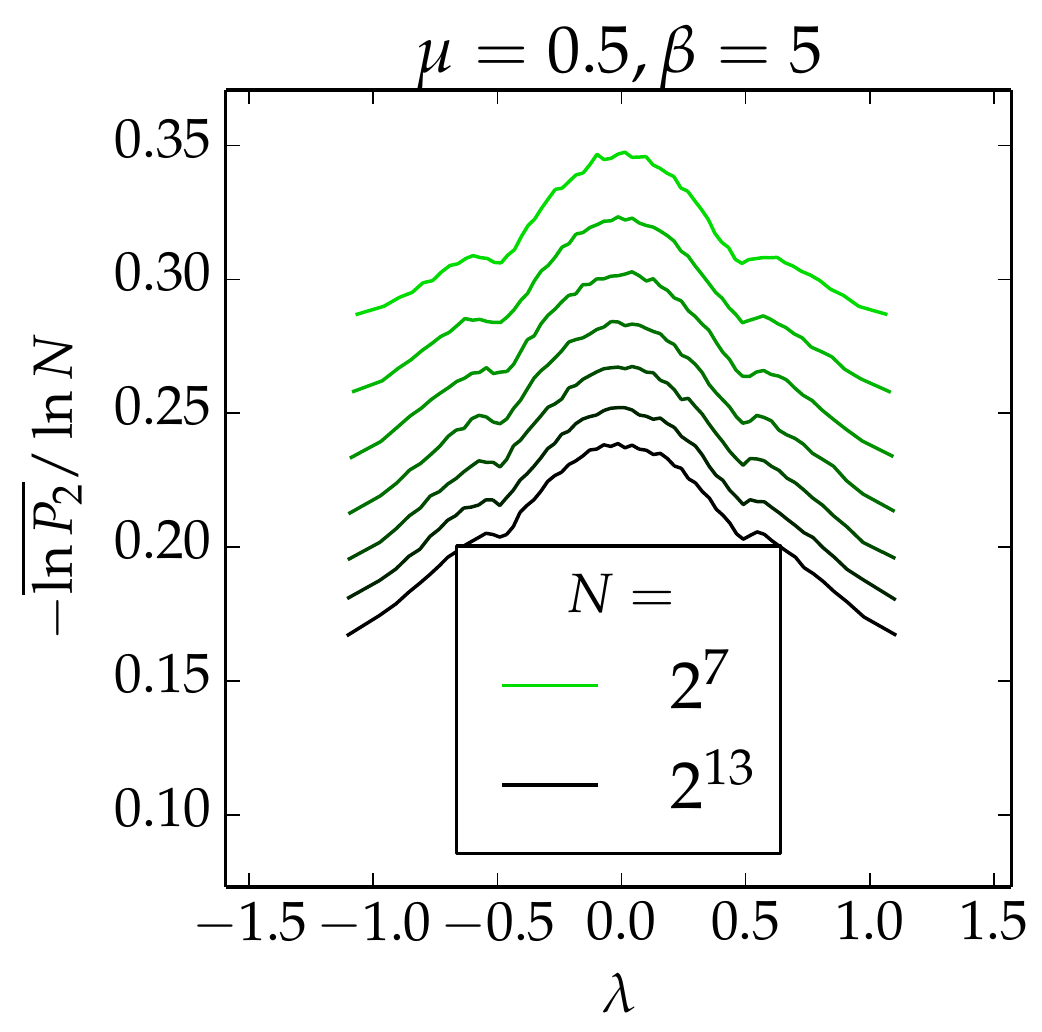}
(d) \includegraphics[scale=.6,valign=t]{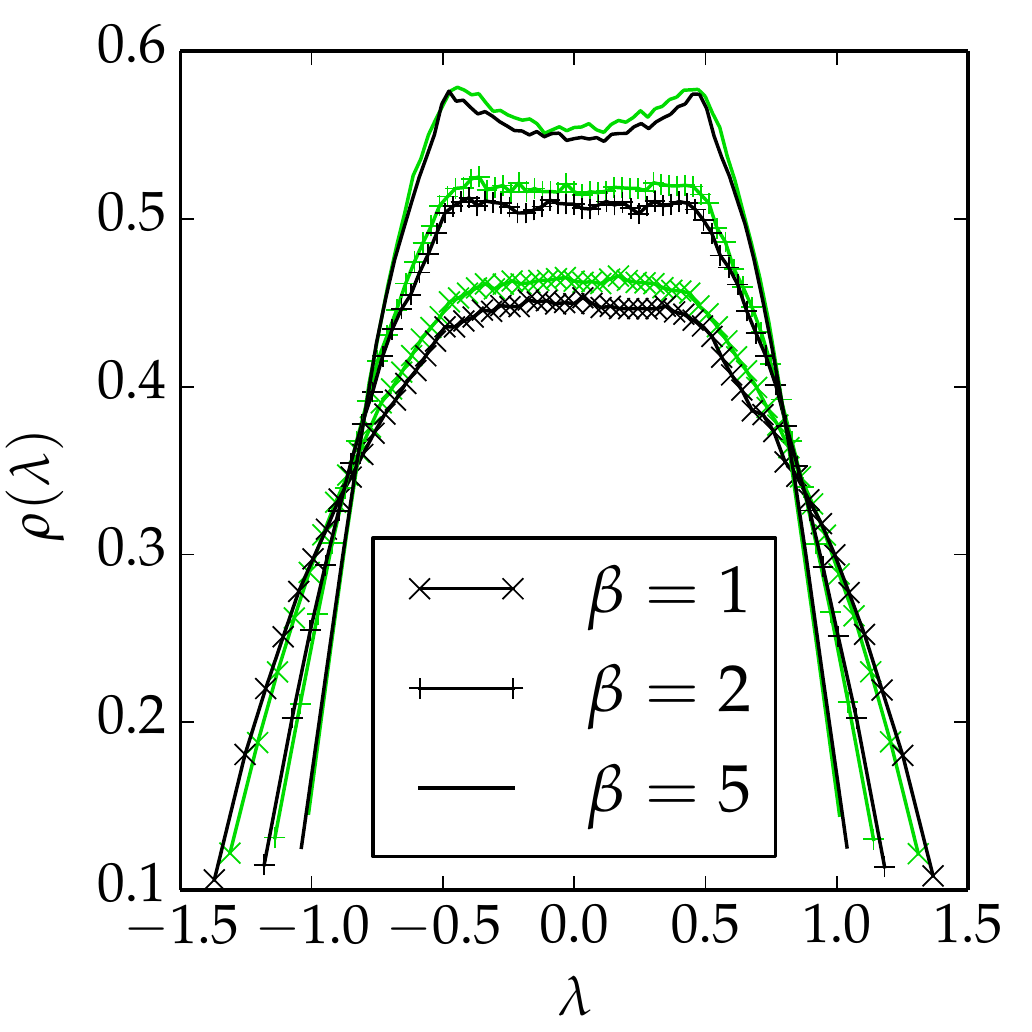}
 \caption{(a--c): The \gls{ipr} of the eigenstates of the randomised sparse matrix defined in eq. \eqref{eq:sparsedef}, plotted for parameters $\mu = 0.5$, $\beta = 1, 2$ and $5$, for matrix sizes $N = 2^7, \dots, 2^{13}.$ The numerical protocol is identical to that of Figure \ref{fig:IPR}. We recall the for ideally extended (localised) states, the $y$ value is $1$ ($0$, respectively). In panel (b), the estimated position of mobility edges $\lambda_c = \pm 1.0(2)$ is also indicated. The estimated \gls{ipr} exponent is $\tau_c = \left[ \overline{\ln P_2} / \ln N \right]_{\lambda_c} \approx 0.36(5)$, comparable to the corresponding \gls{bbrm} value, see Figure \ref{fig:IPR}. (d): The \gls{dos} for the same parameters of $\mu$ and $\beta$, with $N = 2^7$ and $2^{13}$. In all plots, lighter colours represent smaller system sizes.} \label{fig:sparse}
 \end{figure}
To demonstrate the above points, let us consider a specific model in the broadly distributed class: a randomized sparse matrix. To define it, let us denote 
\begin{equation} \tilde{g} = \tilde{g}_{mn} = \begin{dcases}
1 + \beta \abs{m-n}^{\mu+1} \,,\, & \abs{m-n} \geq 2 \,, \\
1 \,,\, & \abs{m-n} = 0, 1 \,.
\end{dcases} \label{eq:gtilde}  \end{equation}
Therefore, $\tilde{g}$ has the same asymptotic decay as $g$ in eq. \eqref{eq:broaddef}). The matrix elements have the following distribution
\begin{equation}
P(Q_{mn}) = \delta(Q_{nm}) \left(1 - \tilde{g}^{-1}\right) +   \tilde{g}^{-1} \theta\left(\frac12 - \abs{Q_{mn}}\right) \,. \label{eq:sparsedef}
\end{equation}
In other words, with probability $\tilde{g}^{-1}$, $Q_{mn}$ is drawn from a uniform distribution in $[-\frac12, \frac12]$; otherwise, the matrix elements vanish $Q_{mn} = 0$. This random matrix ensemble is clearly a member of the broadly distributed class. The modification of $\tilde{g}$ (with respect to $g$) at small distances ensures that in the $\beta \to \infty$ limit, the model tends to a 1D short--range hopping model (in fact, a modified Anderson model, in which the hopping becomes also random). Note that the near--diagonal behaviour of the present ensemble is very different from the \gls{bbrm}. In particular, the diagonal elements are uniformly distributed in the present case (they are $0$ for \glspl{bbrm}). As a consequence, the \gls{dos} of the present model has no singularity, in contrast to \gls{bbrm} (see Figure \ref{fig:DoS}). In fact, as we observe numerically in Figure \ref{fig:sparse} (d), $\rho(\lambda)$ is roughly constant in the interval $[-\frac12, \frac12]$ for the range of $\beta \in [1, 5]$ plotted. As $\beta$ increases further, this description will break down and two \gls{dos} peaks will grow at $\lambda = \pm \frac12$, as one can expect from the \gls{dos} of the 1D Anderson model. 

Now, using the same numerical method as in section \ref{sec:LT} (Figure \ref{fig:IPR}), we look at the localization properties of the eigenstates through their \gls{ipr}. The results, for $\mu = 0.5$ and $\beta = 1,2,5$ are shown in Figure \ref{fig:sparse} (a--c). For a weak disorder $\beta = 1$, all the eigenstates tend to be extended (as the matrix size increases), except those near the edges of the spectrum, whose behaviour are not clear. For a strong disorder $\beta = 5$, all the eigenstates tend to be localised: there are no mobility edges, in contrast to the strong disorder \gls{bbrm} model (for $\mu \in (0,1)$). 

For $\beta = 2$, however, there appear to be mobility edges separating the extended states in the middle of the spectrum from the localized ones near the edges. Therefore, the qualitative picture is identical to the \gls{bbrm} case. At a more quantitative level, we observe a similarly strong correction--to--scaling effect: the crossing point moves towards to edges as $N$ increases. Taking this into account, our estimate for the mobility edges is $\lambda_c \approx \pm 1.0(2)$ (in this model there is $\lambda \to -\lambda$ symmetry), with critical \gls{ipr} exponent $\tau_c = \left[ \overline{\ln P_2} / \ln N \right]_{\lambda_c} \approx 0.36(5)$, which is in agreement with the \gls{bbrm} model for the same $\mu$. (see Figure \ref{fig:IPR}). Although the agreement is preliminary and subject to more careful numerical analyses (for both models), it is tempting to conjecture that the critical behaviour of the broadly distributed class may depend only on the ``effective dimension'' $\mu$. 

In summary, the preliminary numerical study of the randomized sparse banded matrix model (defined in eq. \eqref{eq:sparsedef}), a member of the broadly distributed class, confirms the existence of predicted localization transition, and the expected mobility edges. Note that the sparseness of this matrix ensemble, which has not been exploited in the numerics presented above, makes this model an advantageous example for the future investigation of the broadly distributed class. In particular, an interesting direction for future numerical work could be inspired by the powerful methods of \cite{garcia2016scaling}.

\section{Summary and perspectives}\label{sec:finalanderson}
The original material of this Chapter can be divided into two parts: the results specific to the particular model of \acrfullpl{bbrm} (section \ref{sec:bbrm}), and their generalization to a larger class of banded random matrices with broadly distributed elements. Recall that the latter class can be characterized by the following: the distribution of the matrix element $Q_{mn}$ depends only on the distance to the diagonal through an algebraic law, $g =\beta \abs{m-n}^{\mu+1}$. With probability $1/g$, $Q_{mn}$ is a ``black swan'' of order unity; while the other typical elements decay at least exponentially fast $Q_{mn} \sim \exp(-c g)$. 

Although these Hamiltonians (matrices) are defined on a regular lattice,  both the model--specific and general analyses point to the importance of the graph formed by the edges where the matrix element is atypically large. By definition, this graph is statistically identical to that of the 1D long--range percolation problem. In the hindsight, the subject matter of this Chapter is rather Anderson localization transition on \textit{random graphs}. Compared to the \textit{regular random graphs} on which the Anderson localization transition attracts a recent interest \cite{altshuler2016rrg}, the random graphs underlying our banded matrices are not mean--field, and retain the notion of spatial distance. 

The comparison to regular--random--graph Anderson model raises another natural question, that of the existence of a \textit{critical}, \textit{i.e.}, de--localized but non--extended (ergodic) phase. Such a phase is a \textit{finite} interval of the disorder parameter (with $\lambda$ fixed, usually at the centre of \gls{dos} in the literature), in which the eigenstates have a non--trivial multi--fractal spectrum; equivalently, the \glspl{ipr} satisfy neither of the cases in eq. \eqref{eq:Pqphases}. Their existence in the regular--random--graph (and the Bethe--lattice)  Anderson model is the primary motivation behind \textit{op. cit.} (and \cite{biroli2012difference,deluca14bethe}, respectively). Another random matrix model with a critical phase is the generalized Rosenzweig--Porter \cite{rosenzweig60rp} model proposed recently \cite{kravtsov2015random}. It is also a mean--field (in the sense of fully--connected) model. This situation makes it very interesting to determine the (non)--existence of critical phase in the broadly distributed class. Unfortunately, we cannot attack this question in a trustworthy manner with the analytical and numerical methods presented in this thesis, and should leave it to future study.

Given the naturalness of the definition of the broadly distributed class, we should not underestimate its potential connections to more experiment--driven models in the literature. The latter should also be a guide for future study and extension of our theoretical toy. We wish to suggest two topics in this respect:
\begin{itemize}
\item The first is the vibration modes of random spring network defined by a sparse random set of spatial points; such models are studied recently in \cite{amir10glass,amir13localisation} in the context of electronic glass. In general, the random matrix (denoted $K$) arising in this context satisfies the semi--positive sum rule:
\begin{equation} K_{mn} \leq 0 \,, m \neq n \,,\, K_{nn} = -\sum_{m \neq n} K_{nm} \,,\,  \label{eq:sumrule} \end{equation}
in addition to being real symmetric. Therefore, the eigenvalues of $K$ are all non--negative, and the lowest eigenstate has constant coefficients (it is the zero--mode reflecting the translation invariance of the system). It is formally straightforward to implement the sum rule eq. \eqref{eq:sumrule} in, \textit{e.g.}, the \gls{bbrm}. However, we expect that the spectral and localization properties will be qualitative changed; in particular, in cases where there are mobility edges, the extended states may move to the low--energy (frequency) edge of the spectrum (rather than in its centre). 

Despite the differences, it is remarkable that the methods of \cite{amir13localisation} to study the localization transition in their models have similar points to ours, including a coarse--graining procedure and a mapping to the long--range percolation model. 

\item The second is the many--body localization transition with \textit{long--range} resonance interaction of type $ \abs{\mathbf{r}}^{-\mu - d} \hat{S}(\mathbf{0}) \hat{S}(\mathbf{r})$, studied in \cite{burin15mbl,burin2006energy,yao14mbldipolar}, where $\hat{S}(\mathbf{r})$ is the spin operator of some electron (for example) at position $\mathbf{r} \in \R^d$, $d$ being dimension. In \textit{op. cit.}, it was shown that in when $\mu \in (0, d)$, single--particle Anderson de--localization is impossible but many--body resonance de--localization is possible. Interestingly, setting $d = 1$, the above ``interesting'' range of $\mu$ is exactly the where the broadly distributed class has localization transitions. Moreover, given the close relation between the broadly distributed class and long--range percolation and \gls{fpp}, we note that many results of these models extend to $d$ dimensions (with $ \abs{m-n}^{\mu+1} \leadsto \abs{\mathbf{r} - \mathbf{r}'}^{\mu + d} $), in which the stretch--exponential regime (section \ref{sec:stretch}) is $\mu \in (0, d)$. These coincidences all suggest some deeper relations, which have not been elucidated. Working them out may lead to an exciting situation where an interacting (many--body) problem becomes related to a one--particle case.
\end{itemize}
To conclude, we point out that in both of the connections discussed, the models we quoted are defined \textit{off}--lattice and in a general dimension. We believe that this is a good approach to extend this Chapter to higher dimensions, so that the short--range Anderson de--localization does not interfere with the long--range phenomena we are interested in.

\chapter{Epilogue}\label{ch:end}
In this thesis, we have applied the formalism of equilibrium statistical mechanics to two prototypical problems with quenched disorder: that of  classical, thermal particles in a logarithmically correlated random potential (\glspl{logrem}) in Chapter \ref{ch:logrem}; and that of a quantum particle in a disordered environment whose structure is close to the graph of the 1D long--range percolation problem (banded random matrices of the broadly distributed class) in Chapter \ref{ch:anderson}. 

Our theoretical motivation behind the study of both problems is the existence of phase transitions: freezing transition, termination point transition, and others in \glspl{logrem}, and the localization transition in the random matrices. In both cases, the low--temperature (strong--disorder) phase is governed by the zero--temperature limit (strong--coupling fixed point in the renormalization group language). For the \glspl{rem}, the frozen phase free energy is dominated by the minima of the potential, while the termination point phase \glspl{ipr} are dominated by the largest Gibbs measure weights (in rare samples). For the \glspl{bbrm}, the decay of the localized phase is governed, by the \textit{ground state} energy (\textit{first}--passage time) of the polymer in disordered media model (\gls{fpp} model, respectively). Most of the analytical treatments in both Chapters are concentrated on the \textit{existence} of transition(s) and the properties of the low--temperature phase: \acrfull{rsb} for the \glspl{logrem}, and the quantum--statistical mappings (as well as mapping--inspired general arguments) in Chapter \ref{ch:anderson}. 

This focus suggests that much is to be done in the future \textit{at or near criticality}, by the development of field theory and renormalization group methods in both contexts. We stress that this question is of \textit{qualitative} importance and is not limited in computing, for example, the critical \glspl{ipr} and the position of mobility edges in the localization transitions of the broadly distributed class. Indeed, as we have discussed in section \ref{sec:finalanderson}, we cannot exclude the existence of a finite critical \textit{phase}. For such questions, exact--diagonalization based numerical study would be never conclusive enough (not only in our context, but also for the models in the literature  \cite{biroli2012difference,deluca14bethe,kravtsov2015random,altshuler2016rrg}; see however recent advances \cite{garcia2016scaling}), while much more predictive power could be obtained by a combination with adapted analytical techniques, \textit{e.g.}, super--symmetric field theory \cite{efetov1983supersymmetry,fyodorov1991band,efetov1999supersymmetry,evers2008anderson}, Wegner flow equation \cite{wegner94flow}, or strong--disorder renormalization methods \textit{à la} Levitov \cite{levitov89loc,levitov90deloc,levitov99pbrm} or \textit{à la} Imry--Ma \cite{imry75ma,Igloi2005rg}.

For the \glspl{logrem}, there are similar qualitative questions that are left open. For example, it is generally recognid that \textit{whole} frozen ($\beta > 1$) phase is critical. However, the theoretical evidences for such a claim are scattered: the selected velocity (extensive free energy) frees at the critical--temperature value; the \gls{1rsb} of \glspl{logrem} is a degenerated full \gls{rsb} solution, which has marginally stable modes in general; the finite--size corrected overlap distribution is a power law in the whole frozen phase. On the other hand, the numerical signature is the pronounced finite--si corrections in the whole $\beta > 1$ phase to the $M \to \infty$ limit predictions. To fit these pieces together requires a deeper understanding of the freezing transition, which we believe can be achieved by creatively combining the \gls{rsb} picture, the insights of the recent \acrfull{lft}--\gls{logrem} mapping, and the \gls{kpp}--functional renormalization group point of view already present in the pioneering work \cite{carpentier2001glass}.

\Glspl{logrem} are a remarkable situation where problems defined in \textit{finite dimensions} can be mapped to mean--field (hierarchical) models, so that the \gls{rsb} method can be applied. Note that the basis of this statement is the asymptotic and emergent ultra--metricity thanks to the log correlation in finite--dimensions. An important question is to what extent the \gls{rsb} is valid, and when this is not the case, how to correct it. Clearly, this question in general is central in the domain of disordered statistical mechanics, and is investigated along many lines of attack, \textit{e.g.}, functional renormalization group \cite{ledoussal03frgrsb,wiese07rev,LeDoussal2010rev,ledoussal08cuspshock}, replica--field theory \cite{mezard1991replica,charbonneau2016nontrivial}, and the study of artificial yet non--mean field models such as \cite{castellana10REM}. Returning to the context of this thesis, we know the model of \acrfull{dpct} can be studied by \gls{rsb}; on the other hand, the directed polymer model in $(1+1)$-d is deeply understood analytically (since it is in the $(1+1)$--d \gls{kpz} universality class and amenable to integrability techniques), \textit{without} referring to \gls{rsb} at all. Now, the polymer model used in in Chapter \ref{ch:anderson} is defined on an effectively high--dimensional space. Furthermore, in the stretch--exponential regime where there are localization transitions, the long--range percolation random graph has an arguably hierarchical structure. This raises the crucial question of whether \gls{rsb} can applied to this regime. In fact, the question is equivalent to that of the relevance of other mean--field spin--glass methods, such as the cavity method \cite{mezard87beyond,dotsenko1995introduction}, which was used in the Lévy random matrix model \cite{cizeau1994levy,tarquini16levy} and Bethe--lattice Anderson model \cite{abou1973selfconsistent,abouchacra74self,altshuler2016multifractal}. One can further argue that, \textit{in fine}, the strong--disorder method of Levitov (see \cite{evers2008anderson,mirlin00pbrm} for applications in critical \glspl{pbrm}) reduces the problem at hand to a variant of the \gls{dpct} model. In summary, the stretch--exponential regime of the broadly distributed class provides a new finite--dimensional laboratory for the disordered systems theory.

We would like to devote the concluding discussion to the \textit{dynamical} aspects. Most of the problems studied in this thesis have dynamical counterparts, although the theoretical framework of this thesis has prevented us from addressing them directly \footnote{Nevertheless, dynamical systems such as the \gls{kpp} equation and Eden/long--range epimedics growth model do appear in our analysis.}. Such questions arise naturally in the setting of Chapter \ref{ch:anderson}, where the eigenvector properties of the random matrices have determining  consequences on the corresponding diffusive/quantum dynamics of the particle, and on the corresponding random spring network (see section \ref{sec:finalanderson} around eq. \eqref{eq:sumrule} for further details). In particular, it would be interesting to elucidate the relation of the above dynamics to the epidemics growth at the other end of the mapping (see section \ref{sec:mappingAL}).

For the \glspl{logrem}, one may consider three types of dynamics: that of the thermal particle(s), that of \textit{quantum} particles, and that of the random potential itself. We have indeed touched upon the former problem when studying the joint minimum--maximum in section \ref{sec:minmax} (see also \cite{cao16maxmin}); indeed, a standard definition of the problem is the one in which the particle follows an over--damped Langevin dynamics in a quenched log--correlated potential. However, besides thermal activations, it would be more interesting to drive the particle, by smoothly changing a deterministic part of the potential. In this case, even the zero--temperature dynamics (the change of the minimum) becomes non--trivial, and the particle is expected to move discontinuously by making sudden jumps, called \textit{avalanches}. In general disordered systems, Their statistical properties have intricate relations with the statics of the problem \cite{ledoussal08cuspshock,franz2016mean}, and in particular can be related to the \gls{rsb} solution. Moreover, as shown in \cite{fyodorov2010freezing}, the same protocol can be used to study the shocks in the decaying Burgers turbulence with log--correlated initial data. For exciting recent mathematical progress on this direction, see \cite{biskup2016return} and references therein.

As we mentioned in section \ref{sec:treetoplane}, localisation transitions in 2D disordered quantum mechanics (more precisely, certain non--conventional symmetry classes) were amongst the historical motivations of \glspl{logrem}. Indeed, these models are known to be abundant of non--analytic behaviours reminiscent of the freezing transition. Some of them (\textit{e.g.}, concerning the multi--fractality of the zero--energy state in the disordered Dirac fermion model) are directly related to the termination point transition and its quenched counterpart. Others, \textit{e.g.}, the divergence of the \gls{dos} at zero--energy in bipartite hopping models \cite{mudry03DoS,motrunich02particle}, can be addressed only in a setting beyond this thesis.  Given the progress on \glspl{logrem} we reported, in particular, the relation to \gls{lft}, it is important to revisit this rich literature equipped with the new insights.

The dynamics of the \textit{random} potential is more widely open: for example, the $(2+1)$-D Edwards--Wilkinson equation \cite{edwards82wilkinson} and the \textit{anisotropic} \gls{kpz} equation have the 2D \gls{gff} as stationary state. How can we describe the evolution of the minima process under these dynamics? Going beyond log--correlated fields, we may ask the same question for the isotropic \gls{kpz} equation in $(2+1)$-D, whose stationary state is algebraically rough and breaks up--down symmetry. Then, even the statics could be the subject of another thesis.
\bibliographystyle{SciPost_bibstyle}
\bibliography{rems}


\end{document}